\documentclass[12pt,letterpaper]{book}
\usepackage[dvips]{graphicx,color}
\usepackage{amssymb,hyperref,rotating}
\usepackage{array,hhline,multicol,multirow,dcolumn} 
\usepackage{amsmath}
\usepackage{amsfonts}

\newcommand{\numu}{\mbox{$\nu_\mu$}}
\newcommand{\numubar}{\mbox{$\overline{\nu_\mu}$}}
\newcommand{\nue}{\mbox{$\nu_e$}}
\newcommand{\nuebar}{\mbox{$\overline{\nu_e}$}}

\parskip=\medskipamount                         
\thicklines                                     
\def\beq{\begin{equation}}
\def\eeq{\end{equation}}
\def\beqa{\begin{eqnarray*}}
\def\eeqa{\end{eqnarray*}}
\def\beqs{\begin{eqnarray}}
\def\eeqs{\end{eqnarray}}
\def\n{\global\advance \eqnumber by 1\eqno(\the\eqnumber)}
\def\puteqno{\global\advance \eqnumber by 1 (\the\eqnumber)}

\def\openone{\leavevmode\hbox{\small1\normalsize\kern-.33em1}}

\begin{document}

\pagestyle{empty}

\begin{centering}
\mbox{}
\vfill
\Large{
A Proposal for a Near Detector Experiment on the Booster Neutrino Beamline:} \\
\Large{FINeSSE: Fermilab Intense Neutrino Scattering Scintillator Experiment}\\
\normalsize{
\vspace{.2in}
\today
\\
\vspace{.2in}
L.\ Bugel, J.\ M.\ Conrad, J.\ M.\ Link, M.\ Shaevitz, L.\ Wang, G.\ P.\ Zeller\\
{\em Columbia University, Nevis Labs, Irvington, NY 10533}\\ 
\vspace{0.1in}
S.\ Brice, B.\ T.\ Fleming$^{*}$, D.\ Finley, R. \ Stefanski \\ 
{\em Fermi National Accelerator Laboratory, Batavia, IL 60510}  \\
\vspace{0.1in}
J.\ C.\ Peng\\
{\em University of Illinois at Urbana-Champaign, Urbana, IL  61801}\\
\vspace{0.1in}
J.\ Doskow, C.~Horowitz, T.\ Katori, H.\ O.\ Meyer, P.\ Ockerse, R.\ Tayloe$^{*}$, G.\ Visser \\
{\em Indiana University, Bloomington, IN 47408}\\
\vspace{0.1in}
C.\ Green, G.\ T.\ Garvey, W.\ C.\ Louis, G. McGregor, R. Van de Water\\
{\em Los Alamos National Laboratory, Los Alamos, NM 87545}\\
\vspace{0.1in}
R.\ Imlay, W.\ Metcalf, M.\ Sung, M.~O.~Wascko\\
{\em Louisiana State University, Baton Rouge, LA 70803}\\
\vspace{0.1in}
V.\ Papavassiliou\\
{\em New Mexico State University, Las Cruces, NM 88003}\\
\vspace{0.1in}
L.\ Lu\\
{\em University of Virginia, Charlottesville, VA 22901}\\
\line(1,0){260}\\
$^{*}$ Co-spokespersons: B.\ T.\ Fleming and R.\ Tayloe}
\vfill
\end{centering}
\clearpage

\clearpage
\begin{centering}
\large
ABSTRACT \\
\vspace*{4ex}
FINeSSE: Fermilab Intense Neutrino Scattering Scintillator Experiment\\
\vspace*{4ex}
\end{centering}

Understanding the quark and gluon substructure of the nucleon has been
a prime goal of both nuclear and particle physics for more than thirty
years and has led to much of the progress in strong interaction
physics.  Still the flavor dependence of the nucleon's spin is a
significant fundamental question that is not understood.  Experiments
measuring the spin content of the nucleon have reported conflicting
results on the amount of nucleon spin carried by strange quarks.
Quasi-elastic neutrino scattering, observed using a novel detection
technique, provides a theoretically clean measure of this quantity.

The optimum neutrino beam energy needed to measure the strange spin of
the nucleon is 1 GeV.  This is also an ideal energy to search for
neutrino oscillations at high $\Delta m^2$ in an astrophysically
interesting region.  Models of the r-process in supernovae which
include high-mass sterile neutrinos may explain the abundance of
neutron-rich heavy metals in the universe.  These high-mass sterile
neutrinos are outside the sensitivity region of any previous neutrino
oscillation experiments.

The Booster neutrino beamline at Fermilab provides the world's highest
intensity neutrino beam in the 0.5-1.0~GeV energy range, a range ideal
for both of these measurements.  A small detector located upstream of
the MiniBooNE detector, 100~m from the recently commissioned Booster
neutrino source, could definitively measure the strange quark
contribution to the nucleon spin. This detector, in conjunction with
the MiniBooNE detector, could also investigate $\nu_{\mu}$
disappearance in a currently unexplored, cosmologically interesting
region.

\clearpage

\pagestyle{plain}
\pagenumbering{roman}
\setcounter{page}{1}

\tableofcontents
\clearpage

\listoftables

\clearpage

\listoffigures

\clearpage

\pagestyle{myheadings}
\markright{}
\pagenumbering{arabic}
\setcounter{page}{1}

\chapter{Introduction}
\label{ch:Introduction}

\thispagestyle{myheadings}
\markright{}

The Fermilab Intense Neutrino Scattering Scintillator Experiment (``FINeSSE'') 
is designed to measure the strange quark contribution to the spin of the proton, 
and to search, in conjunction with 
the MiniBooNE experiment, for $\nu_\mu$ disappearance.  These
measurements will employ a novel detection technique, and will examine
a kinematic region inaccessible to any existing or
presently-planned experiment.  FINeSSE will be located 100~m from the Booster
neutrino beamline production target, and 441~m upstream of the 
currently-running
MiniBooNE experiment.  The low energy Booster neutrino beam is
crucial to achieving this experiment's goals; they can only be realized 
on this Fermilab
beamline.  The number of protons on target (POT) needed to reach the
FINeSSE physics goals is 6$\times 10^{20}$, attainable in two
years of running.  The detector is designed to resolve both short, low
energy proton tracks, and longer muon tracks from $\nu N$
interactions.  The FINeSSE detector will cost \$2.25M (\$2.8M with
contingency); the FINeSSE Detector enclosure, \$800K (\$1.6M
with contingency, escalation, and EDIA).

\section{Outline}
This proposal sets forth the experiment's goals, design, costs, and schedule
in the following sections:
\begin{itemize}
\item Chapter~\ref{ch:PhysicsMotivation} provides the physics
  motivation for FINeSSE;
\item Chapter~\ref{ch:TheNeutrinoBeam} describes the flux and event
  rate at FINeSSE produced by the Booster neutrino beamline;
\item Chapter~\ref{ch:TheFINeSSEDetector} details the detector
  design, construction, and installation, as well as the readout and
  trigger systems;
\item Chapter~\ref{ch:EventSimulationandReconstruction} examines
  event kinematics, efficiencies, and backgrounds for
  the FINeSSE physics measurements;
\item Chapter~\ref{ch:OtherPhysics} points out additional physics
  measurements the FINeSSE experiment can perform;
\item Chapter~\ref{ch:CostandSchedule} provides a breakdown of costs
  for the detector, the electronics, and an enclosure for both, as well as
  a timeline to first beam in mid-2006.
\end{itemize}

\section{FINeSSE Physics, Detector, and Neutrino Beam}
The fundamental question of the flavor dependence of the spin of the
proton is not understood.  The still unresolved ``spin
crisis''~\cite{vassili} points to the fact that the proton's spin is
not carried, as was expected, by the valence quarks. How much is
carried by the light quark sea has been the subject of much
controversy.  In addition, measurements of the spin carried by the
strange quarks in the nucleon have been plagued by model assumptions
and experimental limitations.  The FINeSSE experiment will measure the
proton's strange spin, $\Delta s$, avoiding the pitfalls of previous
measurements; our approach will be described in detail in
Chapter~\ref{ch:PhysicsMotivation}.

FINeSSE, in conjunction with the MiniBooNE experiment, is sensitive to
$\nu_{\mu}$ disappearance in an as-yet unexplored, astrophysically
interesting region.  Incorporating oscillations
to ~1 eV sterile neutrinos into the r-process in supernovae 
can explain the abundance of neutron-rich
heavy metals in the universe~\cite{georgewasfirst}.  
Oscillations between these sterile
neutrinos and muon neutrinos are expected over short baselines for neutrino energies
around 1~GeV.  The combination of FINeSSE and MiniBooNE, functioning as
near and far detector, enables a $\nu_{\mu}$ disappearance
search sensitive to these oscillations.  This sensitivity exceeds that of 
any existing or planned experiment, and permits exploration of the full allowed 3+1 region.

\begin{figure}
\centering
\includegraphics[bb=14 500 556 748,width=6.in]{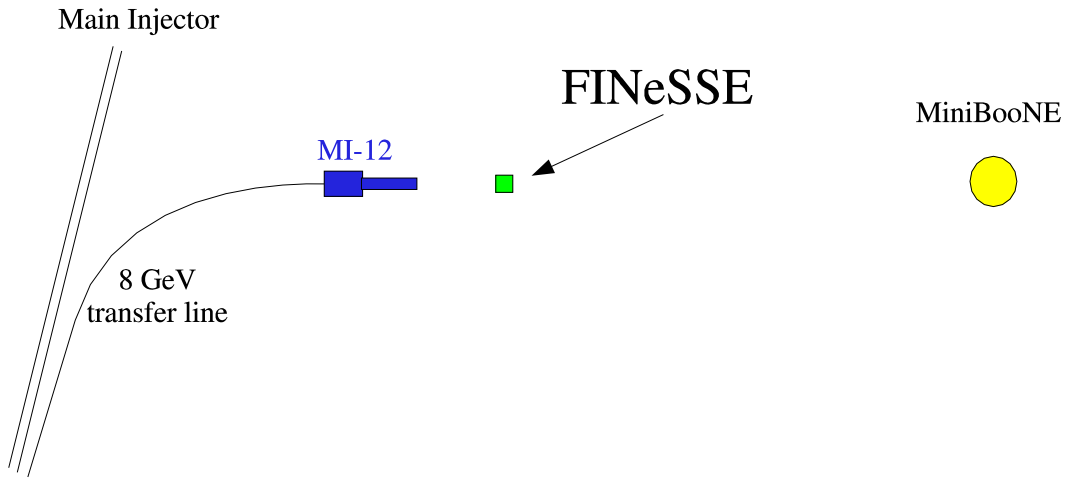}
\caption{\it Proposed FINeSSE location with respect to the existing 
neutrino target building (MI-12) and MiniBooNE detector. North is to the right in 
this drawing.}
\label{fig:transferline}
\end{figure}

The physics goals of FINeSSE can be achieved using a novel, relatively
small, tracking detector placed 100~m from the neutrino
production target on the Booster neutrino beam line (Fig.~\ref{fig:transferline}).  
The detector is comprised of two subdetectors.
The upstream Vertex Detector is a highly-segmented,
liquid scintillator ``bubble chamber'' that tracks particle
interactions via scintillation light read out on a grid of Wavelength
Shifting (WLS) fiber strung throughout the volume. The Muon Rangestack
downstream of the Vertex Detector ranges out high energy muons
produced in neutrino interactions. 

The Booster neutrino beam provides an intense source of muon neutrinos
(with a small background of electron neutrinos) in the energy range of 0.5-1.5 GeV 
and a mean energy of $\sim$ 700 MeV.  This spectrum is ideal for 
the $\nu p$ elastic scattering measurement as
well as for the $\nu_{\mu}$ disappearance search.  Using the currently
estimated MiniBooNE neutrino flux~\cite{mb-runplan}, and assuming $3.0
\times 10^{20}$ protons on target per year~\cite{psourceteamreport},
there would be approximately 360k neutrino scattering events in a
detector of 9~ton active volume during the FINeSSE run of 2 years.  This would
provide a neutrino event sample of unprecedented size in this energy range.

\section{The FINeSSE Collaboration}
FINeSSE is currently a collaboration of 29 scientists 
from six  universities and two national
laboratories.  Once approved, the collaboration is expected to grow.  
Both university groups and national labs have already
incorporated post-docs, graduate students, undergraduates, and a high
school teacher into FINeSSE physics projects.  FINeSSE has organized an
executive committee comprised of senior scientists from each group to
oversee the project.  The collaboration is diverse yet balanced in a
number of ways:
\begin{itemize}
\item Collaborators are drawn from both the nuclear and particle physics
communities.  This will help the collaboration to attain its physics goals, which span
both of these communities, and will encourage cross-disciplinary
interactions between nuclear and particle physicists, as well.

\item FINeSSE is comprised of both MiniBooNE and non-MiniBooNE scientists.
This balance helps to ensure a good understanding of a neutrino beam
already well studied by MiniBooNE, and at the same time bring in new ideas and
perspectives on physics with this beam.

\item FINeSSE scientists carry with them a broad spectrum of
  experience.  The new perspectives brought by the FINeSSE
  spokespeople are tempered by a well-seasoned executive committee
  comprised of experienced scientists with, again, both nuclear and
  particle physics backgrounds.
\end{itemize}

FINeSSE's diversity in background and experience are already a great
asset to its physics program, and will help to sustain it
throughout its physics run.

\section{FINeSSE as part of the Fermilab program}
FINeSSE brings timely and important physics to the Fermilab program.
In addition, FINeSSE takes advantage of the investment Fermilab has
already made in the Booster neutrino beamline, provides an excellent
training ground for young Fermilab scientists, and already  actively
contributes to the growing Fermilab neutrino program.

To achieve its physics goals, FINeSSE requires $6.0\times 10^{20}$
total protons on target, received over the course of a two year run on
the Booster neutrino beamline.  This is within the Booster's
capability in the era of NuMI and Run II running, as described in
Chapter~\ref{ch:TheNeutrinoBeam} and
Reference~\cite{psourceteamreport}.  It also takes advantage of a
running beam, and another running experiment, adding value to already
committed resources.

FINeSSE is a small, focused collaboration.  Such groups are proven
training grounds for graduate students and post-docs, vital to the
future of Fermilab and of high energy physics.  It is on such
small-scale experiments that young scientists
are guaranteed to get their hands on almost every aspect of design,
construction, data taking, and data analysis.

In summary, FINeSSE represents an important addition to Fermilab's program: it provides an 
extraordinary opportunity for physicists from a number of subfields and a variety
of levels of experience to work together; it makes advantageous use
of an existing beamline; it increases the physics reach of an existing experiment; 
and it uses a novel detection technique to address significant and interesting physics.

\section{Requests to the PAC}
Please consider the following specific requests with respect to 
approval and funding for FINeSSE:
\begin{itemize}
\item Grant this experiment ``stage 0'' or ``stage 1'' approval at 
this time 
(approval pending response to any outstanding questions).  
This will allow us to submit an NSF proposal by a January 2004 
deadline.
\item Recommend to the Fermilab directorate to support FINeSSE 
for the first stages of detector enclosure design work.
\item Recommend to the Fermilab directorate to provide FINeSSE 
with office and lab space.
\end{itemize}

\clearpage

\chapter{Physics Motivation}
\label{ch:PhysicsMotivation}

\thispagestyle{myheadings} 
\markright{} 

Two physics measurements form the foundation of the FINeSSE program:
the measurement of the strange spin of the proton, $\Delta s$; and the
search for $\nu_\mu$ disappearance in an astrophysically interesting
region.  Both topics are compelling, and can only be addressed with the
Booster neutrino beam design.  Along with these studies, a complement
of other measurements and searches are open to FINeSSE.  These other
physics projects are addressed in Chapter~\ref{ch:OtherPhysics}.  In
this chapter we concentrate on the two main physics goals, which make
FINeSSE unique.

\section{Strange Quark Contribution to Nucleon Spin}
From the time that the composite nature of the proton was discovered,
physicists have sought to understand its constituents.  The study of
nucleon spin has grown into an industry experimentally, and opened new
frontiers theoretically.  Deep Inelastic Scattering (DIS) measurements
with polarized beams and/or targets have given us a direct measurement
of the spins carried by the quarks in the nucleon.  A central mystery
has unfolded: in the nucleon, if the $u$ and $d$ valence quarks carry
approximately equal and opposite spins, where lies the remainder? 

One key contribution that has eluded a definitive explanation is the
spin contribution from strange quarks in the nucleon sea.  A large
strange quark spin component, extracted from recent
measurements~\cite{Filippone:2001ux}, would be of intense theoretical
interest, since it would require significant changes to current
assumptions.  Is this large value of the strange spin due to chiral
solitons~\cite{Brodsky}, a misinterpretation of the large gluon
contributions coming from the QCD axial anomaly~\cite{Bass,axanom}, or
incorrect assumptions of SU(3) symmetry~\cite{zhu}?  In addition, an
understanding of the nucleon spin structure is a key input to dark
matter searches, since the couplings of supersymmetric particles and
axions to dark matter depend upon the components of the spin.

It has been known for some time that low energy (and low $Q^2$)
neutrino measurements are the only theoretically robust technique (as
robust as, e.g., the Bjorken sum rule) for isolating the strange quark
contribution.  New low energy, intense neutrino beams now make it
possible to take greater advantage of this method.  The FINeSSE
experiment, using these newly-available beams along with a novel
detection technique, will resolve the presently murky experimental
picture, providing results which are interesting both to particle
physics and astrophysics.

FINeSSE will measure $\Delta s$ by examining neutral current
neutrino-proton scattering; the rate of this process is sensitive to
any contributions from strange quarks (both $s$ and $\bar s$) to the
nucleon spin.  Specifically, $\Delta s$ is extracted from the ratio of
neutral current neutrino-proton ($\nu p \rightarrow \nu p$) scattering
to charged current neutrino-neutron ($\nu n \rightarrow \mu^- p$)
scattering.  The measurement will be made at low momentum transfer
($Q^2 \approx 0.2$ GeV$^2$), in order to unambiguously extract $\Delta
s$ from the axial form factor, $G_A$.  FINeSSE will improve on the
latest measurement of neutral current neutrino-proton scattering (BNL
734) by measuring this process at a lower-$Q^2$, with more events,
less background, and lower systematic uncertainty.

In the following sections, we describe some of the previous and
current experiments relevant to the question of strange quarks in the
nucleon.  We then describe why neutral current neutrino-nucleon
elastic scattering is sensitive to the strange-quark contributions to
the nucleon spin.  We conclude with a summary of the sensitivity of
FINeSSE to $\Delta s$ (detailed more completely in
Chapter~\ref{ch:EventSimulationandReconstruction}).

\subsection{Experimental Results: Strange Quarks in the Nucleon}
The role that strange quarks play in determining the properties of the
nucleon is an area of much experimental and theoretical interest, and
is not well understood.  Deep inelastic scattering of neutrinos
on nucleons indicate that strange quarks constitute a substantial
fraction (20\%) of the nucleon sea~\cite{sssea}.  Nevertheless, the
latest results from parity-violating (PV) electron scattering show
that the contribution of strange quarks to the charge and magnetic
moment of the nucleon is consistent with zero~\cite{SAMPLE, HAPPEX}.
However, results from DIS of polarized charged leptons indicate
non-zero contributions of strange quarks to the nucleon spin.  In the
sections that follow, we describe the some of the experimental and
theoretical issues which FINeSSE will help to address.

\subsubsection{Parity-Violating Electron Scattering}
There is a large and continuing effort to investigate the nucleon
structure via Parity-Violating (PV) electron scattering.  Electron
scattering is sensitive to the strange (and anti-strange) quark
contributions to the nucleon charge and magnetic moment distributions.
The recently-completed MIT/Bates SAMPLE experiment~\cite{SAMPLE}
measured the strange-magnetic form factor, $G_M^s$, at a momentum
transfer $Q^2=0.1$~GeV$^2$, to be consistent with zero (albeit with
large errors).  The ongoing Jefferson Lab HAPPEX
experiment~\cite{HAPPEX} measures a PV scattering asymmetry sensitive
to a combination of the strange-electric and strange-magnetic form
factors ($G_E^s$ and $G_M^s$) at $Q^2=0.48$~GeV$^2$.  This combination
is also consistent with no strange-quark effects in the nucleon.
HAPPEX continues to search, and will make a measurement at lower $Q^2
(=0.1$~GeV$^2$) on a helium target in the near future~\cite{HAPPEX2}.

Upcoming PV electron experiments looking for strange quark effects are
the PVA4 experiment~\cite{PVA4} at Mainz, which will measure a
combination of $G_E^s$ and $G_M^s$ at $Q^2$ from 0.1 - 0.48~GeV$^2$; and the
G0 experiment~\cite{G0}, which will cover a large $Q^2$ (0.1 - 1.0
eV$^2$) range.

There is a large effort to look for strange quark effects via PV
electron scattering.  Unfortunately, these measurements are not
sensitive to the strange axial vector form factor ($G_A^s$, related to
the spin structure and $\Delta s$).  This is because the parity
violating contribution to the axial vector form factor of the nucleon
couples via the very small vector form factor of the charged lepton
($1- 4 \sin \theta_W$).  As a result, this contribution to the PV
asymmetry is dominated by a large PV radiative correction known as the
anapole moment~\cite{SAMPLE}.

PV electron scattering combined with neutrino scattering would be a
powerful approach to the study of strange quark effects in the
nucleon.  As will be explained below, neutrino scattering is very
sensitive to the strange axial-vector form factor $G_A^s$, and much
less so to the strange electric and magnetic form factors $G_E^s$ and
$G_M^s$.  The opposite is true in PV electron scattering.  The
measurements from FINeSSE could be combined with those from this
thorough program of PV electron scattering to obtain accurate
knowledge of all three strange form factors $G_A^s$, $G_E^s$, and
$G_M^s$~\cite{Pate}.

\subsubsection{Polarized Lepton Deep Inelastic Scattering}
Results obtained in DIS by polarized leptons from polarized nucleons
have been interpreted (from measurements of the polarized structure
function, $g_1^p$) as evidence for a non-zero and {\em negative}
$\Delta s$.  The SMC experiment, e.g., has reported $\Delta
s=-0.10\pm0.05$~\cite{SMC}.  The method of extracting $\Delta s$ from
$g_1^p$, however, has been subject to much debate, due to
model-dependent assumptions of $SU(3)$ symmetry~\cite{zhu,gluck} and
to extrapolation of the spin structure function to $x \rightarrow
0$~\cite{Bass}.  In addition, recent controversial results from
HERMES~\cite{HERMES} semi-inclusive DIS indicate a zero or small 
{\em positive} value for $\Delta s$ 
($\Delta s = +0.03\pm0.03\pm0.01$ where the first error is statistical
and the second is systematic).
It should be noted that the HERMES measurement is only sensitive to
$\Delta s$, as opposed to the SMC measurement of $\Delta s + \Delta
\bar{s}$~\cite{HERMES}.  The errors reported in these current
``state-of-the-art'' DIS measurements of $\Delta s$ are typically in
the range of $0.02-0.05$.
 
As will be shown in the sections below, the interpretation of 
neutral current neutrino-nucleon scattering suffers from none
of the theoretical uncertainties inherent in the DIS measurements.

\subsubsection{Neutral-Current Neutrino Scattering}
\begin{figure}[tbh]
\centering
\includegraphics[width=4.in]{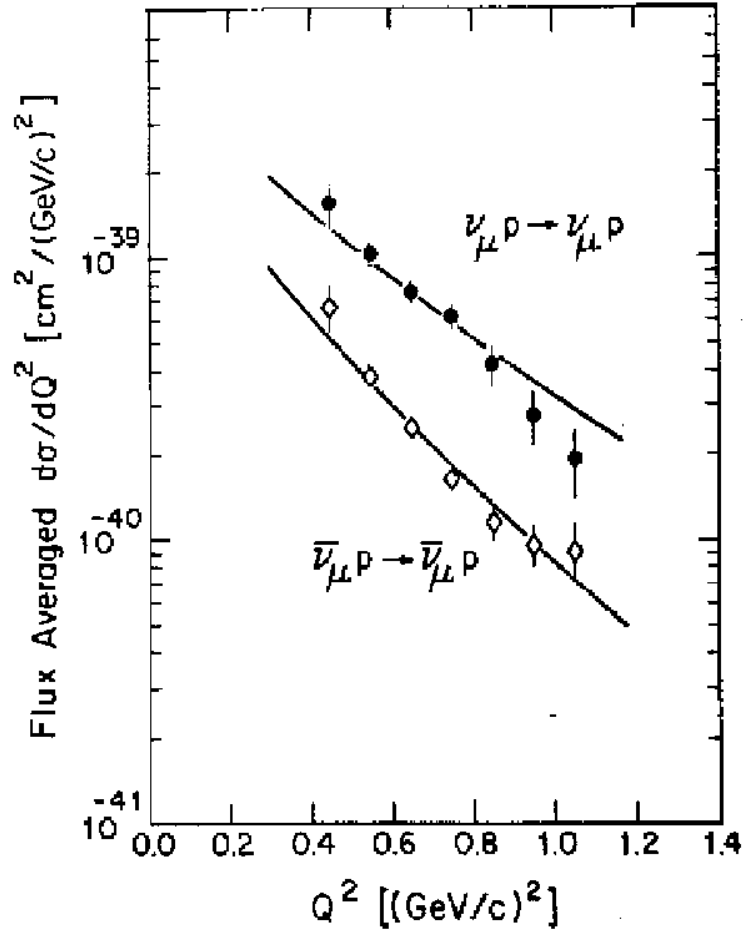}
\caption{\em BNL E734 results on $\nu p \rightarrow \nu p$ and 
$\bar{\nu} p \rightarrow \bar{\nu}p$~\cite{BNL734}.
The solid lines are fits to the data.}
\label{fig:BNL734_dsdq2}
\end{figure}

\begin{figure}[tbh]
\centering
\includegraphics[width=4.in]{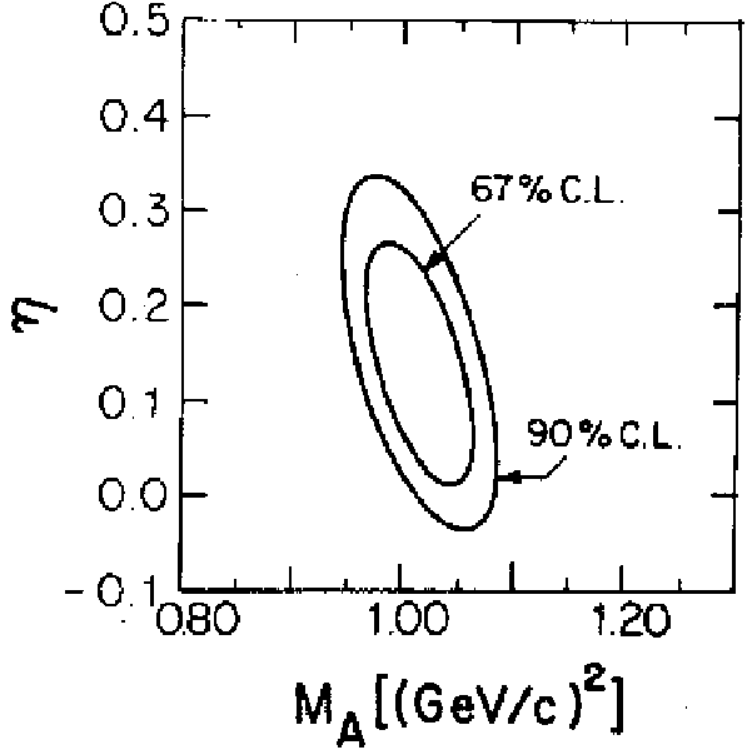}
\caption{\em Results from a fit of the BNL E734 
  $\nu p \rightarrow \nu p$ and $\bar{\nu} p \rightarrow \bar{\nu}p$
  scattering data~\cite{BNL734} indicating the preferred values of
  $\eta$ ($=-\Delta s$) and $M_A$.  In this fit, $M_A$ was constrained
  to the world-average value at the time $M_A=1.032\pm0.036$ GeV.}
\label{fig:BNL734_gasma_fit}
\end{figure}

The best measurement to date of neutral current neutrino scattering
is Experiment E734 at Brookhaven National Laboratory (BNL). They
measured neutrino-proton ($\nu p \rightarrow \nu p$) and
antineutrino-proton ($\bar{\nu} p \rightarrow \bar{\nu}p$) elastic
scattering~\cite{BNL734} using a 170~t tracking detector in the BNL
wide-band neutrino beam ($\bar{E_{\nu}}=1.3$ GeV).  From a sample of
951 $\nu p$ and 776 $\bar{\nu} p$ elastic scattering events, they
extracted differential cross sections ($d \sigma/d Q^2$) for $0.4 <
Q^2 < 1.1\, (\rm{GeV}/c)^2$ (Fig.~\ref{fig:BNL734_dsdq2}).  These data
were fit to a description of $\nu p \rightarrow \nu p$ and $\bar{\nu}
p \rightarrow \bar{\nu}p$ to obtain the results shown in
Figure~\ref{fig:BNL734_gasma_fit}.  The results from this fit were
often cited to support the claims from the DIS experiments at the time,
that $\Delta s$ was non-zero and negative.

This experiment also simultaneously measured the neutrino and
antineutrino neutral current to charged current ratios:
\begin{equation}
R_\nu = \frac{\sigma(\nu p \rightarrow \nu p)}
{\sigma(\nu n \rightarrow \mu^- p)}
= 0.153 \pm 0.007 \pm 0.017
\label{eq:rnu}
\end{equation}
and
\begin{equation}
R_{\bar{\nu}} = \frac{\sigma(\bar{\nu} p \rightarrow \bar{\nu} p)}
{\sigma(\bar{\nu} p \rightarrow \mu^+ n)},
= 0.218 \pm 0.012 \pm 0.023~,
\label{eq:rnubar}
\end{equation}
calculated over the interval  $0.5 < Q^2 < 1.0\, (\rm{GeV}/c)^2$.
The dominant error in these ratios was an 11\% systematic.

\subsubsection{Reanalysis of BNL E734}
The non-zero value for $\Delta s$ as obtained by BNL E734 was later
reexamined~\cite{BNL734_garvey}.  This reanalysis more carefully
considered the effects of strange contributions to the vector form
factors and the $Q^2$ evolution of the axial form factor in the
differential cross sections for $\nu p$ and $\bar{\nu} p$ elastic
scattering.  A value of $\Delta s = -0.21 \pm 0.10$ was extracted.  An additional form factor uncertainty of $\pm 0.10$, determined from
fits to the data with different assumptions about the vector
form factors and evolution of the axial form factor (via the axial vector
mass, $M_A$), should be assigned.

Another group has reexamined the BNL E734 result on the neutral
current to charged current ratios, $R_\nu$ and
$R_{\bar{\nu}}$~\cite{BNL734_Alberico}.  These ratios are sensitive to
the axial form factor, and avoid the systematic uncertainty of the
neutrino flux and nuclear model corrections.  While these ratios hold
the promise of a superior method to extract $\Delta s$, the
experimental errors from BNL E734 were too large to provide a
definitive answer, and the conclusions of this analysis were
consistent with the previous reanalysis of the
data~\cite{BNL734_garvey}.

Another reanalysis of these data has been recently
performed~\cite{Pate}.  This approach is interesting in that it uses
the latest data from HAPPEX on the electric and magnetic strange form
factors in an effort to reduce the systematic errore from these form
factors on this meausurement.  The results show the viability of this
method of combining the PV electron and neutrino scattering into one
analysis, but are limited by the large systematic errors on the BNL734
data sample.

Note that the BNL E734 data have been reanalyzed at least three times
since being published in 1987; this points to growing appreciation of
the fact that neutral current neutrino scattering is an excellent
probe of $\Delta s$.  Unfortunately, though, the precision of the BNL
E734 data limits the conclusions that can be drawn from it.

\subsubsection{How FINeSSE Will Improve These Measurements}
The neutrino neutral current to charged current ratio as measured by
BNL E734 (Eq.~\ref{eq:rnu}) is reported with a statistical error of
5\% and a systematic error of 11\%.  This ratio was made by
integrating over the $Q^2$ interval $0.5 < Q^2 < 1.0$ GeV$^2$.

FINeSSE will improve upon this measurement in several ways:
\begin{itemize}
\item The neutrino neutral current to charged current ratio
  measurement will be made as a function of $Q^2$ in the interval $0.2
  < Q^2 < 1.0$ GeV$^2$.  FINeSSE's measurement will be made at a
  lower-$Q^2$ value, where form factors are more easily interpreted.
  In addition, the shape of the ratio as a function of $Q^2$ holds
  additional information about the evolution of the form factors.
\item With the proposed FINeSSE detector and run plan, the statistical
  errors in the $0.2 < Q^2 < 0.4$ GeV$^2$-bin alone will amount to a
  relative error on the ratio of only 2\%.
\item Systematic errors have been estimated with detailed simulations
  of the detector to be 5\%.  Much of this reduction in systematic
  error is due to the greatly reduced background in FINeSSE.  The
  background to the $\nu p \rightarrow \nu p$ reaction is estimated at
  26\% (and is likely to be further reduced) compared to 40\% in BNL
  E734~\cite{BNL734}.
\end{itemize}

In summary, FINeSSE will be able to make a measurement of the
neutral-current to charged-current ratio with a 6\% total error down
to $0.2$ GeV$^2$.  BNL E734 made a measurement with 12\% total error
down to only $0.5$ GeV$^2$.  This will allow a significant improvement
to the uncertainty on the extracted value of $\Delta s$ as described
in Section~\ref{sc:NCCCR2} and
Chapter~\ref{ch:EventSimulationandReconstruction}.
 
\subsection{Relevance to Searches for Dark Matter}

Understanding of the spin contribution to the nucleon of the strange
quarks is important for certain searches of dark
matter~\cite{Ellis:2001pa}.  In $R$-parity-conserving supersymmetric
models, the lightest supersymmetric particle (LSP) is stable and
therefore a dark-matter candidate; in certain scenarios, the relic LSP
density is large enough to be of cosmological interest. Experimental
searches for cosmic LSPs can be competitive with accelerator-based
searches~\cite{Olive:2003xc}.

In the case where the LSP is the neutralino, cosmic LSP can be
detected either directly, through elastic neutralino scattering in an
appropriate target/detector, or indirectly. The indirect method
involves detection of high-energy neutrinos from the center of the
sun, where the heavy neutralinos accumulate and subsequently
annihilate. If the neutralino mass is larger than the $W$ mass,
annihilation into gauge bosons dominates and this gives rise to
high-energy neutrinos that can be detected on earth. The expected rate
for this process also depends on the elastic neutralino-nucleon
scattering cross section.

The neutralino-nucleus elastic-scattering cross section contains a
spin-dependent and a spin-independent part. The spin-dependent part is
given by \[
\sigma=\frac{32}{\pi}G_{F}^{2}m_{r}^{2}\Lambda^{2}J(J+1),\] where
$G_{F}$ is the Fermi constant, $m_{r}$ the reduced neutralino mass,
$J$ the nucleus spin, and \[
\Lambda\equiv\frac{1}{J}\left(a_{p}\left\langle S_{p}\right\rangle
  +a_{n}\left\langle S_{n}\right\rangle \right);\] here $\left\langle
  S_{p(n)}\right\rangle $ is the average proton (neutron) spin in the
nucleus and \[ a_{p(n)}=\sum_{i}\frac{\alpha_{i}}{\sqrt{2}G_{F}}\Delta
q_{i}^{p(n)},\] where the sum is over quark flavors and the
coefficients $\alpha_{i}$ are functions of the composition of the
neutralino in terms of the supersymmetric partners of the gauge
bosons. The factors $\Delta q_{i}^{p}$ and $\Delta q_{i}^{n}$ are the
quark contributions to the proton or neutron spin.

It is established~\cite{SMC,Filippone:2001ux} that
$\Delta u$ and $\Delta d$ have opposite signs. From the above, it
should be clear that knowledge of $\Delta s$ not only is important for
the interpretation of any limits from such dark matter searches, but it
could also influence the choice of detector material for direct
searches~\cite{Bednyakov:2003wf}, making nuclei with either proton- or
neutron-spin excess optimal, depending on its value and sign.

\subsection{A Measurement of $\Delta s$ via Neutral-Current 
Neutrino Scattering} 

In neutral current elastic (NC)
neutrino-nucleon scattering ($\nu N \rightarrow \nu N$), any isoscalar
contribution (such as strange quarks) to the nucleon spin will
contribute to the cross section. This is in contrast to the charged
current quasi-elastic (CCQE) scattering process ($\nu n \rightarrow
\mu^- p$) where only isovector contributions are possible.

FINeSSE will use this feature of neutrino scattering to measure any
contribution of strange quarks to the spin of the nucleon.  The fact
that NC neutrino scattering is sensitive to strange quark (isoscalar)
spin in the proton, and CCQE neutrino scattering is not, will be
exploited by measuring a ratio of these two processes; this will
eliminate a number of experimental and theoretical errors.

\subsubsection{The Neutral Weak Axial Current of the Nucleon} 
\label{sec:NCcur}
The axial part of the weak neutral current may be
written~\cite{BNL734_garvey,Garvey_dels},
\begin{eqnarray}
\left\langle N \left| A_\mu^Z \right| N \right\rangle 
& = & 
- \left[\frac{G_F}{\sqrt{2}} \right]^{\frac{1}{2}}
\left\langle N \left| \frac{\bar{u}\gamma_\mu \gamma_5 u - 
                            \bar{d}\gamma_\mu \gamma_5 d - 
                            \bar{s}\gamma_\mu \gamma_5 s}{2} \right| N \right \rangle \\
& = & 
- \left[\frac{G_F}{\sqrt{2}} \right]^{\frac{1}{2}}
\left\langle N \left| 
- \frac{G_A(Q^2)}{2} \gamma_\mu \gamma_5 \tau_z +
  \frac{G_A^s(Q^2)}{2} \gamma_\mu \gamma_5  
\right| N \right \rangle,
\end{eqnarray}
where $G_A$ is the axial form factor and $\tau_z = \pm 1$ for protons (+) or
neutrons (-).  The strange axial-vector form factor, $G_A^s$, is
identified with the $\bar{s}\gamma_\mu \gamma_5 s$ term which is
$\Delta s$, the spin carried by the strange quarks.  So, the
non-strange ($u$ and $d$) quark axial current is accounted for in
$G_A$, known at $Q^2=0$ from neutron beta decay to be
$G_A(Q^2=0)=1.2673\pm0.0035$~\cite{Bernard}.  The (unknown)
strange quark axial current is subsumed in $G_A^s$.  A similar
decomposition may be obtained for the vector part of the neutral weak
current~\cite{BNL734_garvey,Garvey_dels} in terms of the two
non-strange vector form factors, $F_1$ and $F_2$, and the
corresponding strange quarks parts, $F_1^s$ and $F_2^s$.

As shown above, the strange axial form factor, $G_A^s$, in the limit
of zero momentum transfer ($Q^2=0$), is identified with the strange
quark contribution to the nucleon spin, $\Delta s$. It is worth
mentioning that the arguments leading to the connection between
$G_A^s$ at low momentum transfer and the strange spin content, $\Delta
s$, that can be extracted from DIS data in the scaling limit, are
essentially the same as those that were used to derive the Bjorken sum
rule, which connects spin-dependent DIS structure functions to the
coupling constants in neutron decay and is considered one of the most
fundamental predictions of QCD. They are both based on the operator
product expansion, which expresses the moments of structure functions
in terms of matrix elements of local operators and perturbatively
calculable Wilson coefficients.

\subsubsection{Neutrino Cross Sections and Form Factors}
The differential cross section, 
$d \sigma/d Q^2$, for NC and CCQE scattering of 
neutrinos and antineutrinos from nucleons can be written as a 
function of the nucleon form factors $F_1$ and $F_2$ (both vector) and
$G_1$ (axial vector)~\cite{barnett,Garvey_dels}:
\begin{equation}
\frac{d \sigma}{d Q^2} = \frac{G_F^2}{2 \pi} \frac{Q^2}{E_\nu^2} (A \pm BW + CW^2)
\label{eq:dsdq2}
\end{equation}
with kinematic factor, $W=4E_\nu/m_p-Q^2/m_p^2$. 
The + (-) sign is for neutrino (antineutrino) scattering. $Q^2$ is
the squared-four-momentum transfer.  
The $A$, $B$, and $C$, contain the form factors:
\begin{eqnarray}
\label{eq:ABC}
A & = & \frac{1}{4}[G_1^2(1+\tau)-(F_1^2-\tau F_2^2)(1-\tau)+4 \tau F_1 F_2], \\
B & = & -\frac{1}{4}G_1(F_1 + F_2), \\
C & = & \frac{1}{16}\frac{m_p^2}{Q^2}(G_1^2+F_1^2+\tau F_2^2).
\end{eqnarray}
Up to this point, this formalism is valid for both NC and CCQE
neutrino- (and antineutrino-) nucleon scattering.  The difference
between the NC and CCQE processes is accounted for with different form
factors ($F_1$, $F_2$, and $G_1$) in each case.

For NC neutrino scattering the axial-vector from factor (as described
in Section~\ref{sec:NCcur}) may be written in terms of the known axial
form factor plus an unknown strange form factor:
\begin{equation}
G_1 = \left[ -\frac{G_A}{2}\tau_z + \frac{G_A^s}{2} \right].
\label{eq:G_1}
\end{equation}
The vector form factors $F_{1,2}$ may also be written in terms of
known form factors plus a strange quark contribution,
\begin{equation}
F_{1,2} = \left[ 
\left[\frac{1}{2}-\sin^2 \theta_W \right]\left[ F_{1,2}^p - F_{1,2}^n \right]\tau_z
- \sin^2 \theta_W \left[ F_{1,2}^p + F_{1,2}^n \right] 
- \frac{1}{2} F_{1,2}^s \right]~,
\label{eq:F_12}
\end{equation}
where $F_1^{p(n)}$ is the Dirac form factor of the proton (neutron)
and $F_2^{p(n)}$ is the Pauli form factor.  The CVC hypothesis allows
us to write the same form factors in these equations as those measured
in electron scattering.  Therefore, the only unknown quantities in
these equations are the strange vector form factors $F_{1,2}^s$.

The differential cross section for neutrino scattering
at low $Q^2$(Eq.~\ref{eq:dsdq2}) is dominated by the axial form factor, $G_1$.
This can be seen be examining Equations~\ref{eq:dsdq2}
and~\ref{eq:ABC}.  In fact, at low-$Q^2$, $d \sigma/d Q^2 \propto
G_1^2$.  This is a crucial point, and it is what makes NC neutrino
scattering the best place to look for the effects of strange-quarks in
the nucleon spin.  It also makes the results less sensitive to the
strange vector form factors, $F_{1,2}^s$.

\subsubsection{$Q^2$ Dependence of the Form Factors}
All of the form factors are, most generally, functions of $Q^2$.  The
values of the non-strange form factors at $Q^2=0$ are known from the
static properties of the nucleon (e.g. charge, magnetic moment, and
neutron decay constant). How the form factors change with increasing
$Q^2$ is less well known and must be addressed.  The form factors are
commonly parameterized with a dipole form, where the dipole mass set
by various measurements.  The evolution of $F_{1,2}^{p(n)}$ is known
via numerous experiments on electron scattering; the vector dipole
mass, $M_V$ is 0.843 GeV/c$^2$.  The $Q^2$ dependence of the axial
form factor, $G_A$, is measured via CCQE neutrino
scattering; the world average data yield an axial
mass $M_A=1.026\pm0.021$~\cite{Bernard}.

Both the strange form factors, $F_{1,2}^s$ and $G_A^s$ and their $Q^2$
evolution are unknown.  It is most common to assume the same $Q^2$
dependence as for the non-strange form factors, using $M_V$ for
$F_{1,2}^s$ and $M_A$ for $G_A^s$.  The uncertainty on the $Q^2$
evolution introduces some uncertainty in the extraction of $G_A^s$
from a measurement.  A measurement at $Q^2 \approx 0.2$ GeV$^2$,
however, would keep this contribution to the uncertainty at a low
level, as the value of the form factor differs by only 20\% from
$Q^2=0$ to $Q^2=0.2$ GeV$^2$.

\subsubsection{Nuclear Physics Corrections}
The expression for the cross section (Eq.~\ref{eq:dsdq2}) is for
scattering from free nucleons. Since FINeSSE will have a target that
consists, in large part, of nucleons bound in carbon, consideration
will need to be given of the effects of this
binding~\cite{horowitz1,kim}.  The corrections can be rather large
when considering the absolute event rate, and can depend greatly on
the model employed, because the amount and quality of available
neutrino data to constrain such models, is lacking.  In principle, the
nuclear models can be constrained with the high-quality electron data
available; this, however, is a work in progress.

These effects become less of a concern when {\em ratios} of cross
sections are
considered~\cite{barbaro,alberico_nufact,alberico_internal,piekarewicz1}.
The initial and final states of the hadrons involved are quite similar
in both NC and CCQE neutrino scattering; as a result, the corrections
employed for either channel should be similar as well.  FINeSSE will
utilize this fact with a measurement of $\Delta s$ as explained in the
following sections.

\subsubsection{The Neutral Current to Charged Current Ratio}
\label{sc:NCCCR2}
The NC neutrino scattering cross section depends strongly on $G_1$ and
therefore on $G_A^s$, the quantity of interest. An absolute
measurement of the cross section, however, is an experimental
challenge; the level of precision achievable for this measurement
would not yield the desired precision for $G_A^s$.  An absolute
prediction is also a challenge from a phenomenological standpoint, as
uncertainties in form factors and nuclear corrections can be large.

It is possible to extract $G_A^s$ to the desired precision by
measuring the {\em ratio} of NC to CCQE event rates.  A
measurement of the ratio of NC to CCQE neutrino scattering event rates
may be measured with greater precision, since many systematic
uncertainties cancel in the ratio such as the neutrino flux and
correlated reconstruction efficiencies.  Theoretical uncertainties are
also reduced in this quantity as many of these uncertainties are
correlated between NC and CCQE scattering.  FINeSSE will use this
method to measure $G_A^s$.

First, consider the ratio of NC neutrino-proton to NC neutrino-neutron 
scattering,
\begin{equation}
R(p/n) = \frac{\sigma(\nu p \rightarrow \nu p)}{\sigma(\nu n \rightarrow \nu n)}.
\label{eq:rpn}
\end{equation} 
This ratio is very sensitive to $G_A^s$~\cite{Garvey_dels}. The NC
neutrino nucleon scattering cross section is proportional to $G_1^2$
as explained above.  A non-zero value for $G_A^s$ will pull the
denominator of this ratio one way, and the numerator the other, due to
the $\tau_z$ factor in Equation~\ref{eq:G_1}.

$R(p/n)$, however, is likely to be very difficult to measure
accurately in a neutrino scattering experiment, because of the
intrinsic difficulties and uncertainties involved with neutron
detection.  For this reason FINeSSE will focus on a measurement of the
ratio of the NC neutrino-proton scattering ($\nu p \rightarrow \nu p$)
to CCQE neutrino scattering ($\nu n \rightarrow \mu^- p$).  This
ratio,
\begin{equation}
R(NC/CC) = \frac{\sigma(\nu p \rightarrow \nu p)}
{\sigma(\nu n \rightarrow \mu^- p)},
\end{equation}
can be more accurately measured and is still quite sensitive to
$G_A^s$ (albeit less so than $R(p/n)$).  The numerator depends upon
$G_A^s$ as explained in the formalism introduced above.  The
denominator does not as the $\nu n \rightarrow \mu^- p$ process is
sensitive only to isovector quark currents and not to isoscalar
currents (such as that from strange quarks).

\begin{figure}
\centering
\includegraphics[width=5in]{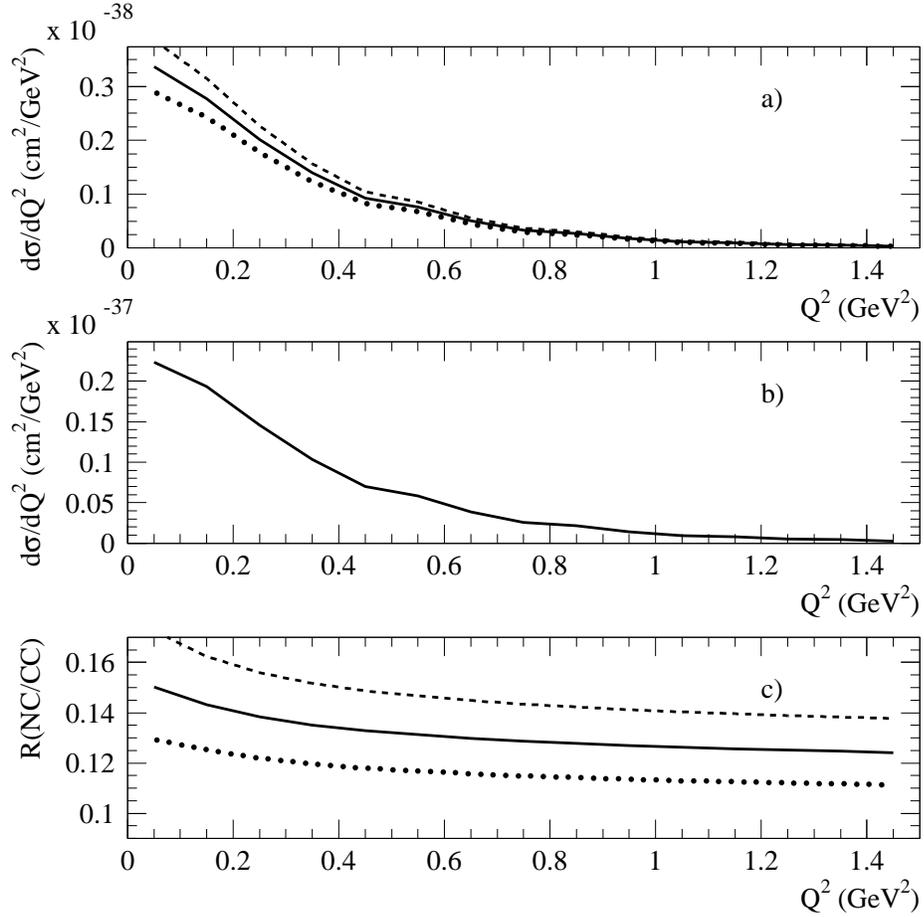}
\caption{\it Flux-weighted differential cross section as a function of 
  $Q^2$ for $\nu p \rightarrow \nu p$ (a) and $\nu n \rightarrow \mu^-
  p$ (b) scattering together with the cross section ratio of the these
  two processes (c).  These quantities are shown for $G_A^s = 0.$
  (solid), $= -0.1$ (dashed), and $= +0.1$ (dotted). The $\nu n
  \rightarrow \mu^- p$ process does not depend upon $G_A^s$.}
\label{fig:dsdq2}
\end{figure}

The $\nu p \rightarrow \nu p$ and $\nu n \rightarrow \mu^- p$
differential cross sections (weighted by the calculated FINeSSE flux)
as calculated with Equation~\ref{eq:dsdq2} are plotted in
Figure~\ref{fig:dsdq2} for $G_A^s = -0.1, 0.0, +0.1$; this shows that
the cross section for $\nu p \rightarrow \nu p$ depends strongly upon
$G_A^s$. The cross section for $\nu n \rightarrow \mu^- p$ is
independent of $G_A^s$, so the ratio of flux-weighted cross
sections (and therefore the event rates) of these two processes
depends upon $G_A^s$.

\begin{figure}
\centering
\includegraphics[width=5in]{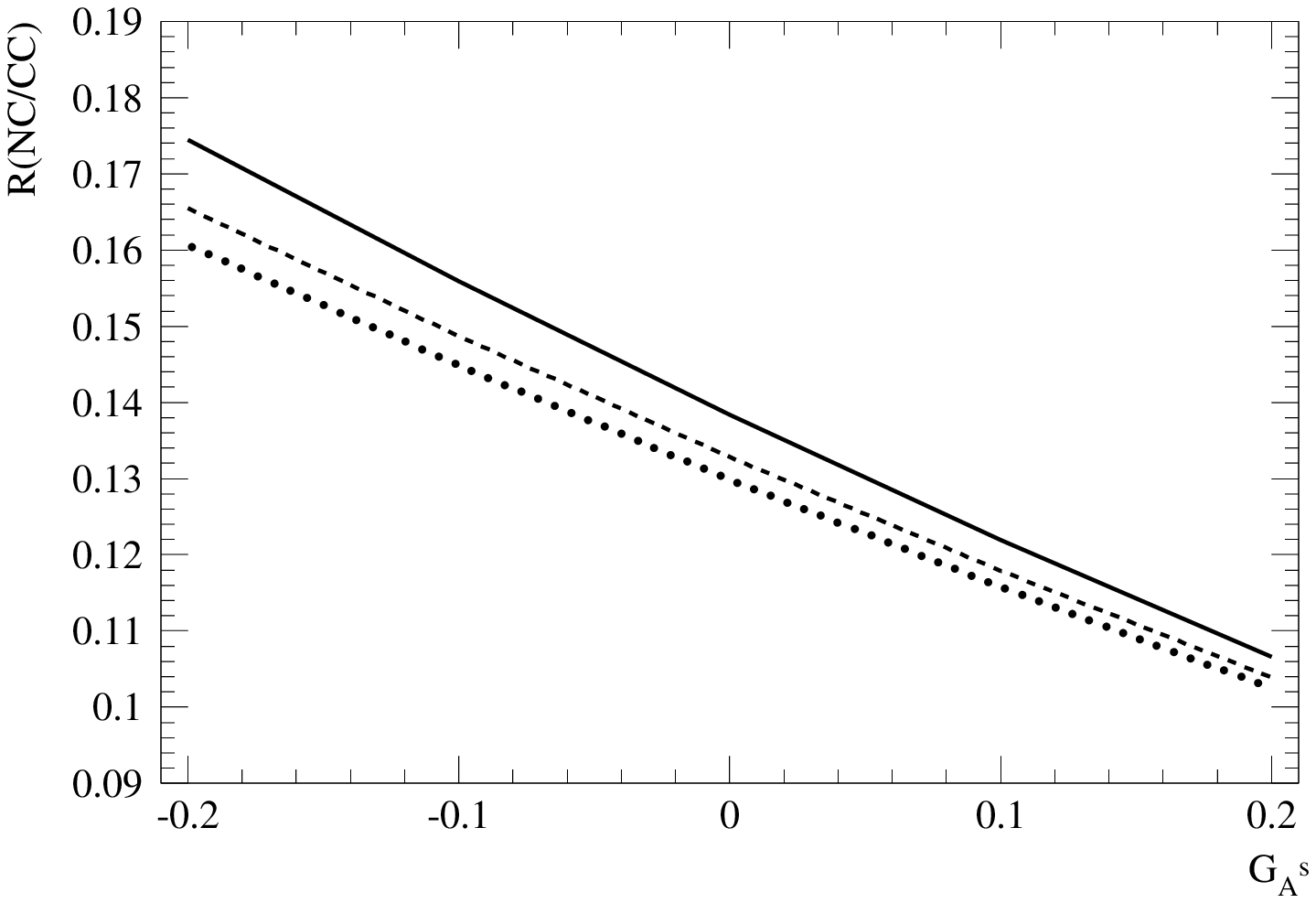}
\caption{\it Ratio of flux-weighted cross sections of the $\nu p \rightarrow \nu p$ and  
$\nu n \rightarrow \mu^- p$ processes as a function of the axial form factor
 $G_A^s$ at $Q^2=0.25$ (solid), 0.45 (dashed), and 0.65 (dotted) GeV$^2$.}
\label{fig:rncvds}
\end{figure}

This dependence is also shown in Figure~\ref{fig:rncvds} as a function
of $G_A^s$ for three different $Q^2$ bins.  In the $Q^2=0.25$ GeV$^2$
bin, the sensitivity of the NC/CC ratio on $G_A^s$ is approximately
1.2.  The relative uncertainty in the NC/CC ratio, $\Delta R/R$ is
related to the absolute uncertainty on $G_A^s$, $\sigma(G_A^s)$ by
\begin{equation}
\sigma(G_A^s) = \frac{1.0}{1.2} \frac{\Delta R}{R}.
\label{eq:siggas}
\end{equation}

Considering this and recalling that $G_A^s(Q^2=0) = \Delta s$ yields
the conclusion that a 5\% relative measurement of the NC/CC ratio at
$Q^2 \approx 0.25$ GeV$^2$ would enable an extraction of $\Delta s$
with an error of 0.04.  This error is comparable to that quoted in the
latest extractions of $\Delta s$ from charged lepton DIS~\cite{HERMES}.

\subsection{FINeSSE Sensitivity to $\Delta s$}
NC, CCQE, and background events in the proposed detector have been
simulated in a detailed manner; a reconstruction procedure has been
performed, to calculate an estimated sensitivity.  All conceivable
effects were considered; systematic errors were estimated.  The
experimental errors considered were those due to statistics and to
various systematics.  The statistical errors were calculated based on
a 9~ton (fiducial) detector running for two years ($6\times 10^{20}$
protons-on-target). The systematic uncertainties considered include those
due to free-to-bound $\nu p \rightarrow \nu p$ scattering rate,
reconstruction efficiencies, background estimation, and $Q^2$
reconstruction.  In addition, theoretical uncertainties from nuclear
model dependence and from form factor estimation were considered.  The
details of this procedure are described in
Chapter~\ref{ch:EventSimulationandReconstruction}.

A fit to the simulated data in the $Q^2$ region where the detector
has reasonable acceptance yields an experimental uncertainty in $\Delta
s$ of $\pm0.039$; the combined uncertainty from the axial
mass, $M_A$, and $F_2^s$ form factor is $\pm0.025$.

Based on these results, it has been determined that FINeSSE, with a
design and plan described in the Chapters below, can make an accurate
measurement of $\Delta s$ at down to $Q^2 \approx 0.2$ GeV$^2$.  This
will enable FINeSSE to answer an important and unresolved question
about the structure of the proton.

\section{$\nu_\mu$ Disappearance}

The FINeSSE detector can be used in conjunction with the MiniBooNE
detector to substantially improve our understanding of neutrino
oscillations at high $\Delta m^2$ by looking for a neutrino 
energy-dependent deficit of $\nu_\mu$ event rates compared to a 
no-oscillation hypothesis.  This search is motivated by
astrophysical models for the production of heavy elements in
supernovae.  If MiniBooNE observes a signal, a combined FINeSSE and
MiniBooNE (``FINeSSE+MiniBooNE'') run represents the next step in
determining the underlying physics model of the oscillation.  These
two motivations are not directly connected: if MiniBooNE does not see
a signal, the astrophysical case still makes this study compelling.

In this section, we first provide a brief overview of the formalism
for neutrino oscillations.  Second, we introduce the LSND signal,
along with theoretical interpretations involving sterile neutrinos.
Third, we describe the astrophysical motivations for the disappearance
search.  Lastly, we describe the capability of a FINeSSE+MiniBooNE
joint analysis of $\nu_\mu$ disappearance.

\subsection{Neutrino Oscillation Formalism}

``Neutrino oscillations'' occur when a pure flavor (weak) eigenstate
born through a weak decay changes into another flavor as the state
propagates in space.  This can occur if two conditions are met.
First, the weak eigenstates can be written as mixtures of the mass
eigenstates, for example:
\[
\begin{array}{l}
\nu _e=\cos \theta \;\nu _1+\sin \theta \;\nu _2 \\
\nu _\mu =-\sin \theta \;\nu _1+\cos \theta \;\nu _2
\end{array}
\]
where $\theta$ is the ``mixing angle.''  The second condition is that
each of the mass eigenstate components propagate with a different
frequency, which can occur only if the masses are different.  We
define the squared mass difference as $\Delta m^2=\left|
  m_2^2-m_1^2\right|>0$.  In a two-component model, the oscillation
probability for $\nu_\mu \rightarrow \nu_e$ oscillations is then given
by:

\begin{equation}
{\rm Prob}\left( \nu _\mu \rightarrow \nu _e\right) = \sin ^22\theta \;\sin
^2\left( \frac{1.27\;\Delta m^2\left( {\rm eV}^2\right) \,L\left({\rm km}%
\right) }{E \left({\rm GeV}\right) }\right)~,
  \label{prob}
\end{equation}
where
$L$ is the distance from the source, and $E$ is the neutrino energy.

Most neutrino oscillation analyses consider only two-generation mixing
scenarios, but the more general case includes oscillations between all
neutrino species.  For the case of the three Standard Model species,
this can be expressed as:
\[
\left(
\begin{array}{l}
\nu _e \\
\nu _\mu \\
\nu _\tau
\end{array}
\right) =\left(
\begin{array}{lll}
U_{e1} & U_{e2} & U_{e3} \\
U_{\mu 1} & U_{\mu 2} & U_{\mu 3} \\
U_{\tau 1} & U_{\tau 2} & U_{\tau 3}
\end{array}
\right) \left(
\begin{array}{l}
\nu _1 \\
\nu _2 \\
\nu _3
\end{array}
\right).
\]
The oscillation
probability is then:

\begin{eqnarray}
P(\nu_\alpha\rightarrow\nu_{\beta})  & = &
|\mbox{Amp}[\nu_\alpha \rightarrow \nu_\beta]|^2 =\delta_{\alpha\beta}
\nonumber  \\
& & \mbox{} - 4\sum_{i>j} \Re\,(U^*_{\alpha i} U_{\beta i} U_{\alpha j}
U^*_{\beta j})
\sin^2[\Delta m^2_{ij}(L/4E)]   \nonumber  \\
& & \mbox{} + 2\sum_{i>j} \Im \,(U^*_{\alpha i} U_{\beta i} U_{\alpha j}
U^*_{\beta j})
\sin[\Delta m^2_{ij}(L/2E)]~,
\label{eq3.2}
\end{eqnarray}

\noindent where $\Delta m_{i\,j}^2=\left| m_i^2-m_j^2\right| $.
Note that there are three different
$\Delta m^2$ (although only two are independent),
and three different mixing angles.   This method can be generalized to
include more neutrino species in Beyond-the-Standard Model Theories.

Although in general there will be mixing among all flavors of
neutrinos, two-generation mixing is often assumed for simplicity.  If
the mass scales are quite different (e.g., $m_3 >> m_2 >> m_1$), then
the oscillation phenomena tend to decouple and the two-generation
mixing model is a good approximation in limited regions.  In this
case, each transition can be described by a two-generation mixing
equation.  However, it is possible that experimental results
interpreted within the two-generation mixing formalism may indicate
very different $\Delta m^2$ scales with quite different apparent
strengths for the same oscillation.  This is because, as is evident
from equation \ref{eq3.2}, multiple terms involving different
mixing strengths and $\Delta m^2$ values contribute to the transition
probability for $\nu_\alpha \rightarrow \nu_\beta$.

From equation \ref{prob}, one can see that the oscillation wavelength
will depend upon $L$, $E$, and $\Delta m^2$.  For short baseline
experiments, sensitivity to oscillations is in a range of $>
0.1$eV$^2$, which will term the ``high $\Delta m^2$ region.  The
oscillation amplitude will depend upon $\sin^2 2\theta$.

\subsection{Experimental Results: The LSND Signal}

One of the most exciting questions in high energy physics, at present,
is whether the ``LSND signal'' is due to neutrino oscillations.  If
the signal is confirmed in the MiniBooNE experiment, then this is an
indication for new physics beyond the Standard Model.  Fermilab will
want to be poised to pursue this result.  A FINeSSE+MiniBooNE run will
be the first step.

The LSND experiment has observed a 4$\sigma$ excess which can be
interpreted as oscillations between muon and electron
neutrinos~\cite{LSNDfinal}.  The beam was produced at LANSCE at LANL,
with 800 MeV protons interacting with a water target, a close-packed
high-Z target, and a water-cooled copper beam dump.  The highest
statistics came from $\bar \nu_\mu$ neutrinos produced by decay at
rest (DAR) of muons, with $20<E<50$ MeV.  However, the lower
statistics decay in flight (DIF) $\nu_\mu$'s were also analyzed.  The
liquid scintillator detector was located 30~m from the beam dump.
Hence the $L/E$ of the experiment was $\sim 1$ m/MeV.  As a result, if
the excess is interpreted as oscillations, the allowed region is
located at high $\Delta m^2$.

\begin{figure}[!tbp]
\centerline{\includegraphics[width=2.in]{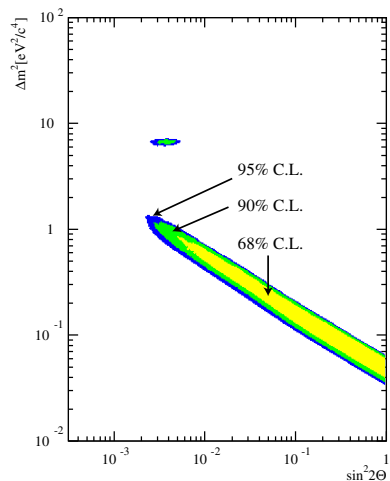}}
\caption{\it LSND and Karmen joint analysis allowed region. }\vspace{0.25in}
\label{fig:joint}
\end{figure}

LSND is the only short-baseline experiment to have observed evidence
for oscillations.  Other short-baseline experiments have searched and
seen no signal.  Those most relevant to this proposal are Karmen~\cite{Karmen} and Bugey~\cite{Bugey}.  Karmen, which ran at the ISIS
facility at Rutherford Labs, was similar in concept to LSND, using a
DAR beam; but the detector was smaller, and the beam of lower
intensity.  Most importantly, it had an L of 17~m; in this way, it can
be thought of as a ``near detector'' for LSND.  Playing the null
signal in Karmen against the observed excess in LSND results in the
allowed regions shown in Figure~\ref{fig:joint}~\cite{joint}.  The
Bugey reactor experiment rules out $\sin^2 2\theta > 0.04$ at 90\%
C.L.  in a search for $\bar \nu_e$ disappearance using a reactor.
Assuming that oscillations respect time reversal, and that only the
three standard model active species are involved, Bugey's result can
be taken as an excluded region for LSND.

The MiniBooNE experiment is a designed to decisively address the LSND
signal.  This experiment will complete its Phase I (neutrino) run in
mid-2005 and is expected to request further (Phase II) running for the
period thereafter.  By the time FINeSSE is on-line, the LSND signal
will be tested in both $\nu$ and $\overline{\nu}$ modes.

If MiniBooNE confirms LSND, and all other oscillation results remain
as they presently stand, then this necessarily implies new physics.
The oscillation signals from solar neutrinos, atmospheric neutrinos,
and LSND cannot be simultaneously fit with the three Standard Model
neutrinos.  A favored method for expanding the theory to allow LSND is
to invoke sterile neutrinos ($\nu_s$).  These are neutrinos which do
not interact via the W or Z, but can couple to the Standard Model
``active'' neutrinos through oscillations.  The most minimal extension
is to introduce a single light sterile neutrino.  This extra neutrino
is not ruled out by cosmology~\cite{cosmo}.  Light sterile neutrinos
can appear in supersymmetry, extra dimensions and GUTs~\cite{theories}.  These can all accommodate more than one light
sterile neutrino, but we will confine our discussion to one for
simplicity.

\begin{figure}[!tbp]
\begin{minipage}{3.in}
\includegraphics[width=3.in]{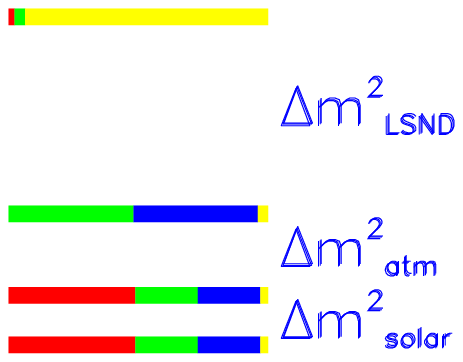}

~~~~~~~~~~~~\includegraphics[width=1.in]{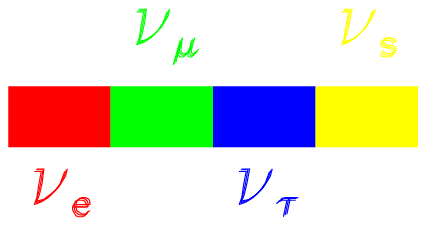}

\end{minipage}
\begin{minipage}{3.in}\includegraphics[width=3.in, bb=60 130 570 740]{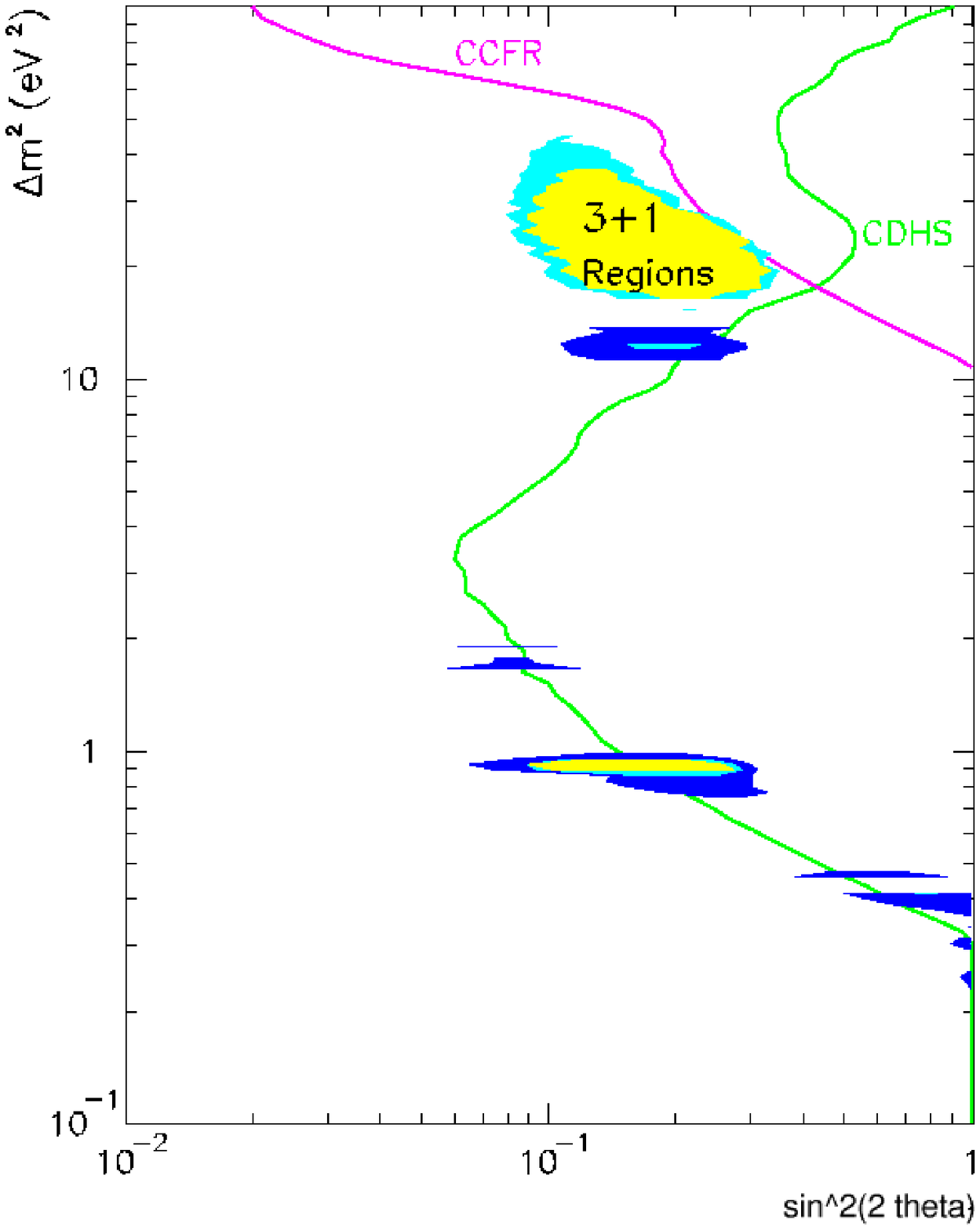}
\end{minipage}
\caption{\it Left: an example of a 3+1 mass spectrum.  Right: allowed region for 
$\nu_\mu \rightarrow \nu_s$ in 3+1 models.  The exclusion regions for 
CDHS and CCFR are also shown.}\vspace{0.25in}
\label{fig:3+1}
\end{figure}

Figure~\ref{fig:3+1}(left) shows a cartoon of how the squared masses
and mixings might be arranged if a single sterile neutrino is
introduced into the theory.  The vertical axis is logarithmic and
arbitrary.  The bars indicate the flavor content of each mass state.
The LSND signal is explained by the largest squared mass splitting,
with the transition $\nu_\mu \rightarrow \nu_e$.  Because there is a
triplet of neutrinos with nearly the same mass, and one large
splitting, this is called a ``3+1'' model.  Note that the transitions
$\nu_\mu \rightarrow \nu_s$ and $\nu_e \rightarrow \nu_s$ must also be
allowed for the same $\Delta m^2$.  The allowed regions for 3+1 for
fits which include LSND, Karmen, Bugey, and two $\nu_\mu$
disappearance experiments (CDHS and CCFR84) are shown in
Figure~\ref{fig:3+1}(right)~\cite{sorel}.  These allowed regions can
be addressed by FINeSSE+MiniBooNE as discussed below.

\subsection{Astrophysical Motivation for High $\Delta m^2$ Disappearance Searches}

The existence of at least one sterile neutrino in the high $\sim 1$ eV
mass range has interesting consequences for the heavy element
abundance in the universe.  In fact, oscillations were predicted on
the basis of this abundance before the LSND signal was presented~\cite{georgewasfirst}.  The allowed range extends beyond the region
constrained by the LSND signal.  Hence, whether or not MiniBooNE
confirms LSND, searches for active-to-sterile oscillations at high
$\Delta m^2$ remain motivated by this astrophysical question.

Active-to-sterile neutrino oscillations in the late time
post-core-bounce period of a supernova will affect the $r$-process, or
rapid neutron capture process.  This presents a mechanism for
producing substantially more heavy elements ($A>100$), solving the
long-standing problem of the high abundance of uranium in the
universe.  The FINeSSE+MiniBooNE search addresses allowed parameters
for this solution.

A favored mechanism for producing heavy elements is through the
$r$-process in the neutrino-heated ejecta of a Type II or Type I/c
supernova.  In this model, during the period of the neutrino-driven
wind, which lasts for $\sim 10$s, ``seed'' elements with $A$ between
50 and 100 capture neutrons to produce the elements with $A>100$.  The
problem is that in most models the neutron-to-seed ratio, $R$, is too
low for production of the heaviest elements~\cite{toolow}.  In fact,
detailed simulation show that a phenomenon called the $\alpha$-effect,
in which neutrons are frozen into alpha particles that do not
recombine to form heavier elements in the requisite time period,
renders the neutron-to-seed ratio downright ``anemic''~\cite{anemic}.

Various solutions have been proposed.  One option is to resort to the
competing theory of neutron star mergers.  The problem with this
scenario is that mergers are too rare to produce the observed abundance
of heavy elements~\cite{starmergers}.  Another alternative is to
introduce physics which adjusts the neutron-to-seed ratio.  This can
be done by modifying the expansion rate, the entropy per baryon or the
$n/p$ ratio -- all of which will affect the neutron-to-seed ratio.
Adjusting any of these three requires invoking new physics in the
model.  We explore the last alternative here: introducing $\nu_e
\rightarrow \nu_{sterile}$ oscillations, which, when combined with
matter effects, enhance the production of neutrons over protons.

The idea~\cite{george, anemic} is that production of neutron-rich
elements requires a neutron-rich environment.  To the level that the
processes $\nu_e + n \rightarrow p + e^-$ and $\bar \nu_e + p
\rightarrow n + e^+$ are in balance during the neutrino-driven wind,
there is no net excess of neutrons.  In its simplest form, neutrino
oscillations between an electron and sterile flavor would not upset
the balance because $\nu_e$ and $\bar \nu_e$ will oscillate at the
same rate.  However, when one introduces matter (or MSW~\cite{MSW})
effects, neutrino and anti-neutrino oscillations are modified with
opposite sign in an electron-rich environment.  Oscillations of
$\nu_e$ to sterile neutrinos are enhanced, while $\bar \nu_e$ are
depressed.  This can produce a substantial neutron excess by removing
the offending $\nu_e$'s. The $\alpha$-process removes some neutrons,
but stalls once the protons are devoured, leaving sufficient neutrons
to produce the high-$A$ elements.

\begin{figure}[!tbp]
\centerline{\includegraphics[width=3.in]{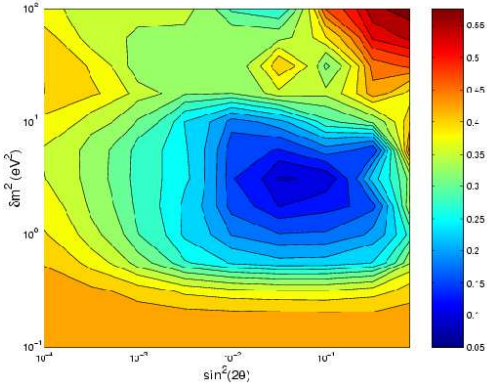}}
\caption{\it Allowed oscillation parameters for the 
  oscillation-enhanced $r$-process.  }\vspace{0.25in}
\label{fig:Ye}
\end{figure}

In this model, the neutron-to-proton ratio (usually characterized as
the function $Y_e = 1/(1+n/p))$), depends upon the choice of $\Delta
m^2$ and $\sin^2 2\theta$.  The condition for a successful r-process is
$Y_e<1/2$.  The smaller the value of $Y_e$, the larger the high-$A$
abundance.  Figure~\ref{fig:Ye} shows $Y_e$ as a function of the
$\nu_e \rightarrow \nu_{s}$ oscillation parameters.  This shows that
there are a wide range of ``robust'' solutions~\cite{anemic}.

One can connect the allowed space for $\nu_\mu \rightarrow \nu_s$ and
$\nu_\mu \rightarrow \nu_e$ to the allowed region for $\nu_e
\rightarrow \nu_s$ within 3+1 and 3+2 models.  If MiniBooNE sees a
signal, this will be a great victory for the oscillation-enhanced
r-process model.  In the case where MiniBooNE does not see a signal,
there remains a large parameter space open to this model.  At present,
$\nu_\mu$ disappearance in FINeSSE+MiniBooNE represents the only way
to access that parameter space.

\subsection{FINeSSE+MiniBooNE Capability for $\nu_\mu$ Disappearance}
\label{sec:oscillations}

\begin{figure}[!tbp]
\centerline{\includegraphics[width=5.in, bb=98 160 525 700]{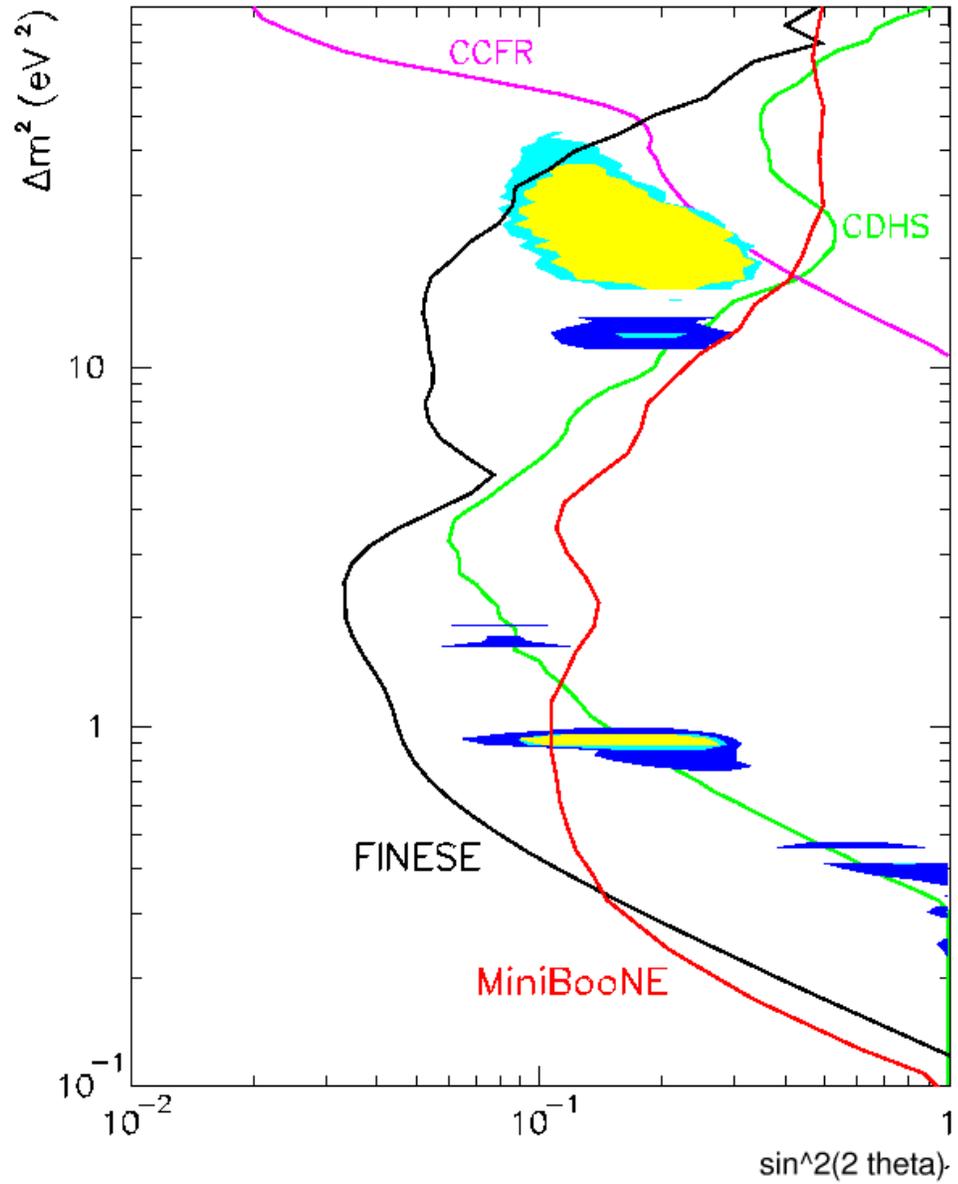}}
\caption{\it The parameter space covered by 
  FINeSSE+MiniBooNE for $\nu_\mu$ disappearance (labeled ``FINESE'').
  Also shown: allowed regions in 3+1 models given the LSND signal
  (solid), existing exclusion regions from CDHS and CCFR, and the expected
  exclusion region for MiniBooNE.}\vspace{0.25in}
\label{fig:bestfinese}
\end{figure}

The FINeSSE detector can be combined with MiniBooNE to explore allowed
regions for oscillations to sterile neutrinos via $\nu_\mu$
disappearance.  In this analysis, FINeSSE serves as a near detector to
accurately measure the flux, and MiniBooNE serves as the far detector
where a deficit may be observed.  This is a unique capability -- there
are no other short baseline $\nu_\mu$ disappearance experiments in the
world.  If MiniBooNE observes a signal, FINeSSE+MiniBooNE will be a
crucial next-step toward understanding the result.  If MiniBooNE does
not observe a signal, this region is still interesting because of 
its relevance to astrophysics.

Figure~\ref{fig:bestfinese} shows the FINeSSE+MiniBooNE expectation
for the default design, with $6\times 10^{20}$ protons on target, in
neutrino mode.  Also shown are the 3+1 allowed regions for fits to
LSND, atmospheric, and solar (as described above); and the expectation
for MiniBooNE prior to FINeSSE running.  MiniBooNE will be able to
address the lower 3+1 allowed regions.  The largest 3+1 allowed
region, however, can only be addressed by the FINeSSE+MiniBooNE
combination. This combination of FINeSSE and MiniBooNE is therefore
very powerful; it alone is able to address the full 3+1 picture.

The``standard'' configuration for FINeSSE and MiniBooNE simultaneous
running places the FINeSSE detector at 100~m from the primary
beryllium target, with the 25~m absorber installed in the beamline.
The angular acceptance from the target to FINeSSE is 25 mrad, and to
MiniBooNE is 10~mrad.  In this analysis, we accept only neutrinos
which traverse both detectors, meaning that we use only the inner 1 m
radius (10~mrad acceptance from target) of FINeSSE.  Event rates are
for a 9~ton fiducial volume and $6\times 10^{20}$ protons in neutrino
mode.

In Chapter 5, we provide details on how the sensitivity shown in
Figure~\ref{fig:bestfinese} was obtained.  We explain why the standard
configuration is best for the analysis.  We also describe the method
for determining the sensitivity, which compares the energy
distribution of events in the near and far detector.  This method
accounts for both statistical and systematic errors.

\section{FINeSSE on the Booster Neutrino Beamline}

The Booster neutrino beamline is the only existing beamline at
Fermilab or around the world where this physics can be accomplished.
The FINeSSE $\Delta s$ measurement requires a clean, low energy neutrino
spectrum, as is produced by the Booster neutrino production target.
The oscillation physics goals of FINeSSE require the energies and baselines
available to experiments on the Booster neutrino beamline.  These
requirements make it impossible to perform this measurement at
Fermilab's other existing neutrino beamline, NuMI.

\begin{figure}[thb]
\centering
\includegraphics[width=3.5in,bb=17 17 771 386]{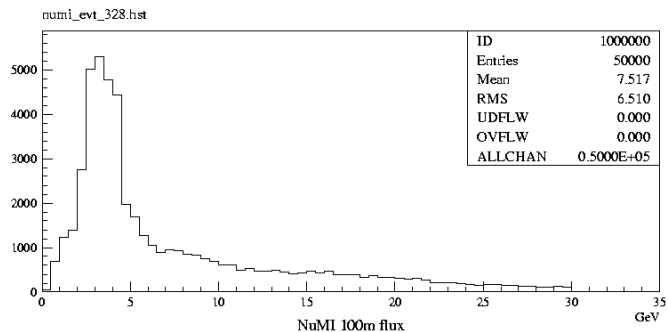}
\caption{\em The flux spectrum of the Booster neutrino beam and the
NuMI neutrino beam at their near detector locations.}
\label{fig:numi-flux}
\end{figure}

Figure~\ref{fig:numi-flux} shows the NuMI beam flux at the MINOS near
detector location, 290~m from the NuMI decay pipe.  The Booster
neutrino spectrum at 75~m from the end of the Booster neutrino
beamline decay pipe with the 25~m absorber in position (100~m from the
production target), is shown in Figure~\ref{fig:flux-100m-25mabs}.  As
indicated, the average energy of neutrinos from the Booster is
700~MeV, with virtually no neutrinos beyond 3~GeV.  This neutrino
energy distribution is excellent for making the $\Delta s$
measurement.  It is an energy large enough to be beyond the region
where low-energy nuclear corrections are significant, yet not so large
where pion production and DIS scattering backgrounds are high.
The NuMI flux, however, has an average energy above 7~GeV and a tail
that extends past 30~GeV.  This energy distribution creates large pion
and DIS scattering rates that increase the background to NC neutrino
elastic scattering considerably.

\begin{figure}
  \centering \includegraphics[width=3.5in]{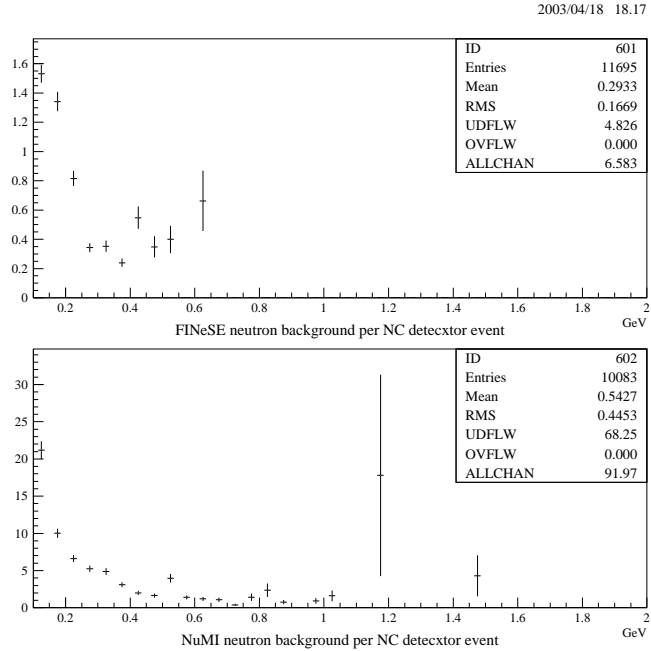}
\caption{\em Ratio of neutron background, from neutrino interactions in the 
  surrounding earth, to NC-like detector events using the FINeSSE
  detector in the Booster neutrino beam (top) and the NuMI neutrino beam (bottom), 
  shown as a function of energy energy deposited in the detector. This plot assumes
  100\% detection and reconstruction efficiency.}
\label{fig:numi-bg}
\end{figure}

In addition, this flux of neutrinos around the MINOS near detector
enclosure will create a large flux of low-energy neutrons from
neutrino interactions in the earth.  This background is much larger in
the MINOS area when compared to that in FINeSSE.  The results of
simulation of this effect are shown in Figure~\ref{fig:numi-bg}.  Note
that in the lowest energy bin, the background is 14 times higher at
the MINOS near detector location.  For these reasons, the $\Delta s$
measurement can not be made in the NuMI beam.

In order to explore the oscillation regions discussed in
Section~\ref{sec:oscillations}, a two detector comparison is required,
and L/E for the far detector must be on the order of 1~m/MeV.  This is
not possible in the NuMI beamline for two reasons.  First, there is no
tunnel for a detector far enough upstream of the MINOS near detector
(equivalent to the FINeSSE detector) to measure the beam before the
neutrinos have oscillated.  Second, the high average neutrino energy
of the NuMI beam means that the baseline for the far detector for an
oscillation experiment would have to be on the order of 800~m, a much
longer distance than is available in the NuMI beamline.  These two
consideration lead to the conclusion that FINeSSE must run on the
Booster neutrino beamline.

\clearpage

%
%

\chapter{The Neutrino Beam and Expected Event Rates}
\label{ch:TheNeutrinoBeam}

\thispagestyle{myheadings}
\markright{}
\section{The Booster Neutrino Beam}

The Booster neutrino beamline presently delivers beam to the MiniBooNE 
experiment;  as the FINeSSE detector will be placed upstream of MiniBooNE, 
the one beamline will provide neutrinos to both experiments. The projections for protons 
on target (POT) in this section are based on a conservative interpretation 
of the Proton committee report~\cite{psourceteamreport}.

\subsection{Beam Intensity Requirements}

During the fall 2003 shutdown, several improvements in the
Booster were made.  These upgrades are expected to provide routine peak
operation with $5 \times 10^{12}$~protons/batch and 5 Hz for the
Booster neutrino beamline.  The efficacy of these improvements will be 
understood prior to FINeSSE running, and there should also be sufficient time 
to implement additional improvements if the goals are not met by the end of 
2004.

By the summer of 2003, the Booster was routinely delivering more than
5$\times 10^{12}$ protons/batch for Stacking for Run II, demonstrating that
Booster can achieve the batch intensity required for FINeSSE.  The
issues are reducing and controlling losses at this intensity, and achieving 
the required repetition rate.  The principal improvements during the Fall
of 2003 were modifications to the doglegs to reduce losses,
installation of two large aperture RF cavities to reduce losses at
these two locations, the installation of collimators to control
losses, and modifications to the RF and magnet subsystems to allow an
increase of the equipment repetition rate to 7.5~Hz.  Once these improvements
are operational, it is expected that the above ground radiation will
be the limit on Booster operation; however, this limit is well above what is 
needed during the FINeSSE era.  In addition, in 2004, Fermilab and Columbia
University are expected to develop a robot for measuring the losses in
the Booster during beam operation, which will
help to understand these losses in detail.

Although the Booster equipment may be able to achieve 7.5~Hz, the MiniBooNE 
horn imposes a limit of 5~Hz.  If the Booster were to achieve 5$\times
10^{12}$ protons/batch at 5~Hz for an hour, the MiniBooNE target would
receive 9$\times 10^{16}$~protons per hour.  This is considered a 
nominal performance level, however, and it is not expected to persist for a
week, (much less for an entire year).

To relate a nominal performance to the number of protons delivered per
year, one can define an annual efficiency.  The analysis used here follows the
same steps given in the Proton committee report~\cite{psourceteamreport}.
The annual efficiency must include factors to account for the number of weeks 
actually scheduled for beam operation in a year; the reliability of the Proton
Source (Linac, Booster, and beam transfer lines) during those scheduled weeks; 
and the operational efficiency for actually achieving 5$\times 10^{12}$~protons/batch and 5~Hz.  
The number of weeks scheduled per year is
determined by the Director's Office and is taken to be in the range
42 to 44 weeks.  The reliability of the Proton Source has been
measured by MiniBooNE to be in the range 0.90 to 0.94.  The operational
efficiency is estimated to be 0.90~\cite{choi}. Combining these factors 
one obtains an annual efficiency of 0.66 to 0.72.

By the time FINeSSE would start to run, however, NuMI will also be 
taking beam. NuMI is expected to use five Booster batches per Main Injector 
cycle.  NuMI is expected to share the same Main Injector cycle as 
Stacking for Run II, and Stacking is expected to take two Booster batches 
per Main Injector cycle.  The Main Injector cycle time is expected to 
be about two seconds.  With these assumptions, NuMI plus Stacking will 
require seven batches every two seconds, which is an average rate of 3.5~Hz.  
At the moment, some of the Booster equipment requires two ``prepulses'' 
with no beam, or 1~Hz.  Thus, the bandwidth required by NuMI, Stacking,
and the prepulses is 4.5~Hz.  This leaves 3~Hz for delivering beam to 
the MiniBooNE target, assuming a total Booster bandwidth of 7.5~Hz.  
This is 60\% of the maximum 5~Hz which the Booster neutrino beamline 
should be receiving in 2004.  If prepulses can be eliminated, then this can 
be used to add 1 Hz to this beamline, but this proposal does not count on that.

Thus, one expects a nominal performance of the Booster neutrino beam for
FINeSSE of 5$\times 10^{12}$ protons/batch and 3~Hz.  Given the range for 
annual efficiency calculated above, one calculates the POT/year for FINeSSE as
5$\times 10^{12} \times$ 3~Hz x 3.15$\times 10^{7}$ sec/yr x (0.66 to 0.72) 
= (3.12 to 3.40) $\times 10^{20}$~POT/yr.

\vspace{0.1in}
This proposal assumes delivery of 3.0$\times 10^{20}$ POT/yr, conservatively 
below the range quoted above.

\subsection{Booster Neutrino Beam Production}
\label{section:flux}

\subsubsection{The Neutrino Flux}
\label{theneutrinoflux}
\begin{figure}[thb]
\centering
\includegraphics[bbllx=19bp,bblly=131bp,bburx=562bp,bbury=698bp,angle=90,width=3.8in]{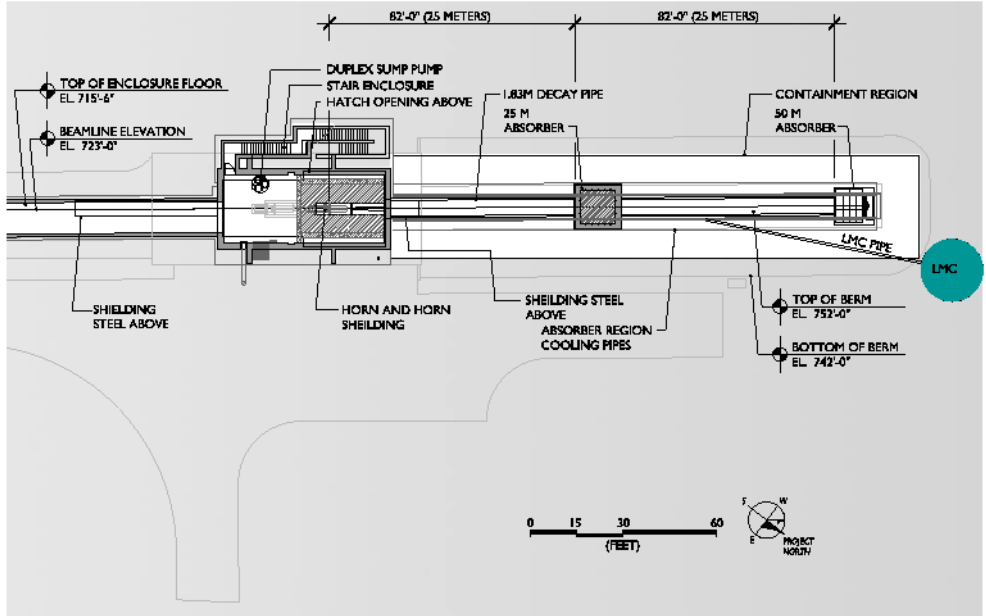}
\caption{\em Schematic of the MiniBooNE target hall and decay region. 
         The 25\,m and 50\,m absorber locations are indicated.}
\label{fig:beam-absorber}
\end{figure}

The neutrino beam is produced by the 8~GeV Fermilab Booster which
currently feeds the MiniBooNE experiment. Protons from the Booster
strike a 71\,cm beryllium target inserted in a magnetic focusing horn.
Protons arrive at this target in 1.6~$\mu$s long Booster spills.  The
timing structure within each spill delivers 84 2~ns wide bunches of
beam, each separated by 18~ns.  Secondary short-lived hadrons (primarily
pions) produced in the target are focussed by the horn and enter a
decay region. In normal MiniBooNE operation, this decay region is
50\,m long, at the end of which region is a beam absorber to stop
hadrons and low energy muons.  Located 25\,m from the proton target is
an intermediate absorber which can be lowered into the beam for use as
a systematic check on the MiniBooNE $\nu_e$ background from $\mu$
decays. It is assumed that the 25\,m absorber will be in place during
the period when FINeSSE is operational to accommodate FINeSSE and
MiniBooNE physics goals.  Figure~\ref{fig:beam-absorber} provides a
diagram of the two possible absorber positions.

\begin{figure}[thb]
\centering
\includegraphics[bbllx=15bp,bblly=135bp,bburx=549bp,bbury=680bp,width=3.5in]{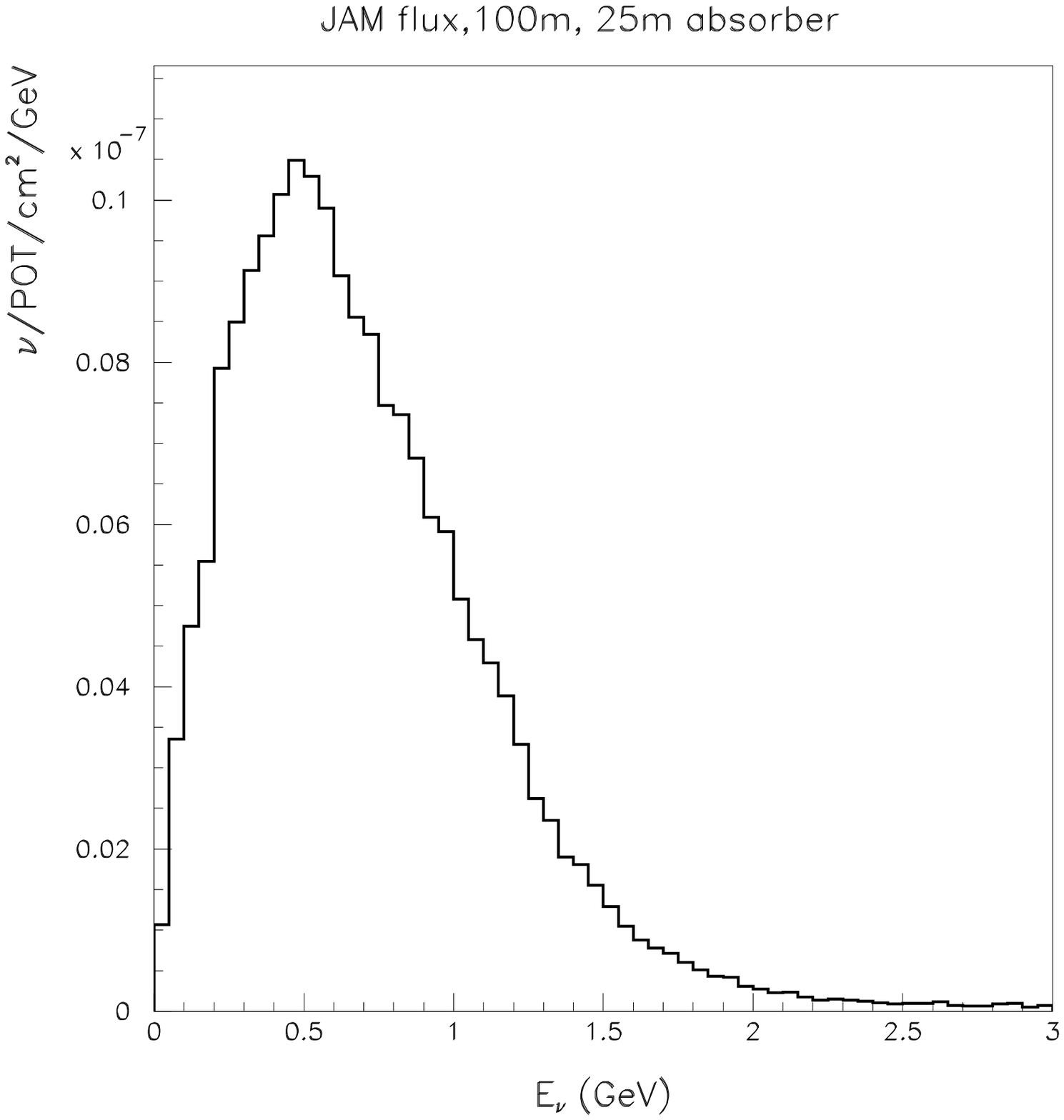}
\vspace{-0.3in}
\caption{\em Expected $\numu$ flux at a 100\,m detector site assuming 
         a 25\,m decay length.}
\label{fig:flux-100m-25mabs}
\end{figure}
%
%
The neutrino flux resulting from this design was simulated using the
same tools currently being employed by the MiniBooNE 
collaboration~\cite{mb-runplan}. The beam simulation utilizes 
GEANT 4 transport code~\cite{geant4}, and the MiniBooNE JAM pion production 
model~\cite{jam-flux} which includes all beamline elements (horn, shielding, 
absorbers, etc.) and $\pi^{\pm}$, $K^{\pm}$, $K^0$ production from proton 
interactions on beryllium. To better reproduce the energy distribution of 
neutrino events observed in the MiniBooNE detector, pion spectra were input 
from a Sanford-Wang-based global fit~\cite{jam-flux} to pion production data 
in the relevant energy range in a procedure similar to that adopted by K2K. 
Figure~\ref{fig:flux-100m-25mabs} shows the resultant muon neutrino flux 
expected from a 25\,m decay length beam produced at the 100\,m FINeSSE 
detector site. In this configuration, $9.55 \times 10^{-9}$ muon neutrinos 
per POT per cm$^2$ are anticipated with a mean energy of $\sim700$ MeV. The 
neutrino flux is roughly 20 times larger than that expected in a comparable
volume at MiniBooNE.  Note that the flux is diminished by about a factor of 1.8 in 
switching from a 50\,m decay length to a shorter 25\,m decay length. However, 
as will be demonstrated, the 25\,m absorber location is ideal for 
optimizing FINeSSE's oscillation sensitivity.


\begin{figure}[h]
\centering
\includegraphics[bbllx=19bp,bblly=156bp,bburx=562bp,bbury=698bp,width=3.5in]{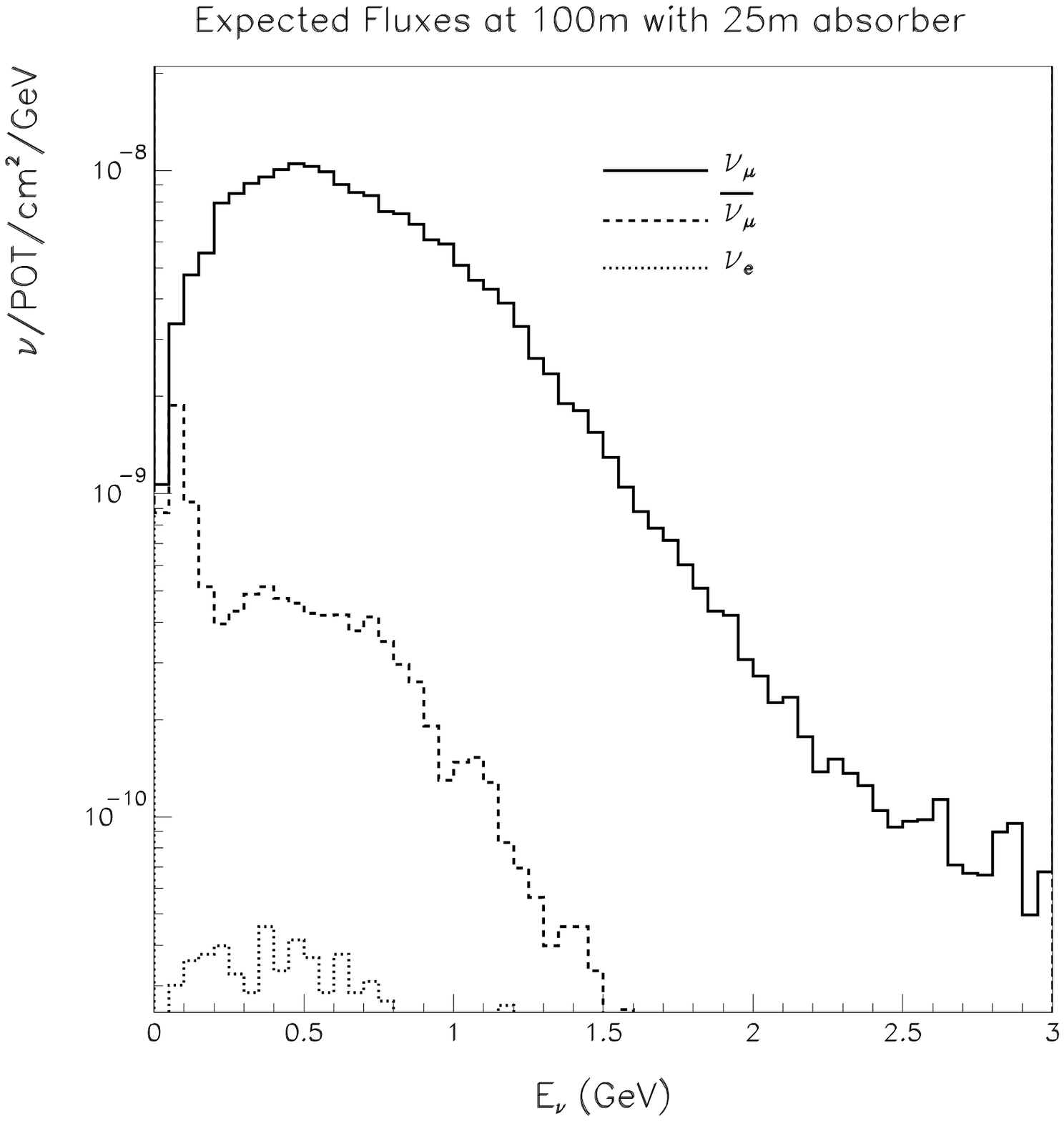}
\vspace{-0.2in}
\caption{\em Predicted flux contributions at a 100\,m detector 
         site assuming a 25\,m decay length. Muon neutrinos comprise 
         $93.5\%$ of the expected neutrino flux.}
\label{fig:flux-contributions}
\end{figure}


Figure~\ref{fig:flux-contributions} shows the individual contributions to
the total neutrino flux expected at FINeSSE. Contaminations from 
$\overline{\nu_{\mu}}$'s and $\nu_e$'s are predicted to be $6\%$ and $0.5\%$ 
of the total $\nu_{\mu}$ flux, respectively. Once the 
``wrong--sign'' $\overline{\nu_{\mu}}$ background events are weighted by their
appropriate cross section, they will comprise less than $1.5\%$ of the total 
events in the FINeSSE detector.

Better knowledge of the incoming neutrino beam flux enables more precise cross
section measurements at both MiniBooNE and FINeSSE. 
The Booster neutrino flux will be much more precisely known than the
fluxes reported in previous low energy neutrino cross section
measurements well in advance of FINeSSE's commissioning.  This
improved knowledge comes from two sources: data from the Brookhaven
E910 experiment~\cite{e910} and from the CERN HARP
experiment~\cite{harp}.  Analysis that is already underway of E910
proton-beryllium data taken at 6, 12, and 18 GeV beam energies will be
instrumental in verifying the extrapolation of the Sanford-Wang
parametrization~\cite{jam-flux} to the 8~GeV Booster beam energy. More
importantly, HARP data taken at 8~GeV on the Booster neutrino
production target slugs will provide a tighter constraint on the flux.
The high statistics HARP data will provide a statistical precision of
$\sim2\%$~\cite{harp-unc} on $\pi^+$ production, which is the main
source of muon neutrinos at both the FINeSSE and MiniBooNE detectors.
Therefore, with these additional inputs, the overall muon neutrino
flux at FINeSSE should be known to roughly $5\%$~\cite{mb-runplan}.

\section{Event Rates}
\label{section:event-rates}

The number of neutrino events expected in the FINeSSE Vertex Detector
is calculated using the NUANCE Monte Carlo~\cite{nuance} to generate
neutrino interactions on $CH_2$. NUANCE is open-source code
originally developed for simulating atmospheric neutrino interactions
in the IMB detector. NUANCE has since been further developed and is
now used by the K2K, Super-K, SNO, MiniBooNE, and MINERvA
collaborations. The neutrino interaction cross sections in NUANCE have
been extensively checked against published neutrino data and other
available Monte Carlo event generators. In addition, the full NUANCE
simulation has been recently shown to provide a good description of
events in both the MiniBooNE detector and K2K near detector ensemble.

For this specific use, NUANCE was modified to include the FINeSSE detector 
composition and geometry, as well as the incident neutrino flux at the 100\,m
detector site (Section~\ref{section:flux}). Using the input neutrino flux 
distribution, NUANCE predicts event rates, kinematics, and final state 
particle topologies that can subsequently feed hit-level GEANT detector 
simulations, or, as in this case, simply estimate the type and number of 
neutrino interactions expected at FINeSSE.

Table~\ref{table:100m-25mabs} lists the NUANCE-predicted event populations 
at the 100\,m FINeSSE detector site with the 25\,m absorber in position.
The table provides the expected $\numu$ rates per ton detector for 
$1\times 10^{20}$ POT as well as the expected backgrounds from the $\numubar$ 
and $\nue$ content in the beam. In all cases, the event rates have been 
normalized to the number of contained neutrino events observed in the 
MiniBooNE detector~\cite{mb-runplan}. Roughly $1.3\%$ ($0.6\%$) of the 
total neutrino events result from $\numubar$ ($\nue$) interactions in the 
detector. The dominant contributions to the total event rate result from 
quasi-elastic and resonant processes: $41\%$ of the $\numu$ events are CC 
quasi-elastic ($\numu \, n \rightarrow \mu^- \, p$), $17\%$ are NC elastic 
($\numu \, N \rightarrow \numu \, N$; $N=n,p$), and $35\%$ resonant single 
pion production ($\numu \, N \rightarrow \mu^- \, (\numu) \, N \, \pi$) 
channels.

\vspace{0.1in}
\begin{table}[h]
\centering
   \begin{tabular}{|l|c|c|c||c|} \hline
               & $\numu$ & $\numubar$ & $\nue+\nuebar$ & $\numu$  \\
$\nu$ Reaction & $10^{20}$ POT  & $10^{20}$ POT  &  $10^{20}$ POT & $6\times10^{20}$ POT  \\
               &  1 ton         &  1 ton         & 1 ton          & 9 ton
\\ \hline \hline
CC QE, $\numu n \rightarrow \mu^- p$                 & 2,715 & 43 & 13 & 146,610 \\ \hline
NC EL, $\numu N \rightarrow \numu N$                 & 1,096 & 18 & 5  & 59,184  \\ \hline
CC $\pi^+$, $\numu p \rightarrow \mu^- p \pi^+$      & 1,235 & 6  & 8  & 66,690  \\ \hline
CC $\pi^0$, $\numu n \rightarrow \mu^- p \pi^0$      &  258  & 3  & 2  & 13,932  \\ \hline
CC $\pi^+$, $\numu n \rightarrow \mu^- n \pi^+$      &  216  & 2  & 2  & 11,664  \\ \hline
NC $\pi^0$, $\numu p \rightarrow \numu p \pi^0$      &  211  & 3  & 2  & 11,394  \\ \hline
NC $\pi^+$, $\numu p \rightarrow \numu n \pi^+$      &  125  & 2  & 0  & 6,750   \\ \hline
NC $\pi^0$, $\numu n \rightarrow \numu n \pi^0$      &  158  & 3  & 2  & 8,532   \\ \hline
NC $\pi^-$, $\numu n \rightarrow \numu p \pi^-$      &   98  & 3  & 0  & 5,292   \\ \hline
CC DIS, $\numu N \rightarrow \mu^- X$                &   80  & 0  & 3  & 4,320   \\ \hline
NC DIS, $\numu N \rightarrow \numu X$                &   37  & 0  & 2  & 1,998   \\ \hline
CC coh $\pi^+$, $\numu A \rightarrow \mu^- A \pi^+$  &  160  & 5  & 2  & 8,640   \\ \hline
NC coh $\pi^0$, $\numu A \rightarrow \numu A \pi^0$  &   98  & 3  & 0  & 5,292   \\ \hline
other                                                &  117  & 2  & 0  & 6,318   \\ \hline 
\hline
total                                                & 6,604 & 93 & 41 & 356,616 \\
\hline
   \end{tabular}
   \caption{\em Number of events expected at 100\,m with a 25\,m 
            decay length for $1 \times 10^{20}$ POT per ton detector 
            and for the full requested FINeSSE running and detector 
            (rightmost column). These predictions do not include final 
            state effects in $^{12}C$ and assume $100\%$ detection efficiency.}
   \label{table:100m-25mabs}
\end{table} 

A total of approximately 360,000 neutrino interactions can be expected at
FINeSSE for the full request of $6\times10^{20}$ POT. This raw estimate
assumes a 9~ton fiducial detector and $100\%$ detection/reconstruction
efficiency.

\subsubsection{Effect of Final State Interactions}

Because a large fraction of neutrino interactions at FINeSSE take place on 
carbon, the number of expected events will depend not only on the predicted 
neutrino cross sections and flux, but also on the final state interactions 
engendered by the local nuclear environment. Particles produced
via neutrino interactions in carbon nuclei will have a chance to reinteract
before exiting the nucleus, and thus can vanish or change identity before 
being detected. Although the initial reaction might be a simple CC 
quasi-elastic interaction ($\numu \, n \rightarrow \mu^- \, p$), the observed 
final state particles might include pions, multiple nucleons, low energy 
photons, or all of these combined. Examples of the types of nuclear 
rescattering that can distort the final state observed at FINeSSE include 
simple absorption, charge exchange ($\pi^+ \, n \rightarrow \pi^0 \, p$, 
$\pi^0 \, p \rightarrow \pi^+ \, n$, $\pi^0 \, n \rightarrow \pi^- \, p$, 
$\pi^- \, p \rightarrow \pi^0 \, n$), and both inelastic and elastic 
scattering. For example, consider the resonant interaction
$\numu \, n \rightarrow \mu^- \, p \, \pi^0$. If the $\pi^0$ is absorbed
before exiting the carbon nucleus, the interaction will appear to be  
quasi-elastic $\numu \, n \rightarrow \mu^- \, p$. Hence, the presence
of such final state interactions demands accounting in our observed event 
rate calculations.

Table~\ref{table:fsi} summarizes the final states expected at FINeSSE
after using NUANCE to simulate re-interactions within carbon nuclei.
The table defines ``QE--like'', ``NC-EL-like'', and ``NC-$\pi^0$-like'' 
event categories, where these classes refer to final states that appear
to be QE, NC elastic, or NC $\pi^0$ events, respectively. Specifically,

\vspace{-0.1in}
\begin{itemize}
   \item CC QE-like: a CC event with a muon and any number of nucleons
                   in the event \\ (no $\pi$ or $K$ in the final state)
   \item NC EL-like: a NC event with any number of nucleons
                     \\ (no $\mu$, $\pi$, or $K$ in the final state)
   \item NC $\pi^0$-like: a NC event with any number of nucleons and a
                          single $\pi^0$ \\ (no other $\pi$'s,
                          $K$, or $\mu$ in the final state) 
\end{itemize}
\vspace{-0.1in}

\noindent
Comparison of Tables~\ref{table:100m-25mabs} and \ref{table:fsi} reveal that
the number of observed QE and NC elastic interactions increases by roughly
$10-15\%$ as a result of final state reinteractions. This is largely a result 
of resonant processes where the final state pion is absorbed. By the same 
mechanism, the overall number of observed NC $\pi^0$ events decreases
by roughly $30\%$ because the final state $\pi^0$ is either absorbed 
or ``charge exchanges''.
%
%
The contributions are further differentiated by the number of
neutrons and protons produced. More than half of the events yield only a 
single nucleon in the final state. The non-negligible rate of NC $\pi^0$ 
events with no final state nucleons results mainly from coherent pion 
production processes where the nucleus remains intact.

\begin{table}[h]
\centering
   \begin{tabular}{|l|c|c|} \hline
final state                & \# $\numu$ events & fraction of total ($\%$) 
\\ \hline
\hline
CC QE-like: $\mu^-$, $1\,p$  &  2136           & $69.5\%$     \\ \hline
CC QE-like: $\mu^-$, $>1\,p$ &   937           & $30.5\%$     \\ \hline
\hline
CC QE-like: total            &  3073           &      \\ \hline
\hline\hline
NC EL-like: $\numu$, $0\,p, 1\,n$   &  361      & $28.4\%$     \\ \hline
NC EL-like: $\numu$, $0\,n, 1\,p$   &  400      & $31.4\%$     \\ \hline
NC EL-like: $\numu$, $1\,p, 1\,n$   &  131      & $10.3\%$     \\ \hline
NC EL-like: $\numu$, $>1\,p, >1\,n$ &   96      & $7.5\%$     \\ \hline
NC EL-like: $\numu$, $1\,n, >1\,p$  &   78      & $6.1\%$     \\ \hline
NC EL-like: $\numu$, $0\,p, >1\,n$  &   71      & $5.6\%$     \\ \hline
NC EL-like: $\numu$, $0\,n, >1\,p$  &   68      & $5.3\%$     \\ \hline
NC EL-like: $\numu$, $1\,p, >1\,n$  &   67      & $5.3\%$     \\ \hline
\hline
NC EL-like: total                   &  1272      &      \\ \hline
\hline\hline
NC $\pi^0$-like: $\numu, 1\pi^0, 1\,p, 0\,n$    & 108  & $35.4\%$    \\ \hline
NC $\pi^0$-like: $\numu, 1\pi^0, 0\,p, 1\,n$    &  43  & $14.1\%$    \\ \hline
NC $\pi^0$-like: $\numu, 1\pi^0, 1\,p, 1\,n$    &  35  & $11.5\%$    \\ \hline
NC $\pi^0$-like: $\numu, 1\pi^0, 0\,p, 0\,n$    &  27  & $8.9\%$     \\ \hline
NC $\pi^0$-like: $\numu, 1\pi^0, >1\,p, >1\,n$  &  22  & $7.3\%$     \\ \hline
NC $\pi^0$-like: $\numu, 1\pi^0, 1\,p, >1\,n$   &  20  & $6.2\%$     \\ \hline
NC $\pi^0$-like: $\numu, 1\pi^0, >1\,p, 1\,n$   &  19  & $6.2\%$     \\ \hline
NC $\pi^0$-like: $\numu, 1\pi^0, >1\,p, 0\,n$   &  17  & $5.2\%$     \\ \hline
NC $\pi^0$-like: $\numu, 1\pi^0, 0\,p, >1\,n$   &  16  & $5.2\%$     \\ \hline

\hline
NC $\pi^0$-like: total                          &  307     &      \\ \hline
\hline
   \end{tabular}
   \caption{\em Number of events for $1 \times 10^{20}$ POT per ton 
            detector after including the effects of final state interactions 
            in $^{12}C$. The event classes are further broken 
            down to indicate the number of nucleons present in the final state
            (either 0, 1, or $>1$ proton or neutron).}
   \label{table:fsi}
\end{table} 
\clearpage

With these definitions, Table~\ref{table:contributions} lists the 
dominant contributions to each final state. Of the observed QE-like events, 
$86.7\%$ are true QE interactions. Of the events appearing to be NC elastic 
in the detector, $85.8\%$ are true NC elastic interactions. Of the events 
appearing to be NC $\pi^0$, $73.5\% + 19.0\% = 92.5\%$ are true NC $\pi^0$ 
resonant or coherent interactions, respectively. Therefore, under the 
assumption of $100\%$ reconstruction and detection efficiencies, the level of 
irreducible backgrounds from final state effects appears to be less than 
$15\%$. However, reconstruction and selection criteria may potentially amplify
or reduce the effect of such backgrounds.

\vspace{0.1in}
\begin{table}[h]
\centering
   \begin{tabular}{|l|r|c|} \hline
final state   & contribution  & fraction ($\%$) 
\\ \hline
\hline
CC QE-like &  QE ($\numu \, n \rightarrow \mu^- \, p$)  & $86.7\%$   \\ \hline
CC QE-like &  CC $\pi^+$ RES ($\numu \, N \rightarrow \mu^- N \pi^+$) 
   & $9.2\%$ \\ \hline
CC QE-like &  CC $\pi^0$ RES ($\numu \, n \rightarrow \mu^- p \pi^0$) 
  & $2.3\%$  \\ \hline
CC QE-like &  CC $\pi^+$ COH ($\numu A \rightarrow \mu^- \, A \, \pi^+$) 
  & $1.3\%$  \\ \hline
CC QE-like &  CC $\eta$ ($\numu \, n \rightarrow \mu^- \, p \, \eta$) 
  & $0.3\%$  \\ \hline
CC QE-like &  CC DIS ($\numu \, N \rightarrow \mu^- \, X$) 
  & $0.2\%$  \\ \hline
\hline\hline
NC EL-like &  NC EL ($\numu \, N \rightarrow \numu \, N$)  
  & $85.8\%$   \\ \hline
NC EL-like &  NC $\pi^0$ RES ($\numu \, N \rightarrow \numu \, N \, \pi^0$)  
  & $7.6\%$   \\ \hline
NC EL-like &  NC $\pi^-$ RES ($\numu \, n \rightarrow \numu \, p \, \pi^-$)  
  & $2.3\%$   \\ \hline
NC EL-like &  NC $\pi^+$ RES ($\numu \, p \rightarrow \numu \, n \, \pi^+$)  
  & $2.1\%$   \\ \hline
NC EL-like &  NC $\pi^0$ COH ($\numu \, A \rightarrow \numu \, A \, \pi^0$)  
  & $1.9\%$   \\ \hline
NC EL-like &  NC $\eta$ ($\numu \, n \rightarrow \numu \, n \, \eta$)  
  & $0.2\%$   \\ \hline
NC EL-like &  NC DIS ($\numu \, N \rightarrow \numu \, X$) 
  & $0.1\%$  \\ \hline

\hline\hline
NC $\pi^0$-like & NC $\pi^0$ RES ($\numu \, N \rightarrow \numu \,N \,\pi^0$)
  & $73.5\%$   \\ \hline
NC $\pi^0$-like & NC $\pi^0$ COH ($\numu \, A \rightarrow \numu \,A \,\pi^0$)
  & $19.0\%$   \\ \hline
NC $\pi^0$-like & NC DIS ($\numu \, N \rightarrow \numu \, X$)
  & $2.4\%$   \\ \hline
NC $\pi^0$-like & NC $\pi^-$ RES ($\numu \, n \rightarrow \numu \,p \, \pi^-$)
  & $1.8\%$   \\ \hline
NC $\pi^0$-like & NC $\pi^+$ RES ($\numu \, p \rightarrow \numu \,n \, \pi^+$)
  & $1.5\%$   \\ \hline
NC $\pi^0$-like & NC EL ($\numu \, N \rightarrow \numu \, N$)
  & $1.3\%$   \\ \hline
NC $\pi^0$-like & NC $\eta$ ($\numu \, N \rightarrow \numu \, N \, \eta$)
  & $0.3\%$   \\ \hline
NC $\pi^0$-like & NC multi-$\pi$ ($\numu \, N \rightarrow \numu \, N \, \pi^0\, \pi^0$)
  & $0.2\%$   \\ \hline
\hline
   \end{tabular}
   \caption{\em Fractional contributions to each observed final state at a
           FINeSSE $100\,m$ detector site. ``RES'' (``COH'') refers to 
           resonant (coherent) pion production processes; ``DIS'' refers
           to deep inelastic scattering.} 
   \label{table:contributions}
\end{table} 

Just as non-QE events can appear to be quasi-elastic in the detector 
(via pion absorption), the reverse can also occur, albeit at a much smaller 
rate. NUANCE predicts that less than $1\%$ of QE (or NC elastic) events 
will fail to appear quasi-elastic. This results from the small probability 
that a proton will rescatter in the carbon nucleus and produce one or more 
pions, for example:

\vspace{-0.3in}
\begin{eqnarray*}
           p + p &\rightarrow& p + n + \pi^+ \\
           p + p &\rightarrow& n + p + \pi^+ \\
           p + p &\rightarrow& p + p + \pi^0 \\
           p + p &\rightarrow& p + p + \pi^0 + \pi^0 \\
           p + p &\rightarrow& p + p + \pi^+ + \pi^- \\
           p + p &\rightarrow& n + n + \pi^+ + \pi^+ \\
           p + p &\rightarrow& p + n + \pi^+ + \pi^0,
           \hspace{0.1in} \mathrm{etc.}
\end{eqnarray*}

\noindent
The situation differs for NC $\pi^0$ events. In this case, roughly $30\%$
of true NC $\pi^0$ interactions will not appear to be NC $\pi^0$ events in the
detector: $\sim20\%$ of the true NC $\pi^0$ interactions have no final state 
$\pi^0$ (that pion is absorbed before exiting the nucleus), and $\sim10\%$ 
instead contain a final state $\pi^+$ or $\pi^-$ due to charge exchange 
processes,

\vspace{-0.3in}
\begin{eqnarray*}
    \pi^0\,p \rightarrow \pi^+\,n  \\
    \pi^0\,n \rightarrow \pi^-\,p.
\end{eqnarray*} 

\noindent
Therefore, because a large number of the neutrino scatters occur on carbon,
it is important that the FINeSSE Monte Carlo simulation include these 
secondary final state interactions. Such a model is provided by the NUANCE 
generator, and is used in all event simulations provided in this document.


\chapter{The FINeSSE Detector}
\label{ch:TheFINeSSEDetector}

\thispagestyle{myheadings}

\markright{}

The FINeSSE detector is a 13~ton (9~ton fiducial) active target,
consisting of a tracking scintillator detector followed by a muon
rangeout stack comprised of scintillator planes interspersed with
iron absorber.  The physics goals of this experiment require the
ability to tag both $\nu_{\mu} p \rightarrow \nu_{\mu} p$ and
$\nu_{\mu} n \rightarrow \mu^- p$ interactions by looking for the
final state protons and muons produced in these channels.  Proton
energy and angle are measured in the first part of the detector,
called the Vertex Detector.  Muon tracks are tagged in both the Vertex
Detector and the downstream Muon Rangestack.  The Vertex Detector is
also ideal for cross section measurements (such as single $\pi^0$
production), which require good final state particle separation and
good energy resolution.  FINeSSE is designed to meet these
requirements with a novel, yet relatively simple, detector.

\section{Detector Design and Construction}
\begin{figure}
\centering
\includegraphics[bb=73 197 539 596,width=\textwidth]{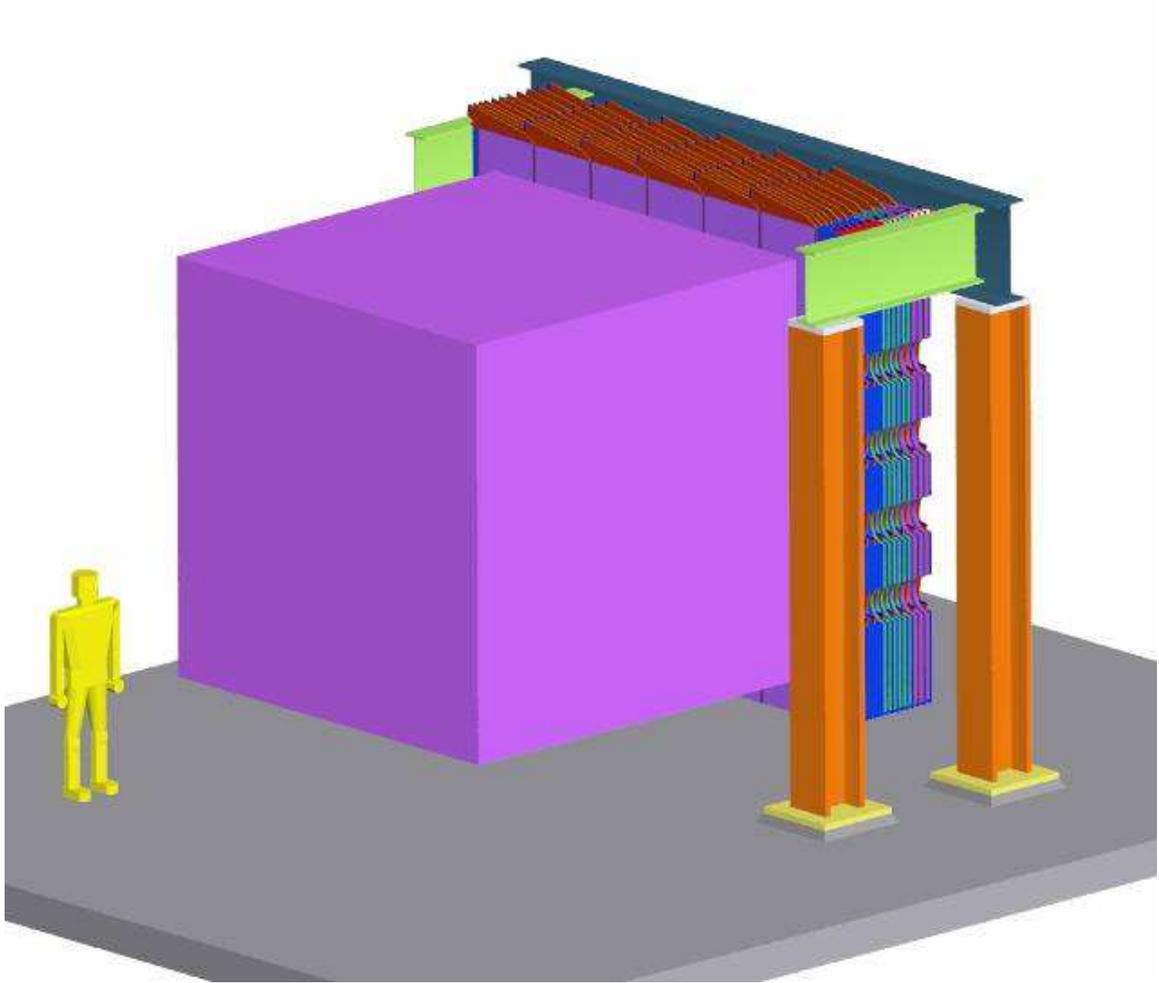}
\caption{\it A schematic drawing of the FINeSSE detector.  The cubic volume is
  the (3.5m)$^3$ Vertex Detector.  It is a (2.4m)$^3$ signal region
  surrounded by a veto, filled with scintillator oil.  The larger
  volume downstream is the Muon Rangestack.}
\label{fig:det_overview}
\end{figure}

The layout of the FINeSSE detector can be seen in
Figure~\ref{fig:det_overview}.  The upstream Vertex Detector contains a
wavelength-shifting (WLS) fiber array situated in a large, open volume
of liquid scintillator.  The downstream Muon Rangestack is comprised
of $4.1$cm $\times$ $1$cm scintillator strips organized into planes in
alternating X and Y orientations, interspersed with iron absorber.  The
Vertex Detector is described below in
Section~\ref{sec:tracking_detector}; this is followed by a description
of the Muon Rangestack in Section~\ref{sec:muonstack}.

\subsection{The Vertex Detector}
\label{sec:tracking_detector}

\begin{figure}
\centering
\includegraphics[width=5.in]{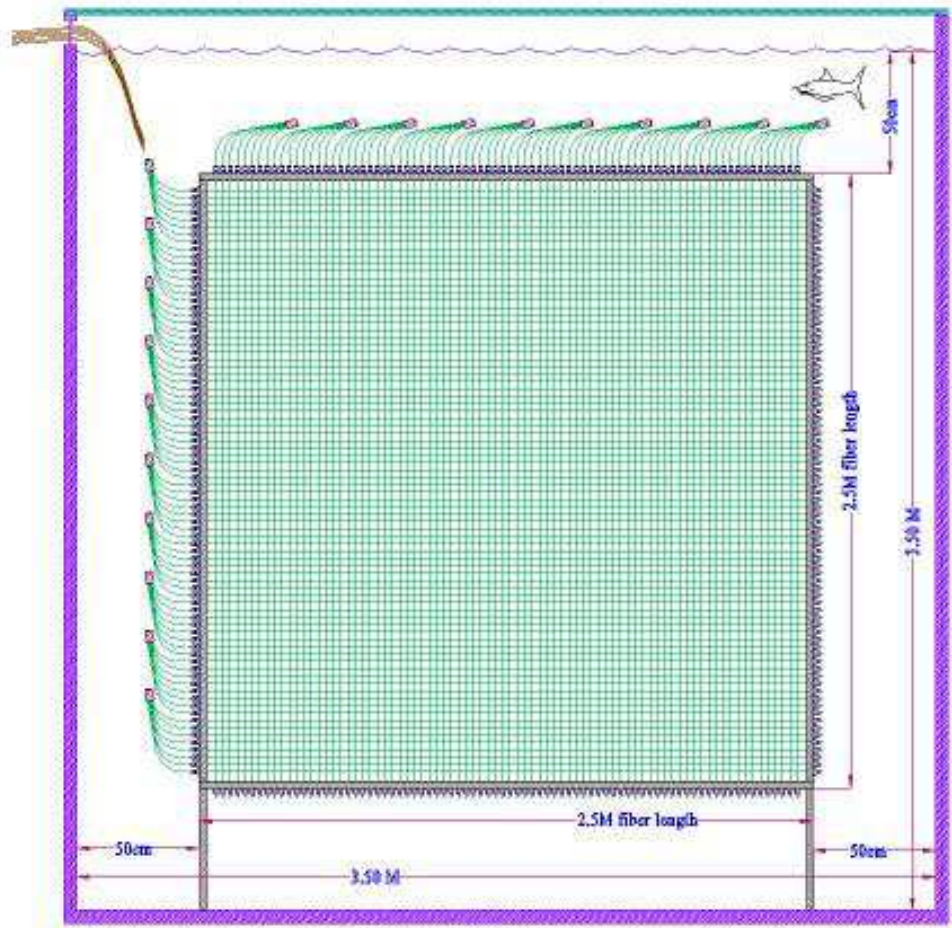}
\caption{\it A schematic drawing of the Vertex Detector shown from 
the side.}
\label{fig:sbathschem}
\end{figure}

The FINeSSE Vertex Detector consists of a cube-shaped volume of liquid
scintillator with dimensions $2.4 \times 2.4 \times 2.4$ m$^3$. Light
generated by ionizing particles traversing the scintillator is picked
up by 1.5~mm diameter, WLS fibers, submerged
throughout the sensitive volume. The fibers are mounted on a support
frame, and are connected on one end to multi-anode photomultipliers,
mounted to the outside of that frame.  The fiber frame,
photomultipliers, and associated electronics form a unit; this unit is
immersed in the liquid scintillator, which is contained in a cubic
tank, 3.5~m on a side. The volume between the fiber structure and the
tank wall is used to monitor charged particles entering and exiting
from the tracking volume ("veto shield"). The photomultiplier signals
are processed {\it in situ} and transmitted by Ethernet to the outside
of the tank, thus minimizing the number of cables that penetrate the
tank wall.  A schematic drawing of the tracking detector is shown in
Fig.~\ref{fig:sbathschem}.  Cables penetrate the tank wall above the
oil level to prevent leaks.

Particle tracks can be reconstructed because the time of arrival of
the light at the end of a fiber from a given source inside the
detector is a known, continuous function of the distance between the
source and the fiber.  The detector will be calibrated using cosmic
rays.

\begin{figure}
\centering
\includegraphics[width=4.in]{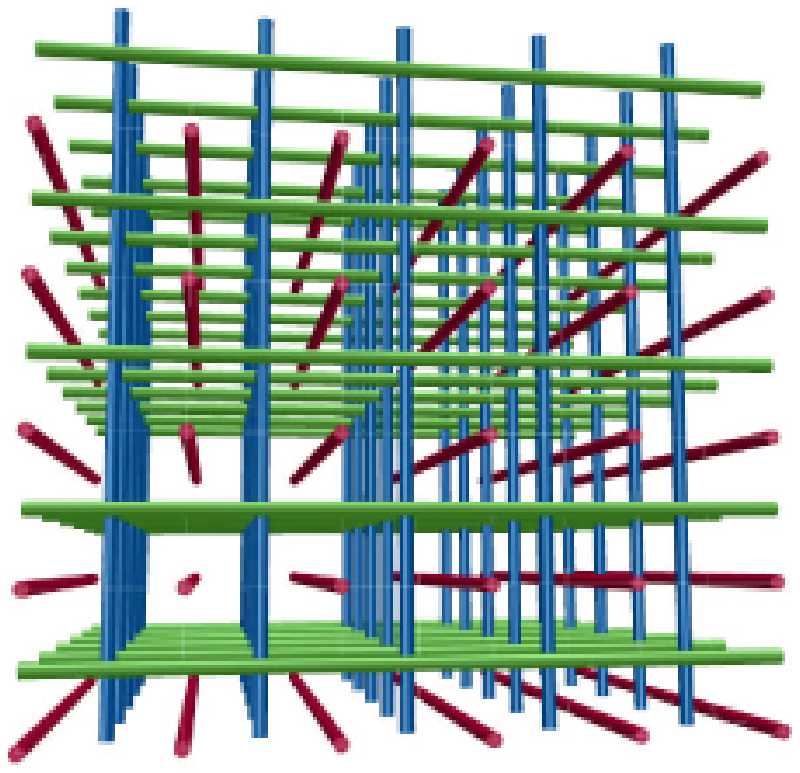}
\caption{\it The geometrical arrangement of WLS fibers inside the
Vertex Detector. The arrangement consists of three orthogonal sets of parallel
fibers. The geometry is symmetric with respect to a rotation by 90$^\circ$ about
any of the three major axes.}
\label{fig:fiberarr}
\end{figure}

The arrangement of the WLS fibers is shown schematically in
Fig.~\ref{fig:fiberarr}. There are three sets of fibers, running
parallel to the axes of a Cartesian coordinate system. Except for a
rotation in space and an offset, the three fiber sets are identical,
consisting of fibers that intercept the wall at the vertices of a
quadrate grid. The distance between grid points is 30 mm. Thus, the
closest distance between any two fibers in the full assembly is 15~mm.
The resulting arrangement is invariant with respect to a rotation by
$90^\circ$ about any major axis. For the given dimensions, there are
a total of $80 \times 80 \times 3 = 19200$ fibers.

\subsubsection{Tracking Scheme}
Consider a point source of light at some distance $d$ from a long, WLS
fiber. A fixed fraction of the light that intercepts the fiber is
wavelength-shifted and transported to the photo detector.  Ideally,
the exiting light is proportional to the diameter of the fiber and
inversely proportional to $d$. In reality, this distance dependence is
faster than $1/d$, because of light attenuation in the scintillator,
and because the fraction of light captured by the fiber depends on the
angle between the fiber and the incident ray.

The light collection efficiency versus distance to the fiber can be
determined by Monte Carlo and test measurements.  This dependence will be
the same throughout the detector volume, with the possible exception
of regions close to the wall, where reflected light may contribute.

The light from the source travels to all nearby fibers through a
completely homogeneous medium. Since the sharing of light by nearby
fibers can be used to localize the source, the position and angle of
tracks can then be reconstructed.  Fibers along a given direction are
only sensitive to the projection of the track onto a plane
perpendicular to that direction. Even if a second, orthogonal set of
fibers is provided, it is still possible that a track may be parallel
to one of two directions. This difficulty is avoided in the FINeSSE
detector by having three orthogonal sets of fibers.

\subsubsection{Light Generation and Transport}
\label{sec:Lightgeneration}
A comparative study of different scintillator fluids used in
conjunction with WLS fibers is given in Ref.~\cite{Bor01}. In general,
an ionizing particle excites ultraviolet fluorescence ($\sim$ 350 nm)
in the scintillator. This light is normally shifted to a longer
wavelength ($\sim$ 430 nm), to avoid problems with absorption in the
scintillator. The shifted light propagates isotropically.  The light
that intercepts a WLS fiber is shifted once more, to typically $\sim$
500 nm to inject light into the acceptance cone of the fiber and to
prevent re-absorption in the fiber. The WLS fiber consists of a
polystyrene core (n=1.60), surrounded by cladding of a lesser index of
refraction. The trapping efficiency of the fiber is significantly
enhanced (to typically 6\%) when two claddings are used. For a 1.5~mm
diameter fiber, the first cladding is 45~$\mu$m thick acrylic (n=1.49),
and the second is a layer of 15~$\mu$m thick fluor-acrylic (n=1.42).
The data given here are for BCF-91A-MC fibers from 
Saint-Gobain~\cite{StGob}. Similar fibers are available from
Kuraray~\cite{Kurar}. The second cladding also provides protection
against possible chemical interaction between the liquid scintillator
and WLS fibers. Long-term tests of fibers in mineral-oil-based
scintillation fluid, carried out in the context of MINOS
R\&D~\cite{Minos, Bor01}, showed no discernible ill effects.
Specifically, the BCF91A fibers used in these studies were not
affected after having been suspended for six months, at temperatures up to
50$^{\circ}$ C, in mineral-oil-based scintillator, BC517L, or in the high
fluor concentration BC517H.  Furthermore, {\it single} clad fibers
suspended in B517L for more than two years also showed no aging or
deterioration~\cite{Bor01}.

The attenuation length of the light propagating in the fiber is given
as 3.5~m by the manufacturer, but a more complicated behavior is
reported in the literature~\cite{Bor01}. The attenuation length is the
same whether the fiber is in liquid scintillator or in
air~\cite{Dou00}. Because of attenuation in the fiber, the light
collected at one end depends on the point of origin along the fiber.
This effect can be reduced by applying an aluminum reflective coating
at the other end~\cite{Chaud}, or even by just painting it white with
TiO$_2$~\cite{Bor01}.  These coated fiber ends will be covered with a
Teflon sleeve and will not be in contact with the mineral oil.  Even
in the case that they would, these coatings are inert and unlikely to
interact with the scintillator oil.

We are currently investigating a number of options concerning the
liquid scintillator. It is true that the present tracking scheme makes
use of the sharing of light from a given source by a number of fibers.
However, it is also true that only fibers in the vicinity of the track
contribute significantly to the determination of the track parameters.
As a result, low-attenuation length oil may be the best choice, which
is an unusual but easily accommodated need.  We are continuing R\&D on
the best choice of scintillator oil.

\subsubsection{Advantages of the Proposed Design}
\label{sec:VDadvantages}
Our design represents a novel approach to the task of tracking
ionizing particles in a large-volume detector. It exploits the fact
that the response of a fiber versus the distance to the light source
inside the detector volume is a universal function, and the three-fold
symmetry makes the tracking sensitivity nearly isotropic. These
properties are particularly important for the physics goals of the
present proposal.  To our knowledge, there is no other scheme that
offers these features.

We investigated an alternative detector design which consists of
planes of bars of solid scintillator, oriented normal to the beam
direction ($z$). Bars in even layers would run in the $x$ direction,
those in the odd layers in the $y$ direction. Scintillating plastic
bars are extruded polystyrene with $2$cm$ \times 1$cm cross section.
Each bar is co-extruded with a TiO$_2$ outer layer for reflectivity
and a hole down the center for a WLS fiber.  This design is very
similar to the K2K scibar detector recently
commissioned~\cite{k2kscibar} and to that employed by the MINOS
experiment~\cite{MINOspdf}.

Such a design for the Vertex Detector has a number of disadvantages
for the FINeSSE physics goals. First, the tracking information would
have to come from light sharing between several intercepted bars. The
track resolution would therefore be limited by the bar dimension.
Second, the amplitude information from a given bar reflects the
overlap of the track with the bar cross section.  Interpretation of
this information is complicated and suffers from irregularities in
stacking, caused by uneven extrusions and variations in the reflecting
layer and in wrapping.  Third, tracks that do not have a sufficiently
large angle with respect to the stacking planes are lost.  These
disadvantages make reconstruction of low energy proton tracks and
therefore low energy $\nu-p$ elastic events particularly difficult.
Details of our studies of this design can be found in
Appendix~\ref{ch:AppendixA}.

\subsubsection{Prototype Setup and Test Measurements}
To demonstrate the viability of the proposed tracking scheme in the
Vertex Detector, we have constructed and tested a prototype.
The following contains a description of the prototype including construction
issues and the results of beam tests.

\begin{figure}
\centering
\includegraphics[width=4.in]{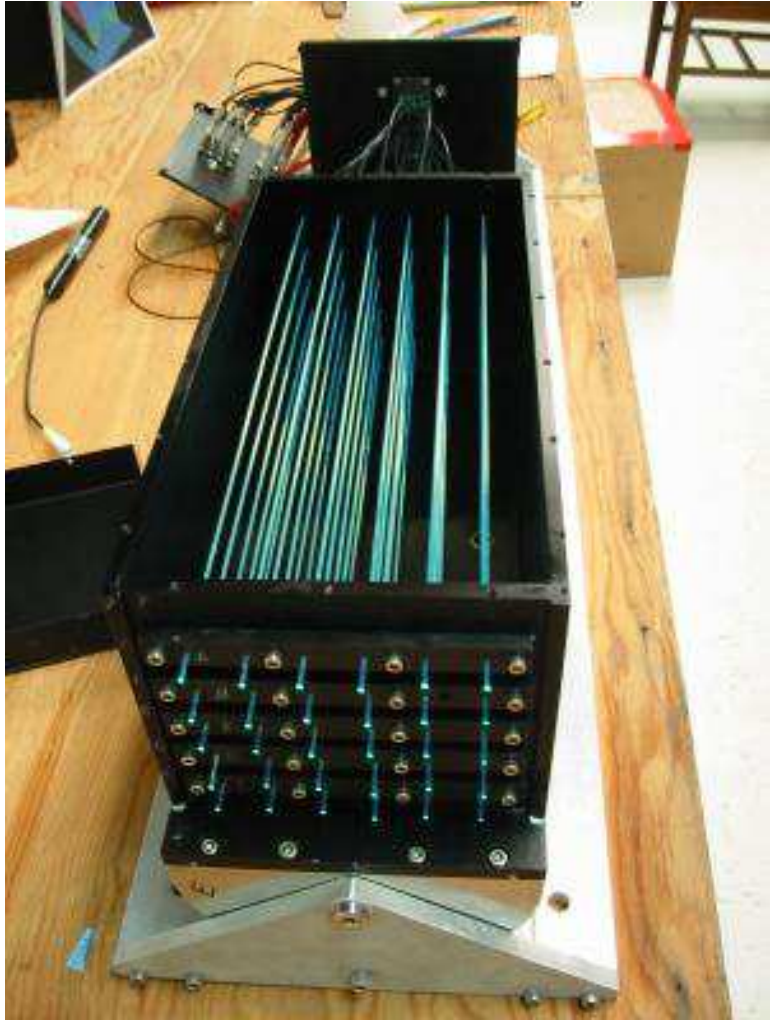}
\caption{\it Photograph of the prototype detector with the mounted WLS fibers. The
  protruding fiber ends were covered with a light shield. The
  photomultipliers were mounted on the plate at the far end. The box
  could be pivoted to study beams that intercept the fiber matrix at
  an angle.}
\label{fig:proto}
\end{figure}

The prototype Vertex Detector consists of a rectangular box made from
anodized aluminum, $16 \times 16 \times 30 $cm$^3$ on the inside
(Fig.~\ref{fig:proto}), with a $6 \times 5$ array of multi-clad,
$1.5$mm diameter, WLS fibers (Bicron BCF-91A-MC,~\cite{StGob}). The
fibers are spaced 20~mm apart, and penetrate the wall of the box.  The
fibers' O-ring seals hold them in place, as will be done in the
full-scale Vertex Detector (see
Section~\ref{sec:mechanicalconstruction}).  Light from the fibers is
detected on one end by two 4$\times$4-anode photomultipliers
(Hamamatsu H8711). The light-tight box is filled with liquid scintillator (Bicron
BCS517H ~\cite{StGob}).

To test this prototype, the box was placed at the end of beam line I
of the Radiation Effects Research Program (RERP) test station at
IUCF~\cite{RERP}. A low-intensity, 200 MeV proton beam with a $6
\times 6$ mm$^2$ profile (defined by two scintillators in coincidence)
was aimed at the center of the box. Each passing proton produced a
track with almost uniform energy deposition.  The box could be moved
vertically (henceforth called x-direction), or horizontally
(y-direction) by remote control, enabling the beam position to be
scanned perpendicular to or along the length of the fibers.  The
amplitude signals from all 30 fibers were digitized by conventional
electronics and stored, event-by-event.

\subsubsection{Light Yield Versus Track-to-Fiber Distance and Angle}
\label{sec:lightyield}
\begin{figure}
\centering
\includegraphics[bb=50 100 480 680,height=6.in]{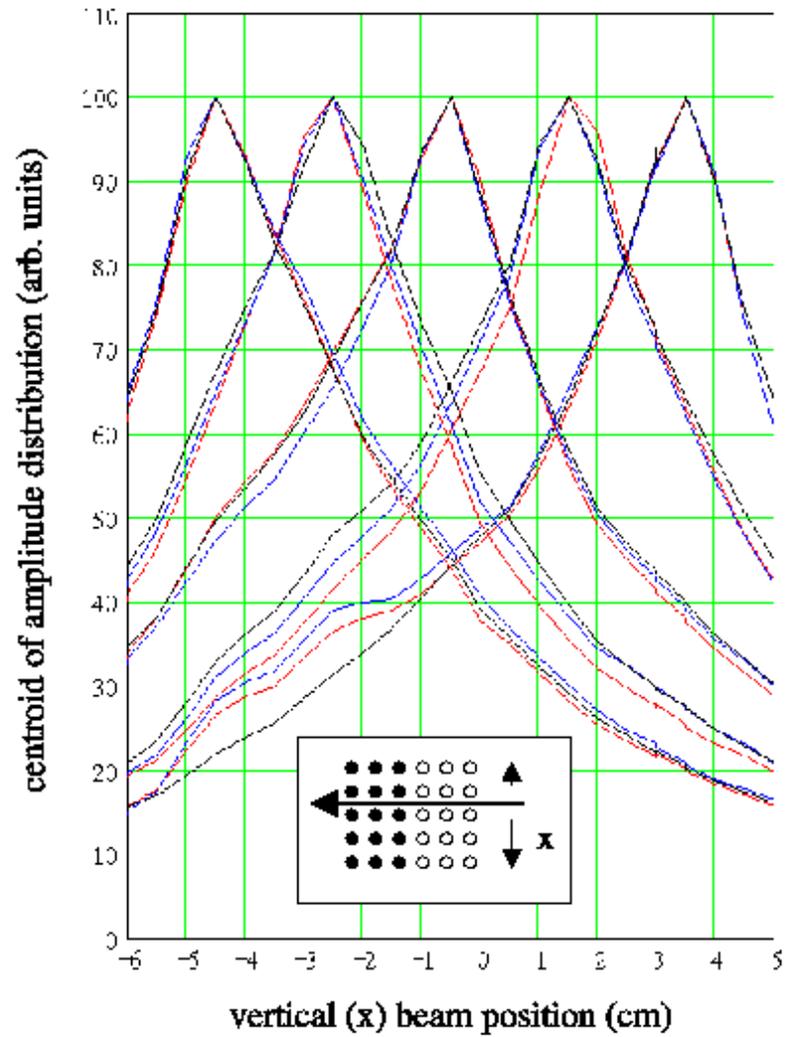}
\caption{\it Normalized centroids of the light yields of 15 fibers as a function of
the vertical beam position. The insert shows the position of the
fibers plotted. The light yield is given in arbitrary units. To convert to 
units of observed photoelectrons, multiply by 0.17 (see text).}
\label{fig:ncentroids}
\end{figure}

With the box axis normal to the beam, a sequence of runs were taken,
moving the beam from top to bottom in steps of 5~mm. For each fiber
and each run, the centroid of the accumulated amplitude spectrum,
corrected by its pedestal value, was determined. The centroid as a
function of beam position is shown in Fig.~\ref{fig:ncentroids}.
Before the centroid values were plotted, each fiber output was scaled
by an individual gain factor such that all peaks lined up at an
arbitrary centroid value of 100. The 30 fiber pedestals and gains thus
determined were retained for the remainder of the test.  This part of
the test demonstrates that there is a universal law that governs the
collected light as a function of the track-to-fiber distance.

In order to study the angular dependence of incident light, 
the box was rotated around a vertical axis by an arbitrary angle
($\beta = 27^\circ$), and the vertical scan, described in the previous
section, was repeated. The result was virtually indistinguishable from
Fig.~\ref{fig:ncentroids}, except for an overall scaling by
$\frac{1}{cos \beta}$.  This shows that the amplitude response of the
detector is indeed sensitive to the projection of the tracks onto a
plane orthogonal to the fiber direction.

\subsubsection{Attenuation of Light in WLS Fibers}
With a normal beam at constant height ($ x = 0.5$ cm), a scan along the
fiber direction was carried out. For all fibers, we observed a light
yield that slowly increased toward the end of the fiber that was
coupled to the photomultiplier, with a sharp fall-off near both ends
of the box.  From the overall increase, we estimated a light
attenuation length in the fiber of 150 to 200 cm.  The fall-off at the
ends is well explained by purely geometric arguments.  

\subsubsection{Absolute Light Yield}
To absolutely calibrate the light yield, it is necessary to relate the 
photomultiplier amplitude to the number of collected photoelectrons (PEs). 
This can be done if the peaks that correspond to events with only one or two
PEs are resolved. In the spectra obtained with the multi-anode 
photomultipliers this was not the case.
We have therefore compared the tubes used in this test with a
high-performance tube, using low-light pulses. We found that a track
that intercepts the fiber (corresponding to 100 units in
Fig.~\ref{fig:ncentroids}) yields an average of $17 \pm 2$
PEs.

\subsubsection{Reconstruction of the Track of a Single Event}
In the following sections we describe an attempt to use a simple
reconstruction algorithm to extract the angle and position of a track
of a single event, and to determine the corresponding uncertainty.
The track reconstruction algorithm used here is primitive but
sufficient to prove the point. For instance, the simple algorithm does
not take into account that the energy loss along the track through our
test setup changes by about 40\%.  The track reconstruction algorithm
used for the Monte Carlo sample for the entire FINeSSE detector,
described in Chapter~\ref{ch:EventSimulationandReconstruction},
accounts for these effects.

The observed dependence of the amplitude, $A$, on the fiber-to-track
distance, $d$, (Fig.~\ref{fig:ncentroids}), is used to construct a
function $A(d)$, chosen so it could be inverted analytically. 
Then, the amplitude data are converted to measured distances. In other 
words, for any given event $i$, the distances $d_{n,i}$ of that track from 
all fibers $(n=1-30)$ are known. 

\begin{figure}[t]
\centering
\includegraphics[bb=123 221 487 568,width=4.in]{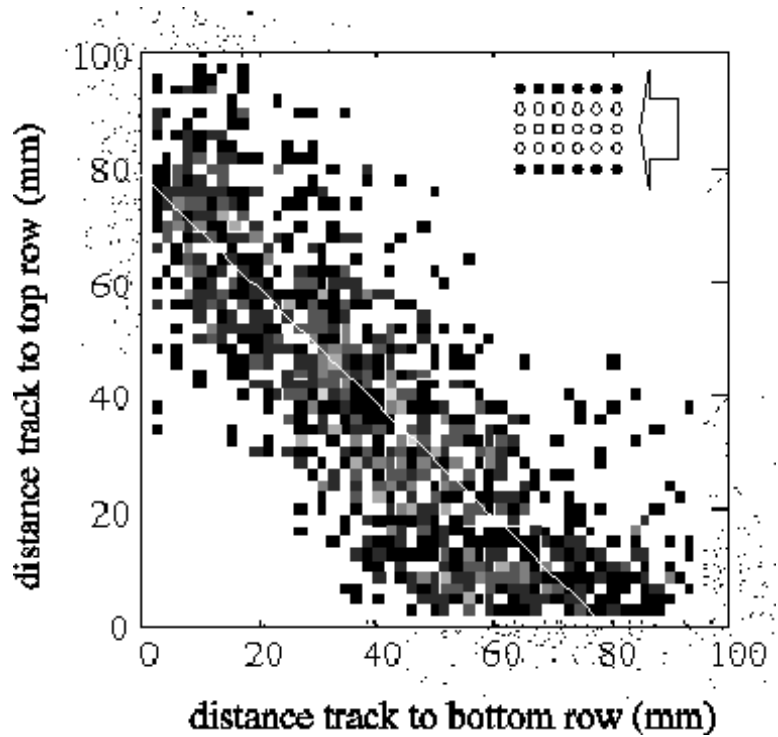}
\caption{\it Distance to top row versus distance to bottom row, averaged
over all six fibers in that row, for horizontal tracks that fill the
space between the two rows.}
\label{fig:disttb}
\end{figure}

Fig.~\ref{fig:disttb} shows the average distance of tracks from the
top row, versus the average distance from the bottom row. Each event
is represented by a point. The plotted events are comprised of 500
events from each of the runs of the vertical scan discussed above.  As
a result, the tracks to be analyzed more or less fill the space
between the top and bottom rows.

\begin{figure}[t]
\centering
\includegraphics[bb=74 365 458 746,width=4.in]{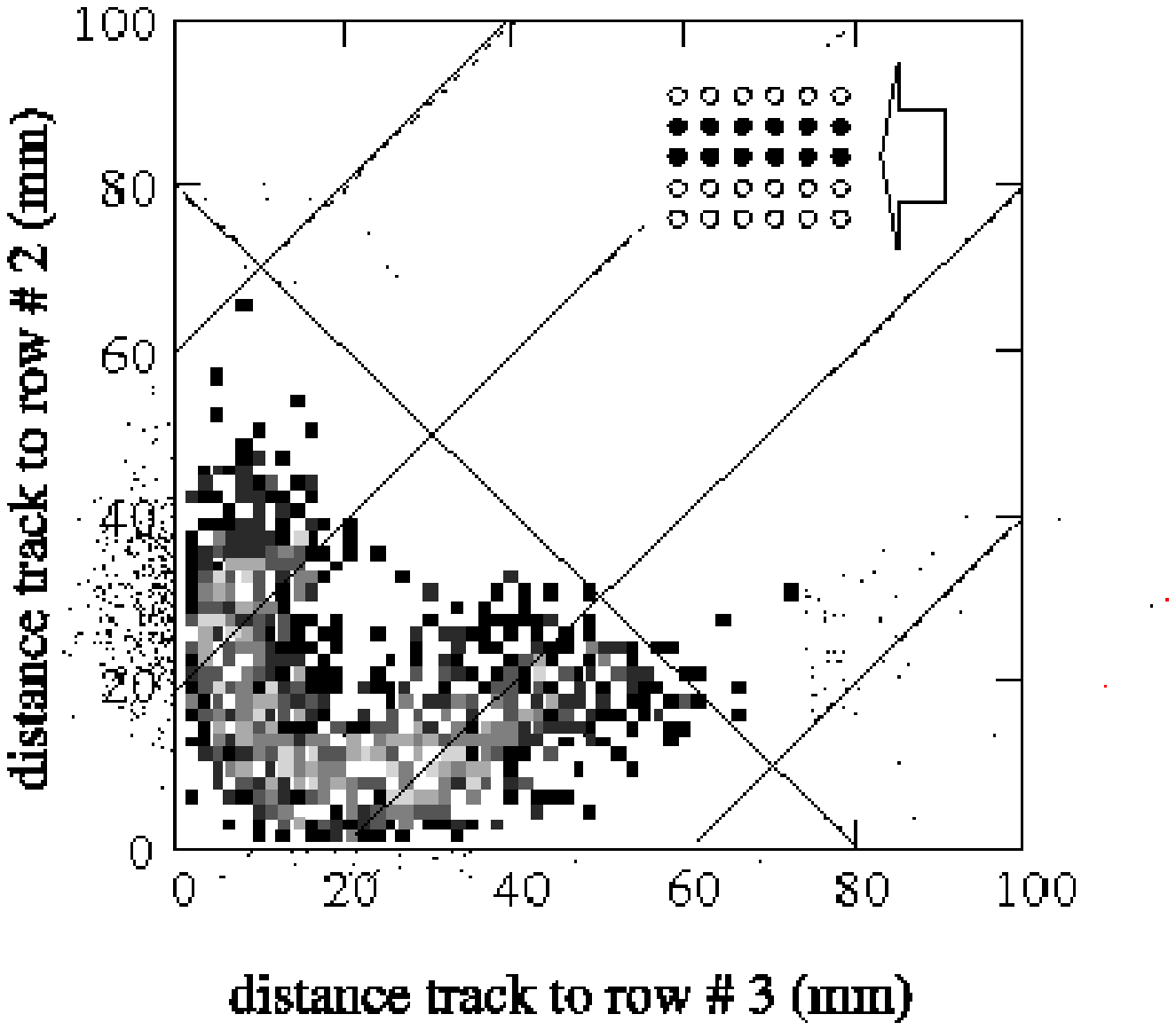}
\caption{\it Distance to row 2 versus distance to row 3, averaged over all six
fibers in that row, for horizontal tracks that fill the space between
top and bottom row.}
\label{fig:dist23}
\end{figure}

Since the sum of the two distances equals the spacing between top and
bottom row (80 mm), the events are expected to lie on a straight line.
For the same events, Fig.~\ref{fig:dist23} shows the average distance
of tracks to the second row versus the average distance of tracks to the third row.
For the tracks between the rows the {\it sum} of the distances is constant, as in 
Fig.~\ref{fig:disttb}  but for those outside the rows the {\it difference} 
must be constant. This can clearly be seen in the figures.

We parametrize a track by its position at the downstream face of the
detector (the intercept with the $x$-axis) and its angle with respect
to the $z$-axis. For a given event (track), the distances to all
fibers are deduced from the measured amplitudes via the function
$A(d)$, mentioned earlier.  To reconstruct the track, one needs to
find the straight line that is most consistent with these distances.
This is accomplished by a regression algorithm minimizing the square
of the difference between the calculated and the measured distance,
weighted by the amount of light, summed over all fibers. In this way,
one obtains the position $x$ and angle $\alpha$ for each individual
track.

\begin{figure}
\centering
\includegraphics[bb=6 226 500 650,width=\textwidth]{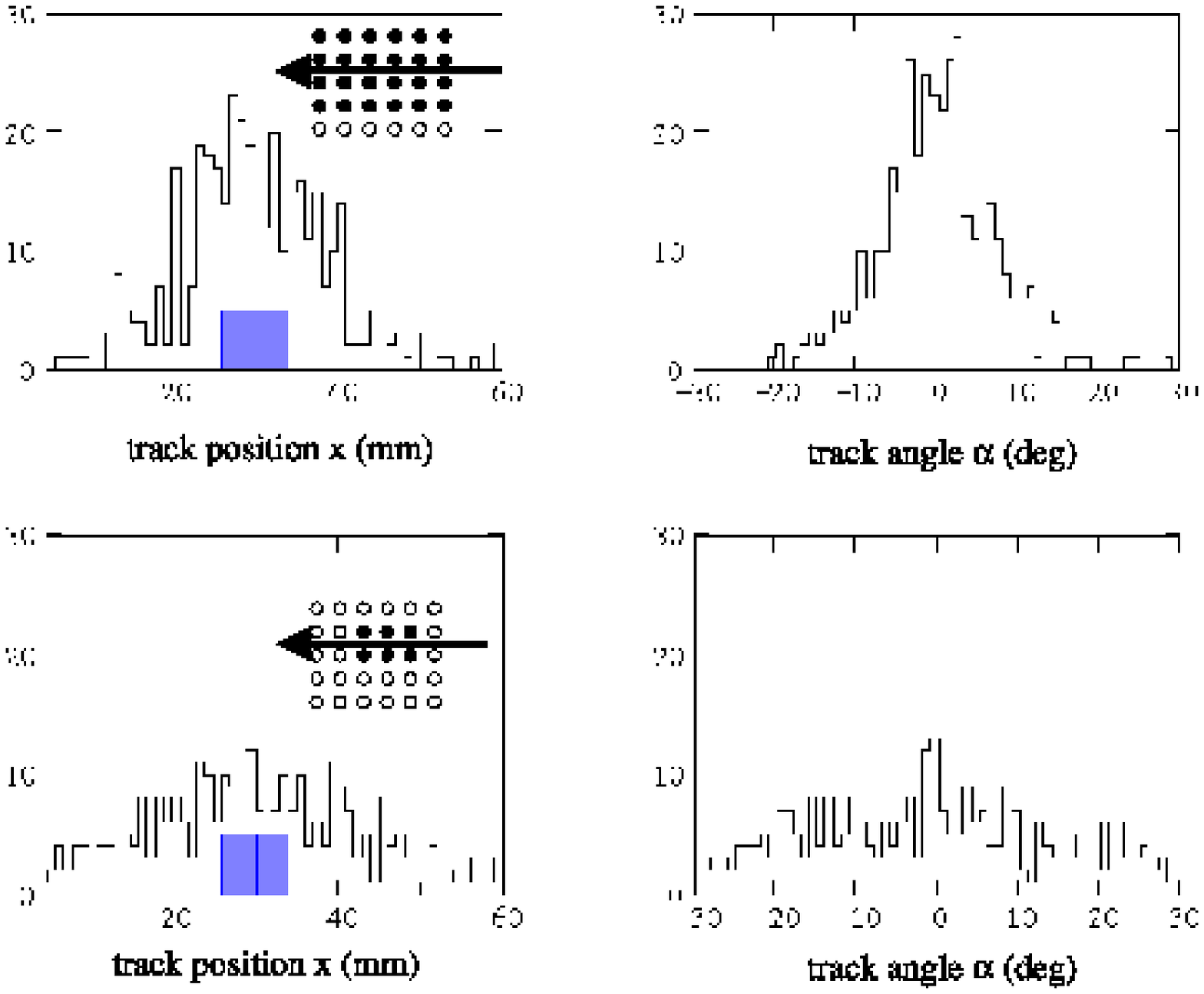}
\caption{\it Reconstructed position and angle of 
tracks for two rows of fibers on
either side of the beam (top plots) and for only three fibers on either
side of the beam (bottom plots). The rectangular box in the left panels
indicates the nominal beam position and width.}
\label{fig:resdata}
\end{figure}

We have applied this procedure to 500 events of a run where the beam
was normal to the fibers, and aimed halfway between rows 2 and 3. The
expected position is therefore $x$=30 mm, and the expected angle is
$\alpha$=0.  In the reconstruction code it was possible to decide
whether a given fiber should be allowed to contribute in the fitting
procedure. The two panels in the top row of Fig.~\ref{fig:resdata}
show the reconstructed position and angle distributions when all
fibers in the two rows on either side of the nominal beam position (a
total of 24 fibers) are taken into account. The distributions
represent a convolution of the intrinsic detector resolution and the
phase space of the ``beam,'' which is defined by the trigger
scintillators. In order to illustrate how this result depends on the
number of participating fibers, we have repeated this analysis, taking
into account only three fibers in each of the rows straddling the beam
(a total of 6 fibers).  The result is shown in the bottom row of
Fig.~\ref{fig:resdata}.

\begin{figure}
\centering
\includegraphics[bb=6 226 500 650,width=\textwidth]{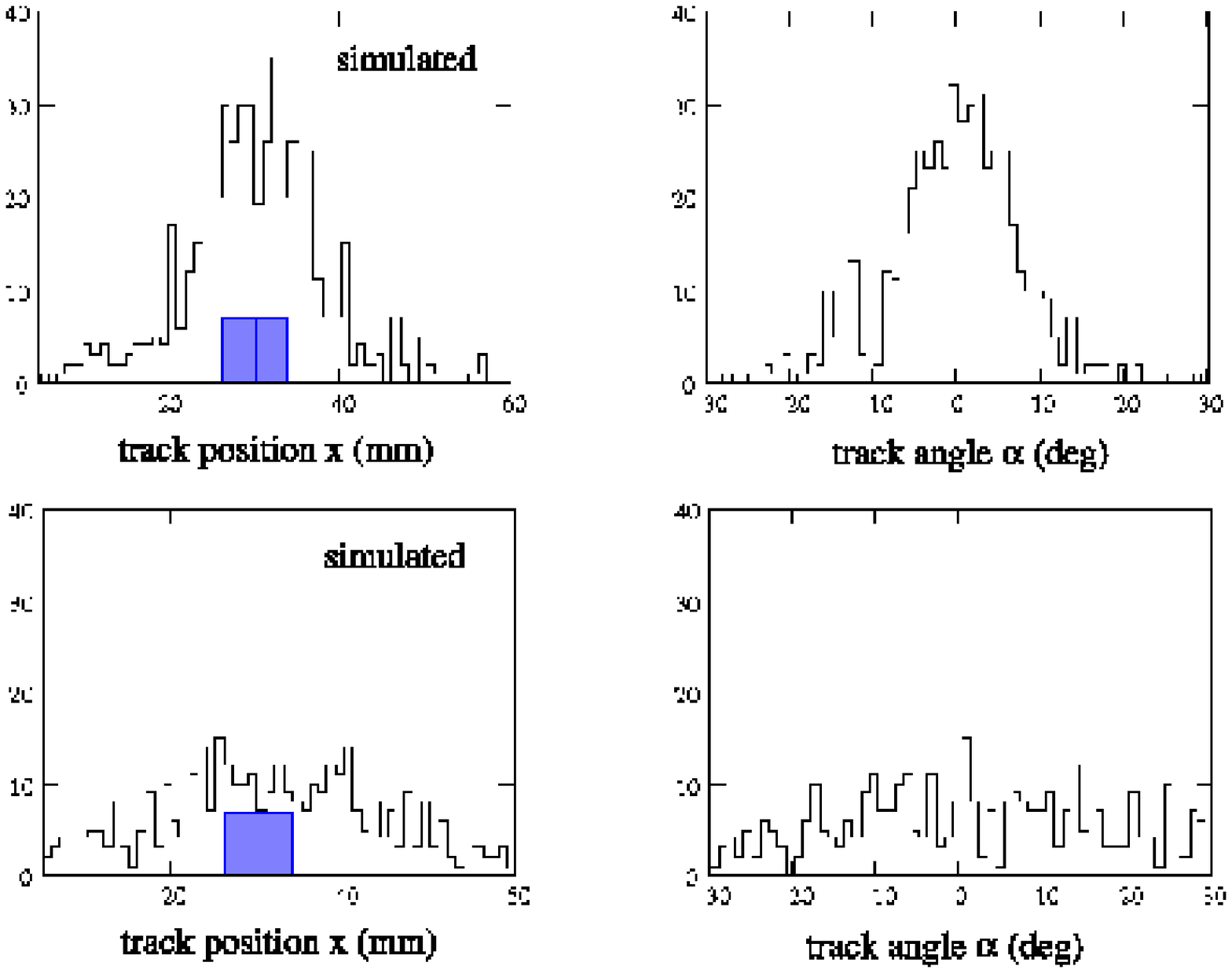}
\caption{\it Simulated events, taking into account
  the Poisson statistics of the observed number of PEs
  (cf. Fig.~\ref{fig:resdata}).}
\label{fig:simdata}
\end{figure}

The measured track position and angle resolutions are dominated by
statistical fluctuations in the number of PEs observed in
the photomultiplier. This can be seen in the following way.  Assuming
a hypothetical track, and using the light-distance relationship $d(A)$
mentioned earlier, the signal amplitude (in arbitrary units) is
calculated for all fibers. From the absolute yield calibration (100
amplitude units correspond to 17 PEs) we get the
corresponding number of PEs.  From a Poisson distribution
with that mean, a random number is drawn to represent the ``observed''
number of electrons. This number is then converted back to amplitude
units for all 30 fibers, and the resulting simulated event is
reconstructed by the same algorithm used for the real events.  As
input, random tracks with $\alpha$ = 0 and 26.5 mm $<x<$ 33.5 mm
(representing the beam width) were used.  The simulated distributions
for $x$ and $\alpha$ are shown in Fig.~\ref{fig:simdata}. Obviously,
they are remarkably similar to the real distributions
(Fig.~\ref{fig:resdata}), and since statistics is the only reason for
smearing in the simulation, we conclude that this is also the case for
the measured data.

Given the good agreement between real and simulated data, we can
deduce the intrinsic detector resolution.  We find that the position
resolution achieved in our test is about 15~mm FWHM, and the angular
resolution about 15${^\circ}$ FWHM.
The difference between using 24 fibers and using only six fibers is
not solely due to counting statistics.  It is also affected by the
fact that the length of the track, that is supported by distance
information from the fibers, changes, thus relaxing the constraint on
the track angles.

\subsection{The Muon Rangestack}
\label{sec:muonstack}
Downstream of the FINeSSE Vertex Detector is the Muon Rangestack.  The
Rangestack is designed to range out muons, in conjunction with the
Vertex Detector, with energies up to 1.5~GeV, and to measure muon
energy to 10\%.  These characteristics are needed to enable the
reconstruction of charged current $\nu_{\mu}$ events, which are used
in both the $\Delta s$ and the $\nu_{\mu}$ disappearance measurements.
The Rangestack is located downstream of the Vertex Detector, because
the muons from high energy CC $\nu_\mu$ events in the Vertex Detector
tend to be produced at forward angles (See Section~\ref{evmurange}).

The iron absorber planes and tracking granularity in the Rangestack
are designed to meet these requirements.  The stack is comprised of $4
\times 4~$m$^2$ planes of scintillator strips and iron absorber with
an overall depth of 0.85 m (0.98 m including support structure) 
in the beam direction (referred to here as the z direction), and a 
weight of 100 t.  Of the 0.85 m thickness, 0.24 m is scintillator and
0.61 m is iron. The design both meets the
physics requirements and minimizes cost and space demands.  The
following sections describe the Rangestack's design, construction, and
readout.

\begin{figure}
\centering
\includegraphics[bb=73 196 539 595,width=4.in]{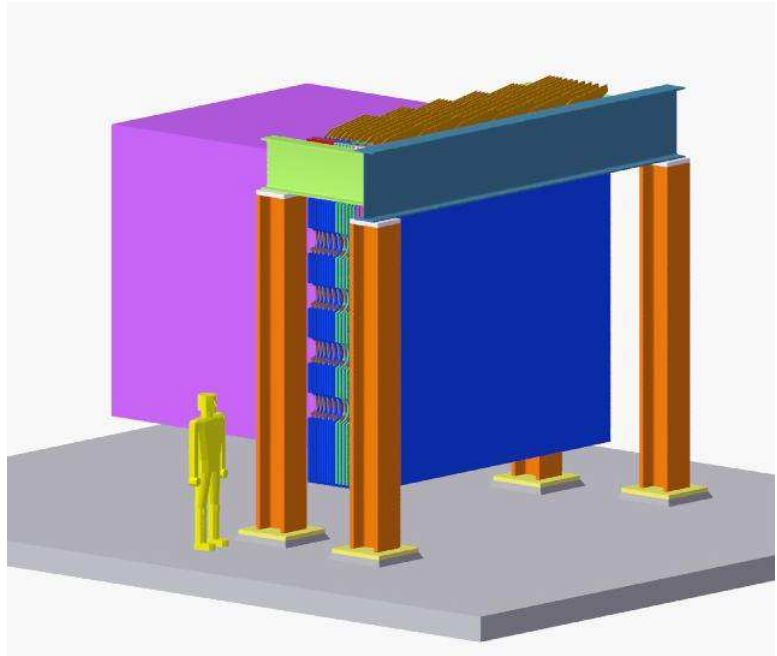}
\caption{\it An isometric rendering of the Muon Rangestack in front of a box 
  bounding the envelope of the Vertex Detector.}
\label{fig:scistack-01}
\end{figure}

Figure~\ref{fig:scistack-01} shows isometric and orthographic views of
the entire detector.  The structural steel shown supporting the stack
is a conceptual design.  No PMT enclosures or clear fiber bundles
surrounding the Rangestack are shown.

\subsubsection{Advantages of Choosing a Rangestack Design}

We have found that a Rangestack is the best choice to achieve our
physics goals.  The Rangestack is designed to range out muons, in
conjunction with the Vertex Detector, with energies between 0.1 and
1.5 GeV.  A stack of plastic scintillator bars interspersed with iron
absorber is a relatively inexpensive and straightforward way to
achieve these goals.  This detector design has been well tested by the
MINOS collaboration.  They have shown that these plastic scintillator
strips have excellent reflectivity, fast timing, simple design, long
term stability, low maintenance and high reliability~\cite{MINOspdf}.
Furthermore, the expertise for construction and assembly of such a
detector already exists at Fermilab.


A magnetic spectrometer was considered and rejected because it is very
expensive compared to the Rangestack.  Its obvious advantage -- charge
identification -- is not necessary for the physics of the experiment.

\subsubsection{Rangestack Design}
\begin{figure}
\centering
\includegraphics[bb=73 196 539 595,width=4.in]{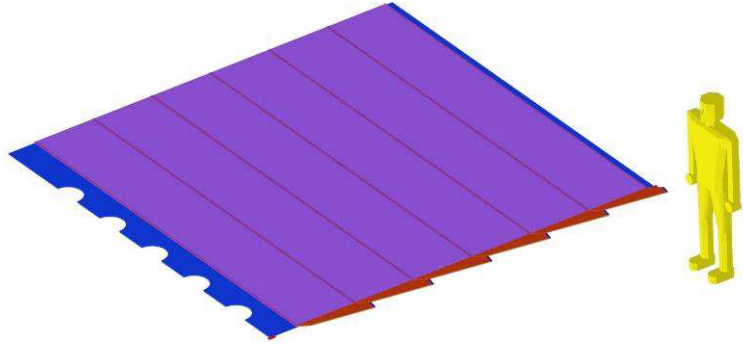}
\caption{\it A completed $x$ layer with six cans of scintillator and a 0.50 inch thick
steel backing.  The assembly weighs 2.2~tons.  Cans are attached to the steel backing by
bolts at the ends of the can, and a strap (not shown) similar to the MINOS design
which prevents the cans from bending away from the backing in the middle.}
\label{fig:scistack-data-07}
\end{figure}

The Rangestack consists of 21 pairs of a scintillating tracking plane
and iron absorber.  Each scintillating tracking plane is made up of
4.1~cm~$\times$~1~cm~$\times$~4~m extruded polystyrene scintillator
strips.  Each plane is oriented normal to the beam, with strips in the
$x$ direction, in even numbered planes, and in the $y$ direction in odd
numbered planes.  The strips are arranged so that their depth in the $z$
direction is 1~cm.  Each 4~m plane, then, contains 96 strips.  The
strips are packaged in groups of 16, surrounded by a 1~mm thick
aluminum can for protection and light tightness.  There are six cans of
strips per $x$ or $y$ layer.  Cans are identical between $x$ and $y$
layers, just rotated 90$^\circ$ with respect to each other.  The six
cans in a layer are attached to a backing layer of 0.50~inch thick
steel which serves to support the 4$\times$4~m$^2$ plane of scintillator and
protect it during crane handling and shipment, and functions as part of the
absorber layer between each $x$ and $y$ layer of scintillator.
Figure~\ref{fig:scistack-data-07} shows an assembled $x$ plane.
Assembled $y$ planes are similar, with scintillator planes rotated by
90$^\circ$.

Scintillator strips in these planes are similar to those used in the
MINOS experiment~\cite{MINOspdf}.  The scintillator is co-extruded
with a TiO$_2$ outer layer for reflectivity.  A groove down one side
of the strip contains a glued-in-place WLS fiber.  Light produced in
the scintillator strips is absorbed by the WLS fiber and re-emitted at
$\sim500$ nm.  Re-emitted light within the acceptance for total
internal reflection is transmitted to the end of the fiber and
detected by the phototubes, which are identical to those used in the
Vertex Detector.  The same type of WLS fiber will be used in the
Rangestack and the Vertex Detector (see Section~\ref{sec:Lightgeneration} for
a description).

Interspersed with the active scintillator planes are iron absorber
planes.  In order to minimize the size of the Rangestack, the
thickness of the iron in the absorber planes increases with increasing
depth into the detector.  This design is chosen to take advantage of
the fact that once lower energy particles range out (leaving only
higher energy particles), the iron absorber thickness can be
increased while maintaining the same relative energy resolution.  This
is achieved easily by dividing the Rangestack into four sections
with the thickness of the beam absorber increasing by 0.5 inches, or
one iron plane, per section.

The upstream end of the range stack (section 1) has six layers, three
$x$ and three $y$, with 0.50 inch total steel between layers.  The
backing plate for the scintillator planes provides this layer of
absorber.  Behind that is section 2, with six layers of $x$ and $y$
planes separated by a total of 1.00 inch of steel between modules.  To
make up the total thickness, there are 0.50 inch thick filler plates
between scintillator modules.  Section 3 has six layers of $x$ and $y$
scintillator planes separated by 1.50 inches of steel.  Section 3
filler modules are 1.00 inch thick.  Finally, section 4 has three
absorber layers 2.00 inches thick, using 1.50 inch thick absorbers.
An $x$ layer sits upstream of the first 2.00 inch thick absorber,
followed by alternating layers of $y$, filler and $x$ layers.  The
stack ends with a $y$ layer.  The final $x$ and $y$ layers in the
stack are triple-thick to assure high efficiency for identifying
stopped muons.  Figure~\ref{fig:scistack-data-09} shows a typical 1.50
inch thick filler module for section 4.  The minimum thickness of 0.5
inches used in the first section of the Rangestack is set so as to
ensure a 10\% energy resolution for the lowest energy muons that enter
the Rangestack.

\subsubsection{Light Output and Readout in Active Rangestack Planes}
\label{sec:StripLightOutput}
Tracking planes in the Muon Rangestack must produce enough
PEs at the readout PMTs to observe a Minimum Ionizing
Particle (MIP) and therefore track muons.  The scintillating strips, WLS
fiber, and readout used in the Muon Rangestack are very similar to
those used in the MINOS far detector.  Light output from a Minimum
Ionizing Particle (MIP) in the FINeSSE Muon Rangestack can be inferred
from prototype testing results from MINOS.

As MINOS studies have shown, the efficiency for observing a muon
crossing a strip is greater than 90\% as long as at least 2.5~PEs are
read out.  In a prototype assembly of part of a scintillator plane for
the MINOS far detector, a MIP passing through the center of a strip
produces, on average, 6~PEs in total, with an average of 3~PEs read
out on either side~\cite{MINOspdf}.  The design of the Muon Rangestack
is such that there will be approximately twice this much light
observed as a result of the differences between the MINOS strips and
the FINeSSE Muon Rangestack strips.
These differences in the design include the overall length of the
strips, the WLS fiber size, and the fact that the readout takes place
from one end only.  The first two factors increase the light
levels in the Rangestack by more than a factor of two as compared to
the MINOS detector.  The fact that readout occurs from only one end
will decrease the light read out, but only slightly as reflective
paint on the terminal end of the fiber will pipe light back toward the
readout end.

WLS fibers are connected to clear fibers at the end of the
scintillator strips as is done in the MINOS detector and as is shown
here in the Muon Rangestack design.  Clear fiber ribbon cable will
then route the signals to the removable floor above the detector where
they will be read out via the same system employed in the Vertex
Detector as described in detail in Section~\ref{sec:SignalReadout}.

\subsection{Signal Readout: Phototubes and Electronics}
\label{sec:SignalReadout}

We propose a common design of signal readout to be used for the Vertex
Detector, its veto shield, and the Muon Rangestack.  Requirements for
the readout system include independent amplitude and timing
measurements for each channel. The amplitude information is used in
the track reconstruction and for $dE/dx$ measurement for the particle
tracks; timing information is used to assemble the hits (rejecting
spurious noise hits and background tracks), to correlate with the beam
spills, and in the analysis of secondary events, such as muon decay
and nuclear decay in the active volume following the capture of a
neutron produced in the primary event.  In the Vertex Detector the
full scale signal range will extend to \mbox{$>50$ PEs},
with an amplitude resolution of \mbox{$<0.1$ pe}.  The timing
resolution is not crucial, but it needs to be \mbox{$\sim$ 10 ns} to
reduce spurious hits and background tracks.

The readout must be sensitive both to the primary interaction events
(which are in time with the beam spill) and to secondary events as
described above. It is also useful that it be sensitive to cosmic ray
muon tracks, for calibration purposes. For these reasons, the front
end electronics must be self-triggering and have a relatively low
deadtime after being hit. Deadtime of the proposed system is about
$1-5\;\mu{\rm s}$, depending on signal amplitude (see discussion
below).  The deadtime applies independently to each channel -- this is
a matter of recovery of the front end electronics only, since there is
zero deadtime associated with the data readout. A external global
trigger will also be implemented, which can be used to force an event
readout and/or a charge pulser event, for diagnostic purposes.

The readout system designed to meet these requirements combines
multianode photomultiplier tubes (MAPMTs) and custom readout
electronics in a 128-channel ``readout module.''  This includes two
MAPMTs, front end electronics, HV bias circuits, charge-injection test
pulsers, and data acquisition and control communications. The module
is a sealed, conductively cooled unit suitable for submerged operation
inside the oil tank (for the Vertex Detector application).  In the
Muon Rangestack the readout modules will be housed above the removable
floor above the detector.  The Vertex Detector requires 150 modules
for readout of the active volume and six modules for the veto shield.
The Muon Rangestack readout requires an additional 17 modules.

\subsubsection{Multianode PMTs}
\label{sec:pmts}
Fibers from the Vertex Detector, veto, and Muon Rangestack are read
out via Hamamatsu R7600-00-M64 MAPMTs.  This multianode
photomultiplier tube provides an $8\times 8$ array of optical readout
in a \mbox{25.7 mm} square by \mbox{20.1 mm} deep metal package. We
intend to operate the MAPMTs at a gain of $8\times 10^{5}$, which
corresponds to a typical cathode supply voltage of \mbox{$-875$ V}.
Channel-to-channel gain variation in the MAPMT is 5:1 maximum.  As a
result, if it is assumed that a channel in the middle of this ratio
is set to the desired gain, the required full scale range of the front
end electronics is $8\times 10^{5} \times\sqrt{5}\times 50q = 14\;{\rm
  pC}$, and the required resolution is $(8\times 10^{5} /
\sqrt{5})\times 0.1q = 5.7\;{\rm fC}$, a dynamic range of 2500.
The R7600-00-M64 MAPMT is similar to the sixteen anode R5900-00-M16
(see in Figure~\ref{fig:star_fee_assy}).

\subsubsection{Front End Electronics}
Continuously running commercial multichannel 10-bit pipeline A/D
converter chips, such as the ST Microelectronics TSA1005-20IF, will
form the heart of the FINeSSE front end electronics (FEE). One such
converter will digitize signals from each anode of each MAPMT, at a
rate of \mbox{20 Msamples/s}. Since this sampling is, of course,
significantly coarser than the MAPMT output pulse width, and
significantly coarser than the desired time resolution of 10 ns,
suitable analog signal processing before digitization, and digital
signal processing after, will be employed.

The analog signal processing must shape the MAPMT output pulse so that
it extends over several A/D converter samples (offering a chance to
measure the amplitude), and so that it does something characteristic
(such as have a peak, or better yet a zero-crossing) at a defined time
after the MAPMT output pulse (offering a chance to measure the time of
occurrence). In addition, the analog signal processing circuit must
present a low and resistive impedance to the MAPMT anode, so that all
the induced charge is swept quickly into the signal processing for
that anode, and does not go into parasitic capacitances to ground or to
other anodes. The simplest circuit which can achieve these things is
an inverting op-amp with an RLC network in feedback. In response to a
$\delta$-function current input, $I(t)=Q\delta(t-t_{0})$, a damped
cosine wave is output,
$V(t)=(Q/C)\cos(\omega(t-t_{0}))\exp(-\lambda(t-t_{0}))$ for $t\geq
t_{0}$.  The amplitude is proportional to the input charge, and the
phase, or equivalently the time of the first zero crossing, is
directly related to the time of occurrence of the input charge.

The required digital signal processing simply undoes the above. If
$R$, $L$, and $C$ are known, then $\omega$ and $\lambda$ are known,
and a simple linear fit of the measured data points to
$V(t)=A\cos(\omega(t-t_{0}'))\exp(-\lambda(t-t_{0}'))
+B\sin(\omega(t-t_{0}'))\exp(-\lambda(t-t_{0}'))$,where $t_{0}'$ is
the time of the first ADC sample found above threshold, determines the
amplitude and the relative time $t_{0}-t_{0}'$ of the input pulse.  Of
course, there will be manufacturing tolerances on the component
values, and potentially a small drift with time or temperature. To
calibrate these, a nonlinear fit involving $\omega$ and $\lambda$ as
parameters is performed occasionally.

The number of raw A/D samples to be used, and the desired value of
$\omega$ and $\lambda$ of the filter, needs to be carefully optimized.
In particular, the response should decay to less than a few percent
within $5\;\mu{\rm s}$.  A following pulse which occurs within
5~$\mu$s can still be measured, provided that the amplitude of the
waveform does not exceed the range of the A/D converter. The actual
deadtime for a full scale amplitude pulse following another full-scale
amplitude pulse may be as large as 5~$\mu$s, but for pulses of smaller
amplitude, e.g., less than half-scale, the deadtime can be as small as
1~$\mu$s or even 800~ns. For pulses separated by less than that, it is
still possible to discriminate that there is a second pulse, by
applying a threshold to the $\chi^{2}$ of the fit, even if it is not
possible to estimate the amplitudes and times separately. The low
deadtime and double-pulse discrimination capability is important to
detect muon decay events, both for particle identification and as a
method of gain calibration.

An important difference from the conventional approach, where a
discriminator and dedicated time-to-digital converter are used, should
be noted. Conventionally, the discriminator is set to respond to the
first, or at least the first few, PEs in the signal, and
thus the measured hit time is insensitive to the later PEs.
As a result, the measured hit time is more accurately related to the
time of the charged particle passing in the detector. In the present
approach, where a timing discriminator is not used, by contrast, the
measured hit time depends on all the PEs in the signal.
This will not yield as good a result; nevertheless, because the
scintillator oil and wavelength shifting fiber both have short decay
times (\mbox{2 ns} for the oil, \mbox{3 to 10 ns} for the WLS fiber),
we can achieve the required \mbox{10 ns} timing accuracy.

The digital signal processing will be performed by a combination of
field programmable gate arrays (FPGAs) and a microprocessor (also used
for communication and control -- see below). The FEE FPGAs, handling
8 or 16 channels each, will monitor the digitized data from each
channel, apply the trigger threshold, capture the required number of
data points after seeing the threshold crossed, and send them to the
microprocessor. There the linear fit calculations are performed, and
the amplitude and time thus determined are queued for transmission to
the DAQ.

This front-end electronics design for FINeSSE is intended to take
maximum advantage of low cost commercial electronics to provide a
highly parallel measurement of pulse amplitude and timing with minimal
deadtime. Although there are ASICs such as the VA-TA series from IDE,
Inc. which have been applied to very high density readout electronics
for MAPMTs, none of the presently available ASICs can match the low
deadtime performance of the system described here when the hit
patterns are non-sparse as in the FINeSSE Vertex Detector.
Furthermore, FINeSSE does not strictly require such very high density
electronics, since each MAPMT covers a 240~mm~$\times$ 240~mm area on
the detector, making almost that much area available for the
readout electronics. In any case, with the large commercial market
driving down the costs of pipeline ADCs and FPGAs, the proposed
approach can be more cost effective.

An independent charge-injection test pulser will be included on each
channel. This diagnostic feature will be useful in testing the front
end electronics and the readout, and debugging the system during
operation.

\subsubsection{The Readout Module}
\begin{figure}
\centering
\includegraphics[width=5in]{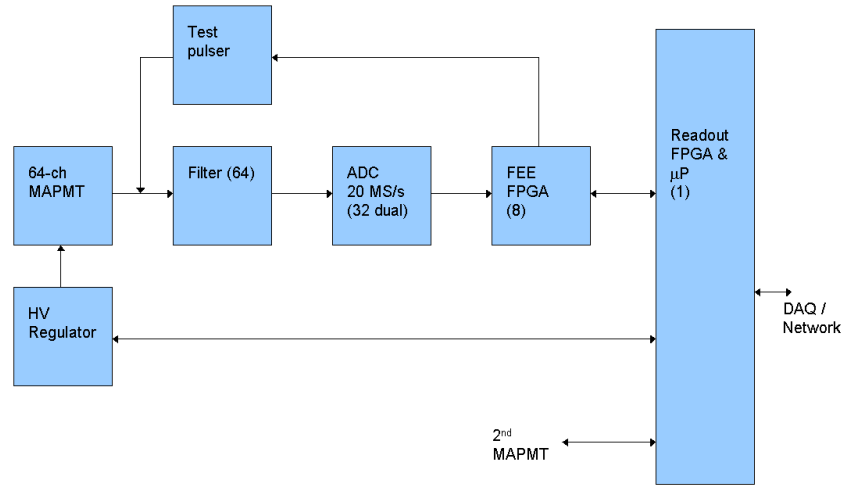}
\caption{\it Block diagram of the readout module.}
\label{fig:readout_blockdiagram}
\end{figure}

The readout module is comprised of two Hamamatsu R7600-00-M64 MAPMTs
described in Section~\ref{sec:pmts}, a high voltage bias generator,
128 channels of front-end electronics, a microprocessor, a data
network interface, a system clock and trigger interface, and a power
supply.  (See the block diagram in
Fig.~\ref{fig:readout_blockdiagram}.)  The microprocessor handles
control and housekeeping tasks, data transmission tasks, and also the
fitting algorithm (described above) to calculate the hit time and
amplitude based on the raw ADC samples captured in the FEE. Normally
only the time and amplitude will be sent to the DAQ, although for
calibration of the algorithm above, the raw samples will be sent on
selected events.

The module will perform zero suppression based on a programmable
amplitude threshold. This threshold is independent of the trigger
threshold used in the FEE to detect and capture an event.  Note that
both thresholds are applied digitally and can easily be set on a
channel-by-channel basis, which can be helpful in dealing with any hot
channels.

The module can also be programmed to suppress data not satisfying a
pattern cut. For instance, a requirement of $\geq 2$ hits within a
20~ns window will almost completely suppress MAPMT dark pulses, even
with the trigger threshold set very low, without adversely affecting
real events. More complex cuts, such as requiring a track through the
contiguous fibers of the module, can be imposed as well, although a
cut requiring data from other modules would need to be imposed in the
DAQ computer, since the event data will not be communicated directly
between the modules.  (Each module reads out an $8\times 16$ array of
fibers on one face of the detector.)  Note that the front-end
electronics will process every hit seen on every channel; both the
zero suppression and any other cuts that may be programmed into the
readout module act on digitized data before transmitting it to DAQ,
and the purpose is only to reduce the volume of unwanted data.

The output data from the readout module will be, for each hit, a
32-bit coarse timestamp from a counter, 8-bit fine time, and 12-bit
amplitude. For data budgeting, we assume here that each of the latter
will be embedded in a 16-bit word, which may include some extra
``status bits.''  (If data volume or data rates become an issue this
could be revisited; also the 32-bit coarse timestamp could be
compressed easily, since the relative time between subsequent
transmitted hit data in general will not require 32-bit
representation.)  At 8 bytes per hit, the worst case event, where all
channels of a module are hit, will generate \mbox{1.02 kbytes} of data
to be transmitted to the DAQ. An elasticity buffer of \mbox{16 kbytes}
or greater will be implemented in the module to allow for bursts in
the incoming data and other network activity such as packet re-tries
and control transactions.

Estimated power dissipation of the readout module is \mbox{20 W}
(\mbox{156 mW/channel}). The power input will be from a single 24 to
\mbox{48 V} DC supply. Each module will have an individual power cable
from a fused power distribution panel outside the tank. In this way a
module with a catastrophic power short can be easily isolated to allow
the others to continue operation.  High-reliability steering diodes
will direct the input voltage to one of two independent fully isolated
DC/DC converters in the readout module, based on polarity. The outputs
of these supplies will be connected through further steering diodes.
In this way we obtain the benefits of two almost fully independent
power systems in each module, with no overhead for additional cables
or connectors. Fully isolated DC/DC converters are required anyway, so
that the modules can be powered from a common DC source and still have
equal ground voltages, which simplifies the data network requirements
(no AC coupling or wide common-mode transceivers are required).  The
internal circuitry of the module will work largely from a single
\mbox{$+2.5$ V} supply (about \mbox{4.5 A}), with an auxiliary
\mbox{$\pm 3.3$ V} supply for some functions, and another auxiliary
\mbox{$-1050$ V} supply for the MAPMT bias circuits.  HV will be
regulated to each MAPMT individually through a high-side pass
transistor driven by an analog optoisolator. This will feed a
conventional resistive voltage divider for the dynode string. A
Cockcroft-Walton type of PMT bias circuit will be evaluated as an
alternative, for reduced power, but reliability is likely to be
greater using the resistive divider circuit.

The microprocessor and the data network interface are also potential
single-point failures affecting an entire module. We will carefully
evaluate a redundant system design around the microprocessor and its
boot EEPROM in particular. We expect to implement a redundant network
(pair of networks) for each of the five logical networks (described
below); in particular an unpowered module shall be positively
disconnected from the data (and clock) networks. Even when powered, it
will be positively disconnected from the primary data network unless a
watchdog timer is kept alive by properly functioning software.
Download of new code shall be possible via either network, even if the
old contents of the boot EEPROM are corrupted. (In particular the
network disconnection watchdog may be defeated until a defined time
after power-up.)

With proper attention to the above details, reliability of the modules
can be assured to a degree which will allow them to be mounted in the
inaccessible interior of the oil tank. The costs involved in this
redundant design are significantly less than would be involved in
transporting either the scintillation light out of the oil tank on
optical fibers, or the anode signals from MAPMTs mounted in the oil
tank to readout electronics outside. Remaining single-point failures,
except in the MAPMTs or the divider resistor string, can affect at
most eight channels.

The power-dissipating electronic components in the readout module will
conduct heat through their leadframes or other thermal bonds into the
extra thick (\mbox{2 oz.}) copper ground plane of the printed circuit
boards. The printed circuit board ground planes are directly tied to
the the outer aluminum case of the module, either by wedge-lock clamps
or simply by screws and aluminum frames. Temperature rise between the
module case and the printed circuit board ground planes will be about
\mbox{5 - 10 K}. Scintillation oil circulating in the Vertex Detector
will cool the readout modules. The total heat load into the oil is
$156\times 20\;{\rm W}= 3.1\;{\rm kW}$, which can be removed by
circulating the oil through an external water-cooled heat exchanger at
at least \mbox{8 gal/min}, assuming a \mbox{5 K} temperature rise.
The 17 readout modules of the Muon Rangestack will be cooled by fans
mounted externally to their cases. All the modules will have internal
temperature monitors and a hardware overtemperature interlock.  

The readout module will be constructed as a black anodized aluminum
box with an oil-tight gasketted lid. Attention will be paid to
minimizing enclosed or recessed spaces in the outer surface of the
box, the screws and other hardware, and the connectors and cables;
furthermore the box will be purged with nitrogen internally after
assembly. These measures should minimize the contamination of the
scintillating oil with oxygen, which can degrade the light yield.

\subsubsection{Network, Data Acquisition System, and Clock and Trigger Distribution}
The 173 readout modules in the system buffer data internally, and
communicate it, as needed, to the data acquisition computer system
over a serial multidrop bus architecture. Although other
implementations, in particular \mbox{10 Mbit/s Ethernet} or CANbus,
will be considered carefully at the design stage, at the time of this
writing we assume a synchronous serial bus (separate data \& clock
lines) implemented with RS-422, running at \mbox{2 Mbit/s,} using HDLC
(High Level Data Link) protocol.  For acceptable bus loading the
readout modules will be arranged in five networks of up to 39 modules
each (four used for the Vertex Detector and a fifth for the Muon
Rangestack).  The DAQ computer will be a Linux box with two
DMA-capable PCI quad RS-422 communication controller boards installed.
These are expected to be commercial off-the-shelf devices. Control and
status communications will run over the same networks as the readout
data.

With 10\% overhead for protocol the worst case module, with \mbox{1.02
  kbytes} of data, will require \mbox{5.6 ms} for data transfer.
Furthermore, assuming that the networks are physically arranged to
balance the data load, each will see typically up to 250 hits per
spill, which is \mbox{2.0 kbytes} of data. Since this is coming from
39 different nodes, assume a 100\% overhead for protocol. (In
actuality the data is buffered in the readout modules, so not every
module will send a data packet on every spill, but for a reasonable
worst-case estimate we ignore that here.)  Then the complete data
transfer on the network requires \mbox{20 ms} (this includes the up to
\mbox{5.6 ms} for the worst case module). Since the spills are
\mbox{66 ms} apart, there is adequate spare capacity to accomplish
this.

The overall data rate into the DAQ computer, \mbox{16.6 khits/s} =
\mbox{133 kbytes/s,} will be quite manageable, for transfer over the
PCI bus, online monitoring software, and local disk storage or TCP/IP
network transmission.

The 32-bit coarse timestamp provided by the readout module rolls over
in 429 seconds. Therefore the DAQ computer will augment this timestamp
with further bits to provide an absolute time measurement without
rollover.

A \mbox{10 MHz} reference clock and global command word are
distributed to the readout modules over separate, but similar, serial
multidrop networks. The global command word is used to synchronize the
timestamp counter of all modules, and to fire the global trigger and
test pulser events as described above. The serial communications
network clocks will also be derived from the same \mbox{10 MHz} clock
for convenience.  The reference clock will be provided from an
oscillator module which may either be a standard crystal oscillator or
an absolute GPS-referenced timebase.

\subsubsection{Heritage}
\label{sec:star}
\begin{figure}
\centering
\includegraphics[width=5in]{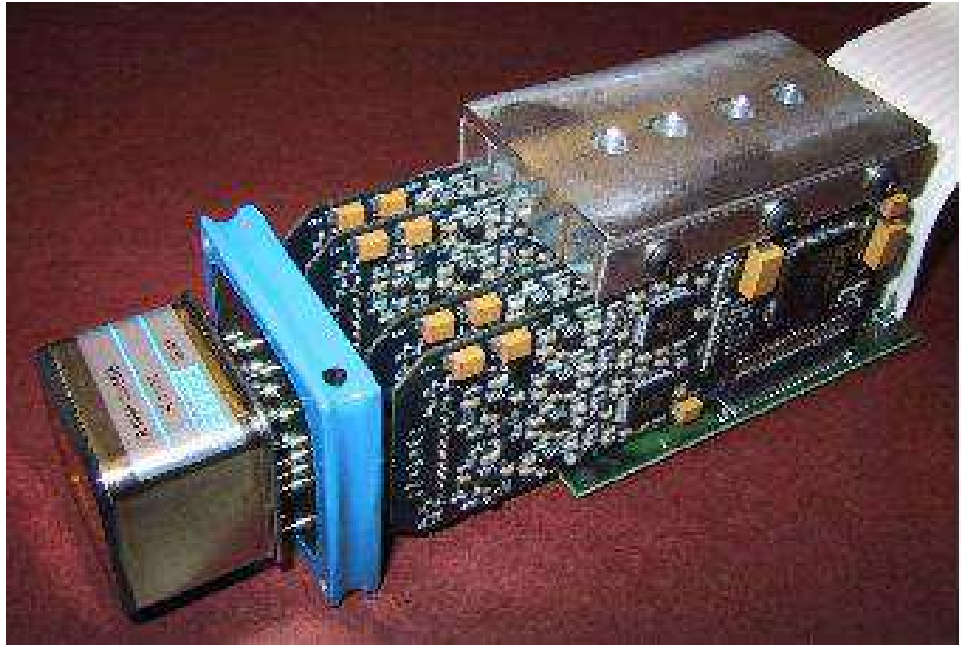}
\caption{\it MAPMT (16-channel), Cockcroft-Walton base, and FEE assembly 
  developed at IUCF for the STAR experiment at BNL.}
\label{fig:star_fee_assy}
\end{figure}

Indiana University Cyclotron Facility personnel have significant
recent experience designing and constructing a 9216~channel MAPMT
readout system for the STAR Endcap Electromagnetic Calorimeter
shower-maximum detector~\cite{StarElectronicsNIM}.
Figure~\ref{fig:star_fee_assy} shows a picture of the R5900-00-M16
MAPMT together with the Cockcroft-Walton base and front-end
electronics. The front-end electronics for STAR uses switched-reset
gated integrators and 12-bit pipeline A/D converters to provide a
12-bit integrated amplitude measurement for every channel for every
bunch crossing time (period of $\approx$~103~ns) in the RHIC ring.
Data is stored until an external trigger decision, and then all 9216
channels are read out in less than 20~$\mu$s. The readout is modular
in groups of 192 channels (12 MAPMTs); the modules being enclosed in
rugged, magnetically shielded, water-cooled steel boxes which bolt to
the STAR magnet poletip. While in functional details the system
proposed here differs considerably from the STAR readout, many aspects
of the design (such as the pipeline A/D converter and FPGA circuits,
the conductively-cooled printed circuit board designs, the choice of
connectors and other components, and test and diagnostic procedures)
will either carry over directly or be informed by the lessons learned
in the STAR readout work.

\subsection{Calibration}
\label{sec:calibration}
The FINeSSE calibration system must calibrate the response of the
MAPMTs of the fiber network to light from charged particles in the
scintillator.  It must also calibrate the energy response of the
tracking detector.  The relative gains of the muon range stack
scintillator strips must also be understood.

To calibrate the response of scintillating fibers and MAPMTs, FINeSSE
will utilize cosmic ray muons and their Michel decay electrons.  The
energy distribution of Michel electrons is well defined and provides a
``standard candle'' for the calibration.

The veto fibers alone can be used to trigger on through-going muons.
By using the veto and tracker sections, a cosmic Michel decay trigger
can be set up.  The cosmic muon events will put a known amount of
light into the detector volume, with the geometry of the tracks
constrained by the veto section, allowing accurate checks of light
production by muons in the detector.  The Michel events will provide
enough low level light to calibrate the charge response of the fibers
and photomultipliers.  They will also provide a good sample for
$\frac{dE}{dx}$ reconstruction calibration.

The muon range stack will also be calibrated using cosmic ray muons.
The primary goal is to calibrate the relative gains and efficiencies
of the scintillator strips.  This should be easily achievable with the
high rate of cosmic muons that FINeSSE will see.  The calibration
procedures using muons and Michels will be developed using Monte Carlo
events.

\section{Detector Location and Enclosure}

\begin{figure}[t]
\centering 
\includegraphics[width=\textwidth,bb=74 225 539 571]{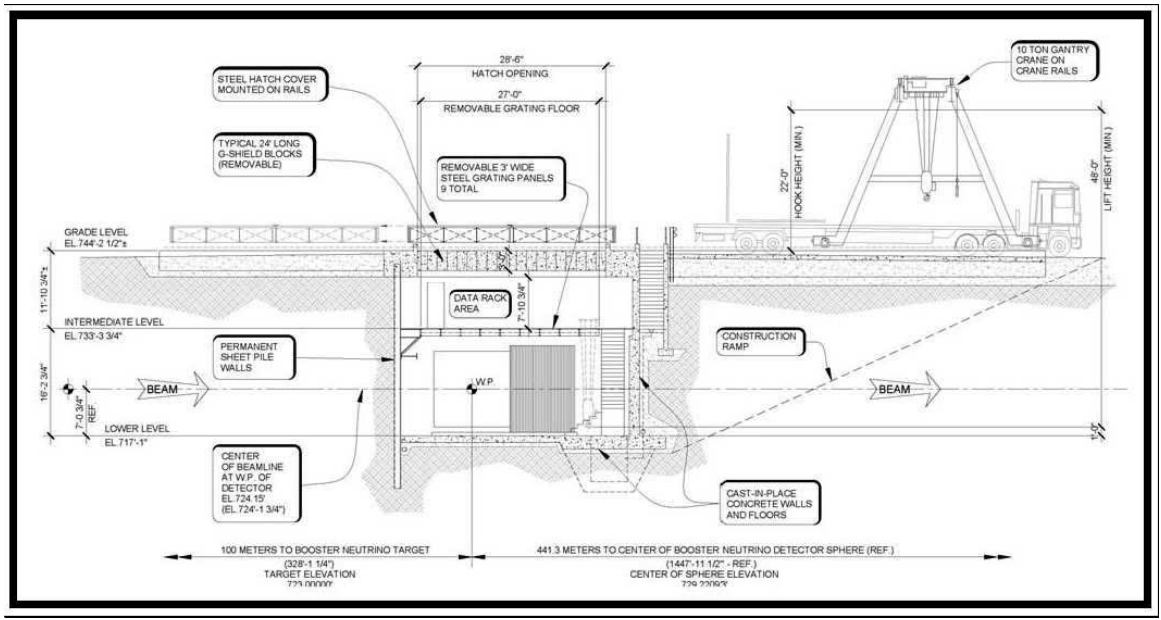}
\caption{\it Side view of the 25~ft below grade detector enclosure to house the FINeSSE detector.}
\label{fig:enclosure}
\end{figure}

\begin{figure}[t]
\centering 
\includegraphics[width=4in.,bb=74 108 541 642]{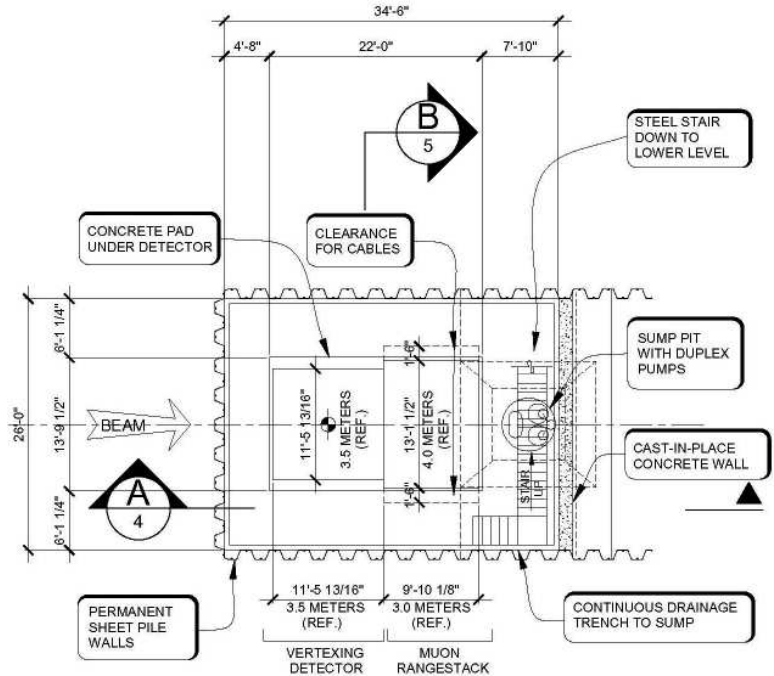}
\caption{\it Top view of the detector enclosure.}
\label{fig:top}
\end{figure}

The FINeSSE detector will be housed in a below ground enclosure 100~m
along the line connecting the Booster neutrino production target and
the MiniBooNE detector.  Figures~\ref{fig:enclosure} and
~\ref{fig:top} show a side and top view of the detector within the
enclosure.  A detailed study, cost estimate, and schedule for this
enclosure, performed by FESS, can be found in
Appendix~\ref{ch:AppendixC}.

The detector is installed using a gantry crane, above the enclosure,
to move detector components from a truck, down through a removable
hatch to the enclosure floor.  The hatch can be rolled in and out of
position on a daily basis to accommodate changes in weather and
therefore in the installation schedule.  Once installation is
complete, shield blocks are installed above this to reduce cosmic ray
rate in the detector.

The detector is accessible via a staircase with landings at a floor
above the detector and at the bottom of the enclosure.  The floor
above the detector holds electronics racks and computing needed for
readout, triggering and monitoring.  Ethernet from both detector
sections is fed through the removable floor to these racks.

Temperature control is minimal as per the requirements of the
detector.  Temperature of the Vertex Detector is controlled by a
standard oil temperature control system like that used in the
MiniBooNE experiment~\cite{miniboonetdr}.  Temperature control of the
electronics racks is done by enclosing them in a caboose and cooling
or heating the caboose as necessary.

This detector enclosure is minimal, but it is practical and fulfills
our needs.  We have studied the feasibility of a below-ground
enclosure in detail and in conjunction with FESS.  A similar enclosure
constructed of sheet piling with a removable roof housed experiments
at PC4.  A senior FINeSSE scientist examined this PC4 enclosure for 
toads, raccoons, and mice, and concluded this design will be sufficient.

\section{Detector Fabrication and Installation}
\label{sec:mechanicalconstruction}

Detector components will be fabricated at Fermilab and FINeSSE
collaborating universities.  Components will be assembled at a staging
area before installation in the detector enclosure.  Lab E or the New
Muon Lab at Fermilab would be a suitable place for this staging area.
Because the enclosure is below ground, the goal is to minimize the
actual construction at the site.  Subdetectors are constructed in
large ``packages'' with outer protection.  What follows is a
description of fabrication and assembly for the Vertex and Rangestack
components of the detector.

\subsection{Vertex Detector Assembly}

The detector assembly is constructed as a unit, at the FINeSSE staging
area on site.  Once assembled, it is moved to the FINeSSE enclosure
and craned into position.  Engineering of the transport is underway.
Finally, the scintillator tank is filled with scintillator oil.
The Vertex Detector assembly is described below.

The main component of the detector is a cubic structure of six
identical grid panels.  Each of these is comprised of a stainless
steel angle frame and an 80$\times$80 array of interlocking thin
stainless steel ribs.  These ribs are notched and positioned at an
equal spacing of 3~cm, welded together at each intersection and to the
frame around the perimeter.  Each grid panel is constructed with the
ribs centered with respect to the frame in one direction, and offset
by 1.5~cm in the other.  Mounting fixtures position the panels so that
alignment may be made to assure that positions of the grid
intersections are accurate and orthogonal throughout the cube.  The
offset directions of the panels are rotated so that any one direction
of grid and the resulting fiber array is centered with respect to the
other two directions.

\begin{figure}
\centering
\includegraphics[width=4.in]{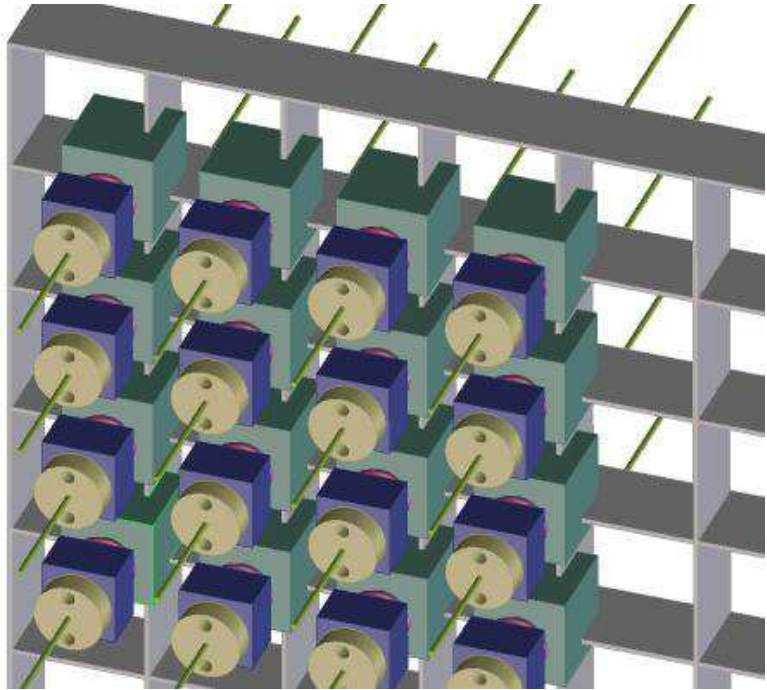}
\caption{\it Supporting grid for the WLS fibers. Shown are the riders (slotted
cubes) and the locking mechanism, which is pushed away from the rider
by a spring.}
\label{fig:supgridout}
\end{figure}

\begin{figure}
\centering
\includegraphics[width=4.in]{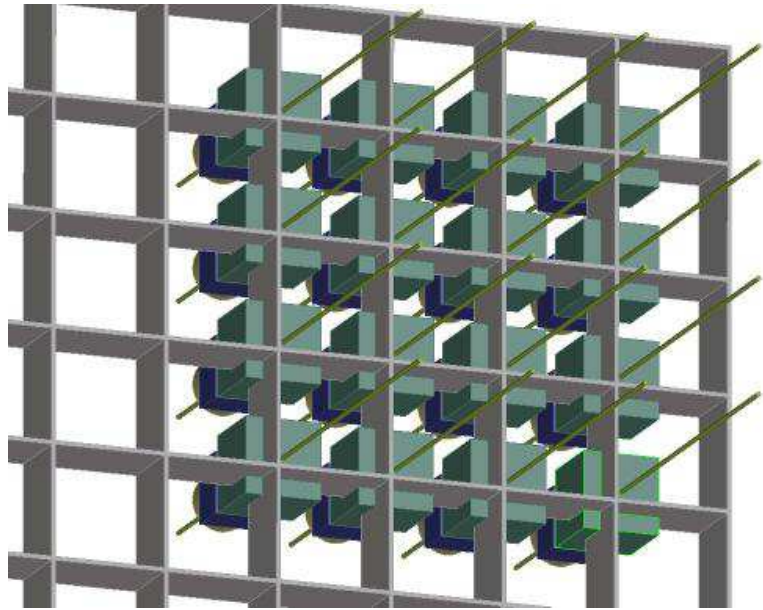}
\caption{\it Same as Fig.~\ref{fig:supgridout}, but seen from the inside of the fiber volume.}
\label{fig:supgridin}
\end{figure}

The fibers are held in place by riders on the intersections of the
grid (see Figs.~\ref{fig:supgridout} and~\ref{fig:supgridin}).  The
riders are spring-loaded, providing a tension of 1~N per fiber. The
deformation by the combined pull of 6400 fibers of a $2.5$m by $2.5 $m
grid plane, supported on the edges, is about $1$mm in the center.

The fibers will be mounted in the completed support structure (after
all six sides of the structure have been joined to form a cube). For
mounting, an insertion tool is needed that can hand a fiber end to the
opposite side of the cube, reaching through a partly occupied grid
cell. It is conceivable to do this with a group of fibers
simultaneously.  Once the fibers are available at both sides, the
riders are put in place. Once set up, we estimate that the mounting
takes on the order of a minute per fiber. This amounts to about 10
weeks (assuming 8 hours/day and 5 days/week) for the fiber assembly.

The weight of the support structure is about 2 tons. This does not
include the 330 MAPMTs with associated
electronics, which are mounted on the outside of the fiber support
structure. When routing the fibers from the riders to the tubes one
must take into account that the minimum recommended bending radius for
1.5~mm diameter fiber is 12 cm. The section of fiber outside the support
lattice and the tubes are encased in light-tight Teflon sleeves.

Once fiber stringing is complete, the electronics module support is
put in place, the modules are mounted and PMTs plugged in, and each of
the 300 'cookies' are locked to their PMT with spring loaded
mountings.  Cables suitable for submersion in scintillator oil are
connected to the electronics modules.

The completed fiber structure will be lowered into a tank which will
eventually contain the liquid scintillator.  This outer tank provides
a cubic volume, 3.5~m on its side. The fibers exit the top of the
tank.  Equipment to detect light from the veto shield, i.e., the
volume outside the fiber cube is mounted along
the walls of the tank.

The assembly is protectively enclosed and transported to the
experimental location by a truck able to move 5~tons with a maximum
tilt of 5$^\circ$ and acceleration of 0.5~g.  The onsite crane removes
the detector from the truck and lowers it in place where fixturing
assures its proper position.  The hatch to the detector hall is then
closed.  The remaining veto area monitoring equipment is installed and
cables from all electronics modules are routed to feedthrough panels.
After all electrical connections are tested, the lid is put in place
and the detector is filled with scintillator oil.

The system for circulation and temperature control of scintillator is
then put into operation, as is the system for maintaining the nitrogen
atmosphere in the tank above the scintillator.  The liquid is
circulated by feeding from the top back to outlets distributed around
the bottom of the tank. The circulation is necessary to control the
temperature of the liquid, and to de-oxygenize the liquid scintillator
initially (resulting in a gain of about 25\% in light yield). The tank
has a gas-tight lid, in order to maintain an oxygen-free atmosphere
above the liquid. The cables from the electronics readout modules in the liquid
exit the tank through gas-tight ports above the liquid level.

Temperature of the submerged PMTs and electronics is controlled by a
system which circulates and cools the oil.  Additional PMTs and
electronics monitor a veto region surrounding the six sides of the
detector active volume.  Electronic cables are routed to feedthrough
panels along the upper surfaces of the walls of the tank above the oil
level, where a clean nitrogen atmosphere is maintained during the
operation of the system.

\subsection{Muon Rangestack Assembly}

For the Muon Rangestack, we must consider fabrication of the steel
absorber and the scintillator hodoscopes.  Fabrication of the pieces
will occur at a Fermilab facility away from the detector site.  Then
the modules will be transported to the detector hall.  

The steel will be procured by competitive bid from commercial machine
shops and steel vendors.  The absorber will have to be welded together
from smaller readily available plate sizes, then cut to final size.
No attempt has been made here to show the plate segmentation, as that
will undoubtedly be determined by each steel vendor and the exact
configuration of weld seams is irrelevant to the operation of the
detector.  Finished plates would be shipped to FNAL by truck.  The
route will have to be planned carefully, as will the fixturing on the
truck, as these will be oversize loads.  Calculation has shown that the
plates can be rotated from a horizontal to a vertical position by
pivoting them on the lower edge without exceeding the yield strength
of the steel.  However, once scintillator cans have been installed, a
lifting/rotating fixture will probably be needed.  Section 4 filler
modules weigh 5.72~tons, and are the heaviest single pieces to lift.
This puts a lower limit on the building crane in the assembly area at
6~tons, probably 10~tons when fixtures are included.

There are 2112 scintillator bars to be extruded, cut to
length and staged.  These could be extruded at
Fermilab~\cite{alanbross}.  After cutting and finishing the ends, WLS
fiber of the appropriate length will be laid in the groove and glued
to the bar.  The fiber gluing operation can be done after the bars
are glued into the can bottoms.  Large tables at least 4.6~m long will
be used for curing the glue between the bars and fibers, and cans
and bars.  With staging areas, the floor space available should be at
least $10\times10$ m$^2$.  Machines for mixing glue, applying it, and
laying the fibers will be needed.

The MINOS TDR showed the aluminum cans which will hold the
scintillator as made on-site from coils of 3003 aluminum sheet 0.040
inches thick~\cite{MINOspdf}.  A rolling machine was used to flatten
the coils, and then to put corners along the edges used for crimping
the lower and upper sheets together.  We will have to examine the
economics of this process, as having finished pieces delivered from a
vendor may be more cost and time effective.  The 4~m long pans are
no problem for delivery by truck.  132 bottom cans and 132 top plates
will be made.

The PVC spacers and fiber routing manifolds are very straightforward
machined parts that can be fabricated at any machine shop equipped
with CNC milling machines.  PVC is stable after machining and
relatively cheap.

Clear fiber bundles with their connectors must also be assembled,
glued, and finished.  Care must be taken to appropriately label fibers
to assure that connections between scintillator cans and PMTs are
correct.

Assembly of the can proceeds as follows.  The set of scintillator
strips are glued to an aluminum skin to hold them fixed, and then
placed within a can.  The WLS fibers are routed through grooves milled
into a PVC plate at the end of the can.  The plate forces the fibers
into their recommended 12~cm bend radius and prevents them from being
damaged during handling.  The plate is glued into the can and
light-sealed with a cover plate of PVC plastic.  An optical connector
is also attached to the grooved PVC plate.  This connector joins the
WLS fiber to clear fibers that go to the PMT enclosures.  The
construction technique follows Chapter 5 of the MINOS
TDR~\cite{MINOspdf}, with the aluminum skins crimped at the edges to
enclose the scintillator.

\begin{figure}
\centering
\includegraphics[bb=20 20 800 450,width=4.in]{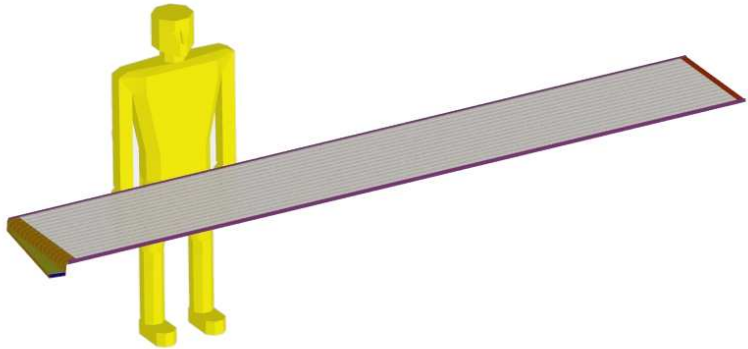}
\caption{\it A can of scintillator strips with WLS fibers and enclosing PVC and
aluminum sheet weighs just under 100 lbs and is approximately 4.3 m long
including the PVC plates for protection of the fibers out to the optical
coupling.}
\label{fig:scistack-data-03}
\end{figure}
\begin{figure}
\centering
\includegraphics[bb=73 160 540 640,width=3.in]{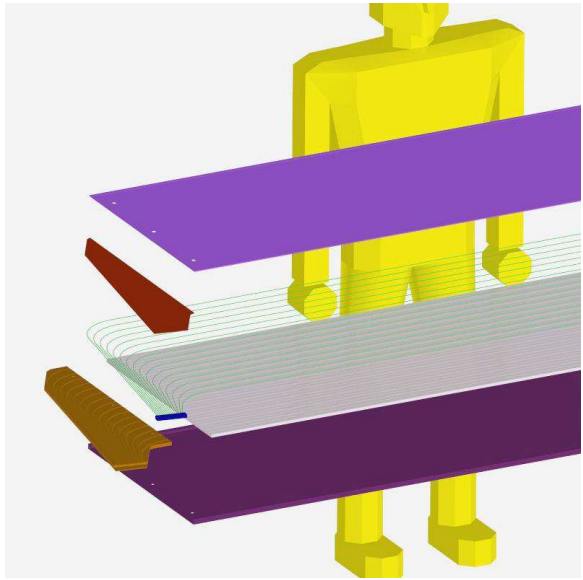}
\caption{\it An exploded view of the end of the can that routes fibers to the optical
coupling.}
\label{fig:scistack-data-04}
\end{figure}

An exploded view of the fiber routing components is shown in
Figure~\ref{fig:scistack-data-04}.  Figure~\ref{fig:scistack-data-03}
shows a rendering of a can without the final light-tighting layer of
aluminum over the strips.  The PVC cover is also not shown.  The
schematics do not show the clear fiber bundles, the PMT enclosures,
and the clear fiber.

\begin{figure}
\centering
\includegraphics[bb=73 160 540 640,width=3.in]{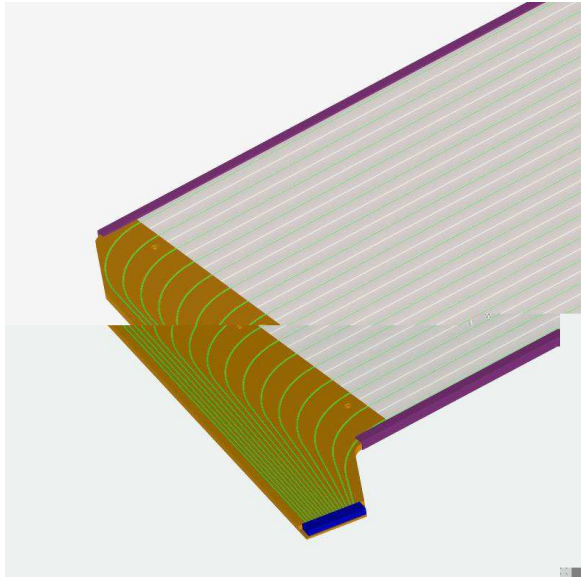}
\caption{\it A close-up of the fiber routing plate at the end of the
scintillator can.  Not shown are the cover plates that light-tight the
assembly.}
\label{fig:scistack-data-02}
\end{figure}

\begin{figure}
\centering
\includegraphics[bb=73 227 540 515,width=4.in]{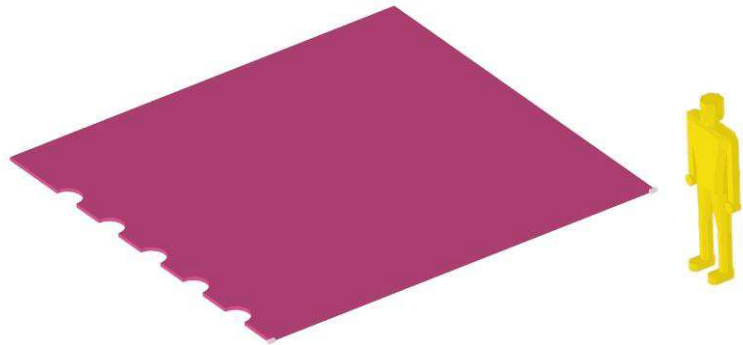}
\caption{\it A typical filler module for section 4.  Not shown on any of the module pictures
are lifting lugs to place the modules in the stack.  The ears that allow the plates to
hang from the support structure are welded on tabs.}
\label{fig:scistack-data-09}
\end{figure}

\begin{figure}
\centering
\includegraphics[bb=73 190 540 640, width=3.in]{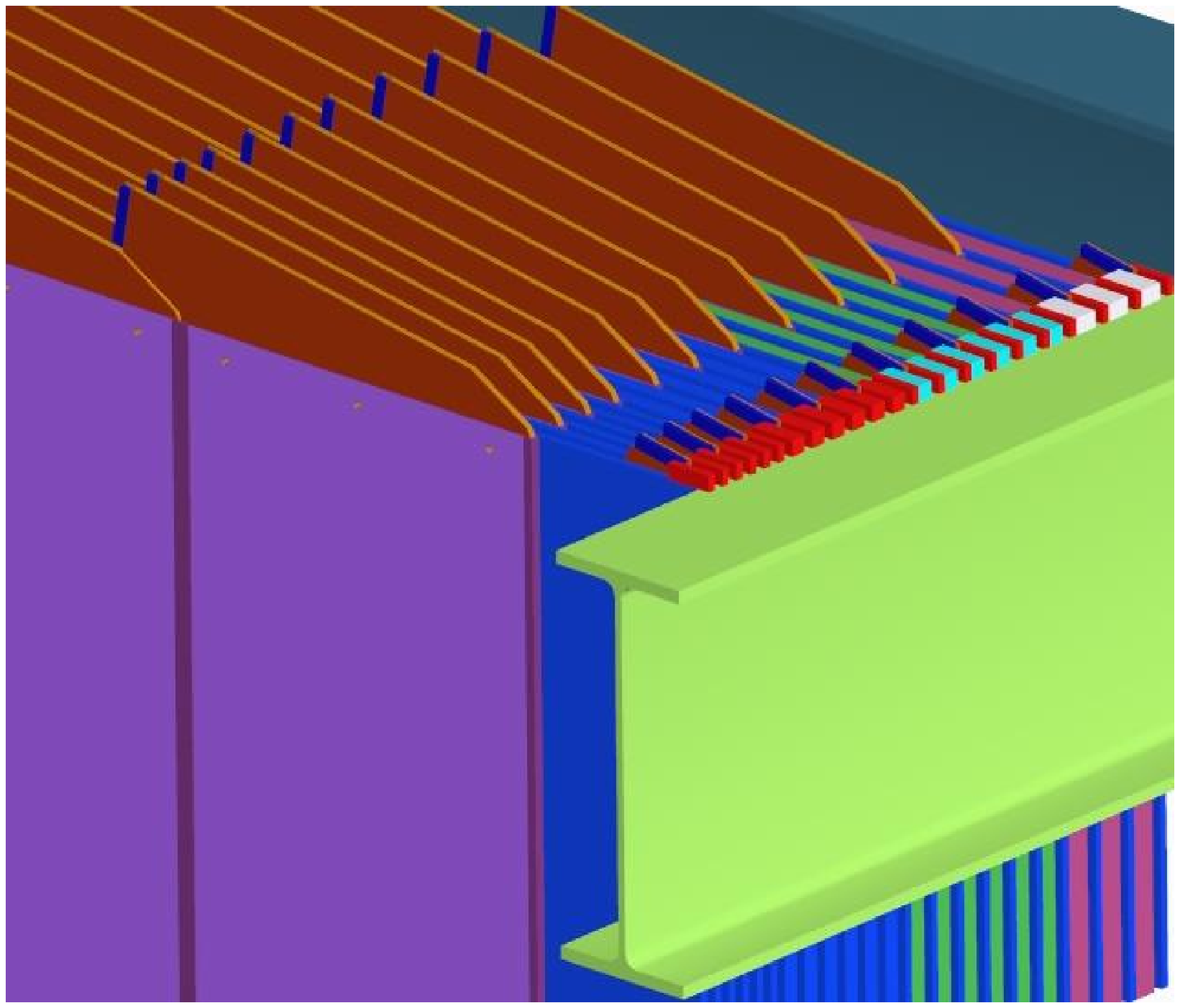}
\caption{\it Close-up of the support rail showing the "ears" on the steel plates that
allow the plates to hang from the rails.  The column and cross member at this corner
were turned off to show more details.}
\label{fig:scistack-02}
\end{figure}

Once scintillator cans are complete, their light tightness and other
performance features may be tested.  This is one of the big advantages
of the 16-fold packaging of the scintillator.  Cans are much easier
units to handle for movement around the assembly area and for testing
than full planes.

Finished cans will be attached to 0.50 inch protective steel backing
plates at the assembly factory.  Fixtures for vertical storage will be
necessary.  Finished scintillator planes weighing 2.2~tons and being
about 15 feet square will be trucked to the detector installation
site.

\begin{figure}
\centering
\includegraphics[bb=73 190 540 640, width=3.in]{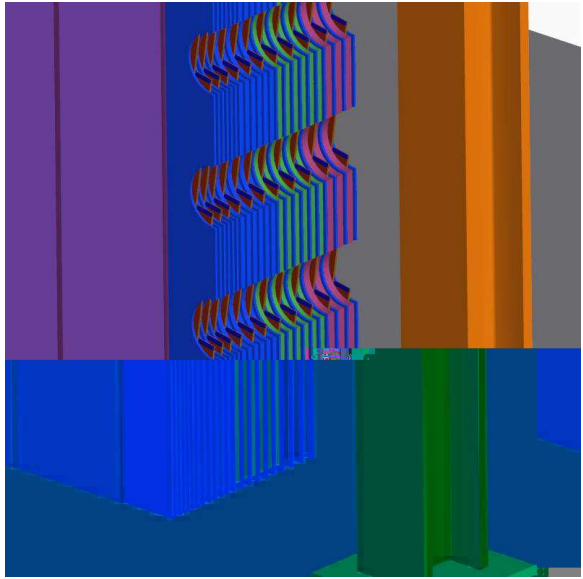}
\caption{\it Close-up view of the changing thickness of the layers from section to section,
  and the scallops allowing access to the optical connectors on the
  $y$ planes.  This face will probably also have structure to support
  PMT boxes and catwalks for maintenance access.}
\label{fig:scistack-data-01}
\end{figure}

At the detector site, each module sits between two support rails upon
which the Rangestack hangs, as indicated by
Figure~\ref{fig:scistack-02}.  The rails must be far enough apart to
clear the ends of the scintillator can optical connectors.  The
protective steel backing, which is wider than the scintillator cans,
is attached to each module.  The steel backing is designed to protect
the optical connectors during lowering into the stack.  These plates
are designed with scalloped edges to allow access to the clear fiber
bundle, connector, and fastener exiting the side of the detector ($y$
view; Figure~\ref{fig:scistack-data-01}) once the module is in place.

Because $4\times 4$ m$^2$ plates are larger than commercially available plate
stock, this design assumes that a 4~m square plate will be welded together from
smaller plates, and water jet or plasma cut to final shape.  Figure~\ref{fig:scistack-02}
shows a closeup of the ears on the steel plates that support the scintillator and
filler modules.  Figure~\ref{fig:scistack-data-01} shows a closeup of the side
of the detector with the $y$ optical connectors and the scalloped steel edge.

Once all scintillator planes and absorber fillers are completed, they
will be transported to the detector site.  MI-12 or MI-8 will be used
for temporary storage at the site.  For installation, each module will
be made vertical (if they are not already on the truck) and lowered
into the detector pit to hang on the two support rails.  The
scintillator planes will require a simple beam lifting fixture to
prevent excessive deflection during crane handling operations.

After installation of planes, the surrounding PMT boxes and support
structure will be assembled.  Clear fiber bundles will then be
connected and testing with cosmic rays can begin.  This can proceed
with the building roof re-installed.

\chapter{Event Simulation, Reconstruction, and Analysis}
\label{ch:EventSimulationandReconstruction}

\thispagestyle{myheadings} \markright{} The detector and physics events have
been simulated and reconstructed to determine the feasibility,
strategy, and sensitivity of the experiment for its two main physics
goals: a measurement of $\Delta s$ and $\nu_\mu$ disappearance.  In
this chapter the general features of the neutrino interactions and
tracks in the detector are described.  Next, the detector simulation
and reconstruction programs and strategies are explained.
Backgrounds, both beam-related and beam-unrelated are then discussed.
In the final part of the chapter, the methods, tools, and estimated
sensitivity are reported for the $\Delta s$ and $\nu_\mu$
disappearance measurements.

\section{Interactions in the Detector}
The event rates expected in the FINeSSE detector are listed in
Table~\ref{table:100m-25mabs}.  These events will need to be
identified, counted, and measured in order to do the desired physics.
The detector as described in Chapter~\ref{ch:TheFINeSSEDetector} will
have the characteristics needed to do that.  The most important and
most prevalent events that the detector will see  are listed in
Table~\ref{tab:evsum}, along with a description of how they will
``look'' in the FINeSSE detector.
\begin{table}[h]
\centering
\begin{tabular}{llcp{2.5in}}
\hline
event name & reaction                    & \#~tracks 
& description                     \\
\hline
CCQE       & $\nu n \rightarrow \mu^- p$      & 2 & $\mu$,p: two-body kinematics  \\
NCp        & $\nu p \rightarrow \nu p$        & 1 & p: two-body kinematics  \\
NCn        & $\nu n \rightarrow \nu n$        & 0 & n: extraneous visible tracks from np scattering \\
CC$\pi$    & $\nu n \rightarrow \mu^- X \pi^{\pm,0}$ & $> \approx 2$ & Not two-body kinematics  \\
NC$\pi$    & $\nu p,n \rightarrow p,n X \pi^{\pm,0}$ & $> \approx 1$ & Not two-body kinematics  \\
\hline
\end{tabular}
\caption{\it Summary and description of event types that the FINeSSE detector will see. 
``\#~tracks'' means typical number of charged particle tracks of significant energy.}
\label{tab:evsum}
\end{table}

\begin{figure}
\centering
\includegraphics[width=4.in]{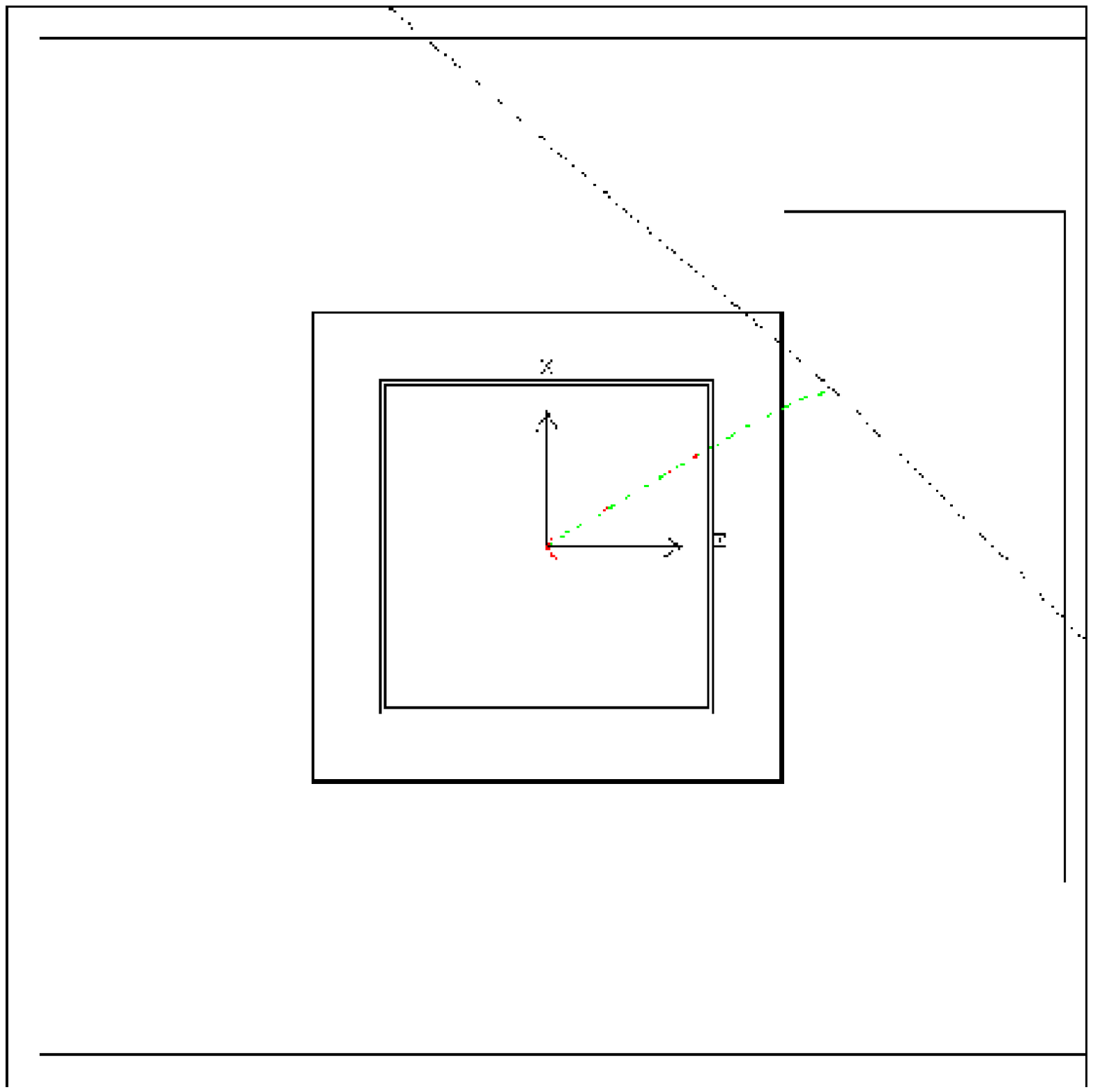}
\caption{\em  A typical $\nu n \rightarrow \mu^- p$ reaction 
in the FINeSSE detector. The event vertex is at the origin of the indicated coordinate system. In this 
event the $\mu^-$ leaves the Vertex Detector, stops in the range stack, and decays.  The two neutrinos
from the muon decay are seen exiting the apparatus.  The short recoil proton track is just 
visible at the origin.} 
\label{fig:qe_1.eps}
\end{figure}

\begin{figure}
\centering
\includegraphics[width=3.in]{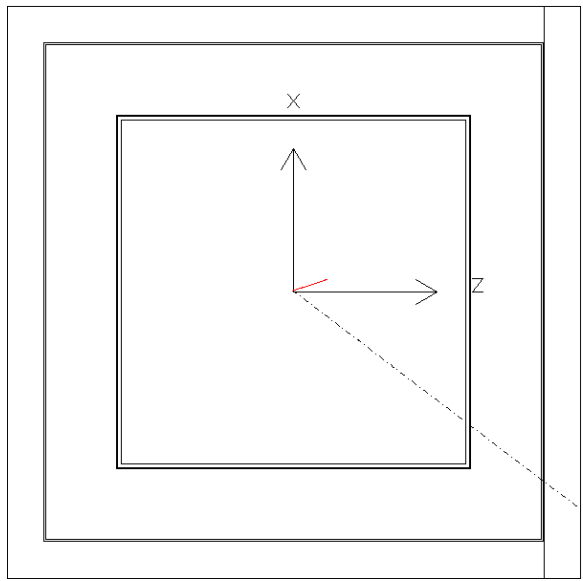}
\caption{\em  A typical $\nu p \rightarrow \nu p$ reaction expected 
  in the FINeSSE detector, generated at the origin.  The short proton
  track can be seen, as can the final state neutrino which exits the
  apparatus. In this view, only the Vertex Detector is shown.}
\label{fig:ncel_p_1.eps}
\end{figure}

\begin{figure}
\centering
\includegraphics[width=3.in]{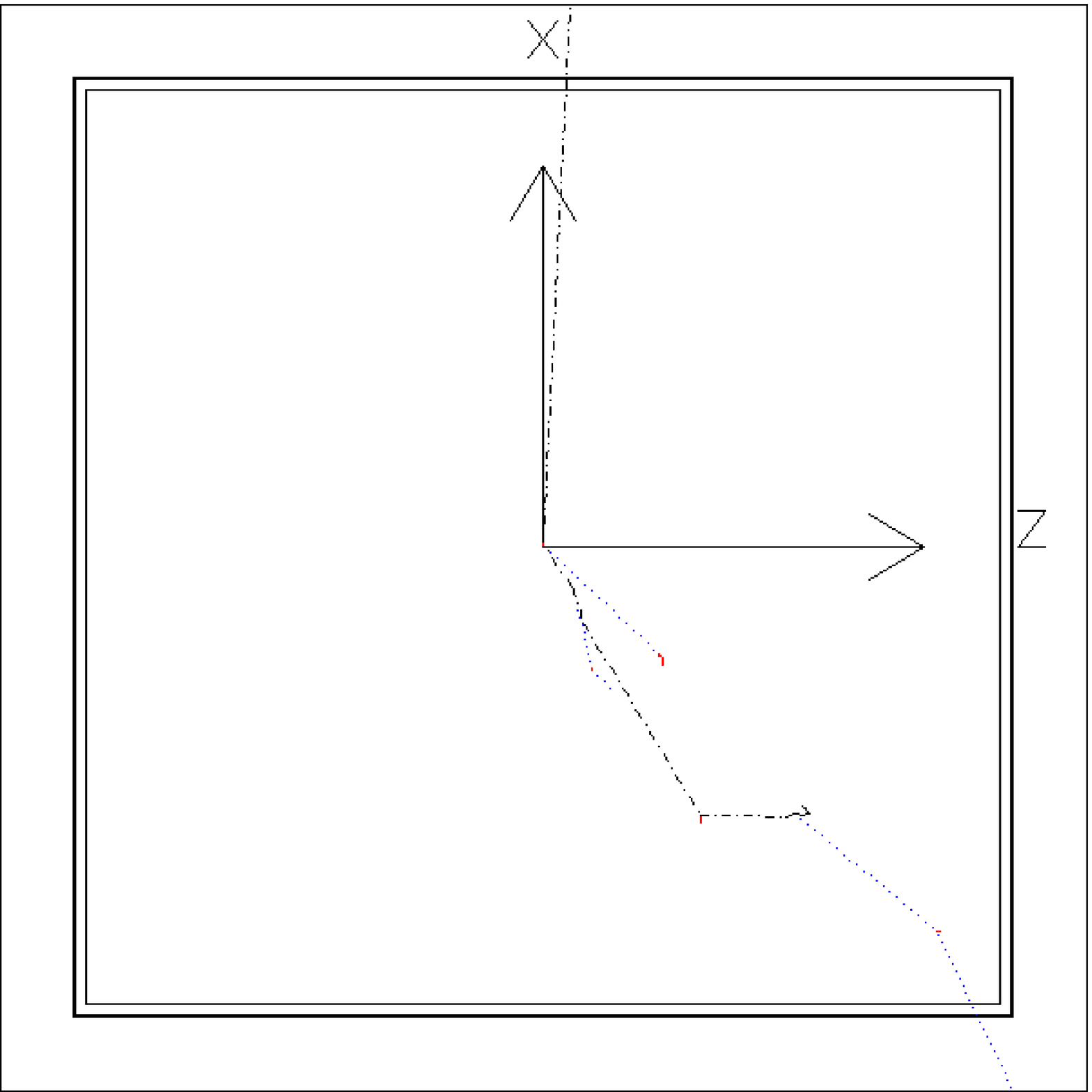}
\caption{\em  A typical $\nu n \rightarrow \nu n$ reaction expected 
  in the FINeSSE detector. Several interactions of the final state
  neutron can be seen.  The final state neutrino exits the apparatus.
  In this view, only the Vertex Detector is shown.}
\label{fig:ncel_n_1.eps}
\end{figure}

The first three reactions are the most important for the main physics
goals of FINeSSE.  Examples of typical events of these reaction types
as simulated with GEANT are shown in
Figures~\ref{fig:qe_1.eps}-\ref{fig:ncel_n_1.eps}.  In these figures,
charged hadrons are shown as solid lines, muons as wide dashed lines,
neutrinos and neutrons as dot-dashed lines, and photons as dotted
lines.  The detector is shown from the side.

The individual particles within a particular event will be identified
via their track length and energy loss density, ``$dE/dx$'', as well
as their decay patterns:
\begin{itemize}
\item $\mu^\pm$: Long tracks with low $dE/dx$.  High-energy $\mu^\pm$
  will exit the Vertex Detector and enter the veto and perhaps the
  range stack. For $\mu^\pm$ that stop in an active area, the decay
  (Michel) electron will be observed.
\item $p$: Short tracks with high $dE/dx$. A 100 MeV proton travels
  approximately 10cm in liquid scintillator.
\item $n$: Extraneous tracks from $\nu p$ scattering.  Occasionally
  transfer enough energy in one collision so as to be misidentified as
  a $p$ track.  Will thermalize and capture in the detector yielding a
  delayed 2.2 MeV $\gamma$.
\item $\pi^\pm$: Longish tracks that look like $\mu^\pm$. For $\pi^\pm$
  that decay in the active area of the detector, the subsequent
  $\mu^\pm$ and $e^\pm$ can be observed.
\item $\pi^0$: 2 hit clusters from the $\pi^0$-decay $\gamma$
  showers.
\item $e^\pm$: 1 ``fat'' track from the $e^\pm$ shower.
\end{itemize}

\subsubsection{Event Kinematics}
\begin{figure}
\centering
\includegraphics[width=5.in]{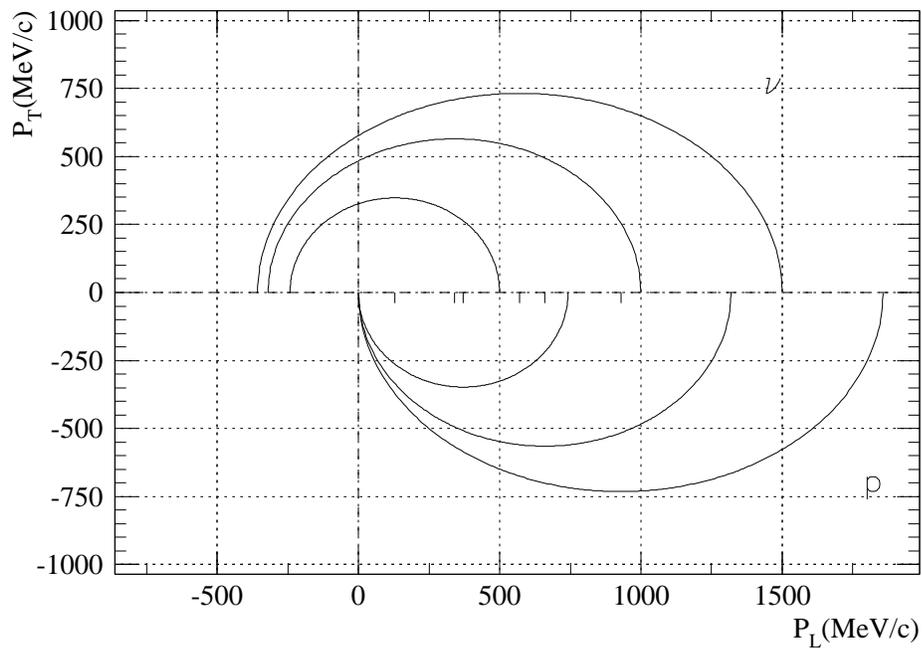}
\caption{\em Kinematic ellipses for the NC elastic 
  ($\nu p \rightarrow \nu N$) reaction. These are also valid for the
  CCQE ($\nu n \rightarrow \mu^- p$) reaction. The circles (ellipses)
  of increasing size indicate the CM (lab) momentum for the event at
  $E_\nu=$500, 1000, 1500~MeV. The longitudinal component (parallel to
  the beam) of the particle momentum is plotted on the x-axis, and the
  transverse component on the y-axis.  In a particular event, the
  particle momentum vector is constrained to lie on the appropriate
  ellipse.}
\label{fig:ncells}
\end{figure}

The two-body kinematics of NC elastic ($\nu N \rightarrow \nu N$)
and CCQE ($\nu n \rightarrow \mu^- p$) interactions are shown
schematically with the use of the kinematic ellipses in
Figure~\ref{fig:ncells}.

Due to the energy of the incident neutrino and the relatively low mass
of the muon, the kinematics of the NCp and CCQE events are almost
identical above $E_\nu \approx 300$ MeV.  This simplifies the analysis
and interpretation of the NC and CCQE event reconstruction.  As can be
seen in Figure~\ref{fig:ncells}, the final-state lepton may have any
angle; the outgoing hadron has a maximum lab angle of 90$^{\circ}$.
The correlations can be seen in this figure as well.  An event with a
lower-energy high-angle lepton is paired with a low-angle high-energy
hadron (and vice versa).

\begin{figure}[tbh]
\centering
\includegraphics[width=5.in]{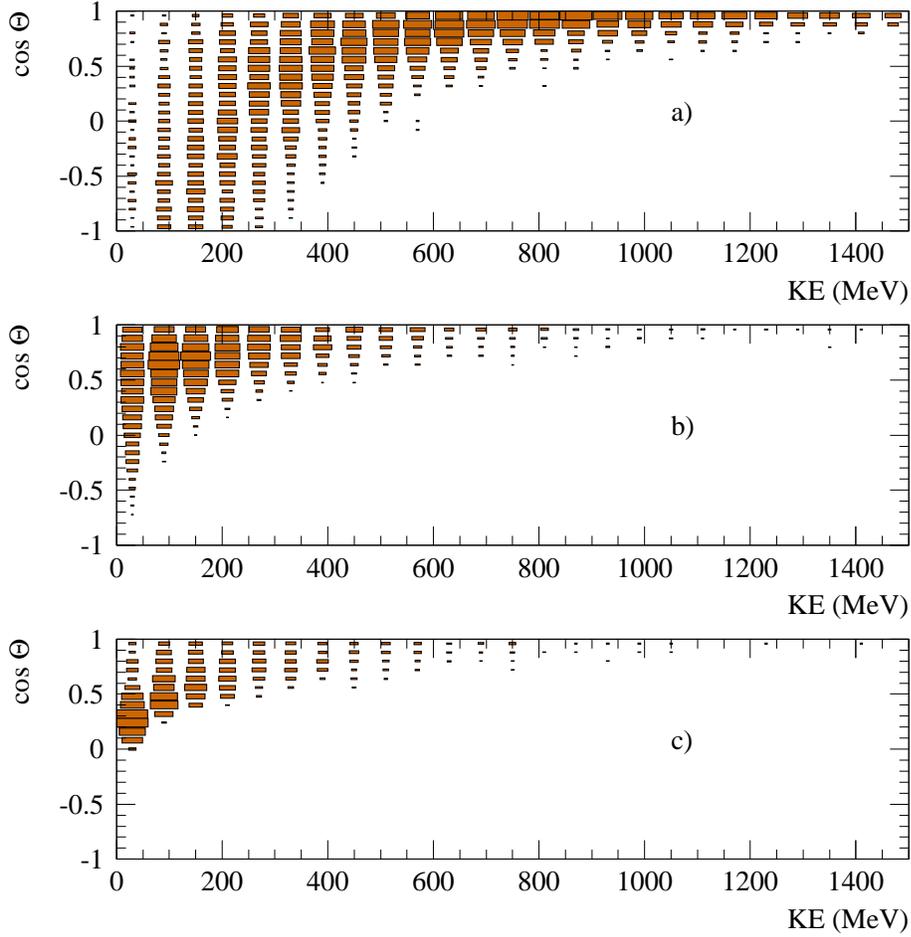}
\caption{\em Kinetic energy vs. $\cos \theta$ for the (a) $\mu$, 
  (b) proton in CCQE scattering from bound nucleons, and (c) proton in
  CCQE scattering from free nucleons. Protons in NCp reactions will
  show the same distribution as that in (b).  }
\label{fig:cckin}
\end{figure}

Plotting energy vs.~angle (Fig.~\ref{fig:cckin}) reveals strong
correlations between the two variables for such two-body reactions.
The events of most interest for the $\Delta s$ analysis are low $Q^2$
events, where the proton has a low energy and a high angle. The lepton
in these events will be in the forward direction at high energy. The
events of most interest for the oscillation analysis are $\nu_{\mu}$
CC interactions with outgoing muons whose kinetic energy lies in the
0.1 to 1.5 GeV range.  Many of these lower energy muons range out in
the Vertex Detector and veto; the most energetic ones, at small angles,
enter and range out in the Muon Rangestack.

The effects of Fermi momentum of the bound nucleons can be seen in
Fig.~\ref{fig:cckin}, which compares the final-state protons produced
in CCQE scattering from nucleons bound in carbon
(Fig.~\ref{fig:cckin}b) to those from free nucleon scattering
(Fig.~\ref{fig:cckin}c).  Fermi momentum widens the angular
distribution of the outgoing proton and suppresses the number of
nucleons at low momentum (``Pauli blocking'').  The effect of this
additional Fermi energy ($\approx 25$MeV) on the energy of the
outgoing proton is small and will have minimal impact on measuring the
$Q^2$ of the reaction (via $Q^2=2m_pT_p$).

The correlations shown in Fig.~\ref{fig:cckin} will be used to reduce
backgrounds from NC and CC single pion reactions by requiring that the
reconstructed tracks obey the illustrated kinematic constraints.

\section{Simulation of the Detector}
\begin{figure}
\centering
\includegraphics[width=5.in]{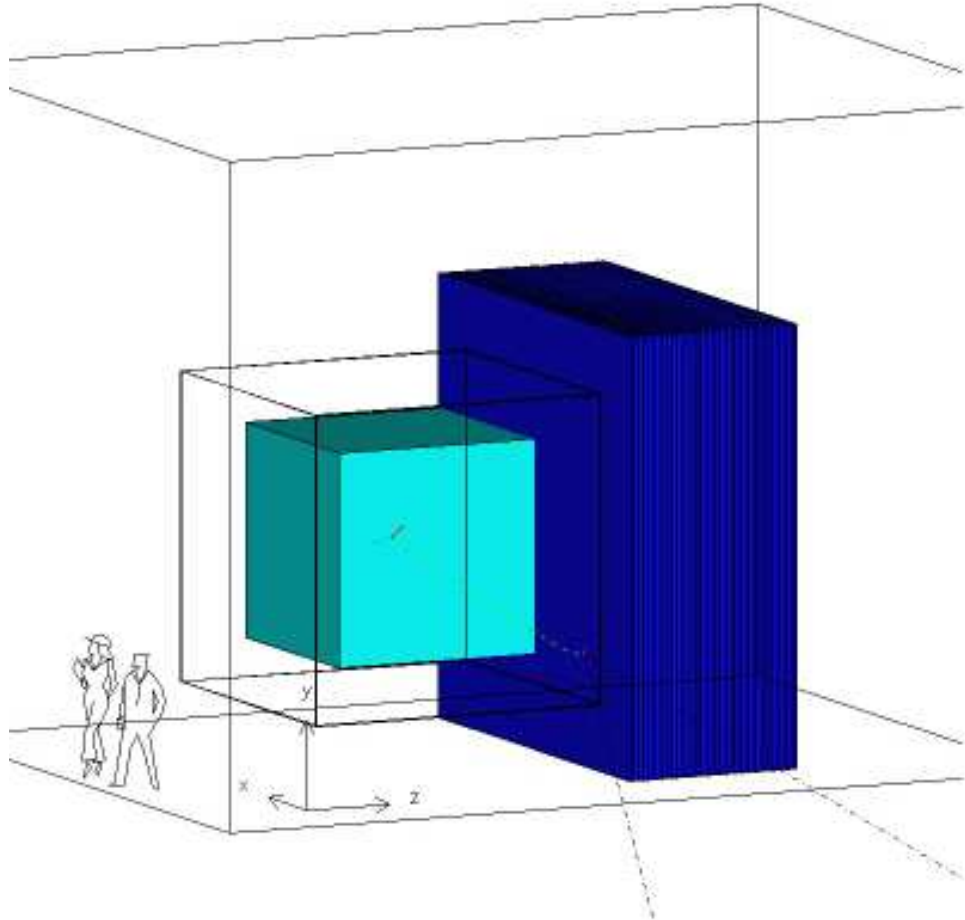}
\caption{\em Diagram of the detector geometry as simulated by GEANT with 
  a superimposed CCQE scattering event.}
\label{fig:GEANT_overview}
\end{figure}

The full FINeSSE detector, including the Vertex and the
Muon Rangestack subdetectors, has been simulated using the the
\texttt{GEANT}~\cite{GEANT3} simulation package. A diagram of the
apparatus as modeled by the program is shown in
Fig.~\ref{fig:GEANT_overview}.

The Vertex Detector is simulated as an (2.4m)$^3$ of liquid
scintillator with a 80 $\times$ 80 grid of embedded 1.5 mm fibers with
a 3.0 cm spacing in three orientations (XY, XZ, YZ;
Fig.~\ref{fig:fiberarr}) for a total of 19,200 fibers.  This inner
volume containing the fibers is embedded in a larger volume of
(3.5m)$^3$ of ($\rho = 0.85$ g/cm$^3$) liquid scintillator.  The fiber
support structure and tank walls are also included in the simulation.  The area outside
of the fiber area in the scintillator tank (the ``veto'') is also
assumed to be active.
 
The Muon Rangestack, downstream of the Vertex Detector, is implemented with
geometry as described in Section~\ref{sec:muonstack}, which consists
of alternating planes of scintillator and iron.

In the inner region, scintillation light and \v{C}erenkov radiation from
passing charged particles is simulated.  Photons thus produced are tracked
until they impinge upon a WLS fiber or the edge of the detector
volume, or are absorbed.  In the outer 50~cm (veto) region of the
liquid scintillator and in the Muon Rangestack, individual photons are
not tracked -- only energy loss is recorded.  However, this is not
important in these regions as photostatistics are not crucial.

In the active region of liquid scintillator, it is estimated that 5000
scintillation photons are produced in the liquid scintillator for
every 1 MeV of energy deposited by a charged particle~\cite{StGob}.
The absorption and capture efficiency of photons intersecting a WLS
fiber is estimated to be 5\%~\cite{StGob}. These fibers have typical
attenuation lengths of 2.5~m, and the quantum efficiency of the PMT is
approximately 20\%. As a result, approximately 10\% of the photons
emitted in the capture cone of the fiber will make it to the PMT and
produce a photoelectron.  Combining these two factors yields an
overall efficiency of 0.5\% that an optical scintillation photon that
strikes a WLS fiber will be detected at the PMT.  To aid the speed of
the simulation, the photon detection efficiency and production was
combined so that 25 ($5000\times0.005$) photons are produced per MeV.
An effective efficiency of 100\% for the photons that hit a 1.5~mm WLS
fiber was then assumed.  In this way, the effects of photostatistics
were properly simulated and the efficiency of the simulation was
kept high.  The attenuation length (5m)~\cite{StGob} of
the scintillator is fairly large compared to the size of the detector
and is not a significant effect.  The saturation due to large
localized energy deposits (``Birks' '' Law) is also modeled.  The
production of \v{C}erenkov photons is simulated but is negligible as
the number of \v{C}erenkov photons is only about 1\% of that for
scintillation.

Using these factors, the simulation predicts that a proton track 
track passing near a fiber will create on average 10 photoelectrons
in the PMT.  This is consistent with the prototype test results
reported in Chapter~\ref{ch:TheFINeSSEDetector} (factoring in the difference in fiber length).  Employing this method of tracking individual
optical photons in the Vertex Detector assures that the photo
statistics (with fluctuations) are properly simulated.

The simulation program can track single-particle events to study the
detector response for each particle type.  Alternatively, it can
accept event descriptions as generated by the NUANCE MC program as
described in Chapter~\ref{ch:TheNeutrinoBeam}.  The later class are
events that were used to predict sensitivities as described below.
The output of the detector simulation includes: a list of all the
``hit'' fibers in the Vertex Detector, a list of all the bars in the
Muon Rangestack that recorded energy loss, and the total amount of
energy and time of deposit in each area of the detector.  These data
are subsequently passed through the event reconstruction program.

\subsubsection{Event Reconstruction}
\label{sec:res}
The simulated event data from the Vertex Detector are analyzed with a
reconstruction program employing a Hough transform
technique~\cite{hough}.  The Hough transform is a global track finding
method that uses the hit fiber information from the XZ and YZ
orientations.  (The information from the XY orientation has not yet
been used.  This information can only improve the reconstruction.)
The coordinates of each fiber (that recorded an amount of light over
an adjustable threshold value) are used to calculate
\begin{equation}
R = X(Y)\sin \alpha + Z \cos \alpha~,
\end{equation} 
where $\alpha$ is a track angle and $R$ is the perpendicular distance
from the track to the origin.  The track angle $\alpha$ is varied in a
loop from $-90^{\circ}$ to $+90^{\circ}$, and the $R$ and $\alpha$
values for each hit used to make an entry (weighted by the amount of
light in the hit) in a histogram.

The task of track finding then reduces to finding ``peaks'' in this
histogram.  Finding single tracks is quite easy with this method.  For
events with multiple tracks, alternate methods had to be developed and tuned to
subtract the light from the first track before the algorithm was
employed to find subsequent tracks.  At present, the reconstruction
program is limited to finding a maximum of two tracks in each of the two
2D-orientations (XZ,YZ).

The 2D-tracks were then combined to form 3D-tracks.  The total
energy and length of each track was also calculated, from which
the $dE/dx$ of the track could be determined. 

\begin{figure}
\centering
\includegraphics[width=5.in]{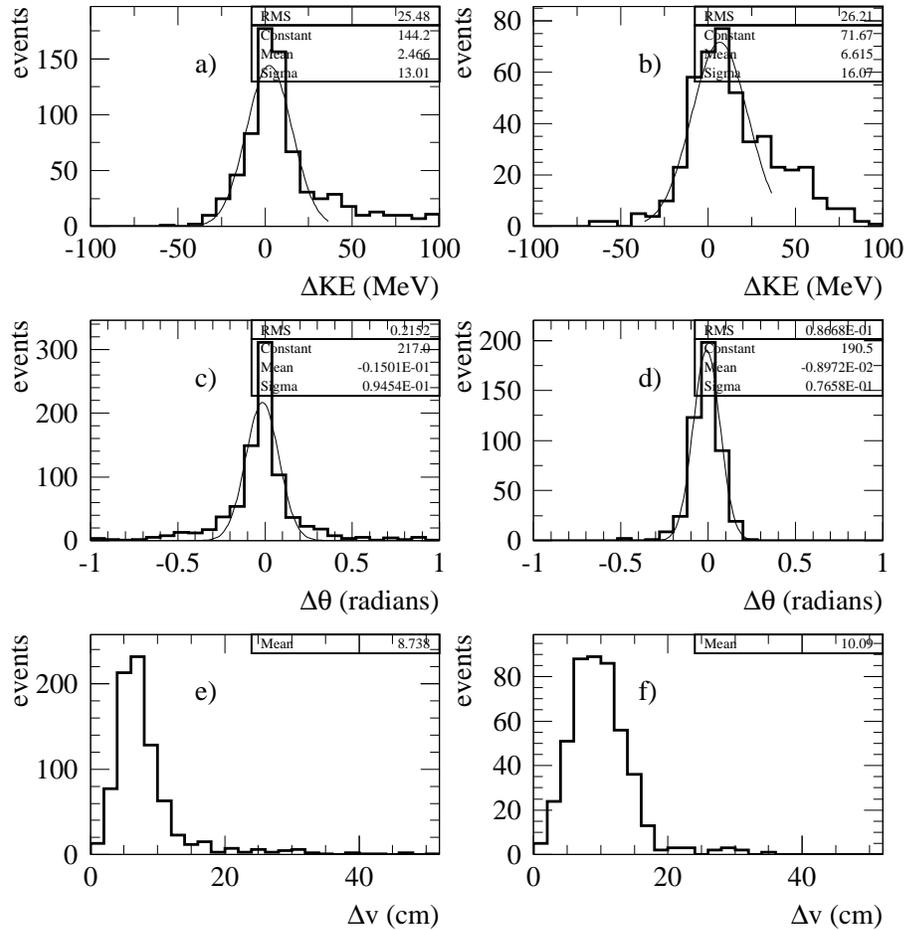}
\caption{\em The energy (a,b), angle (c,d), and position (e,f) resolution of the
  Vertex Detector as simulated and reconstructed for a sample of 1000
  single particle events.  The plots on the left (a,c,e) are for
  50-500~MeV KE protons, those on the right (b,d,f) for 50-500~MeV KE
  muons. Only tracks that were fully contained in the Vertex Detector
  were selected.  This effectively limited the upper muon KE to
  $\approx 300$~MeV.}
\label{fig:etvrec}
\end{figure}

The simulated detector energy, angle, and position resolution for
50-500 MeV KE protons and muons are shown in Figure~\ref{fig:etvrec}.
These particle energies are representative of the tracks that will be
contained in the Vertex Detector for physics events.  These proton
kinetic energies correspond to $Q^2$ values of 0.1-1.0 GeV$^2$ in NCp
and CCQE reactions.  A Gaussian fit to the energy and angle resolution
yields $\Delta E = 13 (16)$~MeV and $\Delta \theta = 100 (80)$~mrad
for protons (muons).  The quantity $\Delta v$ plotted in
Figure~\ref{fig:etvrec} is the distance from the calculated track
origin from the true origin.  The simulations of single particles
predict a mean $\Delta v = 9 (10)$~MeV for protons (muons).  The
distribution is slightly wider for muons as they are longer tracks.
These results indicate detector performance that will meet the physics
goals of the experiment. The effect of detector resolution on the
physics distributions will be shown in the following sections.

\subsubsection{Comparison with IU Cyclotron Facility Prototype Tests}
The results of these resolution studies are consistent with the
prototype tests performed at the IUCF (described in
Chapter~\ref{ch:TheFINeSSEDetector}), factoring in the differences in
the two studies.  For example, the prototype detector was much smaller
(and so did not fully contain tracks); had only one fiber orientation;
and used a slightly different fiber spacing.  Nevertheless, the tracks
reconstructed from the prototype test data yielded an angular
resolution of approximately 6$^{\circ}$ (15$^{\circ}$ FWHM).  The
resolution quoted in Section~\ref{sec:res} for protons is also
6$^{\circ}$.  Similarly, the amount of light detected in the IUCF
tests was $17\pm2$ photoelectrons for a 200 MeV passing adjacent to a
fiber.  These simulations, with light collection and detection
efficiencies estimated from the individual detector properties, are in
agreement with these results.

\subsection{Events in the Vertex Detector}
\begin{figure}
\centering
\includegraphics[width=4.in]{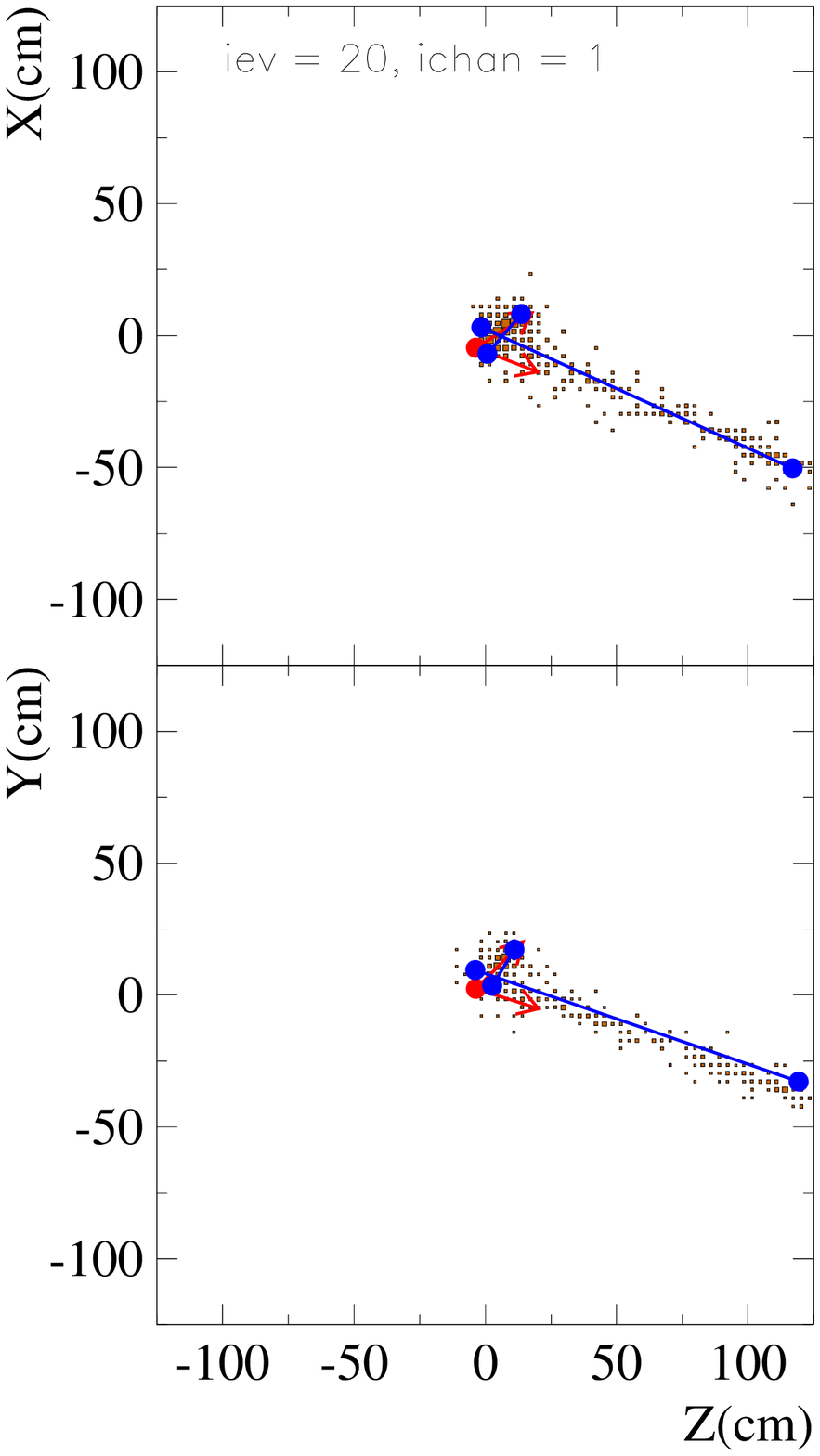}
\caption{\em A XZ (top) and YZ (bottom) projection view of a 
CCQE event in the simulated Vertex Detector with the reconstructed
muon (long line) and proton (short line) tracks superimposed. In this
event, $T_\mu=820$~MeV and $T_p=150$~MeV.}
\label{fig:p5p_1_20}
\end{figure}

\begin{figure}
\centering
\includegraphics[width=4.in]{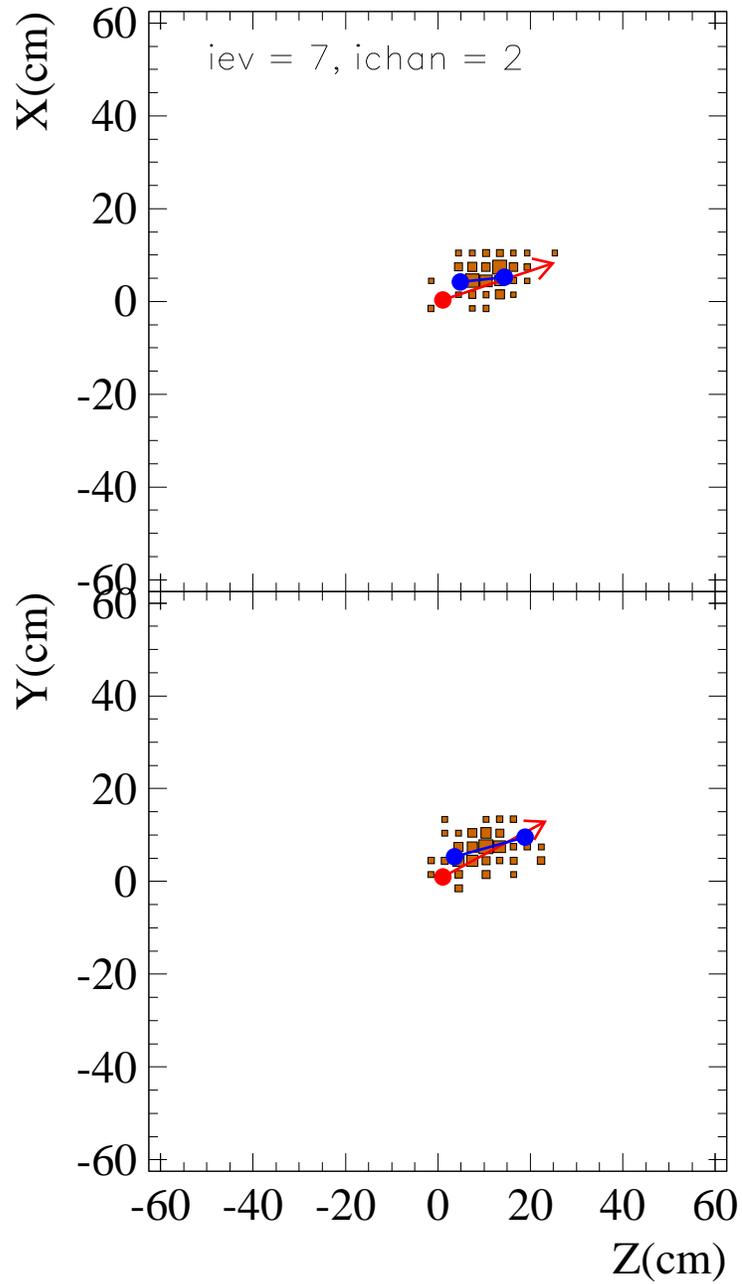}
\caption{\em A XZ (top) and YZ (bottom) projection view of a NCp event 
in the simulated Vertex Detector with the reconstructed proton track superimposed.
In this event, $T_p=100$~MeV.}
\label{fig:p5p_1_7}
\end{figure}

Several examples of the tracks obtained with the Hough transform
reconstruction method from simulated data are shown in 
Figures~\ref{fig:p5p_1_20} and~\ref{fig:p5p_1_7}.  In these figures,
the particle directions are indicated by the light-colored arrows.
The reconstructed tracks and endpoints are indicated by dark
lines and dots. As indicated, this method results in accurate
reconstructed tracks for muons and protons down to kinetic energies of
100~MeV.

\subsection{Events in the Muon Rangestack}
\label{evmurange}
\begin{figure}
  \centering \includegraphics[width=4.in,bb=30 30 540
  380]{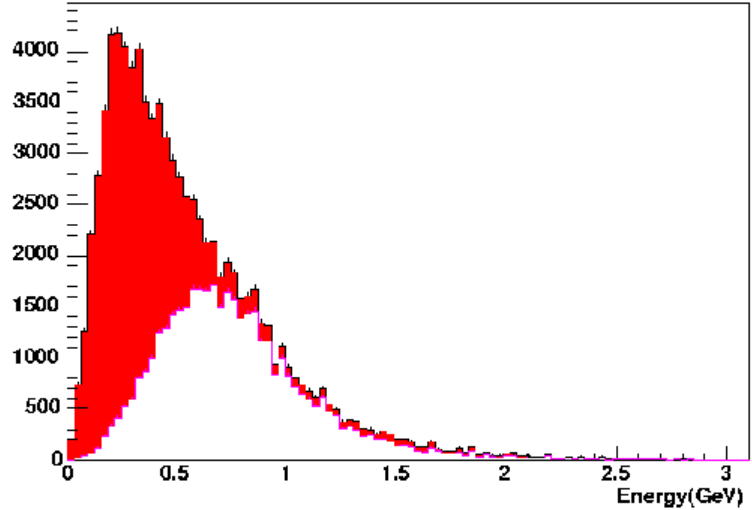}
\caption{\it The energy distribution of muons from CCQE interactions 
within the Vertex Detector 
fiducial volume, that enter and range out in the Muon Rangestack(white) 
overlaid on the distribution of all muons(red).}
\label{fig:muonenergydist}
\end{figure}

\begin{figure}
  \centering \includegraphics[width=4.in,bb=76 248 600 600]{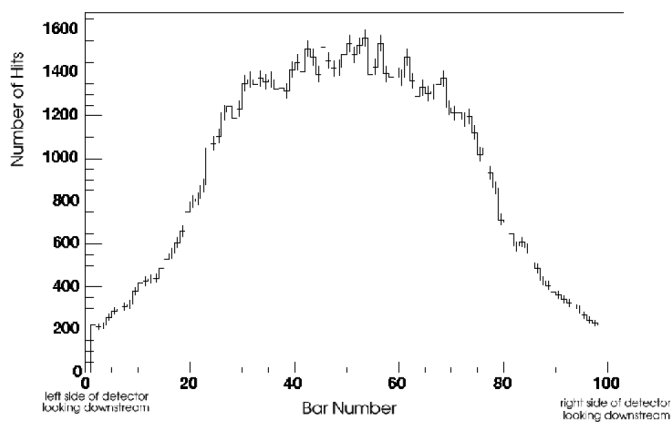}
\caption{\it Number of hits in scintillator strips in the Muon
Rangestack from CCQE muons 
that originate in the Vertex Detector fiducial volume.  The ``Bar Number'' measures the
hit coordinate transverse to the beam direction.}
\label{fig:hitbars}
\end{figure}

\begin{figure}
\centering
\includegraphics[width=4.in,bb=76 248 600 600]{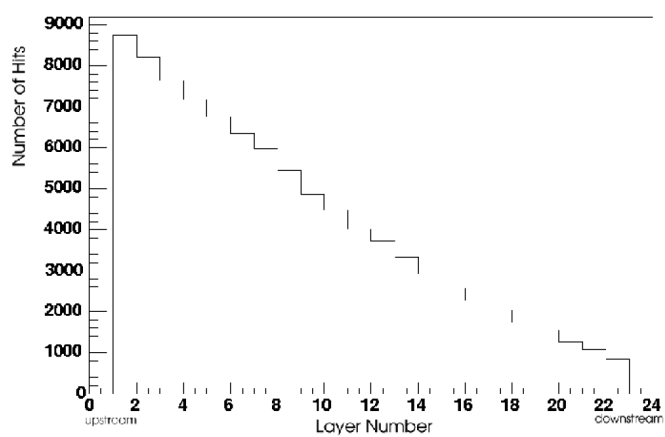}
\caption{\it Number of hits in each scintillator layer in the Muon Rangestack
from CCQE muons that originate in the Vertex Detector fiducial volume. The ``Layer Number'' 
measures the coordinate along the beam direction.}
\label{fig:hitlayers}
\end{figure}

Events in the Muon Rangestack considered here, have reconstructed
vertices in the Vertex Detector fiducial volume, exit the Vertex
Detector, and enter the Rangestack.  These higher energy muons tend to
be emitted at small angles as indicated by Figure~\ref{fig:cckin}a,
which shows the outgoing angle of the muon as a function of the muon
energy.  The energy distribution of these muons events that enter and
stop in the Muon Rangestack is shown in
figure~\ref{fig:muonenergydist}, superimposed over the distribution
for all muons.  The distribution of hits as a function of transverse
position in the range stack created by these higher energy muons is
shown in Figure~\ref{fig:hitbars}.  Note that there are few hits near
edges of the Rangestack.  A muon is considered to have ranged out in
this subdetector if its last hit is not at an edge of the detector,
either in the last plane, or one of the edge strips of an earlier
plane for higher angle muons.  Most muons below 1.5 GeV range out
within the Rangestack as indicated by hits deposited in each rangeout
layer (Figure~\ref{fig:hitlayers}).

\section{Backgrounds}
\label{sec:evt-bg}
We recognize two major classifications of backgrounds for FINeSSE: beam
related and beam unrelated.

FINeSSE can expect to see background events from neutrino interactions
in the detector through channels other than $\nu_{\mu}$ CC and NC
events, as described in Section~\ref{section:event-rates}.  Some of
these will be misidentified as signal channel interactions.  There
will also be neutrino interactions in the dirt around the detector,
sending charged particles and neutrons into the detector.  These are
the most important beam related backgrounds.

FINeSSE will also see a large flux of cosmic ray muons and neutrons.
These events can be used for detector calibration, as described in
Section~\ref{sec:calibration}, but they can also be misidentified as
beam events and therefore represent a potential background.  However,
the small probability of an interaction in time with the short beam 
spill of the booster neutrino source (1.6~$\mu$s) reduces these backgrounds
greatly.  In addition, they can be accurately measured and subtracted.

\subsection{Beam Related Backgrounds}
\label{sec:beam-bg}

The NC and CC measurements will have very different backgrounds, due
to differences in their event topologies.  The CC events have an
outgoing muon and a recoil proton.  These two particles emitted from
the interaction vertex will usually provide a very clean signal.
However, at low $Q^2$, where $\Delta s$ is measured, the recoil proton
will have very little energy.  In these cases, charged particles from
other interactions with two final states could mimic this short,
proton-like track.  The NC events have only the recoil proton.
Interactions producing only low energy short tracks can mimic this
recoil proton.  These will be the hardest events to identify.

Tables~\ref{table:100m-25mabs}~and~\ref{tab:simeffs} list the expected
event rates from the neutrino beam.  FINeSSE expects to see 24,435
CCQE events and 9864 NC elastic events per $1\times10^{20}$ POT,
assuming 100\% detection and reconstruction efficiencies.  Using the
present selection cuts, the expected event totals are 2990 CCQE events
and 1604 NC elastic signal events per $1\times10^{20}$ POT.  Because
the $\Delta s$ analysis will use a ratio to eliminate many systematic
uncertainties, background events become a significant challenge to the
measurement. Table~\ref{tab:simeffs} summarizes the contributions to
the signal from beam related backgrounds that originate in the
detector.

Neutrons from beam neutrino interactions in the material around the
detector are an additional potential background. These events arise
from neutrino NC and CC interactions in the material around the
detector that create secondary particles that impinge on and interact
in the detector.  The majority of the events cannot be eliminated by
taking advantage of the Booster neutrino beamline timing structure, as
they will be in time with the beam.  The majority of the particles
that mimic a signal are neutrons that are produced in the surrounding
dirt.  They have a fairly large range and can travel into the central
area of the detector without leaving a signal in the edge.  Muons, can
not do this and, therefore, are not a background.

Initial Monte Carlo studies, employing the cuts described in 
Section~\ref{sec:simres}, indicate
that the number of neutrons that interact in the dirt around 
FINeSSE and pass the particle ID cuts are insignificant compared to 
the signal. 

An additional background to consider are neutrons directly produced by
protons in the booster neutrino source beam dump.   These neutrons
will constitute a negligible background at the FINeSSE detector
enclosure 50m downstream from the dump.  There is no significant 
neutron flux from the beam dump after 25m downstream due to absorption
in the earth. This was determined in shielding assessment studies 
for the Booster neutrino source~\cite{shieldass}.

\subsection{Beam Unrelated Backgrounds}
\label{sec:non-beam-bg}

The main beam unrelated backgrounds stem from cosmic rays.  After
selection cuts, these are negligible in comparison with the NC elastic
signal.  The cosmic ray muon rate in the FINeSSE detector fiducial
volume, beneath 1~m of concrete and 0.5~m of veto region scintillator,
will be about 900~Hz.  The muons are a potential background both for
CC and for NC events.  The relative rate of coincidences with the beam
timing window is 0.0014.  This can be reduced to $1.4\times10^{-6}$ by
factoring in a conservative veto efficiency of 0.999 for muons. At
0.5~m thick, the veto for the Vertex Detector will be more efficient
than the 0.3~m thick MiniBooNE veto, which has an efficiency better
than 1 in $10^4$ for through-going muons.

Unlike the beam related backgrounds, the number of background events
considered here depends on the Booster performance.  If we assume that
the Booster delivers $4\times10^{12}$ protons per pulse, then we
expect to see 36 non-vetoed cosmic ray muon events per
$1\times10^{20}$ POT in coincidence with the beam.  
Event analysis will reduce this background down further; since there 
will be no recoil proton and the muons will all be headed downward. 
Therefore, cosmic ray muons will present a negligible background 
to CCQE or NC events.  

Cosmic ray neutrons have also been considered as a background.  Their
average rate above 50 MeV during periods of normal solar activity at
sea level and $\sim 40^{\circ}$ geomagnetic latitude is approximately
$9\times10^{-3}$sec$^{-1}$cm$^{-2}$~\cite{moraal,clem}.  They have a momentum
spectrum that falls very steeply with energy.  Fewer than 5\% of these
can traverse the concrete shielding and the veto region.  We therefore
expect a cosmic neutron rate of $<1$Hz in the FINeSSE Vertex Detector,
for neutrons above 100 MeV.  This rate will be reduced to a 
negligible level because of the small coincidence probability with 
the short beam spill. 

\subsection{Charged and Neutral Current Event Identification}

Neutral-current elastic scattering events ($\nu p \rightarrow \nu p$)
will be identified in the FINeSSE detector by looking for single
proton tracks consistent with elastic scattering kinematics. A track
is identified as a proton by a large $dE/dx$.  Charged-current
quasi-elastic scattering events ($\nu n \rightarrow \mu^- p$) will be
identified by looking for events with two tracks each consistent with
the expected $dE/dx$.  In addition, other cuts are employed to reject
backgrounds.  The strategy will be to maintain a compromise between
large efficiency for low-$Q^2$ events while keeping backgrounds as low
as possible.  The squared four-momentum transfer, $Q^2$, will be
determined event by event, by measuring the energy of the proton in
both NC and CC events.  $Q^2$ is determined from the energy via
$Q^2=2m_pT_p$.

\subsection{Simulation Results}
\label{sec:simres}

The \texttt{GEANT} simulation of the FINeSSE detector and reconstruction
program as described above is used to estimate efficiencies and
backgrounds for the NC/CC ratio measurement.  It is also relied upon to
determine the experimental error on a measurement of $G_A^s$.

\subsubsection{Cuts}
The experimental cuts employed for $\nu p \rightarrow \nu p$ (NCp) and 
$\nu n \rightarrow \mu^- p$ (CCQE) event identification are summarized 
in Table~\ref{tab:simcuts}.  A reasonable compromise
of efficiency and background rejection could be obtained out to within 15~cm
of the edge of the active Vertex Detector volume.  This results in a
fiducial volume of (2.2~m)$^3$ = 10.6m$^3$ or 68.4\% of the total (2.4~m)$^3$
``active'' volume of the Vertex Detector.  With $\rho=0.85$~g/cm$^3$ liquid
scintillator, a fiducial mass of 9.1~t is obtained.  

\begin{table}[h]
\centering
\begin{tabular}{cll}
\hline
cut \# & NCp cuts & CCQE cuts \\
\hline
0 & edge distance $<$ 15cm       &   edge distance $<$ 15cm           \\
1 & \# 3d tracks = 1             &  \# 3d tracks.eq.2               \\
2 & $dE/dx(p)>2.5$               &  $dE/dx(p)>2.5, dE/dx(\mu)<2.5$  \\
3 & $\theta(p)>0.5$              &  $\theta(p)+\theta(\mu)>1.5$     \\
4 & no ``late'' light in vertex det. &   no ``late'' light in vertex det. \\
5 & no veto or muon stack energy &  low ``remaining'' energy        \\  
\hline
\end{tabular}
\caption{\it Cuts used to identify  ``NCp'' and ``CCQE'' events in the simulated
         sample.}
\label{tab:simcuts}
\end{table}

Cuts 1 and 2 identify and separate NCp and CCQE based on the number and
type of charged particle tracks in the event.  Cut 2 accepts events with
tracks of larger angle, a feature of the two-body kinematics of these
events, and rejects background events which tend to have tracks at smaller 
angles.  Cut 4 lowers background by eliminating pions that stop and decay
in the Vertex Detector into muons (which decay ``late'').  This also has
a slight effect on the efficiency for accepting CCQE events, but it is not
large since most of the muons in CCQE scattering leave the Vertex Detector.
Cut 5 lowers backgrounds for both channels.  For NCp events, it further
lowers the pion (charged and neutral) backgrounds by cutting those events
where the pions leave little signal in the Vertex Detector.  For CCQE, events
cut 5 also reduced pion backgrounds as a cut on low ``remaining'' energy
(energy not assigned to tracks after tracking is complete) reduces the
number of events with extra energy (due to pions).

\subsubsection{Efficiencies and Purities}

A sample of 215k NUANCE-generated events were tracked through the detector 
with the \texttt{GEANT} simulation with vertices evenly distributed within the 
nominal volume of the Vertex Detector (2.4~m)$^3$.  These events are then 
reconstructed using the algorithms described above.  The results are 
summarized in Table~\ref{tab:simeffs}.

\begin{table}[h]
\centering
\begin{tabular}{|r|rrrrr|}
\hline
                  & \multicolumn{5}{c|}{reaction channel}      \\
\hline
\textbf{NCp cuts} &   NCp  &   NCn & NC$\pi$&   CCQE  &  CC$\pi$ \\
\hline
raw events        & 21219  & 20487 & 19062  & 100102  & 54107 \\
passed events     &  3929  &  1162 &   167  &     48  &     4 \\
efficiency (\%)   &  18.5  &   5.7 &   0.9  &    0.0  &   0.0 \\
fid.~eff.~(\%)   &  27.1  &   8.3 &   1.3  &    0.1  &   0.0 \\
purity (\%)       &  74.0  &  21.9 &   3.1  &    0.9  &   0.1 \\
\hline\hline
\textbf{CCQE cuts}&   NCp  &   NCn & NC$\pi$&   CCQE  &  CC$\pi$\\
\hline
raw events        & 21219  & 20487 & 19062  & 100102  & 54107 \\
passed events     &  165   &    76 &   581  &   7323  &  1322 \\
efficiency (\%)   &  0.8   &   0.4 &   3.0  &    7.3  &   2.4 \\
fid.~eff.~(\%)   &  1.1   &   0.5 &   4.5  &   10.6  &   3.6 \\
purity (\%)       &  1.7   &   0.8 &   6.1  &   77.4  &  14.0 \\
\hline
\end{tabular}
\caption{\em Summary of events that passed the NCp and CCQE cuts along with 
efficiencies and purities:  ``efficiency'' is the reconstruction efficiency
throughout the (2.4~m)$^3$ volume. ``fid.~eff.'' is the reconstruction 
efficiency within the (2.2~m)$^3$ fiducial volume.  The simulation data
set contained 215k events.  To scale to the total expected FINeSSE event
count, multiply these numbers by 2.45.} 
\label{tab:simeffs}
\end{table}

\begin{figure}
\centering
\includegraphics[width=5.in]{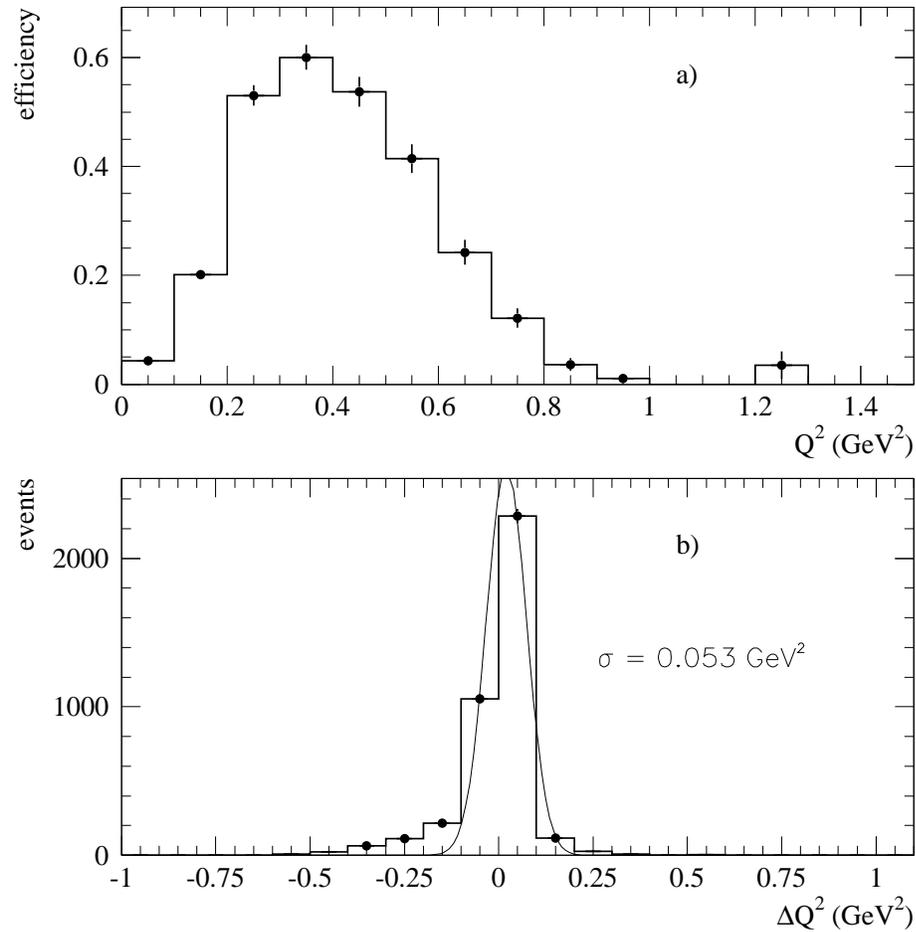}
\caption{\em (a) Efficiency for reconstructing true NCp events with the NCp 
  cuts within the fiducial volume as a function of generated $Q^2$ and
  (b) the distribution of the difference between generated and
  reconstructed $Q^2$ for this sample.  The $\sigma$ of this
  distribution is 0.053~GeV$^2$}
\label{fig:ncpeff}
\end{figure}

\begin{figure}
\centering
\includegraphics[width=5.in]{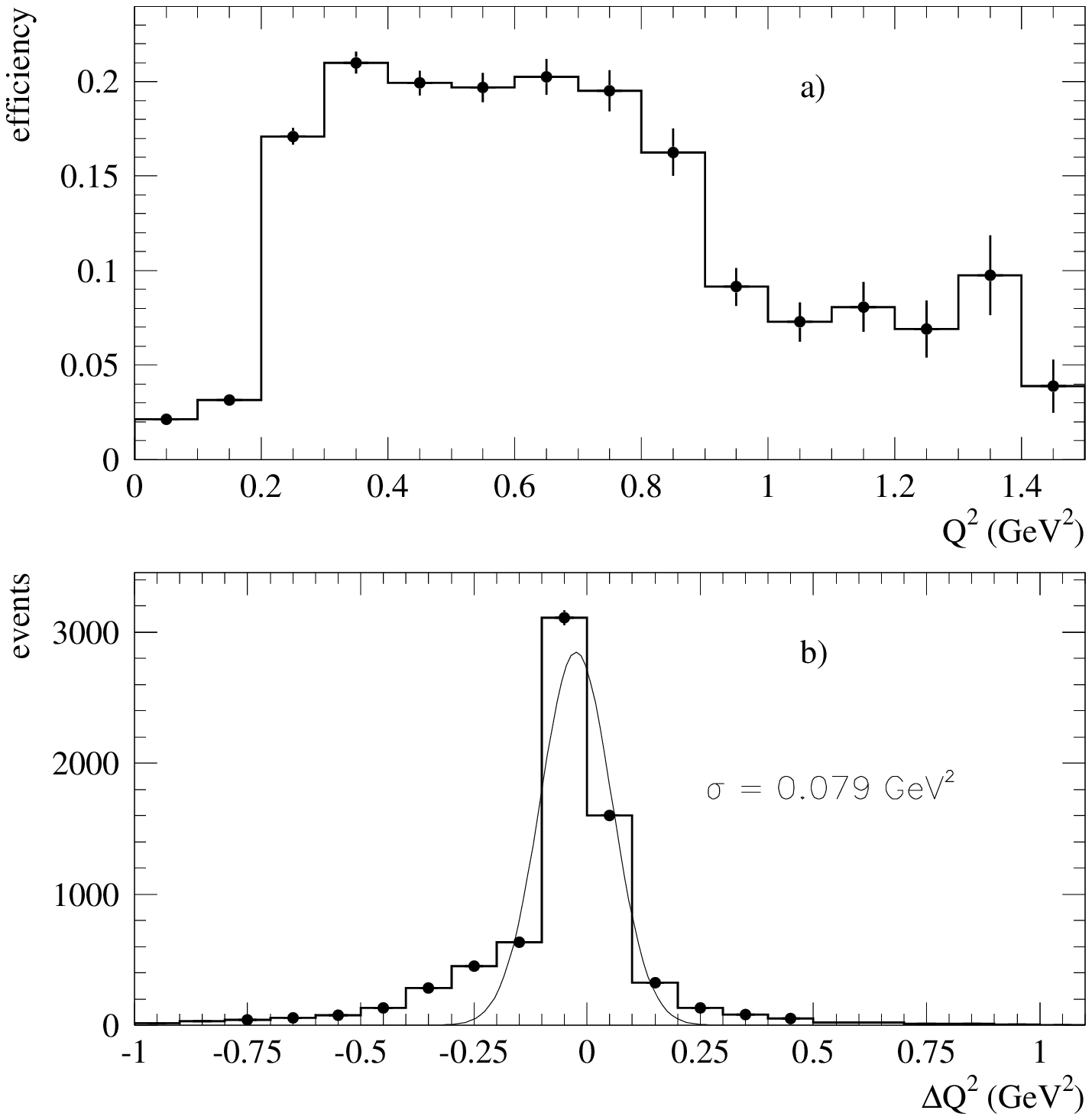}
\caption{\em (a) Efficiency for reconstructing true CCQE events with the CCQE 
  cuts within the fiducial volume as a function of generated $Q^2$ and
  (b) the distribution of the difference between generated and
  reconstructed $Q^2$ for this sample.  The $\sigma$ of this
  distribution is 0.079~GeV$^2$}
\label{fig:ccqeeff}
\end{figure}

The efficiencies and $Q^2$ resolutions for the NCp and CCQE ``signal''
samples are shown in Figures~\ref{fig:ncpeff} and \ref{fig:ccqeeff}.
Note the large efficiency for reconstructing NCp, and reasonable
efficiency for CCQE, in the $Q^2=0.2-0.4$ GeV$^2$ region.  At low
$Q^2$, proton tracks from NCp are short and difficult to reconstruct,
causing a fall off in the efficiency. At high $Q^2$, final state
particles tend not to be contained for both the NCp and CCQE samples,
causing a fall of in the efficiency in these regions also. In addition
to this, very forward events in the NCp sample are cut to remove
backgrounds from pions.  This further decreases the NCp efficiency at
high $Q^2$.

\subsection{Method to Extract $\Delta s$ from the Data} 

To extract $\Delta s$ (or $G_A^s$) from the data, the NC to CC ratio must be 
formed and compared to predictions obtained from Eq.~\ref{eq:dsdq2}.  This 
ratio must be corrected for backgrounds and detector efficiency and may be 
written as:

\begin{equation}
  R(NC/CC) = \frac{\epsilon_\mathrm{CC}}{\epsilon_\mathrm{NC}}
  \frac{(N_\mathrm{NC}-\sum_i \epsilon_{\mathrm{NC},i} N_{\mathrm{NC},i})}
  {(N_\mathrm{CC}-\sum_j \epsilon_{\mathrm{CC},j} N_{\mathrm{CC},j})}~,
\label{eq:Rexp}
\end{equation}

\noindent
where $\epsilon_\mathrm{NC}$ ($\epsilon_\mathrm{CC}$) is the NC (CC)
reconstruction efficiency, and $N_\mathrm{NC}$ ($N_\mathrm{CC}$) is
the number of NC (CC) identified events in the detector.  The
quantities $\epsilon_{\mathrm{NC},i}, \epsilon_{\mathrm{CC},i},
N_{\mathrm{NC},i},$ and $ N_{\mathrm{CC},i}$ are the background
efficiencies and calculated number of background events for the NC and
CC channels, respectively.

Each of the terms in Equation~\ref{eq:Rexp} will contribute to the
total error on $R(NC/CC)$. The relationship between the error on $G_A^s$
($\sigma(G_A^s)$) and R is quantified in Eq.~\ref{eq:siggas}, so we
can estimate the error on $G_A^s$ due to each of the terms in
Eq.~\ref{eq:Rexp}.  These errors are discussed and quantified in the
sections below, assuming that the measurement is performed in the
$Q^2=0.2-0.4$ GeV$^2$ bin, and that an estimate of $\sigma(G_A^s)$ is
obtained.  After this treatment, a fitting procedure that uses the
entire range of $Q^2$ is used, to illustrate an improved method for extracting
$G_A^s$.

\subsubsection{Statistical Errors}

The event sample of 215k NUANCE-generated events is 0.408 of the total
number of neutrino-scattering events expected in the (2.4~m)$^3$ active
volume of the Vertex Detector with $6\times10^{20}$ POT.  So the number of events in
Table~\ref{tab:simeffs} may be multiplied by $1/0.408=2.45$ to obtain
the total number expected.  Taking these scaled numbers, and
considering the efficiencies for the NCp and CCQE cuts, yields 7341
(9936) accepted events in the NCp (CCQE) sample in the $Q^2=0.2-0.4$ GeV$^2$ bin.
Statistical errors from these two sources will contribute 1.2\% and
1.0\% to the relative error on R(NC/CC).

\subsubsection{Neutral Current Neutrino Neutron Scattering}

One of the backgrounds in the numerator of Equation~\ref{eq:Rexp} is
that due to the NC process $\nu n \rightarrow \nu n$ (labeled ``NCn''
in Table~\ref{tab:simeffs}). This reaction will need to be treated
separately, since it is the largest background to the NCp channel, and
also since the rate depends upon $G_A^s$.

The dependence of cross section for $\nu n \rightarrow \nu n$ on
$G_A^s$ is different than that for $\nu p \rightarrow \nu p$, due to
the isospin factor $\tau_z$ (cf.~Equation~\ref{eq:G_1}).  This will
``dilute'' the experimental sensitivity.  For this reason it is
important to keep the efficiency of the $\nu n \rightarrow \nu n$
channel as low as possible.  This is a difficult task as neutrons in a
liquid scintillator detector often undergo charge-exchange reactions
which create energetic protons.  These reactions can be
indistinguishable from the $\nu p \rightarrow \nu p$ reaction.
Neutrons, though, often create two or more protons of lower energy and
these topologies will not be misidentified as $\nu p \rightarrow \nu
p$.

As can be seen from Table~\ref{tab:simeffs}, the reconstruction
efficiency for the NCn background is 8.3\% and comprises 21.9\% of
the NCp sample.  It is non-negligible.  It would be desirable to improve
the detector and reconstruction to lower this background. It is likely
that this is possible by observing the delayed light due to neutron
capture.  The detector as currently
designed would have the ability to observe this delayed light, and
this would improve the neutron background rejection considerably.
This technique, however, has yet to be investigated fully, and any
potential gains are not assumed in this analysis.

The effect of NCn events in the NCp sample dilutes the sensitivity of
the measured NC to CC ratio to $G_A^s$ by a factor
$(1-2\epsilon_\mathrm{NCn}/\epsilon_\mathrm{NCp})$.  The impact of
this on the analysis can be determined using the ratio of
efficiencies, $\epsilon_\mathrm{NCn}/\epsilon_\mathrm{NCp}$ = 0.19, in
the $Q^2=0.2-0.4$ GeV$^2$ bin. Applying this efficiency factor, the
sensitivity of R(NC/CC) is reduced from 1.2 to 0.75.  This is a fairly
substantial reduction but not large enough to prevent a
measurement.  The uncertainty in $G_A^s$ becomes:

\begin{equation}
  \sigma(G_A^s) = 1.3 \Delta R/R.
\label{eq:siggas2}
\end{equation}

\noindent A 5\% relative measurement of the NC/CC ratio at $Q^2=0.25$ GeV$^2$ would 
enable an extraction of $G_A^s$ with an error of 0.07.

\subsubsection{Scattering from Free Protons}

The FINeSSE neutrino scattering target is mineral-oil-based liquid 
scintillator (mainly CH$_2$).  The hydrogen will provide two additional 
free protons for $\nu p \rightarrow \nu p$ scattering.  This entails a 
correction to the measured NC to CC ratio.  The measured ratio (after 
background subtraction) will be:

\begin{equation}
\frac{N_\mathrm{NC}}{N_\mathrm{CC}} =
\frac{N_\mathrm{NC}^\mathrm{C}+N_\mathrm{NC}^\mathrm{free}}{N_\mathrm{CC}}~,
\end{equation}

\noindent
where $N_\mathrm{NC}^\mathrm{C}$ and $N_\mathrm{NC}^\mathrm{free}$ are the 
number of events from $\nu p \rightarrow \nu p$  on bound (in carbon) and free
protons. This can be written:

\begin{equation}
\frac{N_\mathrm{NC}}{N_\mathrm{CC}} =
\frac{N_\mathrm{NC}^\mathrm{C}}{N_\mathrm{CC}} \times (1+\mathcal{F})~,
\end{equation}

\noindent
where:

\vspace{-0.3in}
\begin{equation}
\mathcal{F} = \frac{N_\mathrm{NC}^\mathrm{free}}{N_\mathrm{NC}^\mathrm{C}}.
\end{equation}

\noindent
Naively, one would expect $\mathcal{F}$ to be equal to the free proton
to bound ratio in the target.  This is $\approx 0.34$ \cite{ej-321}
for mineral-oil-based liquid scintillator. In actuality, $\mathcal{F}$
will be slightly larger than these numbers, as the the cross section
per nucleon is slightly suppressed (by $\approx 10\%$) for bound
nucleons.  In addition, the value used for $\mathcal{F}$ will depend
upon the $Q^2$ region accepted for the analysis.  So for the purposes
of a conservative estimate here, $\mathcal{F}=0.34*1.10=0.38$ will be
used.

\noindent
The desired corrected ratio (that on pure carbon) can then be written:

\begin{equation}
R(NC/CC) = 
\frac{N_\mathrm{NC}^\mathrm{C}}{N_\mathrm{CC}} =
\frac{1}{(1+\mathcal{F})} \frac{N_\mathrm{NC}}{N_\mathrm{CC}}~; {\mathrm{and}}
\end{equation}

\begin{equation}
\frac {\Delta R}{R} = 
\frac{\mathcal{F}}{1+\mathcal{F}} \frac{\Delta \mathcal{F}}{\mathcal{F}} = 
0.27 \frac{\Delta \mathcal{F}}{\mathcal{F}}~.
\end{equation}

\noindent
The number of bound and free protons in the target will be known, so
the relative uncertainty on $\mathcal{F}$ will be dominated by the
uncertainty in the ratio of the bound to free NC cross sections. An
estimate of this relative uncertainty is 5\%.  This will add a
systematic uncertainty in the NC to CC ratio of $\approx 1.4\%$.

\subsubsection{Systematic Error in Efficiencies}

The flux and target density is same for both NC and CC reactions and
does not enter as an additional source of uncertainty in a measurement
of the ratio, R(NC/CC).  However the reconstruction efficiencies will
not be the same for both reactions, and the systematic error in the
$\frac{\epsilon_\mathrm{CC}}{\epsilon_\mathrm{NC}}$ term of
Eq.~\ref{eq:Rexp} must be considered.

At a given $Q^2$, the proton in the NC and CC reactions has the same
kinematics.  Because of this, some of the systematic error is highly
correlated and the dominant uncertainty is in the efficiency to detect
the muon in the CC reaction.  This is a well-understood process and it
should be possible to determine it with high accuracy.  For these
reasons, a relative error of 3\% is estimated for
$\frac{\epsilon_\mathrm{CC}}{\epsilon_\mathrm{NC}}$ and will
contribute at this level to the relative error on R(NC/CC).

\subsubsection{Systematic Error in Backgrounds}

The background subtraction in both the numerator and denominator of
Equation~\ref{eq:Rexp} also contributes systematic error to R(NC/CC).  This
error has a contribution from both the absolute rate of the background
and the efficiency to be reconstructed as a NCp or CCQE event.  The
relative error on R(NC/CC) due to these uncertainties is diluted to
the reasonably small background subtraction.  This dilution factor is
equal to the ratio of background to (true) signal for NCp and CCQE
events.  From the numbers in Table~\ref{tab:simeffs} this dilution
factor is 0.35 for NC events and 0.29 for CC events.

It is estimated that the relative error on the product of efficiency
times background can be kept at or below 10\% for both NC and CC
reactions.  This will be possible by means of detailed simulations and
studies of the detector, and having developed a good model for the
background processes.  These backgrounds have been measured in other
neutrino experiments, and likewise, will be measured in FINeSSE. In addition,
relevant electron scattering data exist.

The relative error of 10\% on the background contributions contributes
3.5\% (2.9\%) via NC (CC) backgrounds to the relative error on
R(NC/CC).

\subsubsection{Systematic Error Due to $Q^2$ Reconstruction} 

A large error in the $Q^2$ reconstruction of events will distort the
overall distribution and add to the error in R(NC/CC).  For example,
if low-$Q^2$ ($< 0.2$ GeV$^2$) CCQE events (of which there are many;
see Figure~\ref{fig:dsdq2}) are reconstructed in the $Q^2 = 0.2-0.4$
GeV$^2$ bin, a larger error will result in R(NC/CC) that is not
quantified in the errors considered above.

From examination of Figures~\ref{fig:ncpeff} and~\ref{fig:ccqeeff}, it
can be seen that the $Q^2$ resolution of this detector is quite good
for both NCp and NCn events.  Also, there is no evidence of a large
skew in these distributions.  The $Q^2$ resolution is smaller than
the bin width and will contribute a negligible error to R(NC/CC).

This hypothesis is further checked in the fitting procedure described
below as the {\em reconstructed} $Q^2$ is used to bin the data.

\subsubsection{Nuclear Model Uncertainties}

The majority of protons and neutrons participating in $\nu p \rightarrow
\nu p$ and $\nu n \rightarrow \mu^- p$ interactions are not free, but are bound
within a carbon nucleus.  Any nuclear effects that are not symmetric
between protons and neutrons can lead to a misinterpretation of the
data, when analyzed using  a free nucleon model.  For this reason, a
model with a realistic treatment of nuclear effects will be employed for
the final analysis.

The sensitivity of the NC and CC cross sections to the specific
nuclear model employed has been investigated by several different
groups~\cite{Garvey_dels,barbaro,alberico_nufact}. The effects on the
absolute cross sections can be quite dramatic.  However, the
distortion in the ratio R(NC/CC) has been shown to be small.  It has been
estimated to introduce an error on 0.005 in
$G_A^s$~\cite{alberico_internal}.

This will be an area of continuing work.  Results up to now, however, indicate 
that R(NC/CC) is insensitive to the details of the nuclear model employed for 
the calculations. 

\subsubsection{Form Factor Uncertainties}

The (non-strange) form factors in Eq.~\ref{eq:dsdq2} have been
measured in a variety of experiments at many different $Q^2$ values.
This is not true for the strange form factors.  Information on the
strange-axial form factor $G_A^s$ is almost completely non-existent.
Accumulating more information on this form factor is one of the main
goals of this experiment.  The strange vector form factors influence
the NC to CC ratio only weakly and they will be precisely measured 
by the upcoming G0 experiment~\cite{G0} to take place a
Jefferson Lab in the near future.

In addition, the contribution from these unknown form factors is
minimized by performing these measurements at low-$Q^2$.  The sensitivity of the
final result to these other form factors is small as will be shown in
the results from the fitting procedure below.

\subsubsection{Summary of Errors on R(NC/CC)}

The experimental systematic errors discussed in the proceeding
sections are summarized in Table~\ref{tab:Rerr}.  These contributions,
added in quadrature, yields a relative error on the ratio, R(NC/CC),
of 5.8\%.  This, combined with Equation~\ref{eq:siggas2}, yields a total
error on $G_A^s$ of 0.075.

\begin{table}[h]
\centering
\begin{tabular}{|cccc|}
\hline
quantity,         &           & relative error,  & contribution to \\
$Q$               & prefactor & $\Delta Q/Q$     & $\Delta R/R$    \\
\hline
$N_\mathrm{NC}$   &    1      &   1.2\%          & 1.2\%           \\
$N_\mathrm{CC}$   &    1      &   1.0\%          & 1.0\%           \\
$\mathcal{F} = \frac{N_\mathrm{NC}^\mathrm{free}}{N_\mathrm{NC}^\mathrm{C}}$
                  &    0.27   &   5\%            & 1.4\%           \\
$\frac{\epsilon_\mathrm{CC}}{\epsilon_\mathrm{NC}}$
                  &    1      &   3\%            & 3.0\%           \\
$\sum_i \epsilon_{\mathrm{NC},i} N_{\mathrm{NC},i}$
                  &    0.35   &  10\%            & 3.5\%           \\
$\sum_i \epsilon_{\mathrm{CC},i} N_{\mathrm{CC},i}$
                  &    0.29   &  10\%            & 2.9\%           \\
\hline
\multicolumn{3}{|c|}{Total experimental error} 
                                                 & 5.8\%           \\
\hline
\end{tabular}
\caption{\em Estimated experimental errors on a measurement of R(NC/CC) at 
$Q^2 = 0.2-0.4$ GeV$^2$ with FINeSSE. The prefactor relates the relative error 
on the quantity to the relative error on R(NC/CC). The details are 
explained in the text.}
\label{tab:Rerr}
\end{table}

This is a conservative estimate. This error will be reduced
considerably by reducing the $\nu n \rightarrow \nu n$ background (via
detection of photons from neutron capture), which both contributes to the
error in R(NC/CC) and lowers the sensitivity of $G_A^s$ to R(NC/CC).

\subsection{A Fit Procedure to Extract $\Delta s$ from R(NC/CC)}

In the error analysis explained above, only one $Q^2$ bin was
considered. This yielded a conservative estimate for the error on
$G_A^s$.  As can be seen in Figures~\ref{fig:ncpeff}
and~\ref{fig:ccqeeff}, high reconstruction efficiency is obtained from
$Q^2=0.2$ to 0.8 GeV$^2$, and a measurement of R(NC/CC) may be made in
this range.  This larger $Q^2$ range will accept more events and will
allow for the systematic study of the behavior of R(NC/CC) and,
consequently, the form factors as a function of $Q^2$.

To utilize the entire data set, a fit procedure was developed to
extract $G_A^s$, to determine the effects of systematic errors, and to
investigate the dependence of the extracted $G_A^s$ on the other form
factors.  In this procedure, a corrected NC to CC ratio is formed
using Equation~\ref{eq:Rexp} for each $Q^2$ bin of width 0.1~GeV$^2$
using the simulated data.  The backgrounds are subtracted and
efficiency corrections are applied.  The predicted ratio is then
calculated using the free nucleon cross section (Eq.~\ref{eq:dsdq2}),
correcting for the NC scattering from free protons and the estimated
contribution from $\nu n \rightarrow \nu n$.  This prediction is
calculated for a range of $G_A^s$ values.  The dependence on the other
form factors ($F_1^s$, $F_2^2$) and the axial mass $M_A$ is also
studied.



\noindent
The sensitivity of the data to $G_A^s$ is determined by calculating:

\begin{equation}
\chi^2 = \sum_i \frac{(R_\mathrm{meas}-R_\mathrm{pred})^2}{\sigma(R)^2}~,
\end{equation}

\noindent
where the sum runs over $Q^2$ bins.   The $1\sigma$ error on $G_A^s$ is
determined by the $G_A^s$ value where the $\chi^2$ value changes by 1. 

This procedure yields an error on $G_A^s$ of 0.039.  This is considerably
smaller than the error quoted above due to the fact that this fit procedure
uses more of the data through a range in $Q^2$.  

\begin{figure}
\centering
\includegraphics[width=5.in]{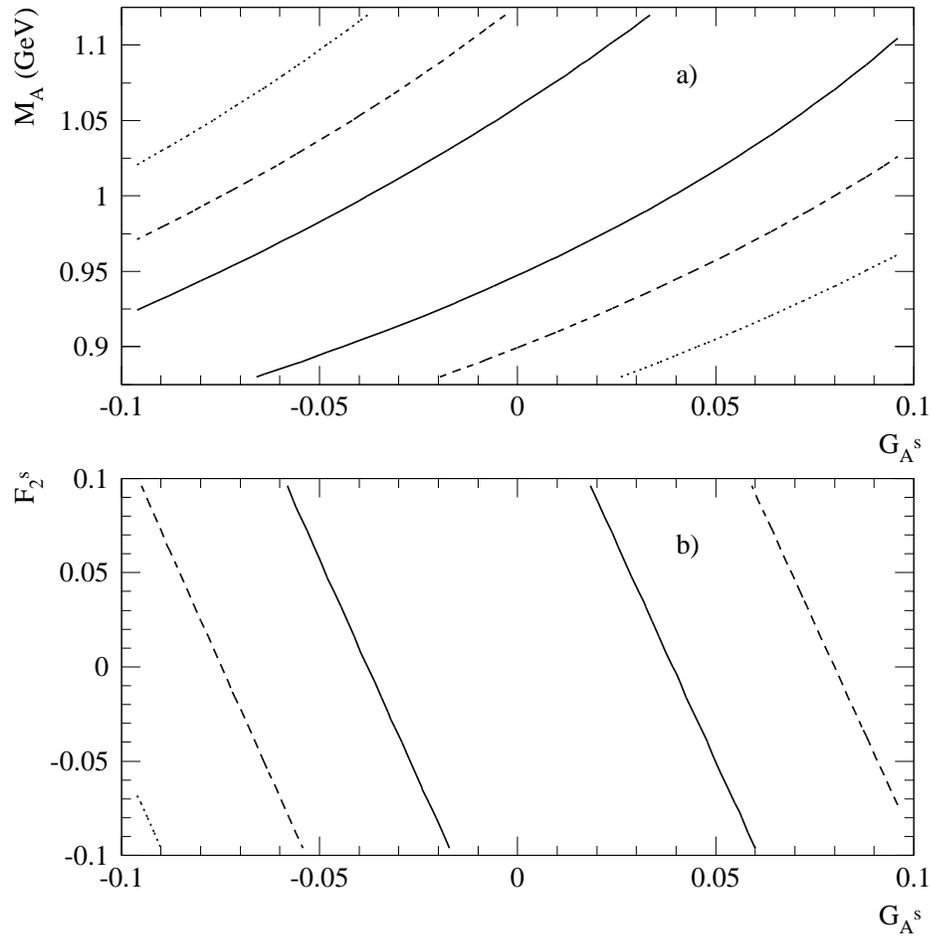}
\caption{\em $\chi^2$ contours corresponding to $1\sigma$ (solid), $2\sigma$ 
         (dashed), and $3\sigma$ (dotted) resulting from the fit procedure 
          described in the text. In (a) $G_A^s$ and $M_A$, the axial vector 
          mass, are varied, in (b) $G_A^s$ and $F_2^s$ are varied.}
\label{fig:chi2ff}
\end{figure}

The sensitivity of R(NC/CC) to other form factors is also studied with
this method.  Figure~\ref{fig:chi2ff} shows how a measurement
of R(NC/CC) depends on other form factors.  An uncertainty in the
axial vector mass, $M_A$, which parameterizes the $Q^2$ evolution of
the $G_1$ form factor, effectively adds a small additional uncertainty
to the extracted value of $G_A^s$. The current world average of $M_A$
as measured in neutrino scattering is $1.026\pm0.021$ GeV~\cite{Bernard}. This
uncertainty will add approximately 0.015 to the uncertainty in
$G_A^s$. The sensitivity of $G_A^s$ on $F_2^s$ is weaker, but this
form factor is not well known directly.  Linear combinations of this
form factor with $F_1^s$ have been measured~\cite{HAPPEX}.  A
conservative estimate on the uncertainty on $F_2^s$ is $\pm0.1$.  An
uncertainty of this size in $F_2^s$ will contribute an uncertainty of
$\pm0.02$ to $G_A^s$.  Combining these yields a $\pm0.025$ uncertainty on
$G_A^s$.

Based on the studies described above, FINeSSE will be capable of a
measurement of $G_A^s$ with small errors. An uncertainty of
approximately 0.04 results from considering statistical and systematic
experimental errors.  Uncertainties in the form factors adds an
additional contribution of 0.025.  This level of accuracy is unprecedented
in $\nu$ scattering will allow for stringent tests with theory. 


\section{Details of $\nu_\mu$ Disappearance Sensitivity Studies}

The method for determining the $\nu_\mu$ disappearance sensitivity
compares the energy distribution of events in the near (FINeSSE) and
far (MiniBooNE) detector.  In this section we justify the choice of
the experimental setup and the estimates of the systematic errors.  We
studied various configurations for FINeSSE+MiniBooNE running in order
to identify the optimal experimental set-up.  Here, we reduce the
comparisons to two representative positions of the near detector (100~m
and 200~m) and the two possible lengths of the decay region (25~m and
50~m).  The event rates assumed in these studies are indicated in
Table~\ref{tab:eventrates}.

\begin{table}[tbp] \centering
\begin{tabular}{|l|c|c|}
\hline
Events for $6\times 10^{20}$ POT &  &  \\
Location & 25~m Absorber & 50~m Absorber \\
\hline
100~m (9~t) & 103,000 & 195,000 \\
200~m (40~t) & 97,000 & 162,000 \\
550~m (650~t) & 205,000 & 323,000 \\
\hline
\end{tabular}
\caption{\it Estimated event sample for $5\times 10^{20}$ POT with
detectors located at various distances with fiducial volumes as
given.  Rates are given for a 25~m and 50~m absorber. 
\label{tab:eventrates}}
\end{table}

\subsection{Tools and Assumptions for This Study}

Determination of the $\nu_\mu$ disappearance sensitivity requires
tools from both FINeSSE and MiniBooNE.  Assumptions related to the
flux in both detector have been described in
Chapter~\ref{ch:TheNeutrinoBeam}.  As discussed there, the NUANCE
Monte Carlo was used to simulate events in both detectors~\cite{nuance}. The
following standard MiniBooNE detector cuts were used: tank hits $>$
100, veto hits $<$ 6, and reconstructed fiducial radius $<$
500~cm.\cite{mb-runplan}.

\begin{figure}
\centerline{\includegraphics[width=4.in, bb=20 20 540 350]{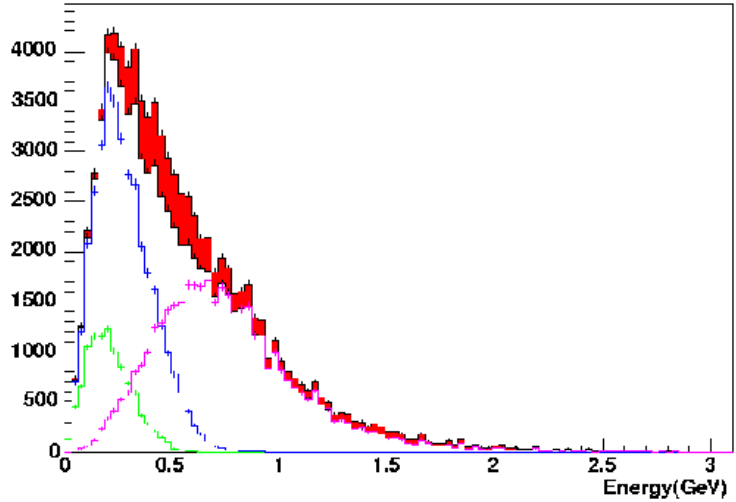}}
\caption{\it True kinetic energy distribution for all muons from 
  CCQE interactions in the Vertex Detector fiducial volume shown in
  red.  Overlaid in white is the sample of those with a contained muon.  
  Also shown are contributions corresponding to muons stopping in the Vertex Detector (green), the Vertex
  Detector veto (blue), and the Muon Rangestack (purple). }\vspace{0.25in}
\label{fig:containedcont}
\end{figure}

\begin{figure}
  \centerline{\includegraphics[width=4.in, bb=20 20 540 350]{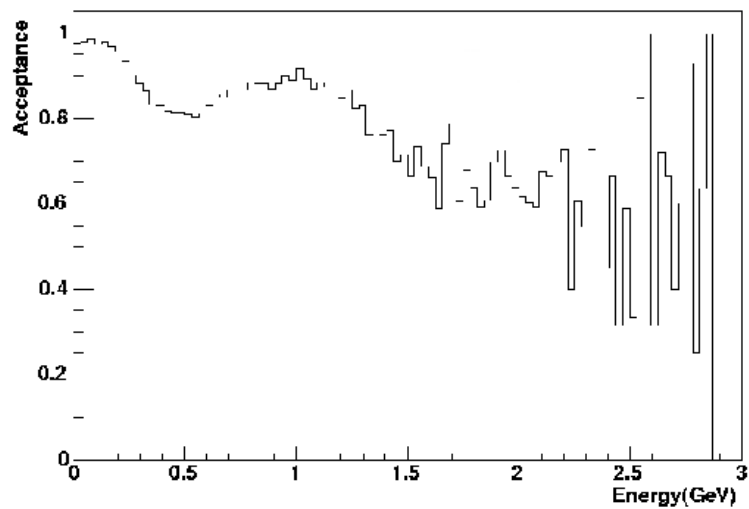}}
\caption{\it Acceptance versus true kinetic energy for muons from 
  the CCQE event sample.}\vspace{0.25in}
\label{fig:wiggles}
\end{figure}

FINeSSE and MiniBooNE are assumed to have the same reconstruction
efficiency. This is conservative because the finely segmented FINeSSE
detector will have much better resolution on the reconstructed
neutrino energy in quasi-elastic (QE) events. Energy resolution for
muons ranging out in the Muon Rangestack will also be known to 10\%
(see Section~\ref{sec:muonstack}).  FINeSSE will also have much
higher purity for the QE sample due to its excellent low energy
reconstruction.  For events to be counted in the FINeSSE data sample
for this analysis, outgoing muons from $\nu_{\mu}$ CC interactions
must be contained entirely in the Vertex Detector, in the Vertex
Detector plus the veto, or in these plus the Muon Rangestack.
Contributions from these three contained samples, as a function of
energy, are shown in Figure~\ref{fig:containedcont}.  Wiggles in this
distribution, seen clearly in Figure~\ref{fig:wiggles}, are due to
contributions from one distribution turning off as another turns on.
Folding this acceptance into overall number of interactions expected
yield the event rates summarized in
Table~\ref{tab:eventrates}.

\subsection{The Motivation for 25~m Absorber Running: Parallax}

\begin{figure}
\centering
\begin{minipage}{2.5in}
\includegraphics[width=2.5in,bb=60 130 570 740]{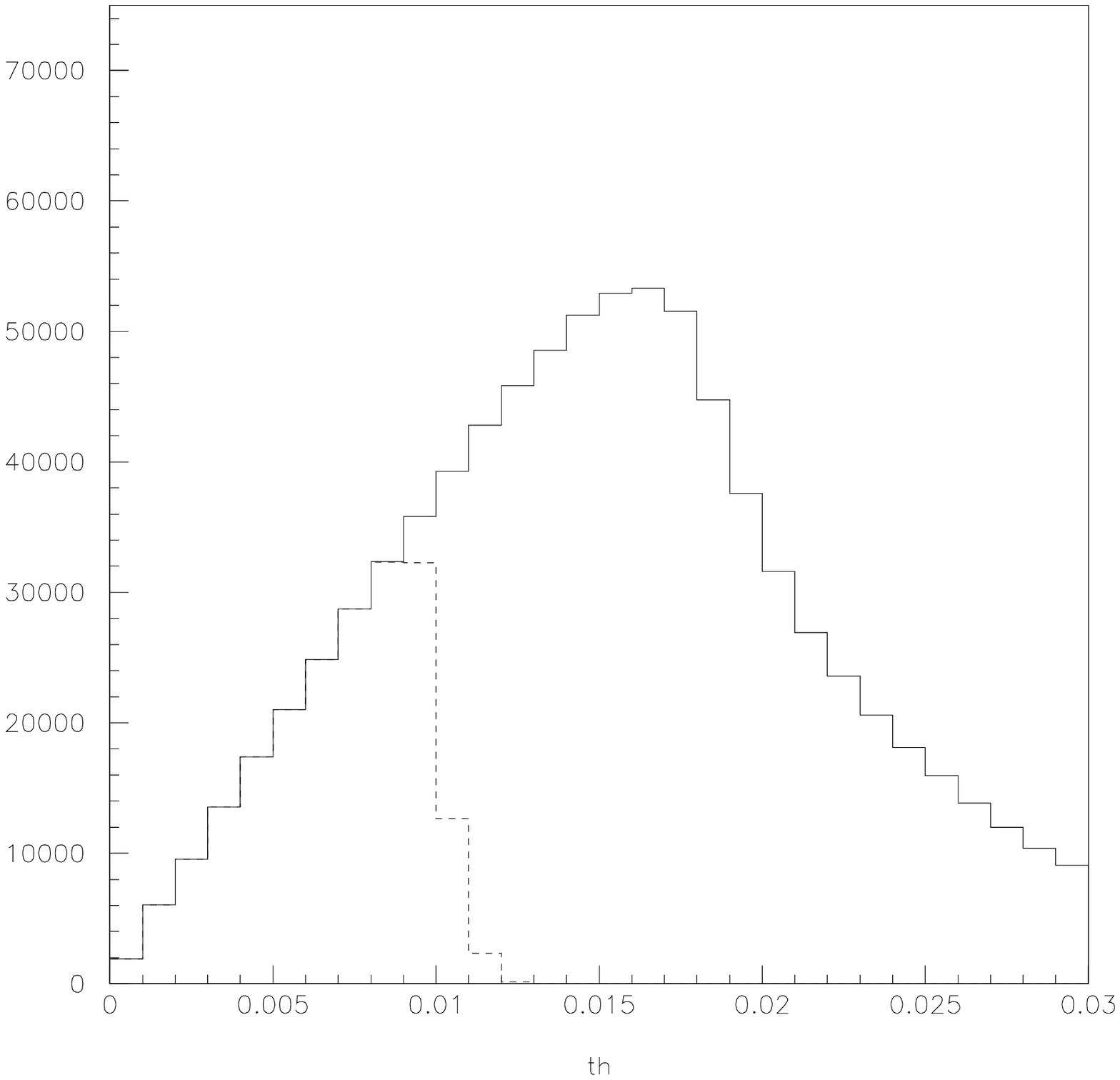}
\end{minipage}
\begin{minipage}{2.5in}\includegraphics[width=2.5in, bb=60 130 570 740]{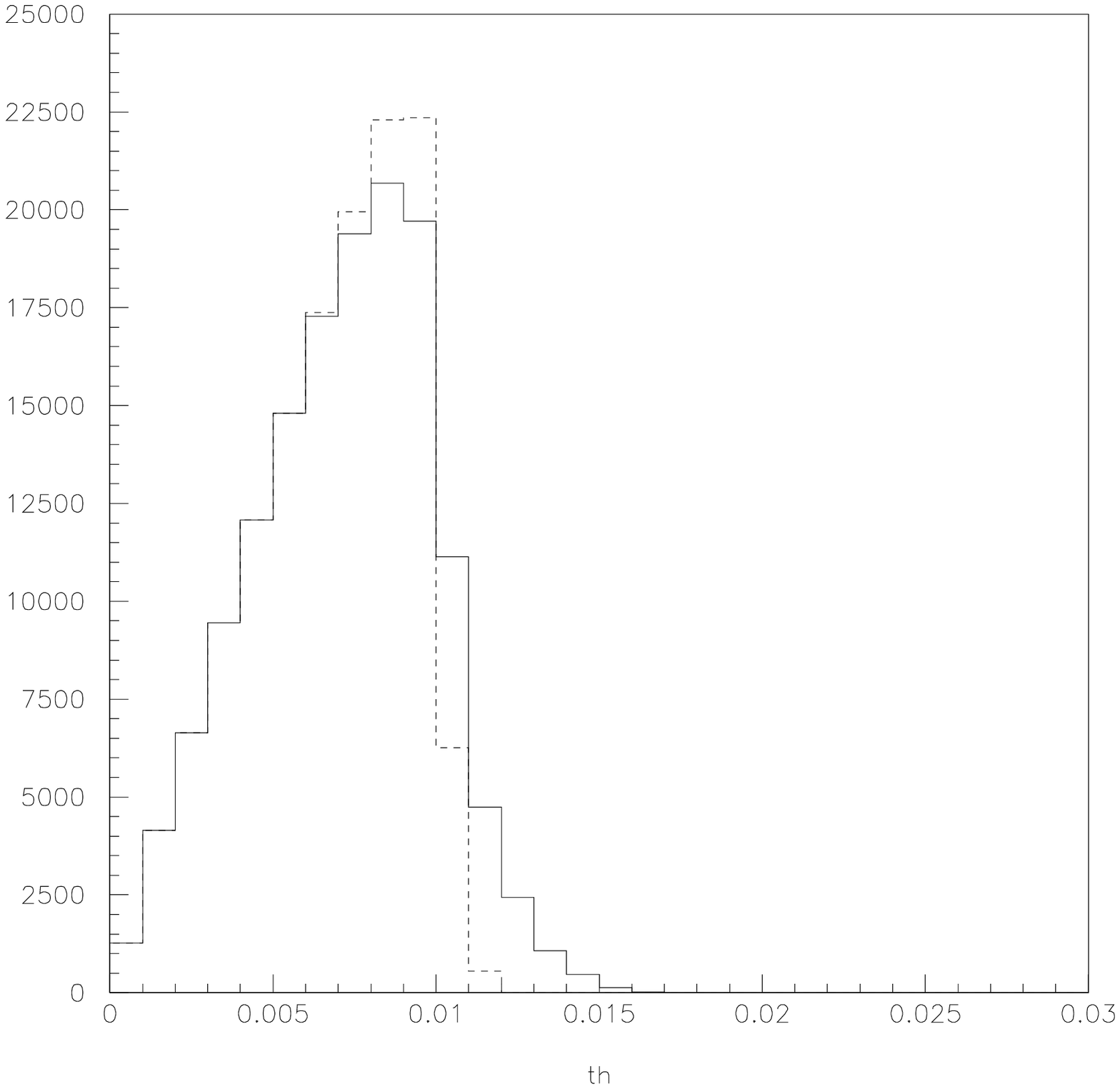}
\end{minipage}
\caption{\it Left, solid: Angular acceptance of the FINeSSE detector with 
  the 50~m absorber in radians; Right, solid: Acceptance with the 25~m
  absorber in radians.  Dashed (both plots): acceptance of the
  MiniBooNE detector in radians.}\vspace{0.25in}
\label{fig:angleaccept}
\end{figure}

\begin{figure}
\centering
\begin{minipage}{2.5in}
\includegraphics[width=2.5in]{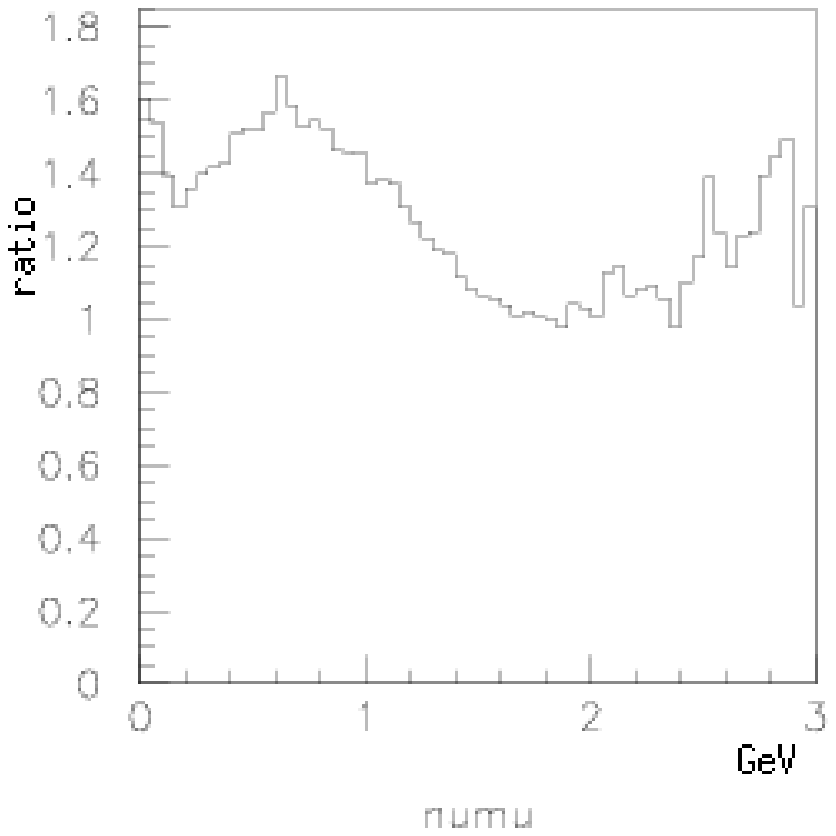}
\end{minipage}
\begin{minipage}{2.5in}\includegraphics[width=2.5in]{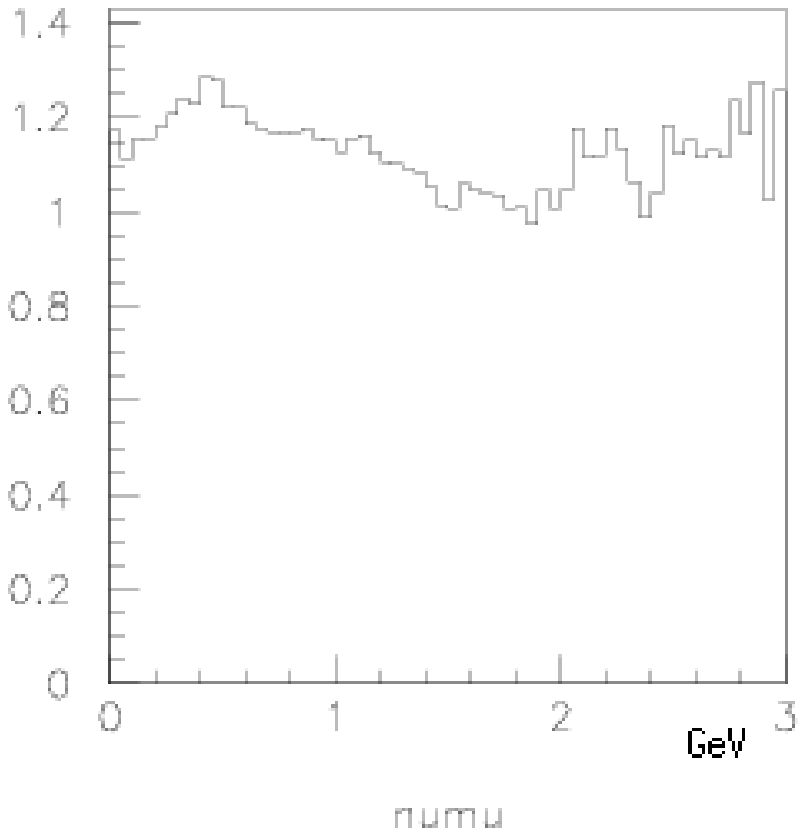}
\end{minipage}
\caption{\it Left: Ratio of the $\nu_\mu$ flux in the FINeSSE to the MiniBooNE detector
  using the 50~m absorber; right: Ratio when 25m absorber is
  installed}\vspace{0.25in}
\label{fig:near2far}
\end{figure}

The main physics driver for the FINeSSE+MiniBooNE standard
configuration was reduction of ``parallax'' in the near detector.  In
this section, we describe the cause of parallax, why it is a
significant issue for disappearance experiments, and how we have
mitigated the effect with our design.

In an ideal world, there are two reasons why disappearance experiments
based on near/far detector comparisons would be performed with
point-source beams.  First, the $L$ of each detector is well
determined.  Second, the near and far detector have the same energy
and angular acceptance for the neutrinos.  Unfortunately, the only
true point sources come from beams created by stopped pion and muon
decay (for example at LANL or SNS). These neutrinos are so low in
energy that a muon cannot be produced in CC interactions.  Hence
$\nu_\mu$ disappearance searches are impossible.

\begin{figure}
\centerline{\includegraphics[width=2.25in, bb=60 130 570 740]{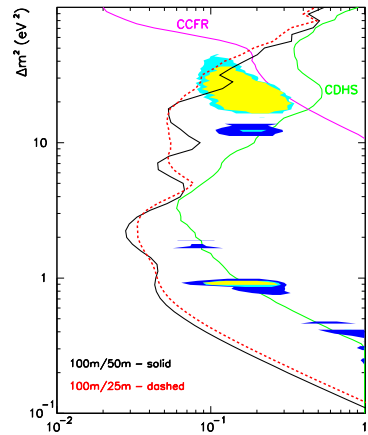}}
\caption{\it Comparison of the parameter space covered by FINeSSE+MiniBooNE with the 25~m absorber (red) and the 50~m absorber (black). The unwanted ``wiggles'' in the 50~m absorber allowed region are caused by the variations in the ratio of fluxes at FINeSSE and MiniBooNE.}
\label{fig:2550}
\end{figure}

\begin{figure}
  \centerline{\includegraphics[width=2.5in, bb=60 130 570 740]{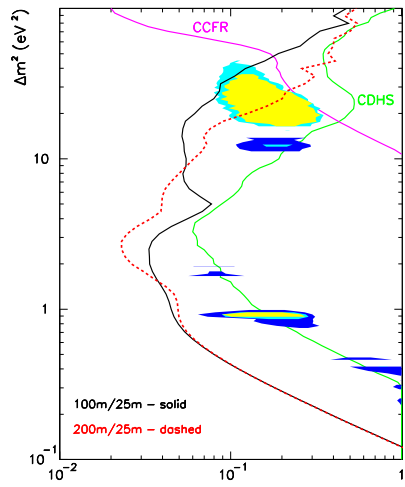}}
\caption{\it Comparison of the parameter space covered by 
  FINeSSE+MiniBooNE for a detector at 100~m (black) from the target
  versus 200~m (red) from the target.  Sensitivity to high $\Delta
  m^2$ is lost at 200~m because neutrinos will oscillate before
  reaching the FINeSSE detector.}\vspace{0.25in}
\label{fig:100200}
\end{figure}

To produce a higher energy neutrino beam, an extended decay path is
necessary.  In this case, the modification of $L$ for the decay length
must be taken into account.  Also, the angular acceptance of the
detector will vary depending on the point along the line of the decay.
For a detector which is relatively far from the beamline, such as is
the case with MiniBooNE, these corrections are negligible.  But for a
near detector, this ``parallax'' can be a large effect.

Because there is a strong correlation between energy and angle of the
neutrinos, differences in angular acceptance between the near and far
detectors is a major concern.  These differences translate directly
into differences in the energy distribution of events.  When taking a
near/far ratio, the result are ``wiggles,'' which are not related to
oscillations.  As long as those wiggles do not mimic an oscillation
signal, this is not a major problem.  But with realistic detector
efficiencies and smearing, these variations can severely limit
sensitivity.  Therefore, a top priority of any near/far experiment
must be to mitigate parallax.

An obvious solution is to move the near detector sufficiently far from
the line that the effect of parallax dies away.  However, for a
short-baseline disappearance experiment, this reduces the reach in
$\Delta m^2$ because some neutrinos have oscillated prior to reaching
the near detector, thereby reducing the apparent flux.

For the FINeSSE+MiniBooNE running, we propose a configuration in which
the 25~m absorber is installed in the beamline.
Figure~\ref{fig:beam-absorber} shows a schematic of the beam design,
including the position of the enclosure for the 25~m absorber.  This
absorber is presently hanging above the beamline.  It can be lowered
(and subsequently raised) on chains.  The process takes approximately
one week.

Installing the 25~m absorber allows the near detector to be placed 100~m
from the target, which would minimize the problem of ``parallax.''  Based
on the JAM flux, 70\% of the pions decay before 25~m, but the
average energy of the neutrinos from those decays is lower.  As a result, the event
rate with the 25~m absorber installed is approximately 50\% of the
event rate using the 50~m decay path.  However, the loss in rate is
more than compensated by the improved systematics.

Figure~\ref{fig:angleaccept} shows the angular acceptance for the
100~m detector with a 50~m (left) and 25~m (right) decay length.  The
25~m decay length has a much narrower angular acceptance. The dashed
line indicates the acceptance of MiniBooNE for the two absorber
lengths.  Introducing the 25~m absorber results in good agreement.
 
As discussed above, the difference in angular acceptance translates
into a difference in the flux spectrum at FINeSSE and at MiniBooNE.
Figure~\ref{fig:near2far}(left) shows the ratio of the FINeSSE to
MiniBooNE $\nu_\mu$ fluxes for the 50~m absorber, while
Figure~\ref{fig:near2far}(right) shows the same ratio for 25~m.  The
25~m option is clearly substantially better.  While some wiggles
remain, these are not resolved as a false oscillation signal (see
discussion of systematic studies below).  Figure~\ref{fig:2550} shows
the parameter space covered by FINeSSE with the 25~m absorber
configuration (black) compared to the 50~m configuration (red).  The
25~m absorber design maintains the better reach.

An alternative method for handling the parallax would be to move the
detector to 200~m.  This has the undesirable impact of reducing the
event rate by more than a factor of four, which limits the other
analyses.  Also, it reduces the sensitivity to the low $\Delta m^2$
allowed regions of 3+1 models, because the neutrinos would have
oscillated before reaching the near detector.  A comparison of the
sensitivity for 100~m and 200~m is shown in Figure~\ref{fig:100200}.
At 200~m, sensitivity to the high $\Delta m^2$ island is lost.
Therefore, this is not an ideal solution.

\subsection{Systematics Errors and Studies}

Systematic errors from the following sources have been included in the
analysis:

\begin{description}
  
\item[$\delta_{eff}$:] The relative error in the total rates between
  the near and far detector.  This is limited mainly by understanding
  of the fiducial volume of both detectors.  The understanding of the
  efficiency of each detector, and the uncertainty in the knowledge of
  $L$ for the 100~m detector, also enter here.  We estimate this from
  Monte Carlo studies of the detectors.  In our studies, we explore
  realistic (5\%) and optimistic levels (2\%) for this systematic.
  
\item[$\delta_{norm}$:] The correlated overall rate normalization of
  near and far detector.  This is limited by knowledge of the number
  of protons on target, and of secondary meson production per proton
  on target.  Since this is not very important for a two detector
  measurement, we have used a conservative estimate of 20\%
  uncertainty for these studies.
  
\item[$\delta_{shape}$:] The systematic error associated with the
  near/far energy difference.  This is the systematic associated with
  parallax.  It was estimated by varying the beam secondary production
  within the allowed errors.  The variation in rate of high angle
  neutrinos in the near detector then translates into this systematic.
  The studies of the variations were done ``blind,'' with one person
  inserting variation and another performing the analysis to identify
  and minimize the effect by changing cuts.  Six studies were
  performed and the range of variation was up to 20\%.
  
  However, we believe that this error can be substantially reduced by
  data from the HARP experiment\cite{harp} and the LMC detector in the
  beamline.  HARP is an experiment studying production of secondary
  mesons by 8~GeV protons on replicas of the MiniBooNE beryllium
  target.  HARP results will easily observe or eliminate the wildest
  variations used in the above study.  The Little Muon Counter (LMC)
  spectrometer also will provide an {\it in situ} test of excess
  wide-angle production.  The LMC is located just beyond the 25~m
  absorber; it consists of a long pipe at 7$^\circ$ degrees, followed
  by a muon spectrometer.  If there is an anomalously high rate of
  high angle secondary production, it will be observed in the LMC
  during the MiniBooNE 50~m running.  Assuming HARP and LMC limit the
  variations, a systematic of 10\% is obtainable.
  
\item[$\delta_{E~scale}$:] The relative energy scale error between the
  near and far detectors.  This will be set using the energy
  distribution of Michel electrons from muon decays as a standard
  candle, along with the $\pi^0$ mass and cosmic ray muon tracks. We
  believe that a 5\% relative error can be straightforwardly achieved.
  We have also looked at the improvement if a 1\% relative calibration
  were to be obtained; this, however, would be very difficult.

\end{description}

In order to cross-compare the sensitivity for various combinations of
the systematic errors, we use the 90\%~CL limit on $\sin^2 2\theta$ for
a null disappearance signal at $\Delta m^2$ of 1~eV$^2$.
Table~\ref{tab:disappearsys} shows the results.  The top section of
the table applies to 100~m running with the 50~m absorber; the
middle section shows the expectation for the standard configuration of
25~m running.  For comparison, the bottom section shows the sensitivity
if all systematic errors are assumed to be negligible.

\begin{table}[tbp] \centering
\begin{tabular}{|l|c|cccc|c|}
\hline
 & protons & $\delta_{eff}$ & $\delta_{norm}$ &
$\delta_{shape}$ & $\delta_{escale}$ & $\sin ^{2}2\theta $ \\
 & on target &  &  &  &  & (90\%CL@1eV$^{2}$)\\
\hline
 100~m+50~m & $1\times 10^{20}$ & 0.05 & 0.20 & 0.20 & 0.05 & 0.1040 \\
& $2.5\times 10^{20}$ & 0.05 &  & 0.20 &  & 0.0821 \\
& $2.5\times 10^{20}$ & 0.02 &  & 0.10 &  & 0.0505 \\
& $5\times 10^{20}$ & 0.05 &  & 0.20 &  & 0.0648 \\
& $5\times 10^{21}$ & 0.05 &  & 0.20 &  & 0.0234 \\
\hline
100~m+25~m & $2.5\times 10^{20}$ & 0.05 & 0.20 & 0.20 & 0.05 & 0.0791 \\
& $2.5\times 10^{20}$ & 0.02 &  & 0.20 & 0.05 & 0.0490 \\
& $2.5\times 10^{20}$ & 0.05 &  & 0.10 & 0.05 & 0.0628 \\
& $2.5\times 10^{20}$ & 0.05 &  & 0.20 & 0.01 & 0.0791 \\
& $2.5\times 10^{20}$ & 0.02 &  & 0.10 & 0.05 & 0.0467 \\
& $2.5\times 10^{20}$ & 0.15 &  & 0.20 & 0.05 & 0.0900 \\
& $5\times 10^{20}$ & 0.02 &  & 0.10 & 0.05 & 0.0418 \\
\hline
Rate Only & $2.5\times 10^{20}$ & 0.02 & 0.20 & -- & 0.05 & 0.0650 \\
(100~m+25~m) & $2.5\times 10^{20}$ & 0.05 &  & -- &  & 0.1595 \\
\hline
Shape Only & $2.5\times 10^{20}$ & 0.02 & 0.20 & 0.10 & 0.05 & 0.0459 \\
(100m+25m) & $2.5\times 10^{20}$ & 0.05 &  & 0.20 &  & 0.0825 \\
\hline
200~m+25~m & $2.5\times 10^{20}$ & 0.05 & 0.20 & 0.20 & 0.05 & 0.0702 \\
         & $2.5\times 10^{20}$ & 0.02 &  & 0.10 &  & 0.0524 \\
\hline
Rate Only & $2.5\times 10^{20}$ & 0.02 & 0.20 & -- & 0.05 & 0.0776 \\
(200~m+25~m) & $2.5\times 10^{20}$ & 0.05 &  & -- &  & 0.1898 \\
\hline
Stat.Only & $2.5\times 10^{20}$ & 0.0 & 0.0 & 0.0 & 0.0 & 0.0067\\
\hline
 other $\Delta$m$^2$& &  &  &  &  & (90\%CL@20eV$^{2}$)\\
 100~m+25~m&$2.5\times 10^{20}$ & 0.02 &0.20  & 0.10 & 0.05 & 0.0650\\
 200~m+25~m&$2.5\times 10^{20}$ & 0.02 &0.20  & 0.10 & 0.05 & 0.1386\\
\hline
\end{tabular}
\caption{\it Comparisons of various experimental setups and assumed
systematic uncertainties. \label {tab:disappearsys}}
\end{table}

Results for a range of protons on target are shown.  From the table,
one sees that statistics are important, and that in any configuration
the results improve with further running.  Henceforth we will consider
only $6 \times 10^{20}$ POT, the full FINeSSE run.

From the table, one can see that either the 50~m or 25~m running
conditions allow for a disappearance search; the 25~m
configuration, however, is optimal.  Installing the 25~m absorber gives a
10\% better limit then the 50~m counterpart.  Assuming the 25~m
absorber, the most important systematic becomes $\delta_{eff}$ and
not $\delta_{shape}$, showing that the absorber is ``doing its
job'' in reducing the parallax.

For the bottom line, we assume the following:  $6\times 10^{20}$~POT,
$\delta_{eff}=2\%$, $\delta_{norm}=20\%$, $\delta_{shape}=10\%$, and
$\delta_{scale}=5\%$.    These are the assumptions for the sensitivity
shown in Figure~\ref{fig:bestfinese}.

\subsection{Method for Determining the Sensitivity}

To obtain the sensitivity, we use a standard method applicable to
near/far detector experiments \cite{Minikata}. A $\chi ^{2}$ is formed
from Monte Carlo estimates of the number of events in the near and
far detectors without and with oscillations for a given $\Delta
m^{2}$ and $\sin ^{2}2\theta $.

\begin{align*}
\chi ^{2}& =\sum\limits_{i}\frac{\left( \left(
N_{i}^{far}-N_{i}^{osc\_far}\right) \left( 1+k_{norm}\right) -\left( \frac{%
N_{i}^{far}}{N_{i}^{near}}\right) \left(
N_{i}^{near}-N_{i}^{osc\_near}+k_{shape}\Delta
N_{i}^{shape}\right) \right) ^{2}}{\left( N_{i}^{far}+\left(
\frac{N_{i}^{far}}{N_{i}^{near}}\right)
^{2}N_{i}^{near}\right) } \\
& +\frac{\left( \left( N_{tot}^{far}-N_{tot}^{osc\_far}\right)
\left(
1+k_{norm}\right) \left( 1+k_{eff}\right) -N_{tot}^{far}\right) ^{2}}{%
N_{tot}^{far}} \\
& +\frac{\left( \left( N_{tot}^{near}-N_{tot}^{osc\_near}\right)
\left(
1+k_{norm}\right) -N_{tot}^{near}\right) ^{2}}{N_{tot}^{near}} \\
& +\left( \frac{k_{eff}}{\delta _{eff}}\right) ^{2}+\left( \frac{k_{norm}}{%
\delta _{norm}}\right) ^{2}+\left( \frac{k_{shape}}{\delta
_{shape}}\right) ^{2}+\left( \frac{k_{_{Escale}}}{\delta
_{Escale}}\right) ^{2}
\end{align*}

\noindent The terms involve the events in the far detector binned in visible
energy, without and with oscillations, $N_{i}^{far}$ and
$N_{i}^{osc\_far}$; and the
events in the near detector binned in visible energy shifted by $%
k_{_{Escale}}$, with and without oscillations,  $N_{i}^{near}-N_{i}^{osc%
  \_near}$. (The terms with a $tot$ subscript are the same but summed
over all bins.) The oscillation event estimate correctly integrates
over the length of the decay pipe since this is an important effect
for the near detector. The first term in the expression is for the
energy dependent shape analysis, and the second two terms provide the
sensitivity to a total event counting technique.

Statistical errors are included along with systematic uncertainties
through the additional parameters, $k_{j}$.  The $k_{j}$ are
systematic error fit parameters and the $\delta _{j}$ are the
systematic errors associated with each of the uncertainties. The final
$(k/\delta )^{2} $ terms constrain the systematic parameters by their
assumed uncertainties. The systematic uncertainties, which are
described in detail in the following section, include: $k_{norm}$, the
overall normalization error common to both detectors; $k_{eff},$ the
relative normalization between the two detectors; $k_{shape}$, a
parameterization ($\Delta N_{i}^{shape}$) of a possible energy shape
difference between the two detectors; and $k_{_{Escale}}$, the
uncertainty in the energy scale between the two detectors. The energy
scale error is included by introducing the fit parameter,
$k_{Escale}$, that scales the energy for the near detector binning by
$E^{\prime }=\left( 1+k_{_{Escale}}\right) E$.

To find the sensitivity of a given experimental setup to oscillations,
the above $\chi ^{2}$ at a fixed $\Delta m^{2}$ is minimized with
respect to $\sin ^{2}2\theta $ and all of the $k$ parameters. For the
results shown in this proposal, the fit uses 50~MeV reconstructed
visible energy bins from 0.0 to 1.5 GeV.  Table \ref{systerrors} lists
the assumed systematic errors for the the sensitivity shown in
Figure~\ref{fig:bestfinese}.

\begin{table}[tbp] \centering%
\begin{tabular}{|l|l|l|}
\hline Source & $k$ Parameter & $\delta $ Values \\ \hline
Overall Normalization & $k_{norm}$ & 0.20 \\
Relative Normalization & $k_{eff}$ & 0.02 \\
Shape Uncertainty & $k_{shape}$ & 0.10 \\
Relative Energy Scale & $k_{Escale}$ & 0.05 \\ \hline
\end{tabular}
\caption{\it Systematic uncertainties used in the energy dependent
sensitivity
fits.\label{systerrors}}%
\end{table}%

\subsection{Coordination with MiniBooNE}

This analysis requires that the MiniBooNE and FINeSSE collaborations
agree on the beamline configuration and on sharing data.  After
discussions, the collaborations have agreed to a Memorandum of
Understanding, which is designed to prevent conflicts and to allow the
two experiments to work together.  In short, conflicts regarding the
beam configuration will be resolved by program planning; conflicts
regarding data-sharing will be resolved by a committee consisting of
the co-spokespersons of MiniBooNE and FINeSSE.

We expect this inter-group collaboration to proceed smoothly for three
reasons.  First and foremost, both the FINeSSE and MiniBooNE
collaborations regard the $\nu_\mu$ disappearance search as a high
priority.  Second, configuring the beam with the 25~m absorber may
have significant advantages for the Phase II physics of
MiniBooNE~\cite{MiniBooNEEOI}.  Thus the MiniBooNE collaboration is
expected to be supportive of this configuration.  However, if there is
strong pressure in favor of 50~m running, the $\nu_\mu$ disappearance
search will still be interesting, if not optimal, as discussed above.
Third, there is substantial overlap in membership between the two
collaborations.  Thus it will be possible for this bi-group analysis
to be performed by members of both groups.

\chapter{Additional Physics}
\label{ch:OtherPhysics}

\thispagestyle{myheadings}
\markright{}

While the physics foundations of FINeSSE are $\Delta s$ and
$\nu_{\mu}$ disappearance studies, there is also a wide range of other
interesting physics topics which FINeSSE can address.  For FINeSSE's
initial run, these include a significant contribution to cross section
measurements and an exploration of neutrino magnetic moments.  Potential
future runs of FINeSSE could also permit a measurement of $\Delta s$ using
antineutrinos, and substantial improvement on a $\nu_{\mu} \rightarrow
\nu_e$ oscillation signal, should one be observed in MiniBooNE.  We
explore these capabilities in this chapter.

\section{Cross Section Measurements}
\label{section:cross-sections}



Currently, oscillation experiments rely on modeling of neutrino
interactions in a regime that is poorly constrained by experimental
data.  Although accelerator-based neutrino beams have existed since
the 1970s, our primary knowledge of neutrino interactions at low
energy comes almost entirely from bubble chamber measurements made
decades ago at ANL, BNL, CERN, and FNAL, all of which were limited
both by low statistics and by large neutrino flux systematics. In
addition to (or perhaps because of) these large uncertainties
(typically $10-30\%$) the experimental results often conflict and are
difficult to interpret, mainly because of nuclear corrections and
exclusive final state ambiguities.

Improved knowledge of low energy neutrino cross sections will become
increasingly important as experiments move from discovery to precision
measurements of oscillation parameters. Consider the following
examples. Present atmospheric constraints on $\Delta m_{23}^2$ and
$\theta_{23}$ are already limited by flux and cross section
systematics.  Current uncertainties on neutral current (NC) $\pi^0$
production cross sections currently restrict the ability to
discriminate between $\numu \rightarrow \nu_\tau$ and $\numu
\rightarrow \nu_s$ transitions in studies of enriched NC samples in
the atmospheric neutrino data.  Furthermore, $\numu \rightarrow \nue$
appearance searches are limited by the statistical and systematic
errors related to background subtraction, again, most notably those
associated with NC $\pi^0$ interactions where the final state is
mis-identified as an electron. Both the kinematics and rate of NC
$\pi^0$ production are less precisely known than most other reaction
channels, because of the need to model resonant and coherent
contributions in addition to potential feed-down from inelastic
channels. More precise cross section measurements are not only
important for ensuring the success of oscillation measurements.
Resonant cross sections are particularly relevant for $p \rightarrow
\nu K^+$ proton decay searches, because poorly measured associated
strange particle production reactions such as $\numu \, n \rightarrow
\mu^- \, K^+ \, \Lambda$ and $\numu \, p \rightarrow \numu \, K^+
\Lambda$ present significant backgrounds and hence large resultant
systematics.
As a result, while present neutrino experiments could clearly benefit
from improved knowledge of low energy neutrino cross sections, such
advancement will become more crucial for the success of future
neutrino experiments.

\subsection{Present Understanding of Quasi-Elastic and Single Pion Cross Sections}
\begin{figure}[h]
\centering
\includegraphics[width=7.0cm,bbllx=42bp,bblly=22bp,bburx=353bp,bbury=475bp,clip=0,angle=270]{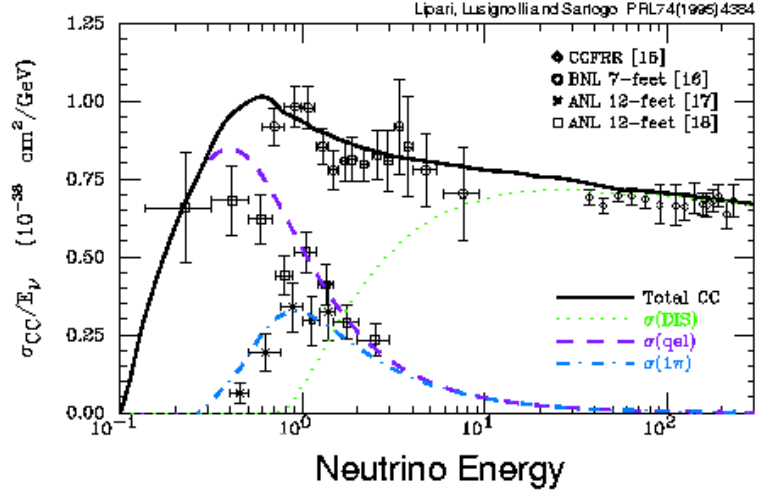}
\vspace{-0.2in}
\caption{\em Charged current neutrino cross section contributions as a 
  function of neutrino energy (in GeV): quasi-elastic (dashed),
  resonant single pion (dot-dash), and deep inelastic scattering
  (dotted) processes. Figure extracted from~\cite{lipari}.}
\label{fig:lipari}
\end{figure}

Figure~\ref{fig:lipari} shows the contributing neutrino cross sections
in the region of interest for atmospheric and terrestrial based
neutrino oscillation experiments.

\noindent
At energies near $\sim1$~GeV, neutrino interactions are predominantly
quasi-elastic (QE) and resonant single pion production; each are known
to the $10-20\%$ level from light target ($H_2$, $D_2$) neutrino data.
Although deep inelastic scattering (DIS) processes have been measured
with impressive precision ($<2\%$) at high energies, it is challenging
both to measure and to model neutrino interactions at low energies,
where there is substantial overlap between various poorly-understood
contributing processes.

Furthermore, because modern neutrino oscillation detectors consist of
heavy nuclei ($C$, $O$, $Ar$, $Fe$, $Pb$), the complex target adds
additional complications. In this case, one must deal with the effects
of Pauli blocking, Fermi motion of the target nucleons, and final
state interactions (i.e. careful accounting for the fact that the
outgoing hadron may reinteract before exiting the nucleus). Final
state effects (nuclear reinteractions, $\pi$ absorption, and charge
exchange) often dominate; they can vary depending on the neutrino
process, and certainly have not been disentangled experimentally.
Nuclear effects significantly impact both the rate and kinematics of
the neutrino reaction, as well as the observed final state event
composition and multiplicity (see, for example,
Section~\ref{section:event-rates}). Although nuclear effects have been
studied extensively using muon and electron beams, no comparable
effort has been made using neutrinos. Neutrino cross sections have
been measured on nuclear targets in the past\footnote{Gargamelle
  ($C_3H_8CF_3Br$), SKAT ($C_3H_8CF_3Br$), FNAL ($Ne$), CHARM and
  CHARM II (marble, glass), and Serpukhov ($Al$) are several
  examples.}, but these experiments suffered from low statistics and
typically published only free nucleon cross sections.  By making
dedicated, high statistics measurements of neutrino interaction cross
sections on a scintillator-based target, FINeSSE could greatly improve
the current experimental situation.

Several efforts are already underway to more precisely measure
neutrino interactions on nuclei at low energy. Measurements of NC
$\pi^0$/QE and inelastic/QE event ratios have been performed in the
K2K water \v{C}erenkov and scintillator-based fine grain near
detectors, respectively~\cite{k2k-near-detectors}. These measurements
exhibit $\sim10\%$ accuracy based on samples of roughly 5,000-10,000
events~\cite{xsect-k2k}.  Although MiniBooNE can additionally offer
improved cross section constraints with increased statistics over the
K2K near detector ensemble, such \v{C}erenkov-based detection methods
are inherently limited in their capabilities. The ability to
disentangle the various channels (QE, resonant, coherent, DIS, etc.)
in a nuclear environment necessitates use of a fine-grained detector
such as that being proposed for FINeSSE.

\subsection{Prospects for Measuring Cross Sections at FINeSSE}

The following subsections outline prospects for several exclusive neutrino 
cross section measurements at FINeSSE. This includes improved constraints on 
NC $\pi^0$ production and strange particle production.

\subsubsection{Neutral Current $\pi^0$ Production}
\label{subsection:xsect-ncpi0}

The dominant backgrounds to $\numu \rightarrow \nue$ appearance searches
result from two principal sources: the intrinsic $\nue$ component in the beam
and NC $\pi^0$ production where the final state is misclassified 
as an electron. Experiments primarily rely on Monte Carlo simulations to 
estimate their $\pi^0$ backgrounds. Such simulations must model
several mechanisms for producing a single $\pi^0$: resonance 
production, coherent single pion production, and deep inelastic scattering 
in which additional hadrons are absorbed in the nuclear medium before being 
detected. The dominant means of single pion production at low energy arises
through this first production mechanism: excitation of baryon resonances ($\Delta, N$)
that decay as:

\vspace{-0.5in}
\begin{eqnarray*}
   \numu \, N \rightarrow l \, &N^*& \\
                               &N^*& \rightarrow \pi \, N'~.
\end{eqnarray*}

\noindent
As a result, there are seven such resonant neutrino reaction channels: three
charged current and four neutral current:

\vspace{-0.2in}
         \begin{minipage}{0.5\textwidth}
             \begin{eqnarray*}
                \numu \, p &\rightarrow& \mu^- \, p \, \pi^+ \\
                \numu \, n &\rightarrow& \mu^- \, n \, \pi^+ \\
                \numu \, n &\rightarrow& \mu^- \, p \, \pi^0
             \end{eqnarray*}
         \end{minipage}\hspace{-1.0in}
         \begin{minipage}{0.5\textwidth}
             \begin{eqnarray*}
                \numu \, p &\rightarrow& \numu \, n \, \pi^+ \\
                \numu \, p &\rightarrow& \numu \, p \, \pi^0 \\
                \numu \, n &\rightarrow& \numu \, n \, \pi^0 \\
                \numu \, n &\rightarrow& \numu \, p \, \pi^-
             \end{eqnarray*}
         \end{minipage}
\vspace{0.2in}

\begin{figure}[h]
\centering
\mbox{
\includegraphics[width=0.6\textwidth,bb=5 5 528 546]{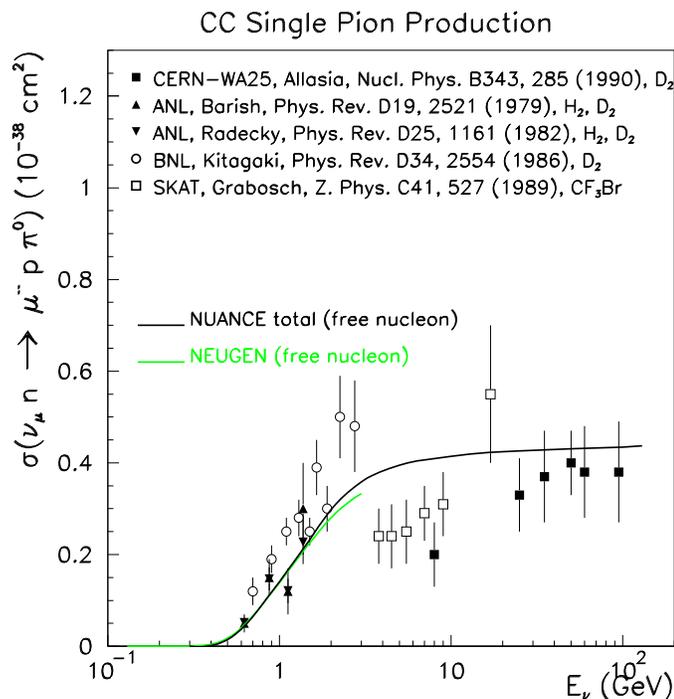}}
\vspace{-0.3in}
\caption{\em Measurements of the CC resonant single $\pi^0$ cross section, 
         $\sigma(\numu \, n \rightarrow \mu^- \, p \, \pi^0)$.
         Also shown are the Rein and Sehgal-based predictions from two 
         publicly available Monte Carlo generators~\cite{xsect-monte-carlos}.}
\label{fig:cc-resonant-pi0}
\end{figure}

\noindent
Traditionally, Monte Carlo simulations covering the low energy region
have used theoretical calculations by Rein and
Sehgal~\cite{rein-sehgal} to predict the rate and kinematics of
neutrino resonance production. Such models are tuned to reproduce
neutrino single pion data, but remain poorly constrained, because of
the limited availability and large uncertainties in existing
experimental data. As an example, Figure~\ref{fig:cc-resonant-pi0}
shows the experimental constraints on single $\pi^0$ production from
charged current (CC) neutrino data.  Note that all of the data at low
energy were collected from light targets.

\begin{figure}[h]
\begin{minipage}{0.5\textwidth}
\centering
\includegraphics[width=7.8cm,clip=0]{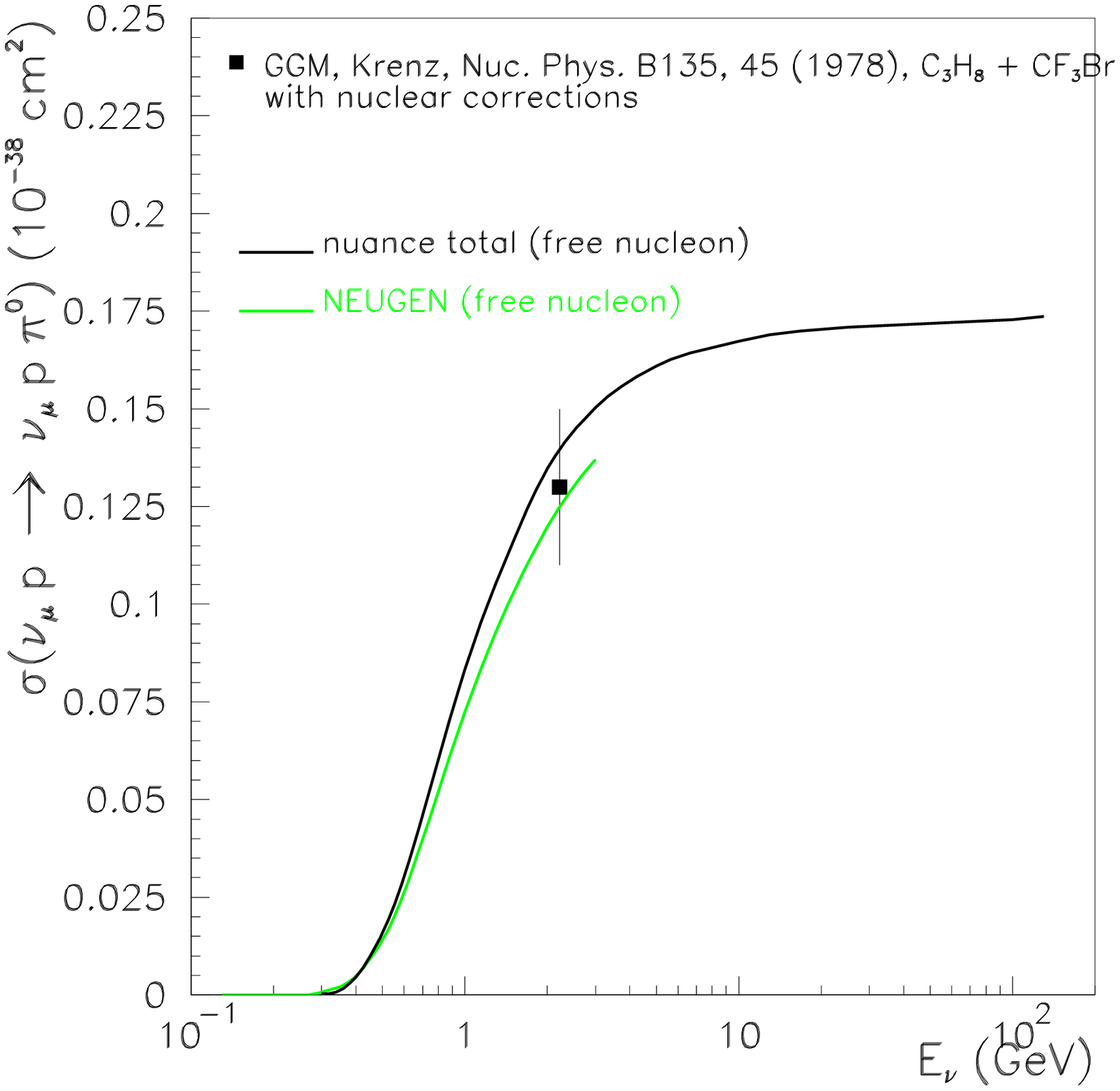}
\end{minipage}\hspace{0.2in}
\begin{minipage}{0.5\textwidth}
\centering
\includegraphics[width=7.8cm,clip=0]{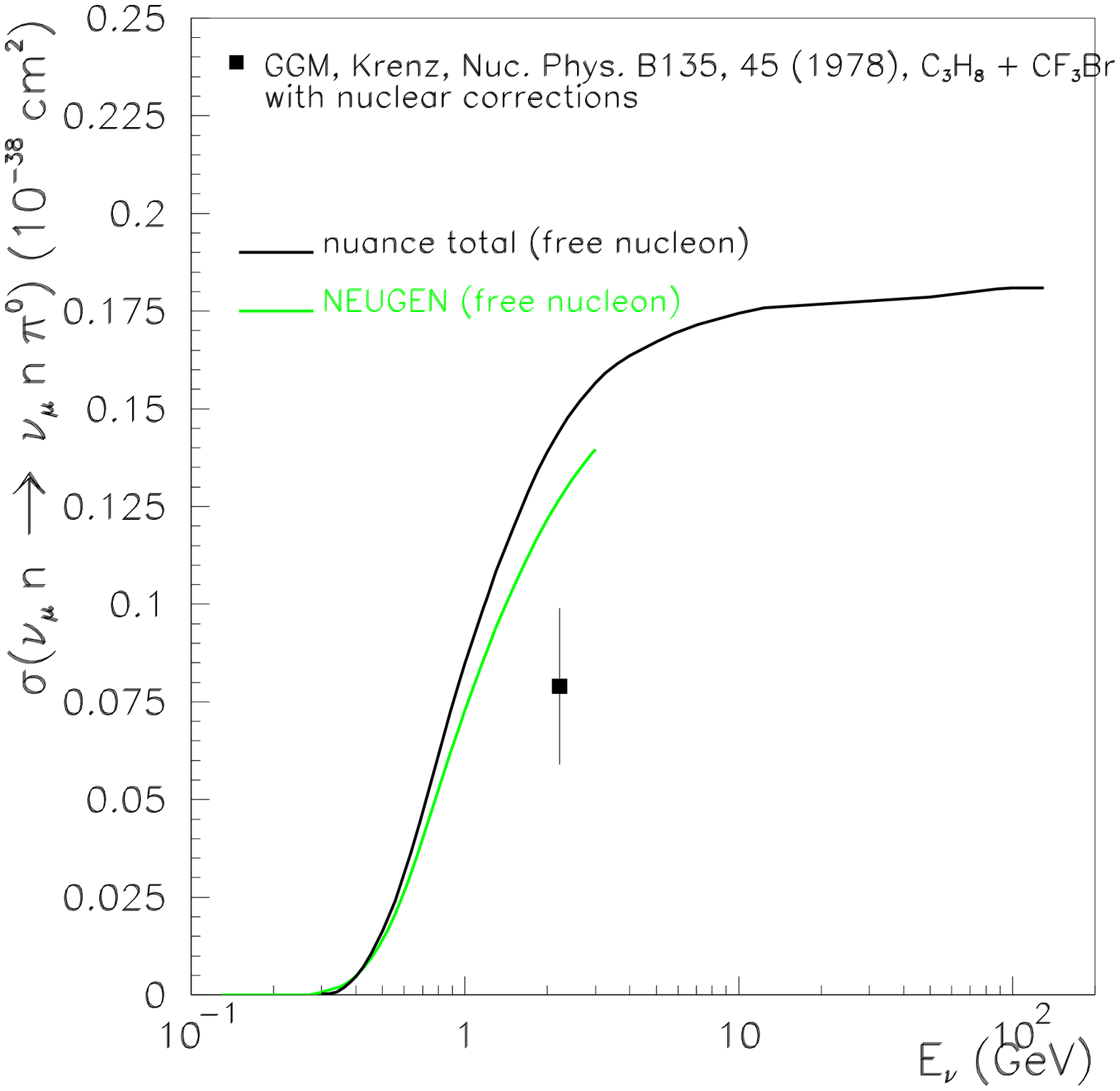}
\end{minipage}

\vspace{-0.2in}
\caption{\em NC resonant single $\pi^0$ cross sections, 
         $\sigma(\numu \, p \rightarrow \numu \,p \,\pi^0)$ (left) and
         $\sigma(\numu \, n \rightarrow \numu \,n \,\pi^0)$ (right).
         Also shown are the Monte Carlo predictions 
         from~\cite{xsect-monte-carlos}.}
\label{fig:nc-resonant-pi0}
\end{figure}

By contrast, Figure~\ref{fig:nc-resonant-pi0} shows the only available
experimental measurement of an absolute resonant NC $\pi^0$ cross
section.  These NC data result from a recent reanalysis of Gargamelle
bubble chamber data at 2 GeV~\cite{hawker-pi0}. Because the NC cross
sections are less well known,  experiments typically in practice
assign large $25-30\%$ uncertainties to NC resonant production
processes.

\clearpage

The data are even more sparse for the other contributing production
mechanism, coherent single pion production. In coherent interactions,
neutrinos scatter off the entire nucleus rather than its individual
constituents.  Because of the negligible energy transfer to the target
nucleus ($A$), such processes distinctly provide a single forward
scattered pion. Like in the resonant case, both NC and CC processes
are possible:

\vspace{-0.3in}
\begin{eqnarray*}
  \numu \, A &\rightarrow& \numu A \, \pi^0 \\
  \numu \, A &\rightarrow& \mu^- A \, \pi^+
\end{eqnarray*}

\begin{figure}[h]
\begin{center}
\mbox{
\includegraphics[width=0.5\textwidth,bb=19 131 562 698]{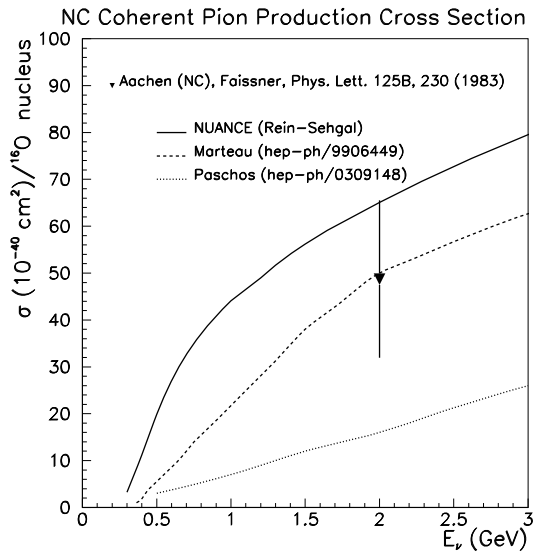}}
\end{center}
\vspace{-0.5in}
\caption{\em Experimental measurement of the NC coherent $\pi^0$ production 
         cross section at low energy compared to current model 
         predictions~\cite{xsect-coherent},\cite{rein-sehgal-coherent}.}
\label{fig:coherent-pi0}
\end{figure}

\noindent
Almost all current Monte Carlo simulations implement Rein and Sehgal's
calculation~\cite{rein-sehgal-coherent} of coherent pion production
cross sections and kinematics. While such predictions have been
constrained by numerous experimental measurements at high 
energy~\cite{vilain}, the lowest energy data available is at 2~GeV
on an aluminum spark chamber target~\cite{aachen}. 
Figure~\ref{fig:coherent-pi0} shows the low energy Aachen measurement 
compared to several model calculations. 

\noindent
The most recent calculations~\cite{xsect-coherent} yield a factor of
two to six less coherent pion production at these energies than the
earlier Rein and Sehgal based predictions~\cite{rein-sehgal-coherent}.
Because of the lack of low energy experimental data and the existence
of several conflicting theoretical predictions, oscillation
experiments typically assign a $100\%$ uncertainty to coherent
processes. This large uncertainty is especially important because
coherent production may comprise up to $20\%$ of the overall NC
$\pi^0$ rate. The ability to further constrain NC $\pi^0$ production
at low energies would thus be of great use in achieving increased
sensitivity to $\numu \rightarrow \nue$ oscillations, and in placing
more stringent limits on the oscillation of standard neutrinos to
sterile states.

\begin{figure}[h]
\mbox{
\begin{minipage}{0.5\textwidth}
  \mbox{
\includegraphics[width=\textwidth,bb=4 4 516 518]{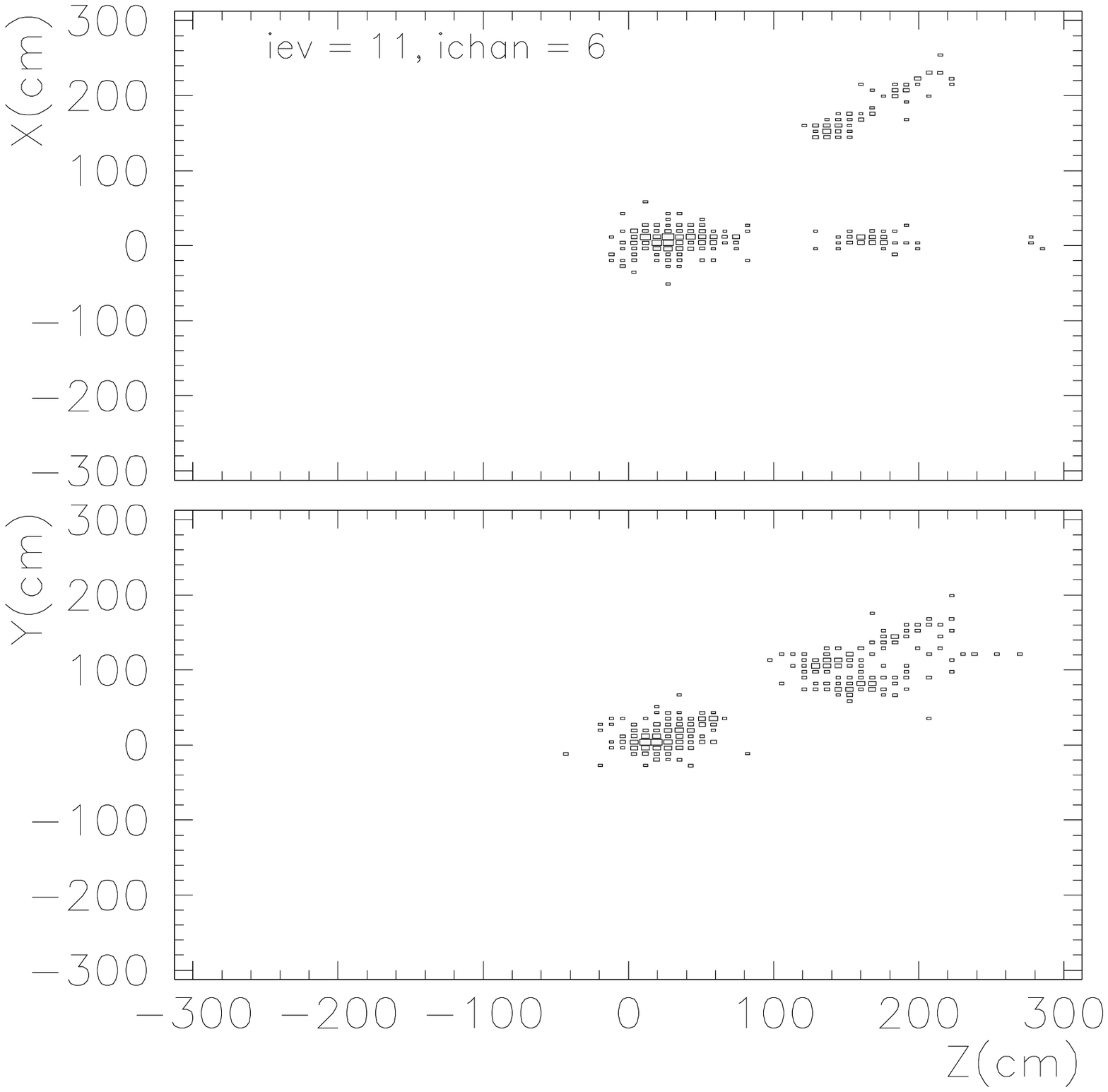}}
\end{minipage}\hspace{0.2in}
\begin{minipage}{0.5\textwidth}
  \mbox{
\includegraphics[width=\textwidth,bb=4 4  516 518]{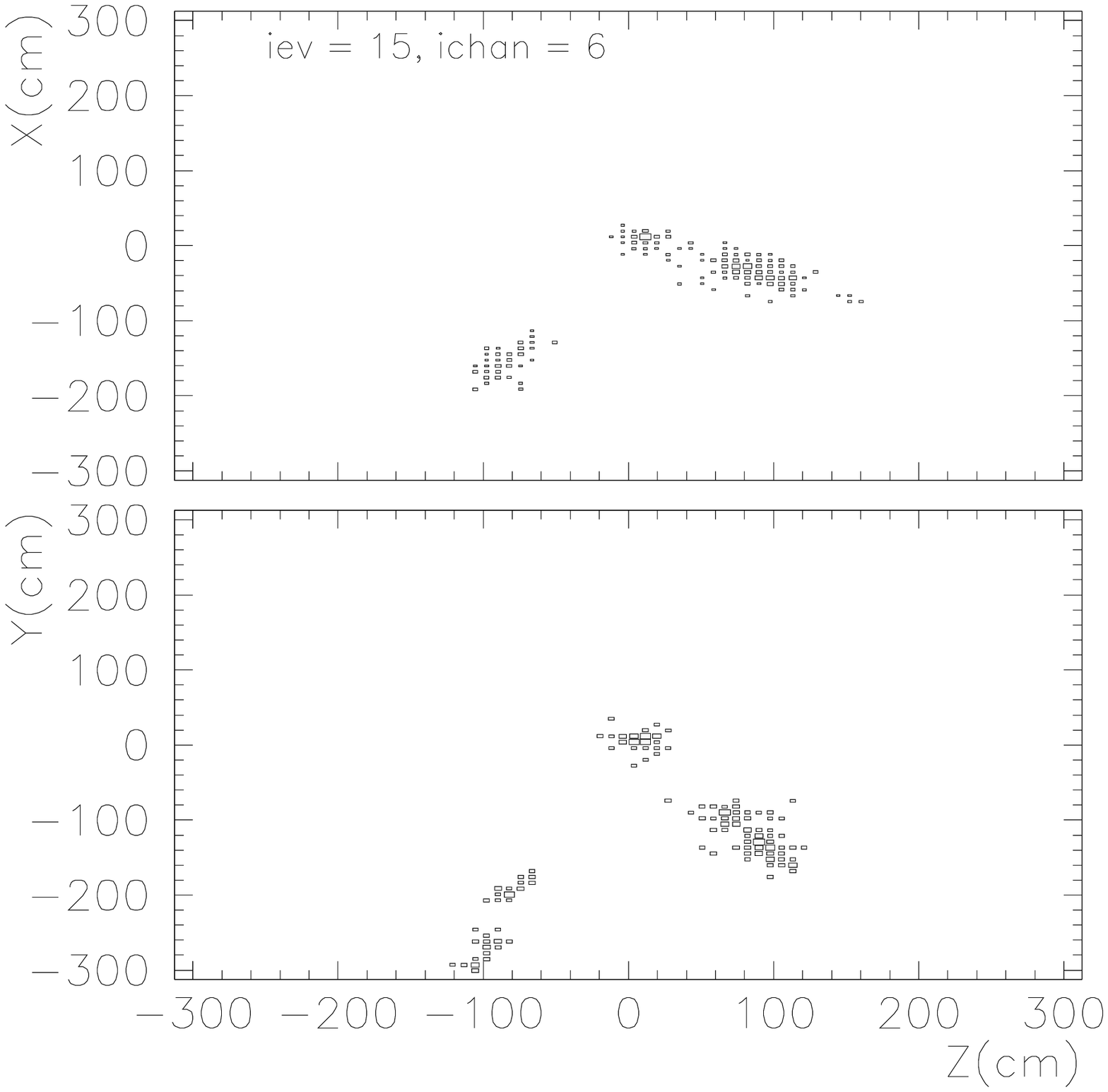}}
\end{minipage}
}
  \caption{\em Hit-level simulations of NC $\pi^0$ interactions in the 
    FINeSSE detector, $\numu \, p \rightarrow \numu \, p \, \pi^0$.
    The three hit clusters correspond to the final state proton and
    two photons from $\pi^0 \rightarrow \gamma \, \gamma$.}
\label{fig:finese-pi0-p-event}
\end{figure}

\begin{figure}[h]
\mbox{
\begin{minipage}{0.5\textwidth}
  \mbox{
\includegraphics[width=\textwidth,bb=2 3 515 518]{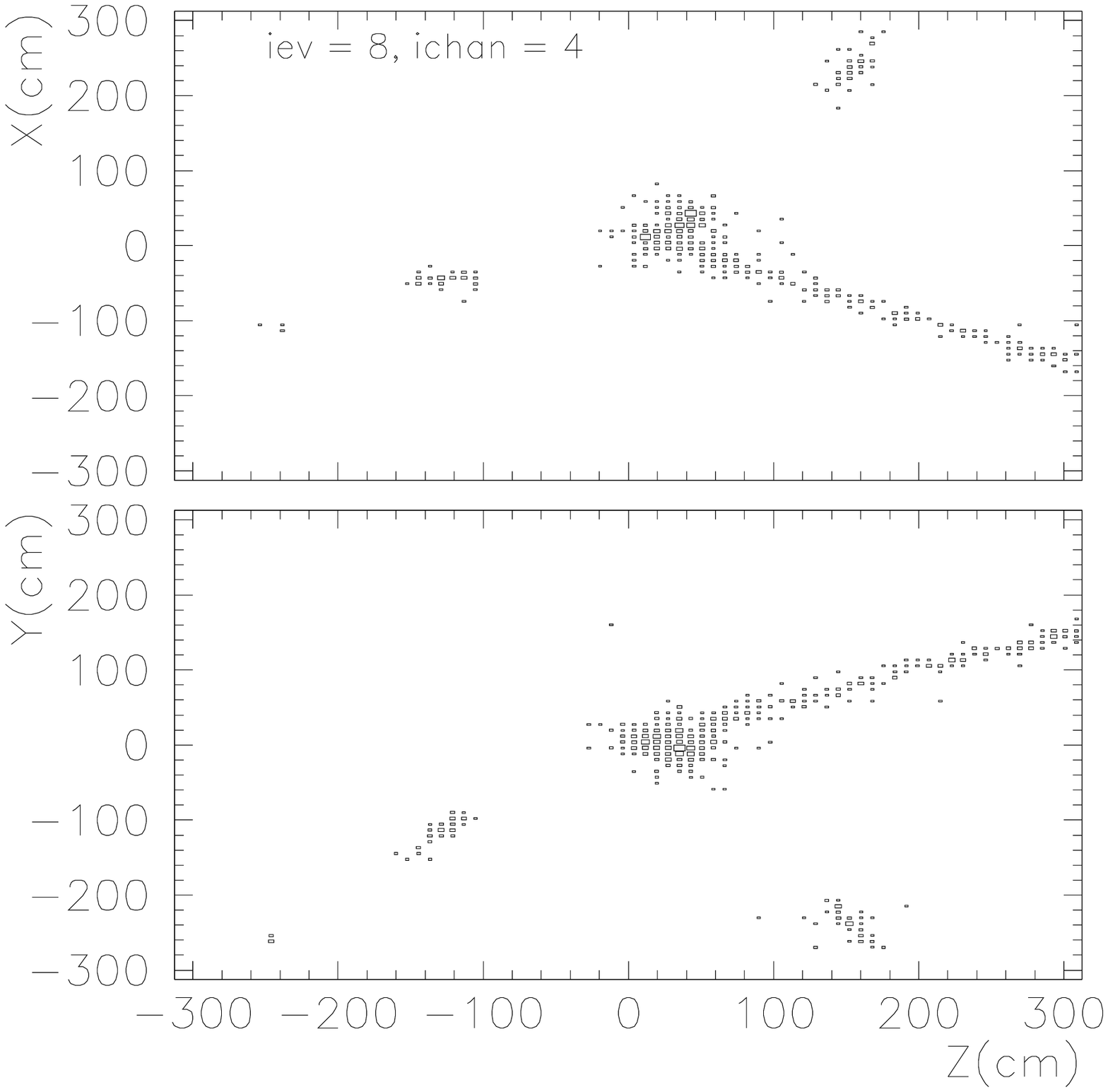}}
\end{minipage}\hspace{0.3in}
\begin{minipage}{0.4\textwidth}
  \mbox{
\includegraphics[width=\textwidth,bb=40 131 588 693]{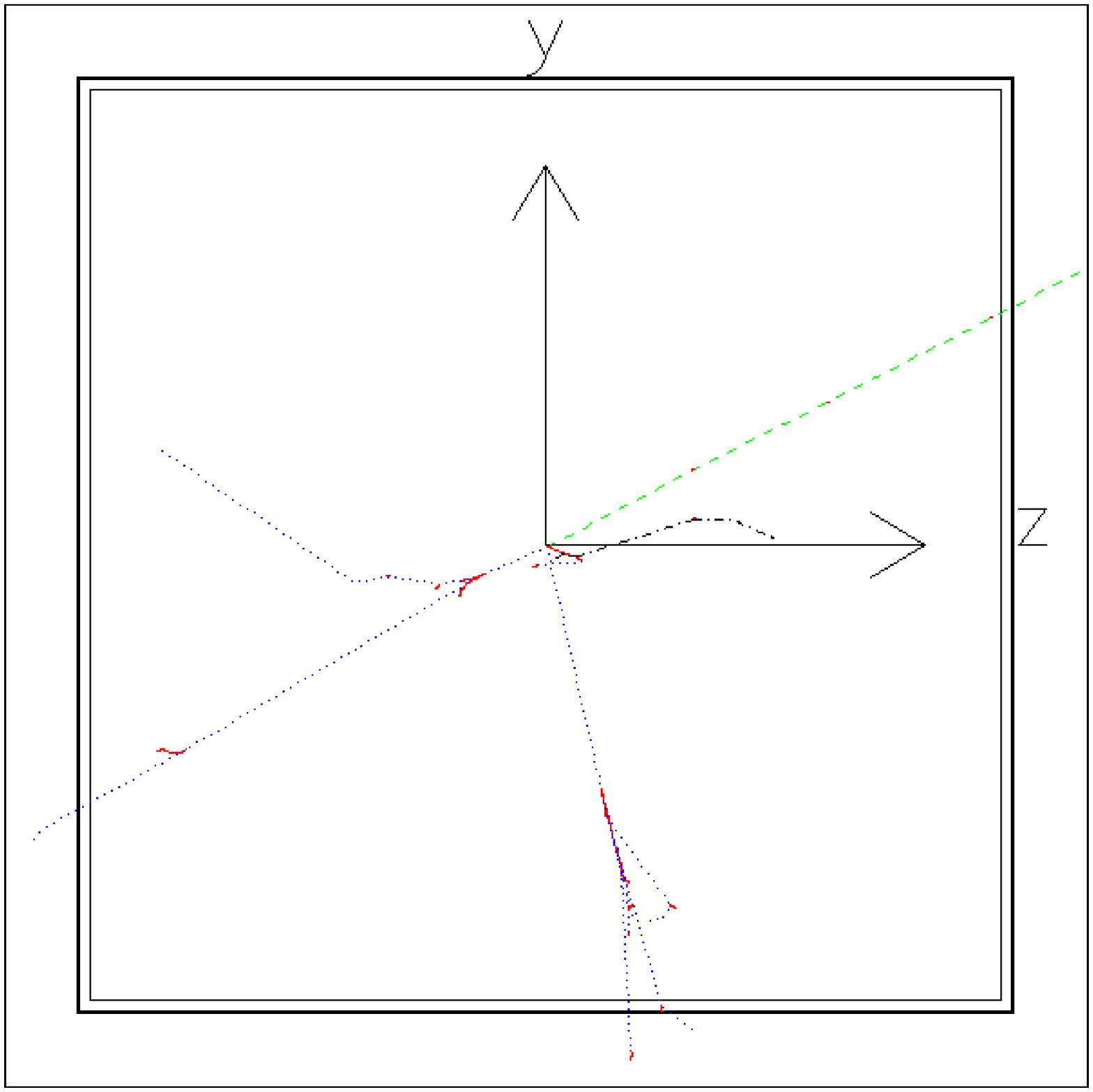}}
\end{minipage}
}
  \caption{\em A $\numu \, n \rightarrow \mu^- \, p \, \pi^0$ CC $\pi^0$ 
           interaction in the FINeSSE detector. The right hand figure 
           shows the true GEANT particle trajectories in the YZ plane.
           In this case, a muon is produced in addition to the three hit 
           clusters from the proton and two photons.}
\label{fig:finese-cc-pi0-event}
\end{figure}

\begin{figure}[h]
\mbox{
\begin{minipage}{0.5\textwidth}
  \mbox{
\includegraphics[width=\textwidth,bb=4 4 516 518]{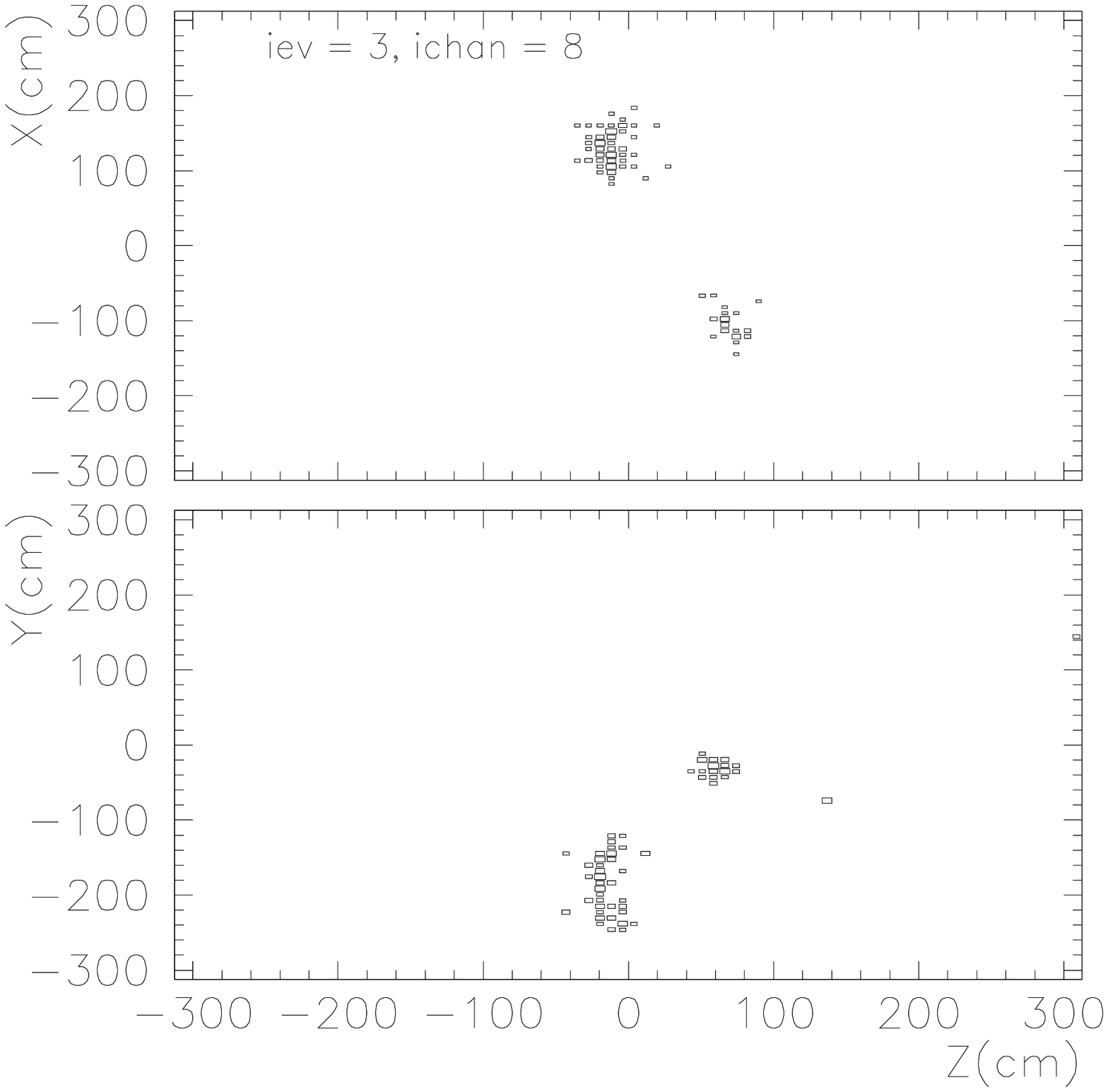}}
\end{minipage}\hspace{0.2in}
\begin{minipage}{0.5\textwidth}
  \mbox{
\includegraphics[width=\textwidth,bb=4 4 516 518]{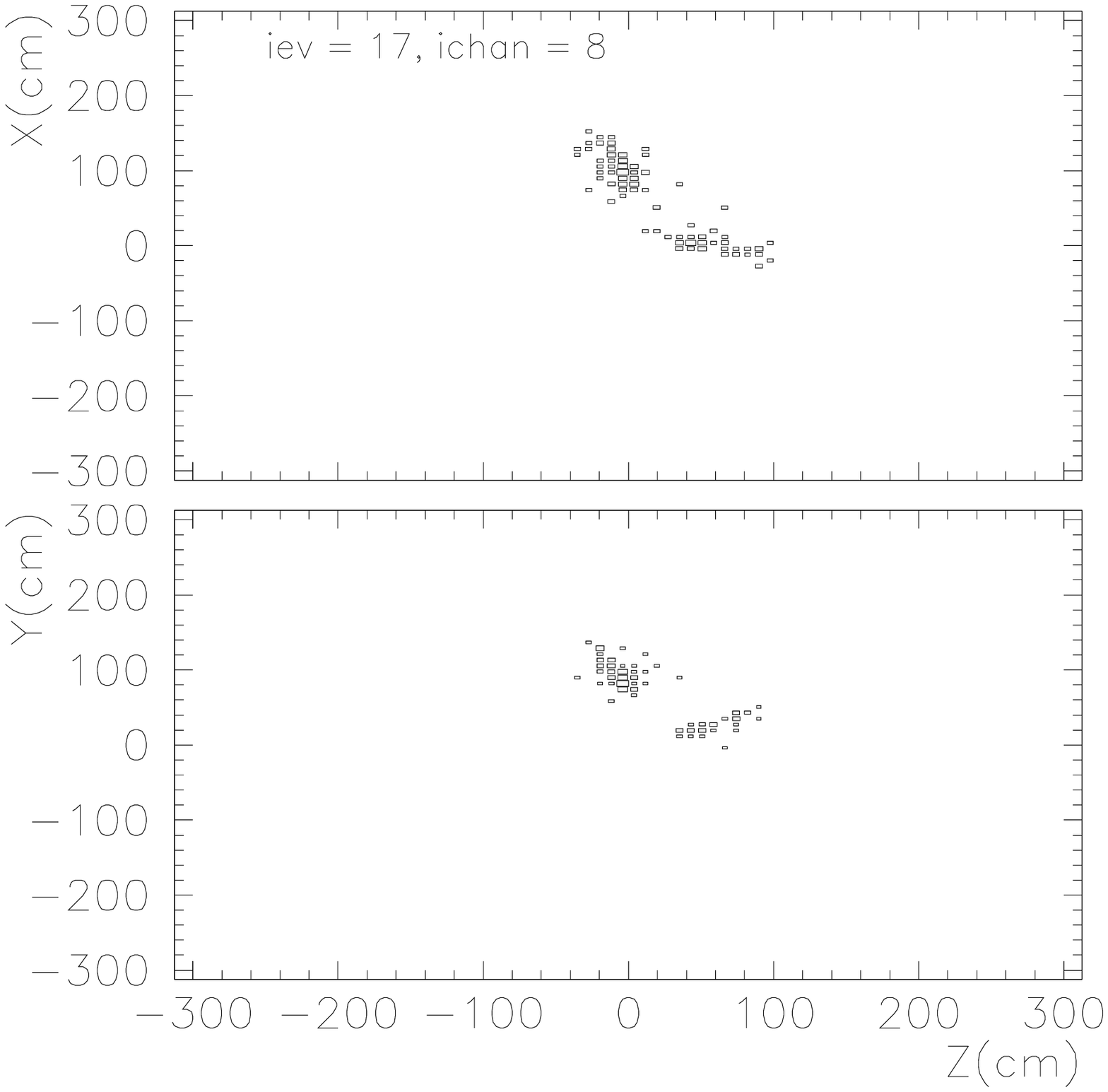}}
\end{minipage}
}
  \caption{\em Two NC $\pi^0$ interactions in the FINeSSE detector,
           $\numu \, n \rightarrow \numu \, n \, \pi^0$. The separated
           hit clusters correspond to the two photons emitted from
           the $\pi^0$ decay.}
\label{fig:finese-pi0-n-event}
\end{figure}

\begin{figure}[h]
\mbox{
\begin{minipage}{0.5\textwidth}
  \mbox{
\includegraphics[width=\textwidth,bb=2 3 515 518]{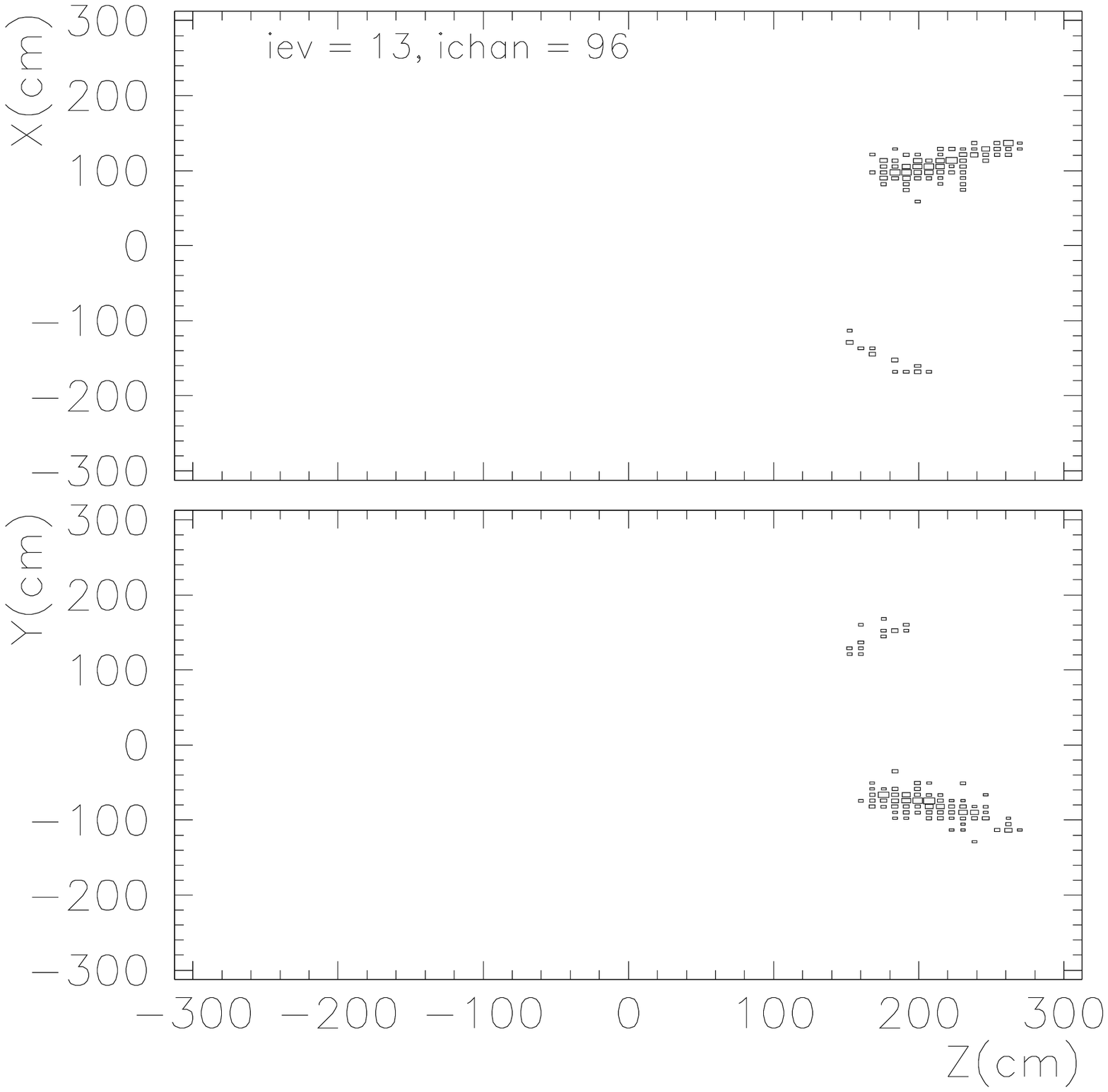}}
\end{minipage}\hspace{0.3in}
\begin{minipage}{0.4\textwidth}
  \mbox{
\includegraphics[width=\textwidth,bb=40 131 588 693]{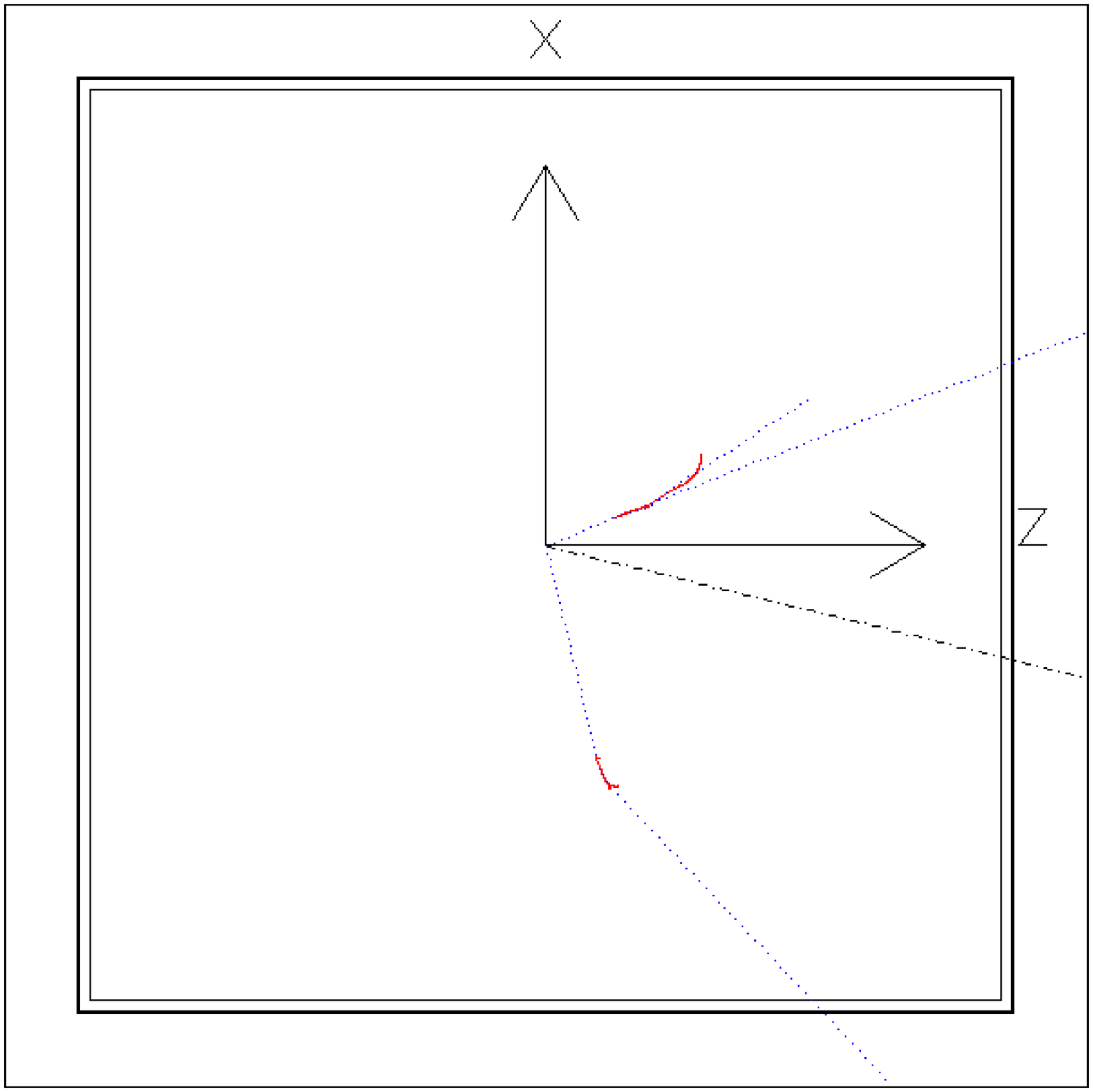}}
\end{minipage}
}
  \caption{\em A NC coherent $\pi^0$ interaction in the FINeSSE 
           detector, $\numu \, ^{12}C \rightarrow \numu \, ^{12}C \, \pi^0$. 
           The figure on the right shows the true GEANT particle trajectories 
           in the XZ plane for the same event. As can be seen, the hit clusters
           correspond to the two photons from the $\pi^0$ decay.}
\label{fig:finese-coherent-pi0-event}
\end{figure}

In two years of running, FINeSSE will collect a total sample of
$\sim20,000$ NC $\pi^0$ resonant events and over 5,000 NC coherent
$\pi^0$ interactions. In addition, FINeSSE's superior energy
resolution and event reconstruction capabilities will greatly enhance
the ability to select $\pi^0$ interactions.
Figures~\ref{fig:finese-pi0-p-event} and \ref{fig:finese-pi0-n-event}
show simulated NC resonant $\pi^0$ events in the FINeSSE detector. A
$\numu \, p \rightarrow \numu \, p \, \pi^0$ interaction
(Figure~\ref{fig:finese-pi0-p-event}) can be distinguished by the
presence of three separated energy deposits corresponding to the final
state proton and the two photons emitted from $\pi^0 \rightarrow
\gamma \, \gamma$.  A charged current $\pi^0$ event, $\numu \, n
\rightarrow \mu^- \, p \,\pi^0$,
(Figure~\ref{fig:finese-cc-pi0-event}) is additionally accompanied by
a final state muon track. In contrast, a $\numu \, n \rightarrow \numu
\, n \, \pi^0$ interaction (Figure~\ref{fig:finese-pi0-n-event}) can
only be produced by scattering from carbon, and will contain only two
clusters of hits corresponding to the two photons from the $\pi^0$
decay. NC coherent $\pi^0$ events
(Figure~\ref{fig:finese-coherent-pi0-event}) are similar in signature
to the former class of events; however, in this case, the energies and
angles of the two final state photons can be used to determine if the
$\pi^0$ angular distribution is more forward peaked as one expects for
coherent scattering.

\clearpage

While such events leave distinct signatures in the FINeSSE detector,
determining the efficiency and accuracy with which FINeSSE can measure
$\pi^0$ rates and kinematics will require development of a cluster
algorithm and tracker designed to located gaps in energy deposits.
While not yet optimized, preliminary studies have demonstrated that
$\pi^0$ efficiencies and purities can be obtained that surpass those
achieved with either the MiniBooNE or K2K \v{C}erenkov--based
detectors. Given that the largest sources of error arise from nuclear
uncertainties, by measuring $\pi^0$ production on carbon, FINeSSE
should therefore be able to reduce the currently assigned NC $\pi^0$
production uncertainty (Section~\ref{subsection:xsect-ncpi0}) by as
much as a factor of two.


In its ability to isolate $\pi^0$ interactions, FINeSSE has an
especially important role to play given that there is currently no NC
$\pi^0$ data on nuclear targets below 2 GeV, and only low energy CC
$\pi^0$ data measured on deuterium targets. FINeSSE is well-positioned
to be the first neutrino experiment to measure NC $\pi^0$ (and CC
$\pi^0$) production on carbon at these energies ($<E_\nu>\sim 700$
MeV), thus providing an important constraint to oscillation
experiments such as MiniBooNE, as well as to other accelerator- and
atmospheric-based $\nue$ appearance experiments employing heavy
nuclear targets. Such efforts at FINeSSE are also complementary to
slightly higher energy ($<E_\nu>\sim 1.3$ GeV) scintillator-based
analyses currently being planned after upgrades to the K2K near
detector site~\cite{k2k-scibar}.

\subsubsection{Strange Particle Production}
\label{section:xsect-strange-prod}

\begin{figure}[h]
\begin{center}
\mbox{
\includegraphics[width=0.5\textwidth,bb=14 145 528 649]{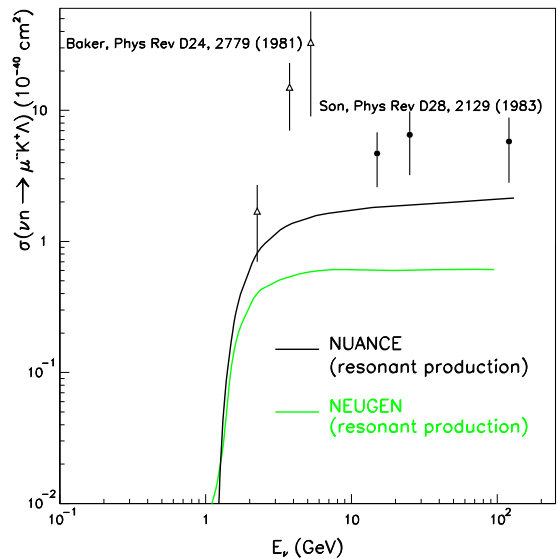}}
\end{center}
\vspace{-0.4in}
\caption{\em Experimental measurements of the
         $\numu n \rightarrow \mu^- K^+ \Lambda$ cross section. Also shown
         are the predictions from two publicly available Monte 
         Carlo generators~\cite{xsect-monte-carlos}.}
\label{fig:kaon-cross-section}
\end{figure}

Proton decay modes containing a final state kaon, 
$p \rightarrow \nu K^+$,  have large branching ratios in
many SUSY GUT models. Because there is a non-zero probability that an 
atmospheric neutrino interaction can mimic a proton decay signature,

\vspace{-0.3in}
\begin{eqnarray*}
   \numu \, n \rightarrow \mu^- \, K^+ \, \Lambda \\
   \numu \, p \rightarrow \numu \, K^+ \, \Lambda, 
\end{eqnarray*}

\noindent 
it is important to reliably estimate this background. Present
uncertainties on this background process are as large as
$100\%$~\cite{p-decay-unc} both because neutrino strange particle
production rates have been measured in only a few experiments, and
because there are not many predictive theoretical
models~\cite{k-prod-models}.  Figure~\ref{fig:kaon-cross-section}
shows the only two experiments which have published cross sections on
the dominant associated production channel, $\numu n \rightarrow \mu^-
K^+ \Lambda$. Both measurements were made on a deuterium target and
based on less than 30 events combined.

\begin{figure}[h]
\begin{center}
\mbox{
\includegraphics[width=0.5\textwidth,bb=4 4 516 518]{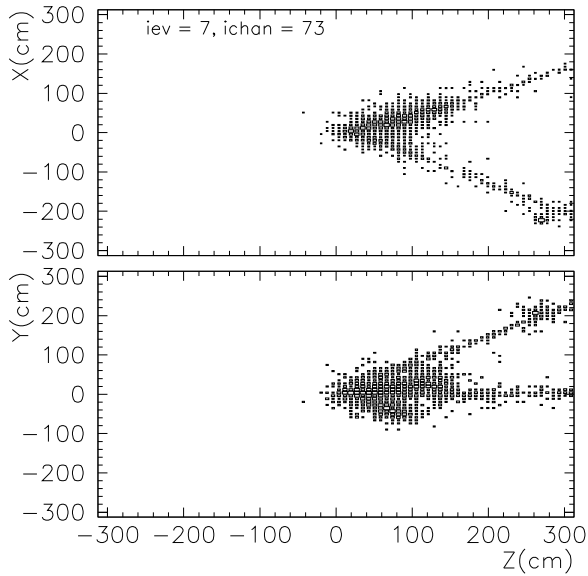}}
\end{center}
\vspace{-0.3in}
\caption{\em Example of a $\numu \, n \rightarrow \mu^- \, K^+ \, \Lambda$ 
         event in the FINeSSE detector. In this event, the associated
         strange particles decayed yielding a hadronic shower and second muon:
         $\Lambda \rightarrow p \, \pi^-$ (BR=$63\%$) and 
         $K^+\rightarrow \mu^+ \numu$ (BR=$64\%$).}
\label{fig:kaon-event}
\end{figure}

Based on NUANCE Monte Carlo predictions, approximately 90 such
interactions are expected at FINeSSE in two years of running. The
ability to identify and isolate neutrino production of strange
particles at FINeSSE would be of even larger importance to proton
decay searches, not only because of potentially large event samples,
but also because these interactions will be taking place at energies
near threshold and on a nuclear target, both of which are particularly
relevant for present and future proton decay studies.  Identifying
such events at FINeSSE will certainly be challenging. Although the
event signature is complex, it is potentially very distinctive.
Roughly $64\%$ of the time, a typical CC kaon event will contain two
oppositely charged muons, one from the primary neutrino interaction
and the other from $K$ decay, $K^+ \rightarrow \mu^+ \numu$
(Figure~\ref{fig:kaon-event}). In addition, since most of the kaons
will stop and decay at rest, the $\mu^+$ is monoenergetic with a
momentum of $\sim240$ MeV, and will appear roughly 12~ns after the
initial interaction.

Further studies must assess FINeSSE's multi-track reconstruction
capabilities and DIS backgrounds before any stringent limits on
strange particle production can be set. However, such a possibility
should be seriously explored as FINeSSE could potentially be the first
experiment to measure this reaction both near threshold and on a
nuclear target.

\section{Neutrino Electron Scattering: Neutrino Magnetic Moment}

In the Standard Model, massive, charged particles have intrinsic
magnetic moments by virtue of their spin.  If we introduce neutrino
mass into the theory, an effective neutrino magnetic moment can arise.
How neutrino mass fits into the Standard Model, be it via a standard
Dirac mechanism, SUSY, or something else, affects the size and origin
of this magnetic moment.  Thus, we can use a neutrino magnetic moment
measurement to tell us about how neutrinos fit into the larger theory.

\begin{figure}[tbh]
\centering
\includegraphics[bbllx=31bp,bblly=606bp,bburx=327bp,bbury=761bp,height=1.5in]{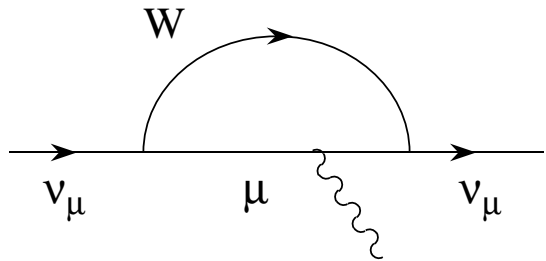}
\caption{\em One loop radiative correction diagram with a massive neutrino.}
\label{fig:nmmdiagram}
\end{figure}

In the minimally extended Standard Model, massive Dirac neutrinos of
mass $m_\nu$ can have a neutrino magnetic moment of:

\begin{eqnarray}
\mu_\nu = \frac{3eG_F}{8 \sqrt{2}\pi^2} m_\nu \sim 3 \times 10^{-19} \mu_B \left(\frac{m_\nu}{1 {\rm eV}}\right)~,
\end{eqnarray}
arising from one loop radiative corrections in diagrams with W-boson
exchange as in Figure~\ref{fig:nmmdiagram}.  

We can place neutrinos with non-zero magnetic moments in the context
of a larger theory such as SUSY and Extra Dimensions.  These
extensions to the Standard Model such as the super symmetric
left-right model predict larger neutrino magnetic moments of the size
\cite{Frank:1999nb}

\begin{eqnarray}
\mu_{\nu_e} \cong 5 \times 10^{-15} - 10^{-16} \mu_B \\
\mu_{\nu_\mu} \cong 1 \times 10^{-12} - 10^{-13} \mu_B \\
\mu_{\nu_\tau} \cong 2 \times 10^{-12} \mu_B. 
\end{eqnarray}
\noindent Theories involving Extra Dimensions produce an effective neutrino
magnetic moment as large as $10^{-11} \mu_B$
~\cite{McLaughlin:1999br}.

A non-zero neutrino magnetic moment would also have important implications
in cosmology in the development of stellar models.  Astrophysical limits
such as plasmon, or photon wave packets,  decay rates from horizontal branch stars and
neutrino energy loss rate from supernovae allow a neutrino magnetic
moment as large as $10^{-11}\mu_B$
\cite{Fukugita:1987uy,Lattimer:1988mf,Barbieri:1988nh}.

The most straightforward way to look for a non-zero neutrino magnetic
moment is via an excess in $\nu-e$ scattering:
\begin{eqnarray*}
   \numu \, e^- \rightarrow \numu \, e^- ~.
\end{eqnarray*}

\noindent A non-zero neutrino magnetic moment will give rise to an
electromagnetic contribution to this process in addition to the Standard
Model Z exchange.  This electromagnetic contribution to the cross
section for neutrino-electron elastic scattering is of the form:
\begin{eqnarray}
\sigma_{EM} = f^2 \pi
r_0^2 \left[ \frac{T_e^{\rm min}}{E_\nu} - ln\left(\frac{T_e^{\rm min}}{E_\nu}\right) -
1\right]~,
\end{eqnarray}
\noindent where $r_0$ is the classical electron radius, $T_e^{\rm min}$ is the
minimum kinetic energy of the recoil electron, $E_\nu$ is the neutrino
energy, and $f$ is the neutrino magnetic moment in units of Bohr
magnetons.

At low $y=T_e/E_\nu$, the electromagnetic contribution to the
neutrino-electron cross section increases rapidly, while the Standard
Model contribution increases only gradually.  The resulting shape
dependence in the differential cross section can be used to look for a
signal in a high statistics experiment like FINeSSE.  The differential
cross section contributions for weak and electromagnetic components of
the neutrino-electron elastic scattering cross section are given by:

\begin{eqnarray}
\frac{d\sigma^{weak}}{dT_e} = \frac{2 m_e G_F^2}{\pi}\left[g_L^2 + g_R^2\left( 1 - \frac{T_e}{E_{\nu}} \right)^2 - g_R g_L\frac{m_e}{E_{\nu}} \frac{T_e}{E_{\nu}} \right]~,~{\mathrm{and}}
\end{eqnarray}

\begin{eqnarray}
\frac{d\sigma^{EM}}{dT_e} = \frac{\pi \alpha^2\mu_{\nu}^2}{m_e^2} \left(\frac{1}{T_e} - \frac{1}{E_{\nu}}\right) ~,
\end{eqnarray}
where $T_e$ is the electron recoil energy and $E_\nu$ is the neutrino energy.  

Present limits set on neutrino magnetic moments are many orders of
magnitude away from the Standard Model prediction including massive
Dirac neutrinos.  They are, however, only one to two orders of
magnitude away from predictions from certain beyond-the-Standard-Model
theories and astrophysical limits.  The most stringent limits set by
experiments for neutrino magnetic moments for each neutrino flavor are
listed in Table~\ref{tab:limits}.  The experimental limit for the
muon-neutrino magnetic moment coming from the LSND experiment is an
upper limit of $\mu_{\nu_\mu} < 6.8 \times 10^{-10} \mu_B$
\cite{Auerbach:2001wg} set by measuring the total neutrino-electron
elastic scattering cross section.  An improved limit by about a factor
of two, on the muon neutrino magnetic moment is expected from
MiniBooNE ~\cite{Beacom:2001}.

\small
\begin{table}
\centering
\begin{tabular}{|p{1.5cm}|p{2.5cm}|p{4.cm}|p{4.cm}|p{1.7cm}|} \hline \hline
$\nu$ Flavor&Limit Set($\mu_B$)&Experimental Data&Process&Reference\\ \hline \hline
$\nu_e$&$1.5 \times 10^{-10}$&Super Kamiokande&shape of differential cross section&\cite{Beacom:1999wx}\\ \hline
$\overline{\nu_e}$&$1.8 \times 10^{-10}$&combination of $\overline{\nu_e}$ reactor experiments&excess in total cross section&\cite{Derbin:1993wy,Derbin:1994ua}\\ \hline
$\nu_\mu$&$6.8 \times 10^{-10}$&LSND &excess in total cross section&\cite{Auerbach:2001wg}\\ \hline
$\nu_\tau$&$5.4 \times 10^{-7}$&DONUT&excess in total cross section&\cite{Schwienhorst:2001sj}\\\hline\hline
\end{tabular}
\caption{\em Limits set on neutrino magnetic moments.}
\label{tab:limits}
\end{table}
\normalsize

\begin{figure}[tbh]
  \mbox{
\begin{minipage}{0.45\textwidth}
  \mbox{
\includegraphics[width=\textwidth,bb=2 3 515 518]{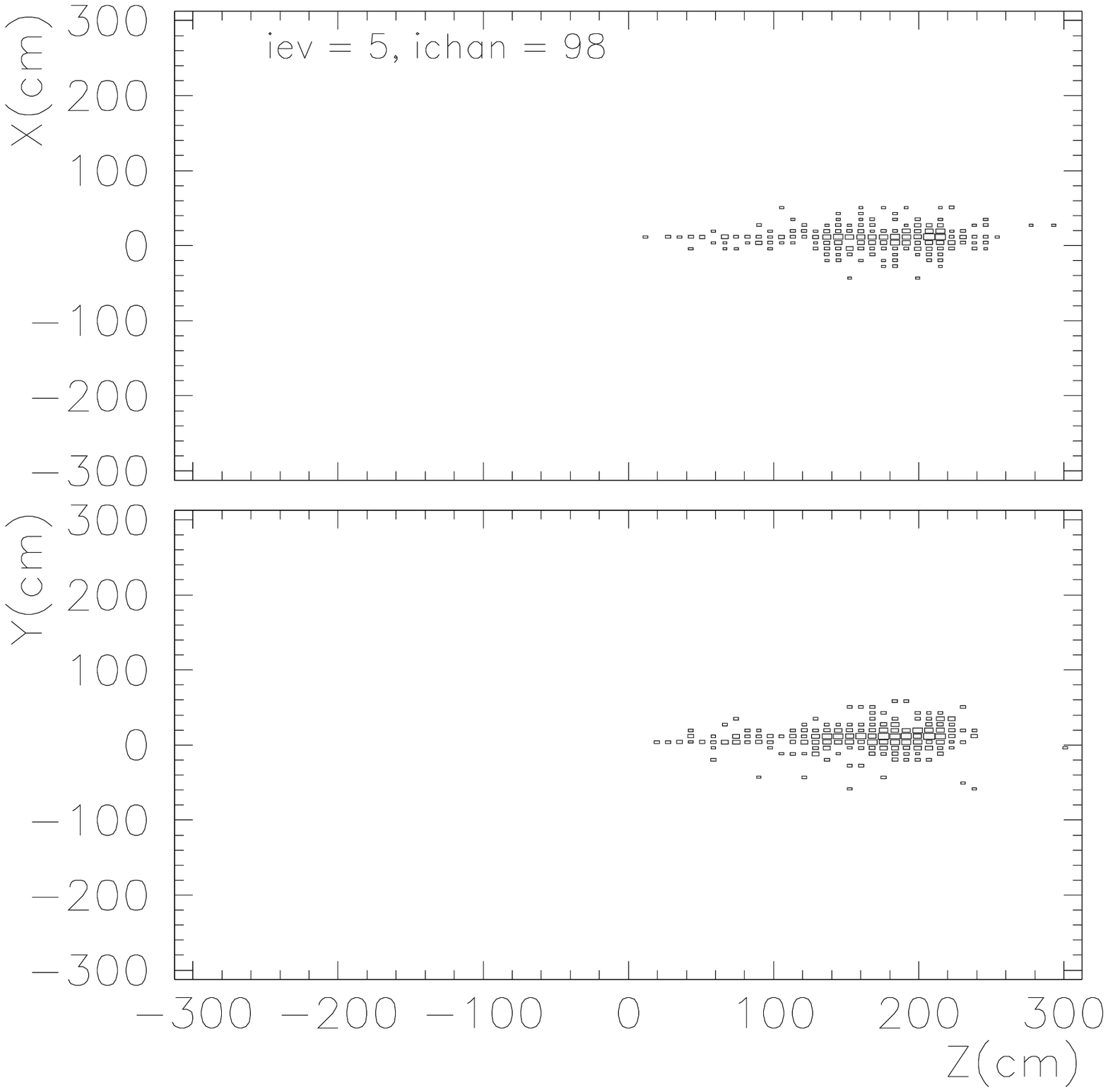}}
\end{minipage}\hspace{0.3in}
\begin{minipage}{0.35\textwidth}
  \mbox{
\includegraphics[width=\textwidth,bb=40 131 588 693]{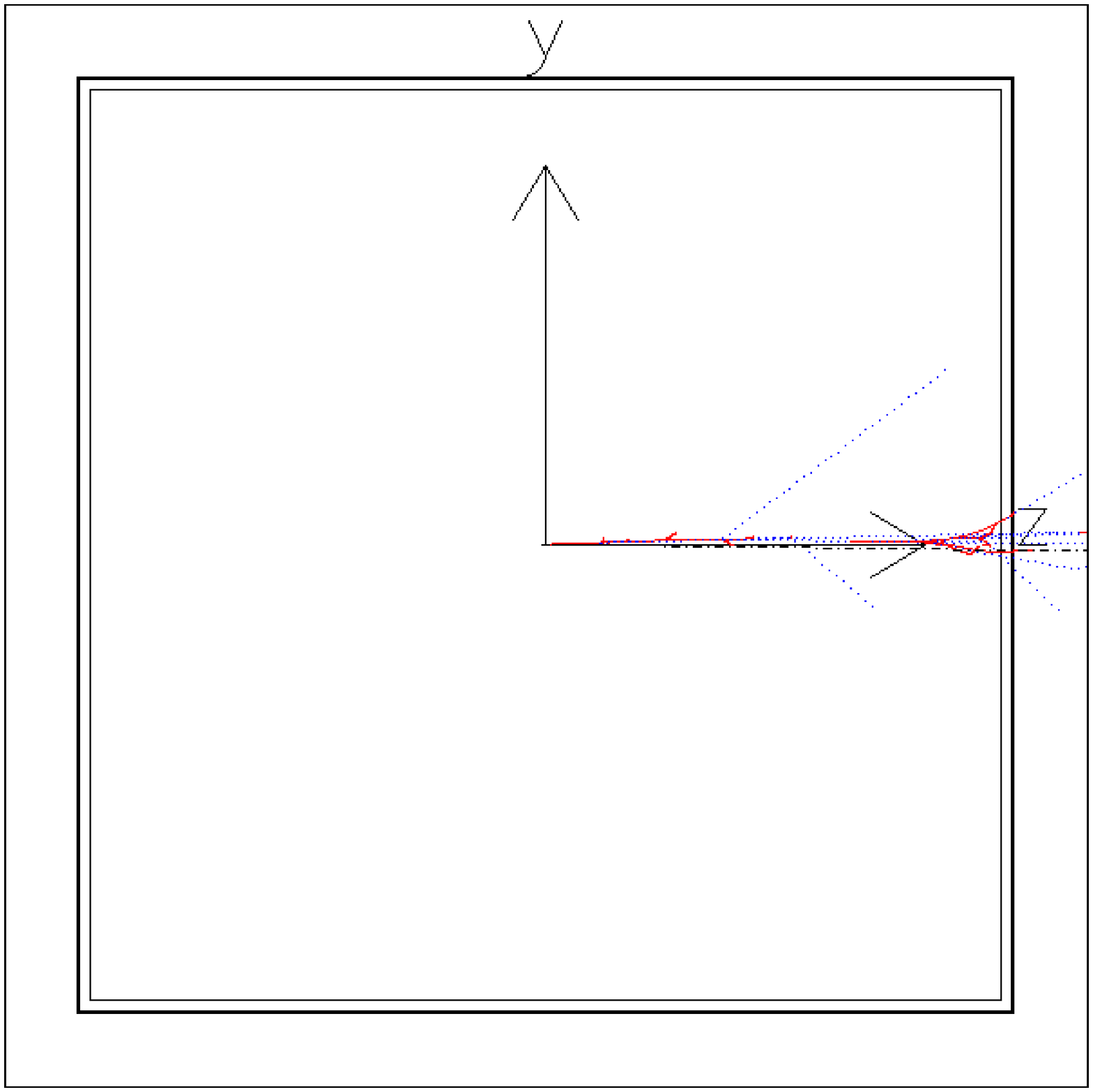}}
\end{minipage}
}
  \caption{\em A $\numu \, e^- \rightarrow \numu \, e^-$ elastic scattering
          interaction in the FINeSSE detector. The figure on the right shows 
          the true electron shower in GEANT for the same event. Such
          events characteristically contain a single forward scattered 
          electron.}
\label{fig:neutrino-electron-event}
\end{figure}

Number of events and, more importantly, low energy threshold for
electron recoil, determine how sensitive an experiment is to a
non-zero neutrino magnetic moment.  FINeSSE expects a number of
$\nu-e$ scattering events (86~events) comparable to the number
MiniBooNE will see (101~events).  FINeSSE can improve upon MiniBooNE's
expected measurement in identifying events at low electron recoil
energy, where the differential cross section is most sensitive to a
non-zero neutrino magnetic moment.
Figure~\ref{fig:neutrino-electron-event} shows a typical $\nu-e$
interaction in FINeSSE's Vertex Detector.  These events are very
forward and have a very distinctive shower shape.  The granularity of
the Vertex Detector as compared to the MiniBooNE detector may allow
FINeSSE to clearly identify these lowest energy events via a shower
shape analysis.  This would provide better sensitivity to non-zero
muon neutrino magnetic moment than any previous experiment.

\clearpage

\section{Antineutrino Running and $\Delta s$}
\begin{figure}[h]
\centering{\includegraphics[width=0.8\textwidth, bb=0 0 430 500]{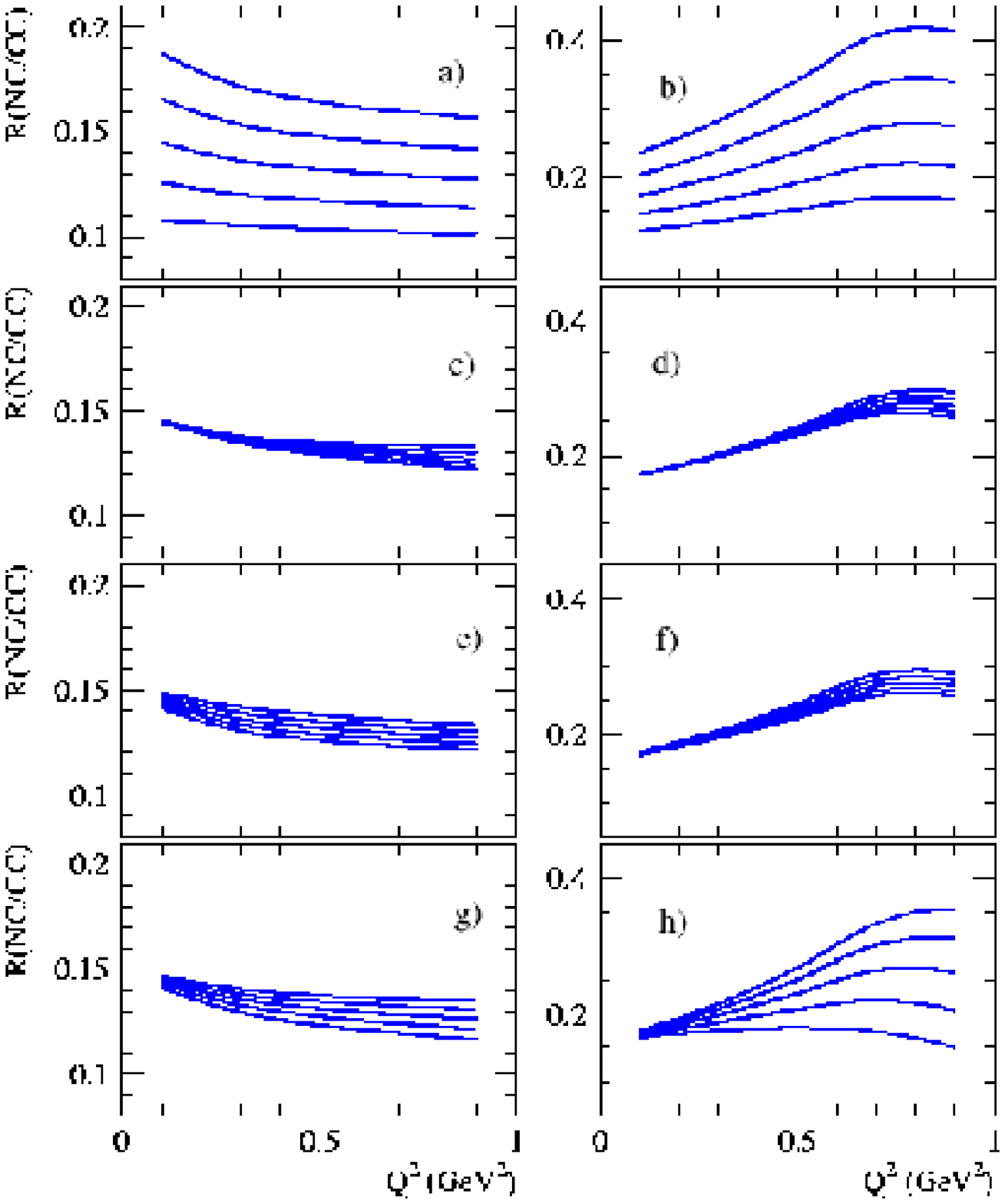}}
\caption{ \it The ratio of NC to CC scattering for neutrinos (left column of plots) and
antineutrinos (right column) as a function of $Q^2$.  In each of the
4 rows one of the form factors is  varied.  In a-b) $-0.2 < \Delta s < +0.2$,
c-d) $-0.2 < F_1^s < +0.2$, e-f) $0.95 < M_A < 1.15$ GeV/c$^2$. }\vspace{0.25in}
\label{fig:rncccq2all}
\end{figure}

If the opportunity arises to run FINeSSE with an antineutrino beam,
there are some features of $\bar{\nu} p \rightarrow \bar{\nu}p$
scattering that would make this channel exciting to investigate.  The
ratio in the antineutrino channel is actually slightly more sensitive
to $\Delta s$ than for neutrinos.  This can be seen in
Figure~\ref{fig:rncccq2all}a-b, where the relative change in the ratio
with the same change in $\Delta s$ is larger for antineutrinos than
for neutrinos.  In addition, the change in the ratio as a function of
$Q^2$ is different for antineutrinos.  This would allow a simultaneous
extraction of $\Delta s$ and $M_A$, since (see
Figure~\ref{fig:rncccq2all}) the ratio has a much different behavior
in $Q^2$ for different values of $M_A$ in $\bar{\nu} p \rightarrow
\bar{\nu}p$.

\section{Long-term Running of FINeSSE+MiniBooNE}

If MiniBooNE sees a signal and it is confirmed by both phase II MiniBooNE
and FINeSSE+MiniBooNE cross checks, Fermilab is likely to
continue the 8~GeV program.  The natural next step is the installation
of BooNE, the multi-detector experiment planned as the upgrade to
MiniBooNE.  Given that BooNE is in the planning stages, it is unclear
when this program would begin.  However, it is safe to say that it is
unlikely that BooNE can be fully constructed and running before 2010.
During this construction period, it would make sense to continue
running FINeSSE+MiniBooNE.

In this section, we consider what can be achieved if FINeSSE+MiniBooNE
receives $3\times 10^{21}$ protons on target before the BooNE detector
system is inaugurated.  This is based on the premise that upgrades to
the Booster can allow the 8~GeV line to return to the level of
$1\times 10^{21}$ protons/year near the end of the decade.  To be
clear, we are not requesting approval for this scenario from the PAC.
We are simply using this to illustrate the long-term value of the
FINeSSE detector.

In this case, the statistics limited measurements described in the
previous section can be greatly improved.  We believe we can reasonably
achieve the systematic errors listed in Table~\ref{tab:systs}.

\begin{table}[h] 
\centering
\begin{tabular}{cc}
  NC $\pi^0$: &   0.03 \\
  Radiative $\Delta$ Decay: &  0.06\\
  $\nu_e$ from  $\mu$: &  0.04 \\
  $\nu_e$ from $K^+$:  & 0.04 \\
  $\nu_e$ from $K^0$: &  0.06 \\
\end{tabular}
\label{tab:systs}
\caption{ \it Systematic errors for the (hypothetical) scenario of 
$3 \times 10^{21}$ POT.}
\end{table}

\begin{figure}[!tbp]
\centering{\includegraphics[width=5.in, bb=60 130 570 740]{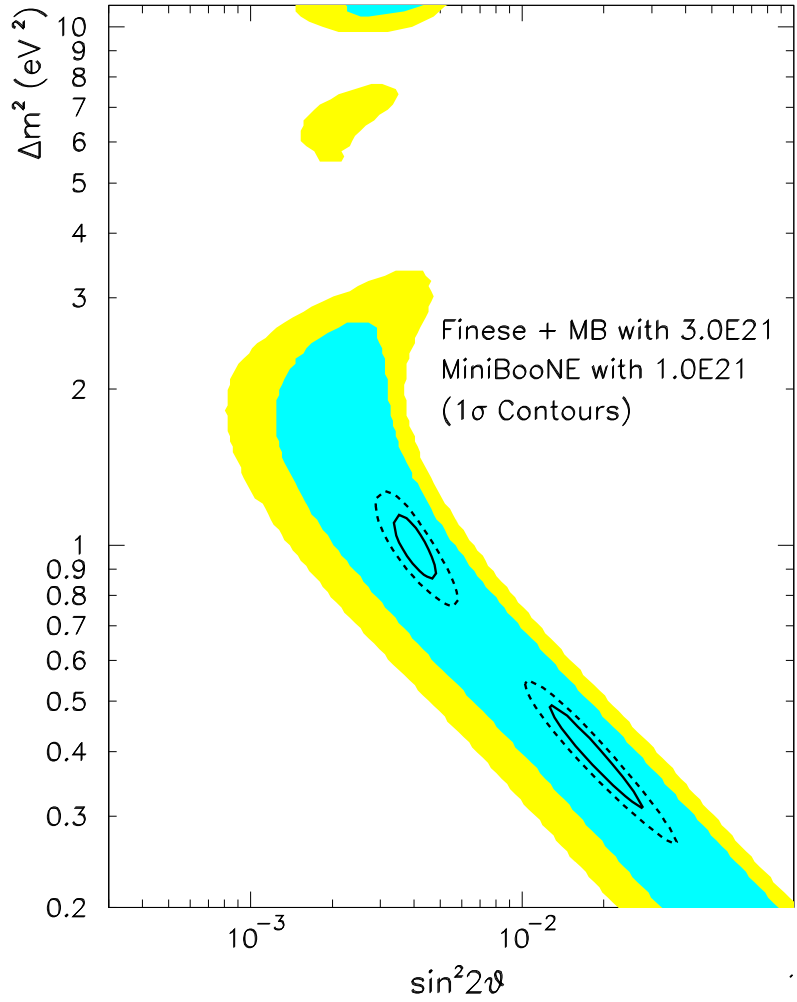}}
\caption{ \it The solid ellipse shows the 1$\sigma$ measurement
  capability for two sets of oscillation parameters for
  FINeSSE+MiniBooNE for $3\times 10^{21}$ POT.  The dashed ellipse
  shows the MiniBooNE Phase I capability. The colored segments
  indicate the LSND-allowed regions.}  \label{fig:wow} \end{figure}

\noindent At this point MiniBooNE will have acquired $4\times 10^{21}$ 
protons on target (25\% with the 50~m absorber and 75\% with the 25~m
absorber).  The capability of such a measurement is shown in Figure
\ref{fig:wow}.  The colored regions show the LSND allowed range.  The
solid ellipses indicates the 1$\sigma$ measurement capability for the
oscillation parameters at two possible true $\Delta m^2$ values.  For
comparison, the capability from MiniBooNE phase I running is indicated
by the dashed ellipses.

\chapter{Overview of Cost and Schedule}
\label{ch:CostandSchedule}

The FINeSSE detector cost is estimated to be \$2.25~M (\$2.8~M with
contingency).  These estimates do not include EDIA or indirect
costs.  Funding for the detector will come from the university funding
agencies. EDIA and indirect costs will be added as appropriate when
the funding structure is in place, and will account for differences in
each university's accounting practices.

The FINeSSE enclosure cost is estimated to be \$800K (\$1.6~M including
contractor O\&H, EDIA, management reserve, and indirect costs) as
determined through a Preliminary Design Report written by the Fermilab
Facilities Engineering Section (FESS).  See Appendix~\ref{ch:AppendixC}
which contains this report.

The dimensions of the enclosure were chosen to comfortably contain the
detector and provide adequate working room around the detector. There
is also adequate room for detector installation. The FINeSSE schedule
incorporates detector and enclosure design, construction, and
installation over the next two years, with data taking
scheduled to begin in mid-2006.

The following sections provide breakdown of detector costs,
enclosure costs, construction and installation schedule, and
an overview of the FINeSSE schedule from now until the beginning of the
FINeSSE physics run.

\section {Detector and Enclosure Costs}

The FINeSSE detector is comprised of two subdetectors
(cf.~Chapter~\ref{ch:TheFINeSSEDetector}): the Vertex Detector, followed
downstream by the Muon Rangestack. Table 7.1 lists preliminary costing
of the components of these detectors, as well as total costs.

The cost of the detector is driven mainly by PMT and electronics costs
that combined total \$900,000. The vertex detector tank that holds the
scintillator, fiber, and PMTs is about \$265,000. The remaining costs
are for fiber, scintillator, and oil.  The cost driver for the
Rangestack detector is the iron planes, costing \$327,000.  Most of
this cost comes from fabrication.  An additional cost of about
\$100,000 is set aside for rigging and installation of the detector.

Contingency is applied to the total detector costs.  EDIA and indirect
costs are not included, as it is expected that the majority of these
costs will come through university groups, where EDIA and indirect
costs are treated differently by each university.  These costs will be
included once the funding profile by universities is established.


The detector enclosure design went through a series of studies within
FESS.  To meet the requirements of detector construction and
operation, several enclosure options were studied. They are summarized
in the Alternate Comparison Matrix given in
Appendix~\ref{ch:AppendixC}.  The options ranged in cost from \$1.48M
to \$2.54M. The study compared the use of a gantry crane to a bridge
crane, a hatch cover to a pre-engineered building, and a sheet pile
pit to a secant pile pit.  The chosen option includes a sheet pile
pit, a hatch cover, and a gantry crane for about \$1.6M.  If a mobile
crane were used, rather than a gantry crane, the construction cost
could be reduced to \$1.48M, but this would be offset by the
operational costs of crane rental during the detector installation
phase.  There will be no need for crane coverage once the detector is
installed and operating. A cross-sectional view of the enclosure is
shown in Figure 4 of the Appendix, and the plan view is given in
Figure 3.

\section{Schedule}
The FINeSSE schedule allows for 2.5 years for enclosure and detector
construction and installation, with the start of the physics run
beginning in mid-2006.  The FINeSSE schedule for enclosure
construction calls for construction of the building to commence in mid
to late 2004.  Total time to construct this enclosure is about six
months, as described in detail in Appendix~\ref{ch:AppendixC}, keeping FINeSSE
on schedule for detector installation beginning in mid-2005.  Detector
construction and installation is expected to take 1.5 years.

FINeSSE is on an aggressive but achievable schedule for a number of
reasons.  First, FINeSSE physics is important and timely and should be
pursued as quickly as possible.  Second, there is an existing, running
neutrino beamline which should be utilized as much as possible.
Finally, the FINeSSE physics run is timed to coincide with the end of
MiniBooNE antineutrino running and fits well with the MiniBooNE
schedule.  The FINeSSE schedule, up to the start of our physics run in
mid-2006, is shown in Figure~\ref{fig:schedule}.

\begin{figure}[p]
  \centering \includegraphics[bb=15 15 627
  807,width=7.in]{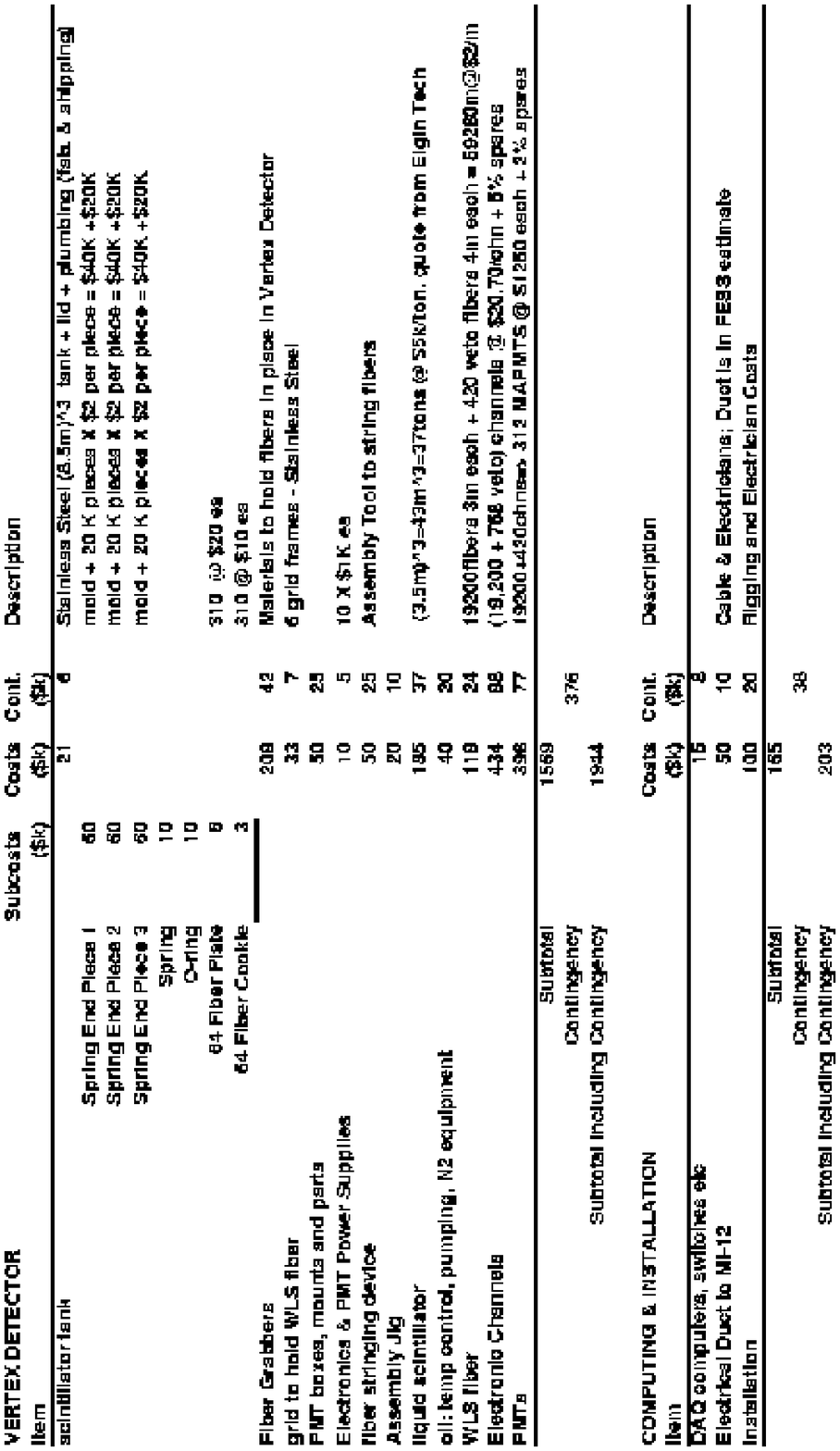}
\end{figure}

\begin{figure}[p]
  \centering \includegraphics[bb=15 15 627
  807,width=7.in]{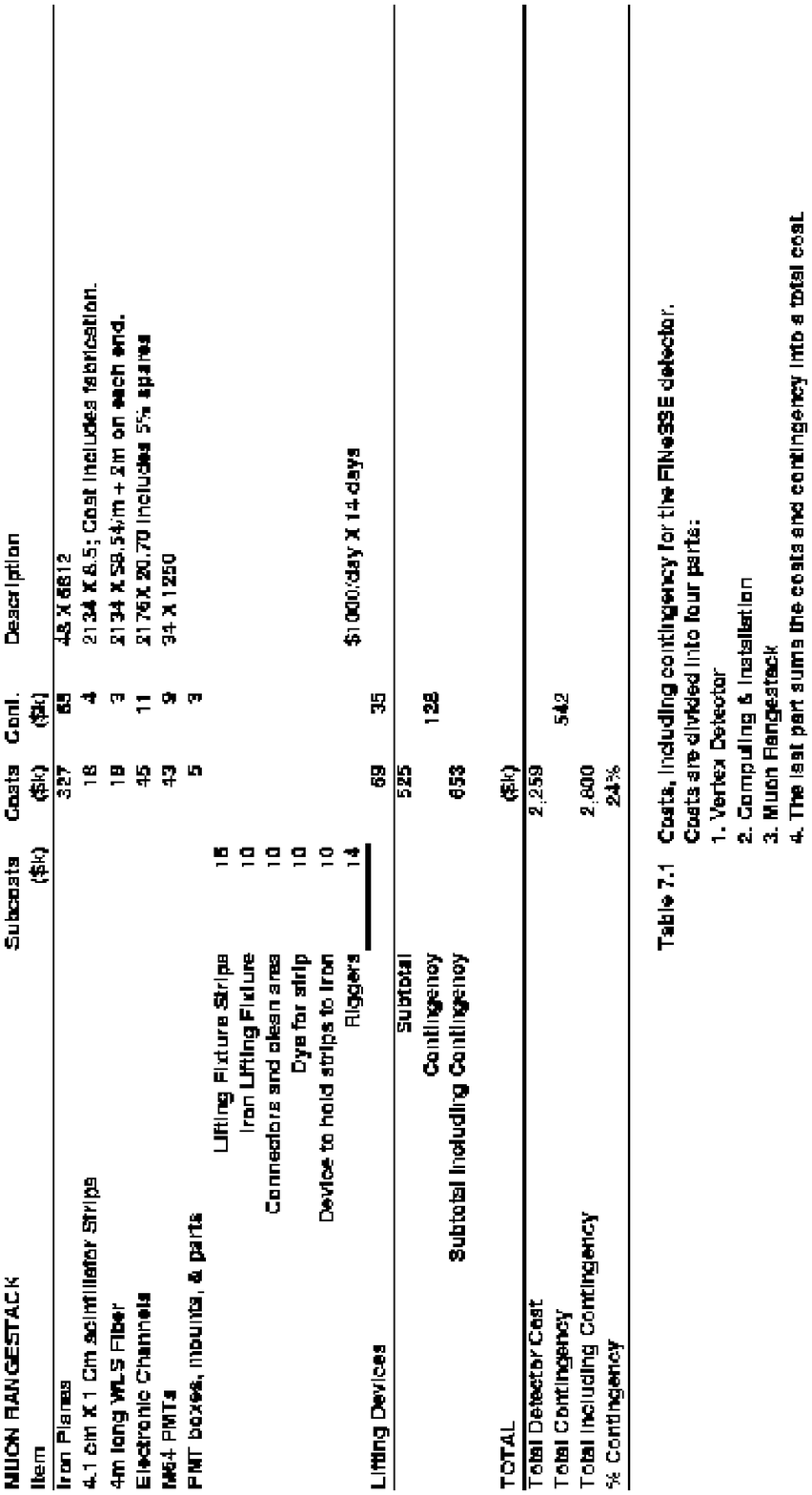}
\end{figure}

\begin{figure}[p]
\centering
\includegraphics[bb=70 70 304 679]{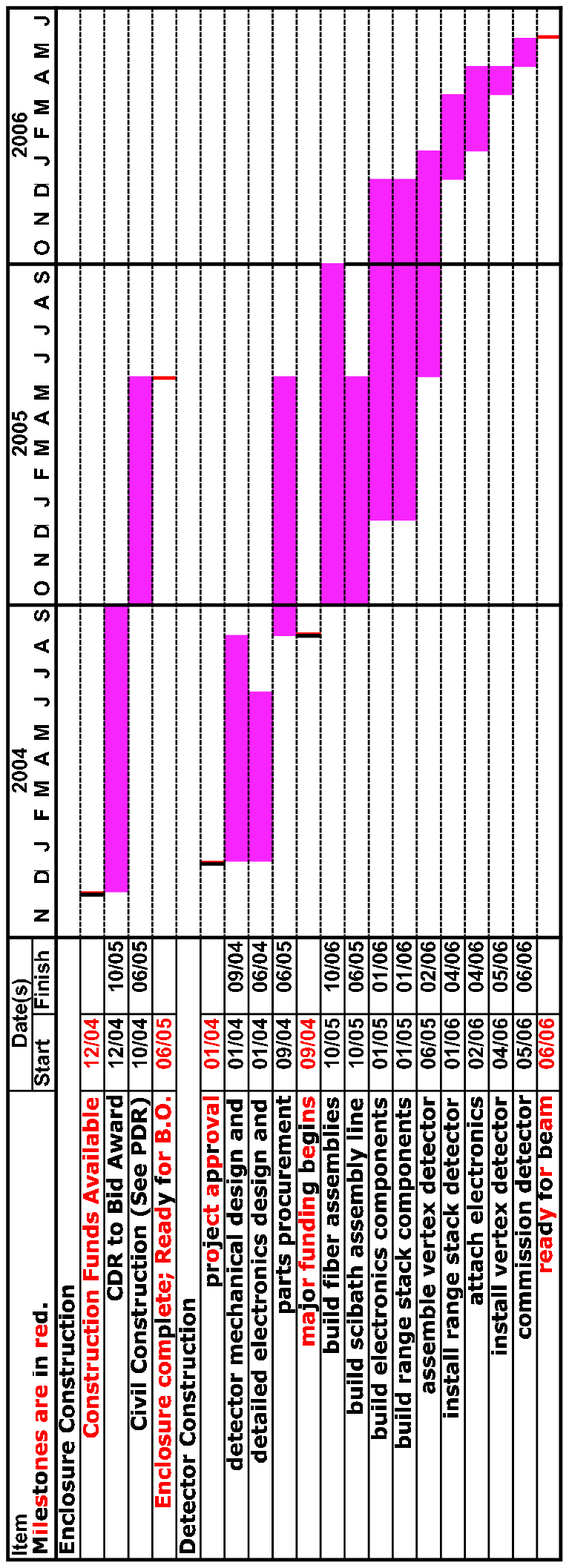}
\caption{\em The FINeSSE schedule, showing the milestones 
  from winter 2004 until beam can be taken in 2006.  Milestones are
  shown in red.}
\label{fig:schedule}
\end{figure}

\chapter{Conclusions}
\label{ch:Conclusions}

\thispagestyle{myheadings}
\markright{}

We propose to construct and operate a novel neutrino detector to run
in the Booster neutrino beamline, 100~m from the Booster neutrino
production target. The physics motivation for this experiment is
twofold.  

First, we will measure the strange spin of the proton $\Delta s$ using
a theoretically robust method, an intense, low energy neutrino beam,
and a novel detection technique.  Second, FINeSSE will look for
$\nu_{\mu}$ disappearance, in conjunction with MiniBooNE, at high
$\Delta m^2$ in an astrophysically interesting region.

FINeSSE will achieve these physics goals using a two-part detector to
measure both short, low energy proton tracks and longer, muon tracks.
Low energy proton tracks will be well measured in a, liquid
scintillator ``bubble chamber''.  Downstream of this Vertex Detector
is a Muon Rangestack designed to range out higher energy muons that
are not contained in the Vertex Detector.

FINeSSE requires 6$\times 10^{20}$ POT total to reach its physics
goals.  In the NuMI era, these protons can be delivered over a two
year period.


\clearpage

\thispagestyle{myheadings}
\markright{}

\clearpage

\appendix

\chapter{Vertex Detector Design Studies}
\label{ch:AppendixA}

\thispagestyle{myheadings} 

\markright{} 

In determining the best detector design to address the FINeSSE physics
goals, a number of options were considered.  In addition to the
``Scibath'' design described in Chapter ~\ref{ch:TheFINeSSEDetector}, we
studied a design called ``Scistack'', much like the K2K Scibar
detector~\cite{k2kscibar}.  We found this detector was not optimal for
identifying low energy proton tracks from $\nu-p$ elastic scatters,
crucial for the $\Delta s$ analysis.  By contrast, the Scibath option
did not have these limitations.  This appendix describes these
Scistack studies including building and testing of a prototype
detector, Monte Carlo studies, and conclusions.

\section{Scistack design, Feasibility and Cost}
The Scistack detector is a 2.5~$m^3$ volume of scintillating plastic
bars.  The 2~cm $\times $1~cm extruded polystyrene bars are co-extruded with
a TiO$_2$ outer layer for reflectivity, and a hole down the center for
a WLS fiber.  Bars are organized in planes normal to the beam
direction with bars in even-numbered planes oriented in the x direction
and bars in odd number planes oriented in the y direction.  This
design has a total of 31250 bars and associated readout channels.
Readout is identical to the Scibath design with light piped out via
WLS fibers, and recorded and digitized as described in
Section~\ref{sec:SignalReadout}.  

The cost of a Scistack detector is comparable to the Scibath detector.
Plastic and liquid scintillator as well as WLS fiber are comparable in
both detectors.  A cost driver for both detectors is channel cost.
The number of channels in this design is larger by a factor of 1.5.
This is offset by employing channel multiplexing which decreases the
channel cost by a factor of eight.  However, channel multiplexing can
be accommodated only with 16 anode MAPMTs (as opposed to the Scibath
64 anode MAPMTs) due to physical size of each anode, driving the cost
per channel back up.  Cost of WLS fiber also increases in Scistack as
there are 1.5 more channels; this, however, is not a significant cost
overall.  Folding in all cost differences, costs for the two detectors
are comparable.

\section{Prototype testing}
\begin{figure}
\centering
\includegraphics[width=4.in,bb=72 221 541 571]{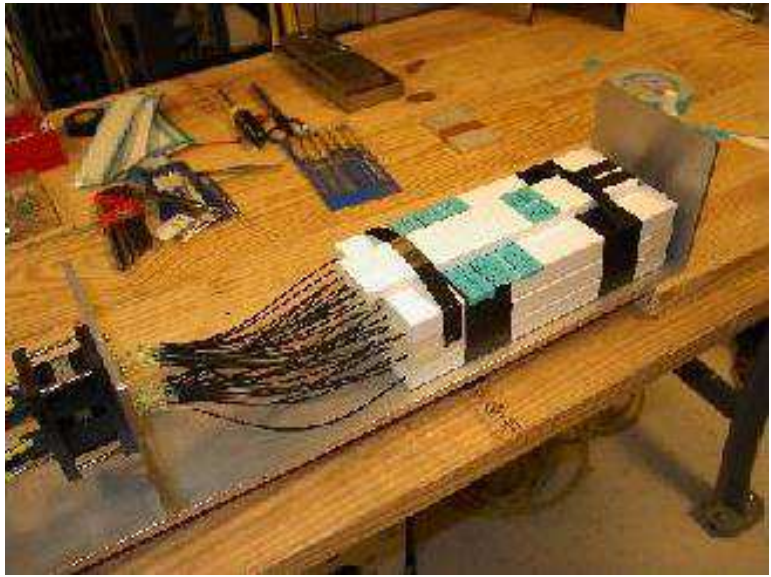}
\caption{\it Photograph of the Scistack prototype.  Scintillator bars 
  are stacked in a 10 x 3 array.  WLS fiber, polished on both ends,
  routed through each bar, is read out on one end by Hamamatsu 16
  anode MaPMTs.  The entire prototype is housed in a light tight aluminum
  box with connections for Lemo and high voltage.  The box is built on
  a rotatable stand to facilitate testing in the IUCF beam.}
\label{fig:scistackprototype}
\end{figure}

A small prototype of the Scistack detector was designed, constructed,
and tested during the summer of 2003.  The prototype design, shown in
Figure~\ref{fig:scistackprototype} consists of a ten by three stack of 2~cm
$\times$ 1~cm scintillator bars, with a WLS fiber inserted down the center of
each.  The fiber is attached to 16 anode MaPMTs and read out through the
same electronics used for the Scibath prototype tests, described in
Chapter~\ref{ch:TheFINeSSEDetector}.  The Scistack
prototype was exposed to the IUCF proton beam, as was Scibath (again, see
Chapter~\ref{ch:TheFINeSSEDetector}).  Scistack performed as
expected.  Light output increased as the proton beam moved, normal to
the bars, and closer to the readout.  Tracks were easily seen, as
expected, as the proton beam traversed the stack both with the stack
oriented normal to the beam and at an angle with respect to the beam.
This detector employs known technology; there were no surprises.

\section{Monte Carlo Studies}
While the detection techniques proved acceptable experimentally,
Scistack Monte Carlo work revealed a feature in the detector geometry
which makes this detector unsuitable for FINeSSE's $\Delta s$
measurement.

The design goal of the FINeSSE Vertex Detector is to be able to
cleanly identify and measure low energy protons from $\nu-p$ elastic
scatters.  Scistack Monte Carlo work shows that these very events,
most important to the $\Delta s$ analysis, are frequently not
reconstructible in Scistack due to limitations in tracking in this
detector.  Specifically, low $Q^2$ events tend to be directed at high
angles with respect to the beam direction.  The Scistack detector
reconstructs these events with less success, since bars are oriented
normal to the beam direction, and therefore in the same direction as
these low $Q^2$ events.  The Monte Carlo and these studies are
described below.

\subsection{Scistack Monte Carlo}
The GEANT-based Monte Carlo for the Scistack option is designed to be
identical to the Scibath option except for the configuration of the
central detector. This includes the output ntuple which has the same
format, with appropriate changes in the meaning of the variables, to
minimize differences in the reconstruction program.

The central detector for Scistack consists of 250 layers, each 1~cm thick,
along the beam axis ($z$). Each layer has 125 scintillator bars, 2~cm in
the transverse direction, with the bars in alternating layers aligned 
along the $x$ and $y$ axes. Each 1~cm by 2~cm by 2.5~m bar has a 1.5~mm~OD
wavelength shifting optical fiber along its axis, to capture and transmit
photons from the scintillator to the PMTs. The outer 0.5~mm of each
bar is a TiO$_2$ wrapper which optically isolates the bars. The highly
reflective inner surface of the wrapper increases the light capture in
the fibers by approximately a factor of two. The scintillator has an
attenuation length of 5.0~m. 

After allowing for fiber capture and transmission efficiencies, assumed
to be the same as Scibath, each photon entering a fiber has its layer
and bar number recorded in the output ntuple. The actual capture 
efficiency as well as the number of photons generated per MeV of energy
loss would be higher in Scistack than in Scibath. But the number of
photons detected is so large that this is not an important factor in the
current studies.

The resulting ntuple is reconstructed by the same Hough transform
method used for Scibath, and the results are used to determine the
angular and energy resolution of Scistack as well as its particle
identification capabilities. The Monte Carlo samples generated for
these studies include single electrons, muons, nucleons, pions, and
kaons with kinetic energies between 0 and 500~MeV, generated
isotropically in direction. We also ran a large sample of neutrino
interactions generated using NUANCE. Each sample was run with vertices
near the center of the detector, and again with vertices spread
uniformly throughout the detector. The results are discussed in the
following section.

\subsection{$\nu-p$ Sample in Scistack}
\begin{figure}
\centering
\includegraphics[width=4.in,bb=40 40 777 531]{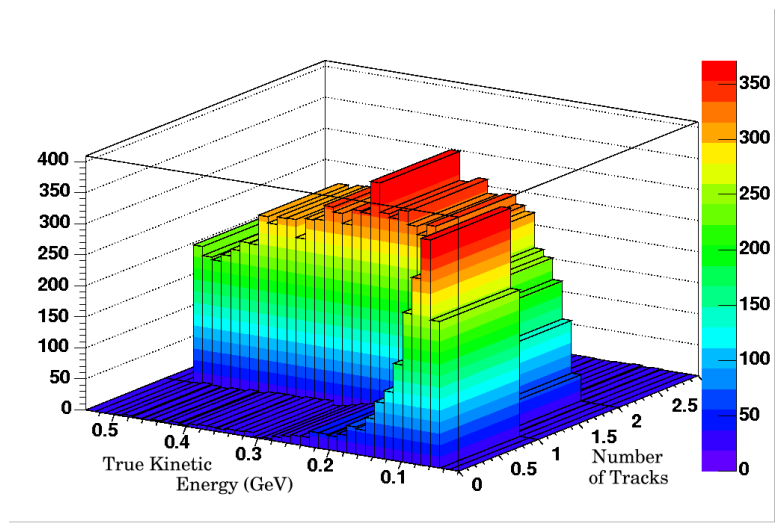}
\caption{\it Number of tracks reconstructed versus true kinetic
  energy for protons generated within Scitrack.  Events with kinetic
  energy below 0.15~GeV are not usually reconstructible (have no
  tracks).}
\label{fig:Ned}
\end{figure}

\begin{figure}
  \centering \includegraphics[width=4.in,bb=120 205 500
  600]{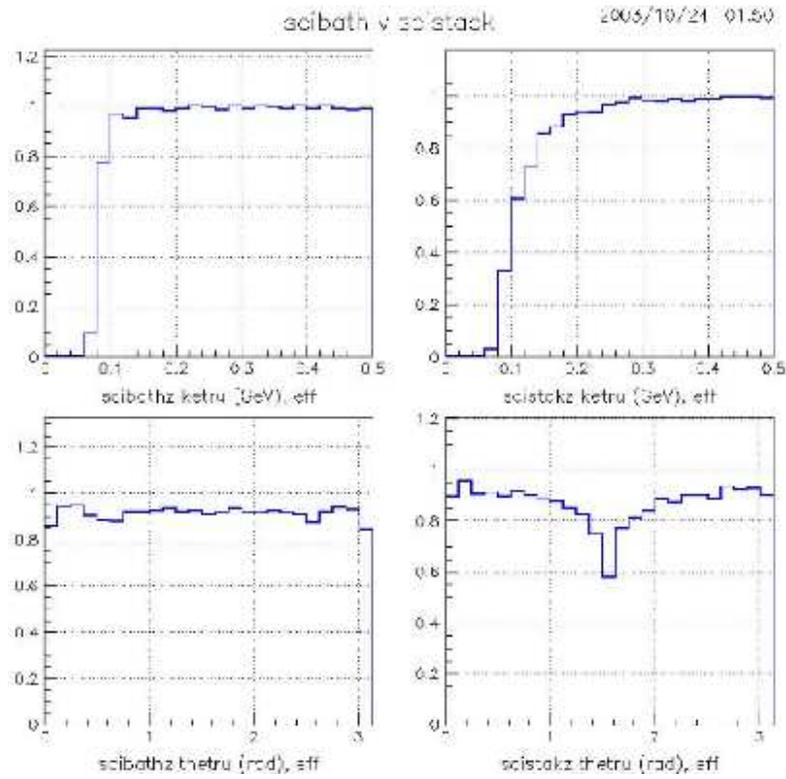}
\caption{\it Reconstructible events versus true angle and 
  energy in Scistack versus Scibath.  Upper left corresponds to true
  kinetic energy of Scibath event; upper right to true kinetic energy
  of Scistack protons.  Bottom left and right are true angle for
  Scibath and Scistack respectively.}
\label{fig:scibathscistack}
\end{figure}

Neutrino-proton elastic scatters at low energies produce short-track,
low energy protons.  Samples of protons generated and reconstructed as
described above were studied to determine if the Scistack detector was
suitable for this measurement.  Figure~\ref{fig:Ned} shows the number
of tracks reconstructed as a function of true kinetic energy for this
sample.  As is indicated, tracks below 150~MeV tend not to be
reconstructible.  These tracks tend to be at high angle and therefore
traversed few bars.  Figure~\ref{fig:scibathscistack} show
reconstructed tracks as a function of true kinetic energy and angle
for both Scitrack and Scibath.

The neutrino interaction sample generated with NUANCE was used to
understand the Scistack efficiency as compared to Scibath.  Similar
cuts as were applied to the reconstructed Scibath events to extract a
NCp sample, were applied to the Scistack sample.  The inability to
reconstruct the proton events described above, translated into a more
than 30\% decrease in $\nu-p$ statistics in the 0.2$<Q^2<$0.4 bin, the
crucial energy bin for the $\Delta s$ analysis.

\section{Conclusions}
While the Scistack detector technology is proven experimentally, it is
not suitable for FINeSSE's $\Delta s$ measurements.  Scibath, by
contrast, has the ability to reconstruct even low $Q^2$ events,
crucial for this analysis.

\clearpage

\chapter{FESS Project Definition Report \\The FINeSSE Detector}
\label{ch:AppendixC}

\thispagestyle{myheadings} 

\markright{}

\clearpage

\begin{figure}
\centering
\includegraphics[bb=0 0 610 800,width=8.in]{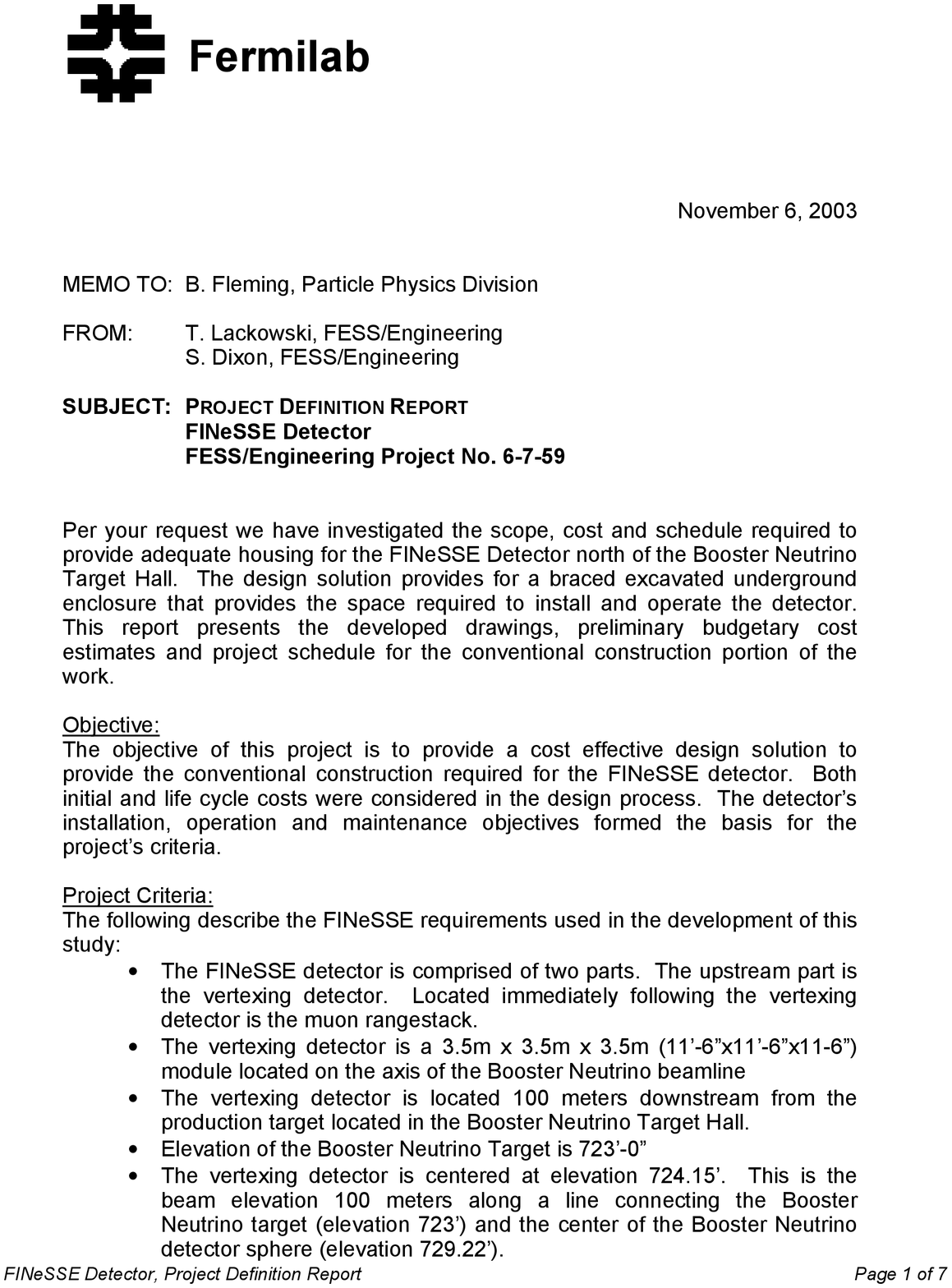}
\end{figure}

\clearpage

\begin{figure}
\centering
\includegraphics[bb=0 0 610 800,width=8.in]{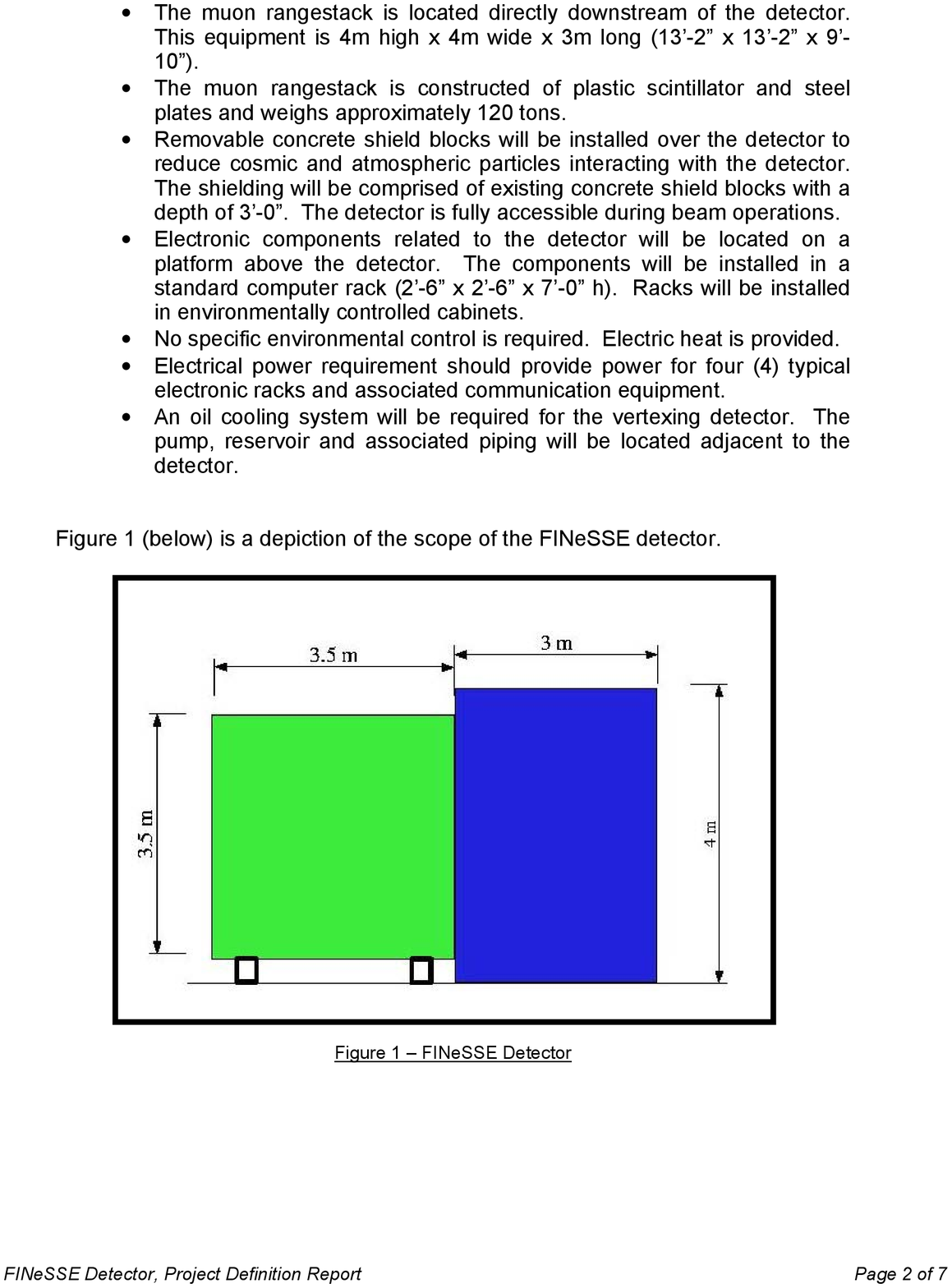}
\end{figure}

\clearpage

\begin{figure}
\centering
\includegraphics[bb=100 100 540 700,width=7.in]{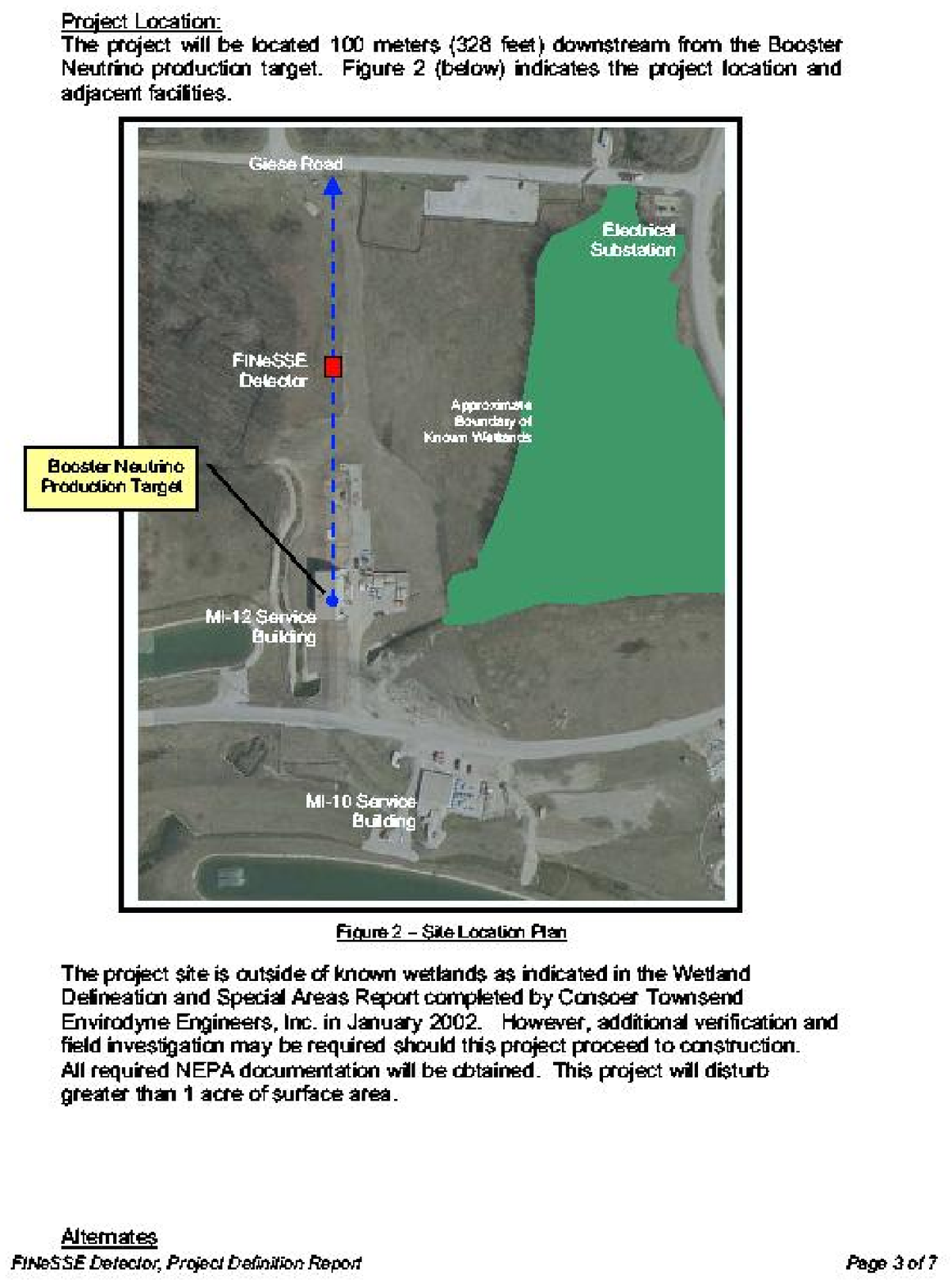}
\end{figure}

\clearpage

\begin{figure}
\centering
\includegraphics[bb=0 0 610 800,width=8.in]{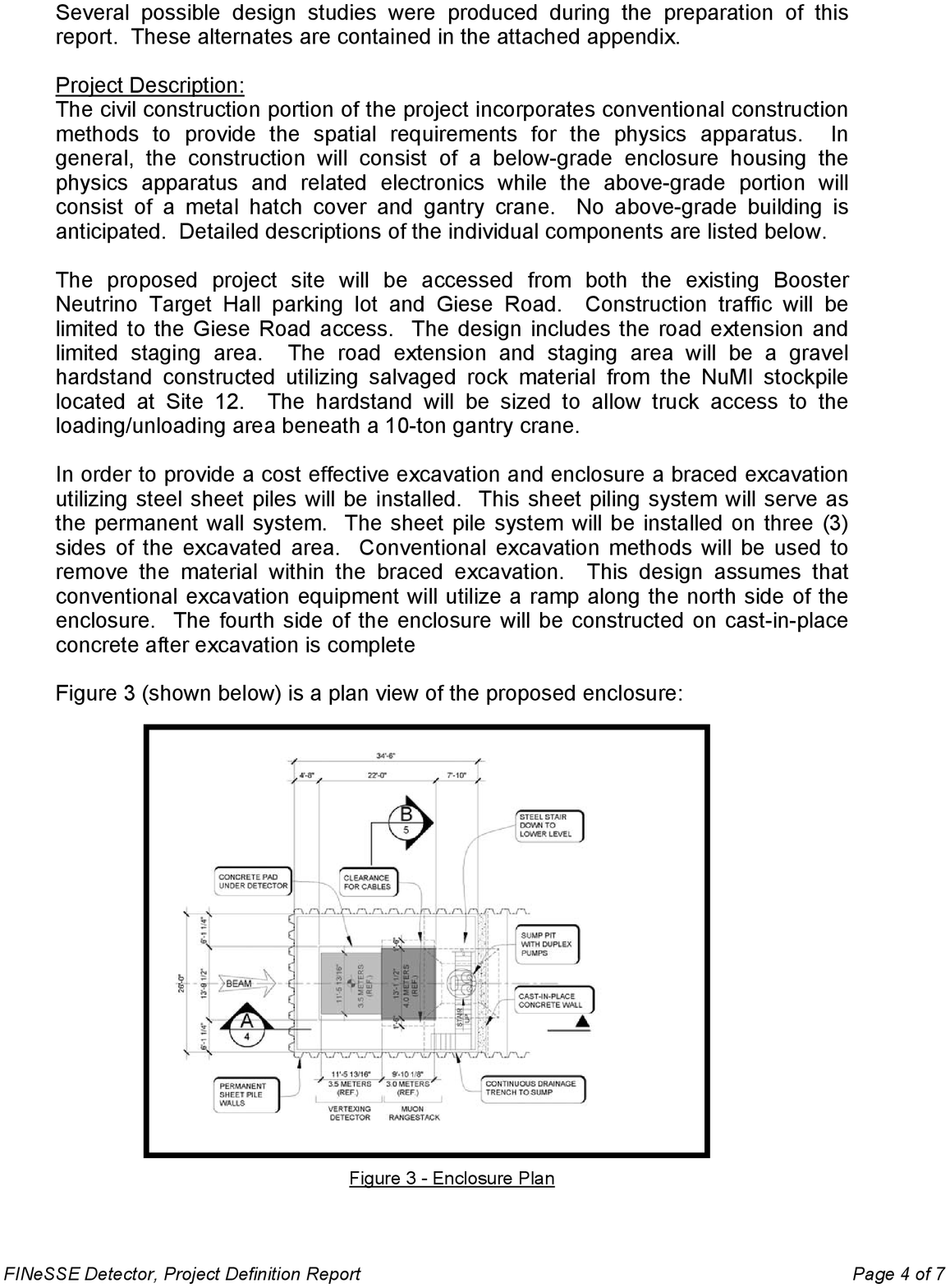}
\end{figure}

\clearpage

\begin{figure}
\centering
\includegraphics[bb=0 0 610 800,width=8.in]{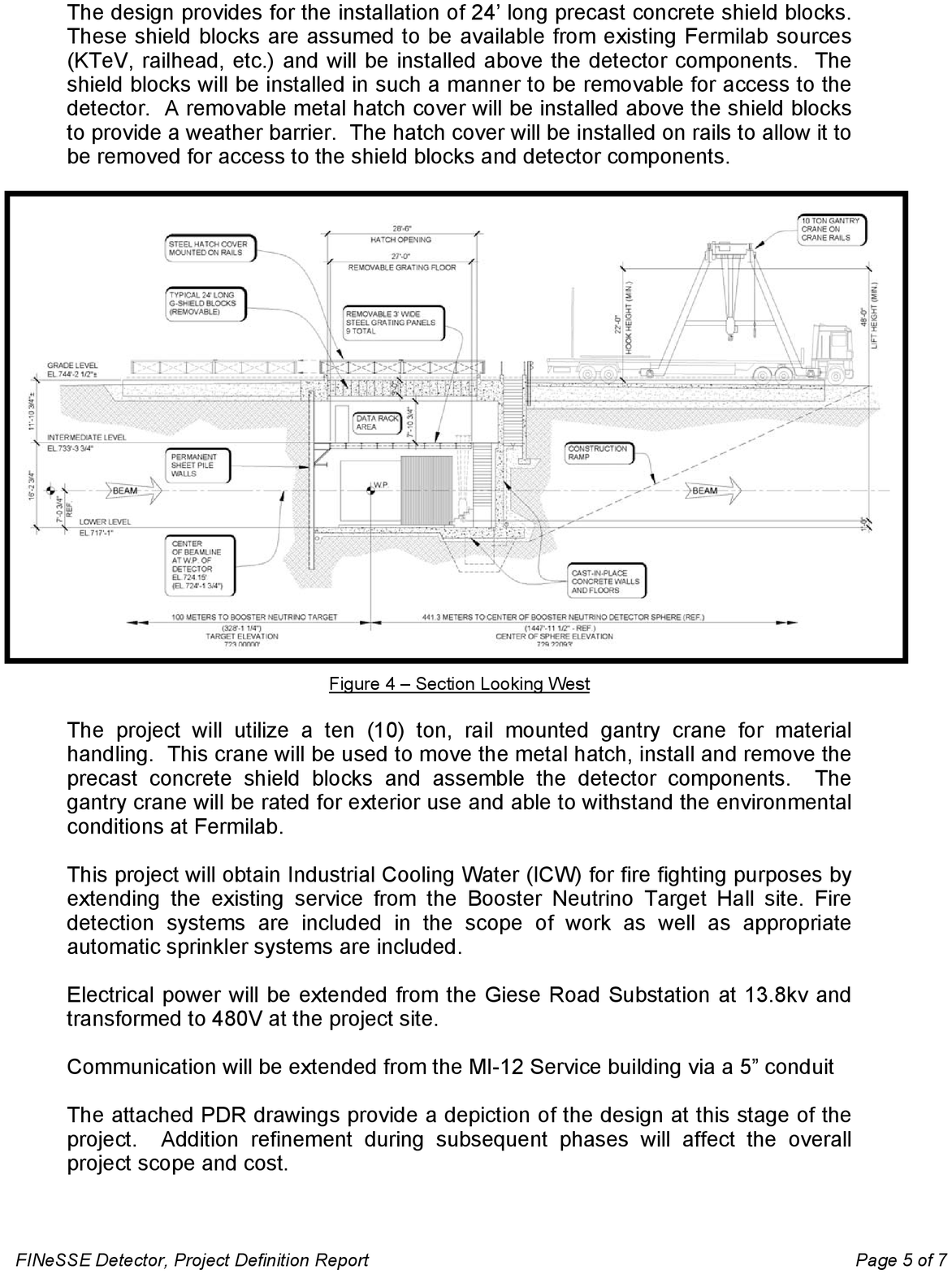}
\end{figure}

\clearpage

\begin{figure}
\centering
\includegraphics[bb=100 100 540 700,width=7.in]{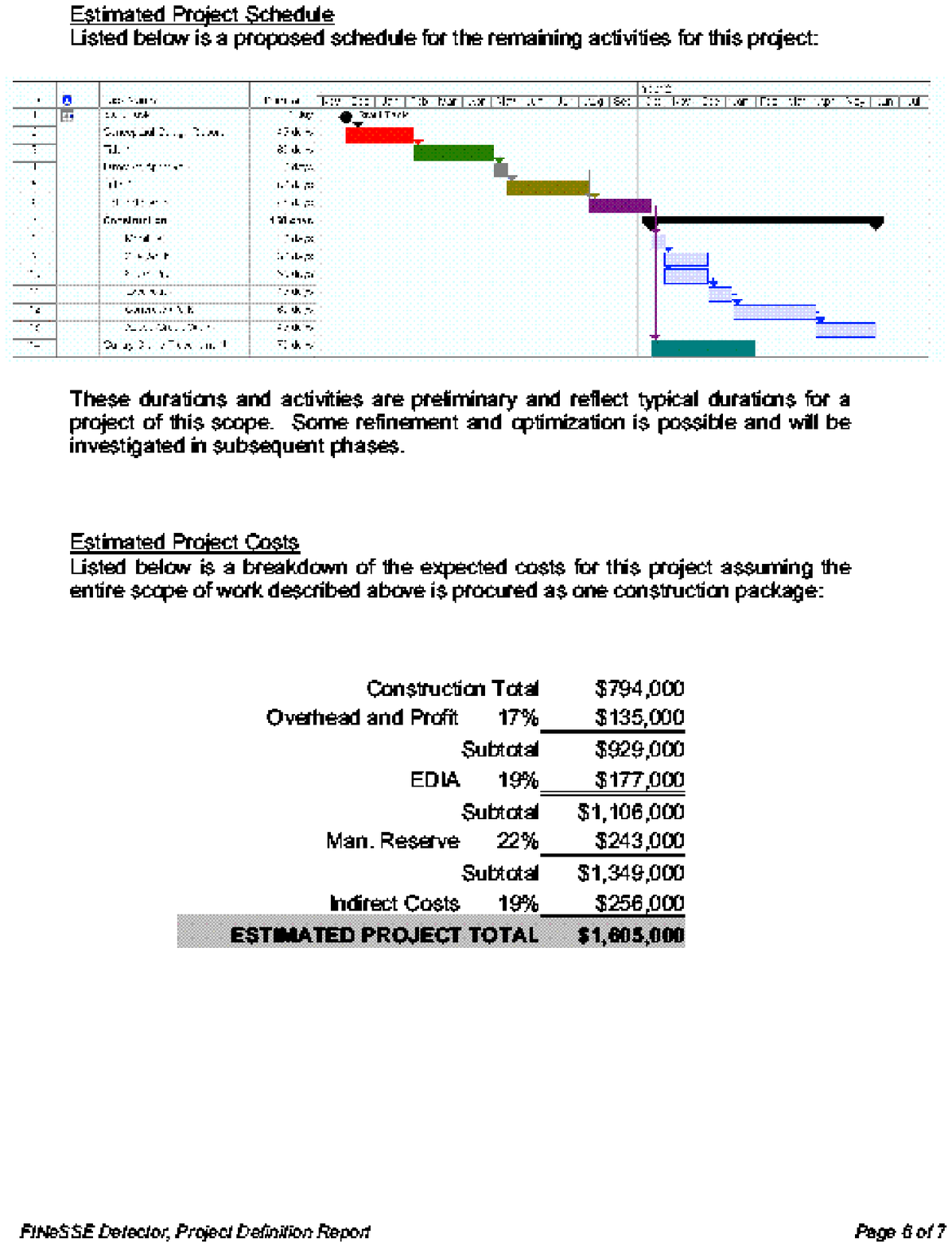}
\end{figure}

\clearpage

\begin{figure}
\centering
\includegraphics[bb=0 0 610 800,width=8.in]{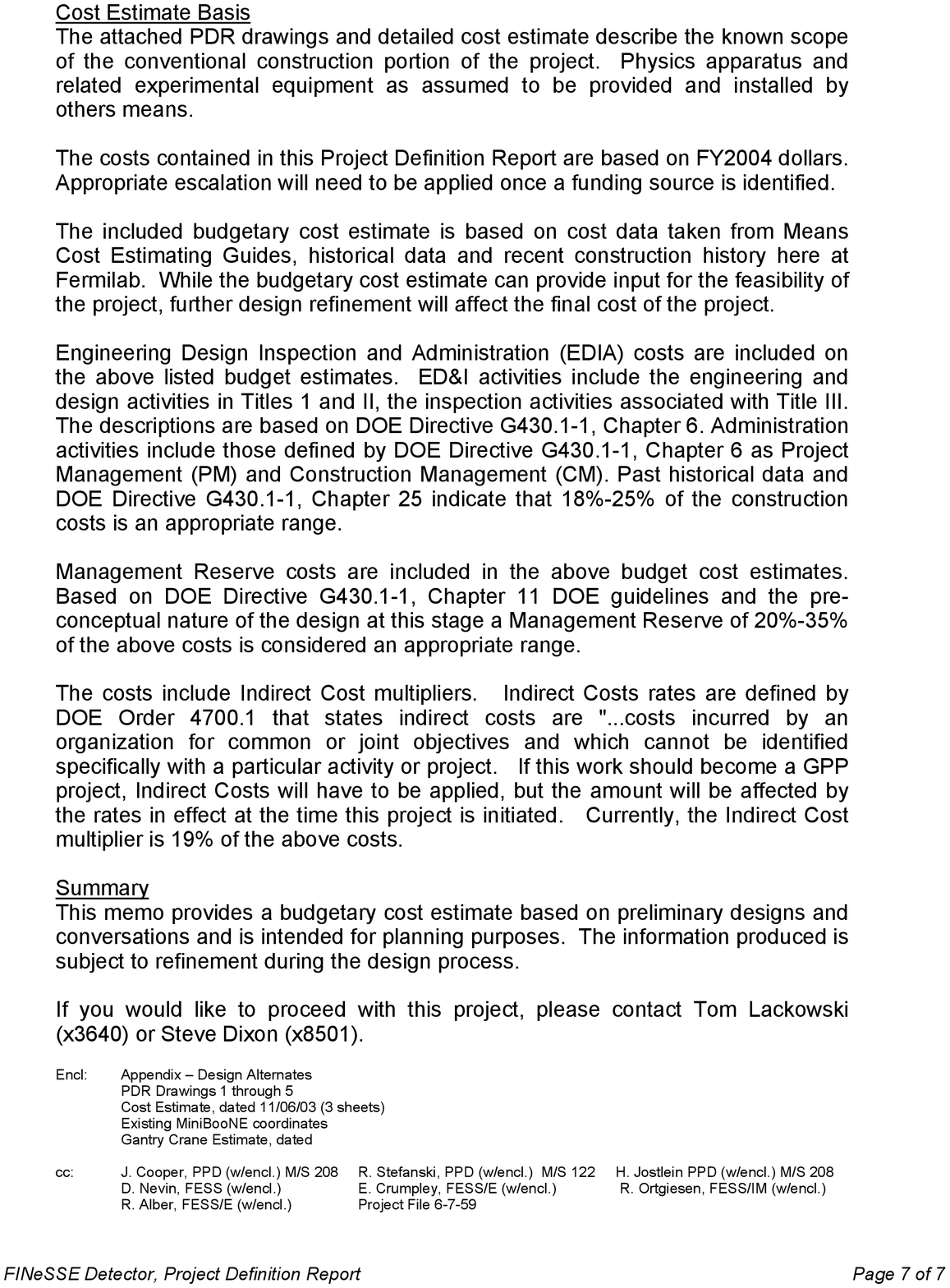}
\end{figure}

\clearpage

\begin{figure}
\centering
\includegraphics[bb=0 0 610 800,width=8.in]{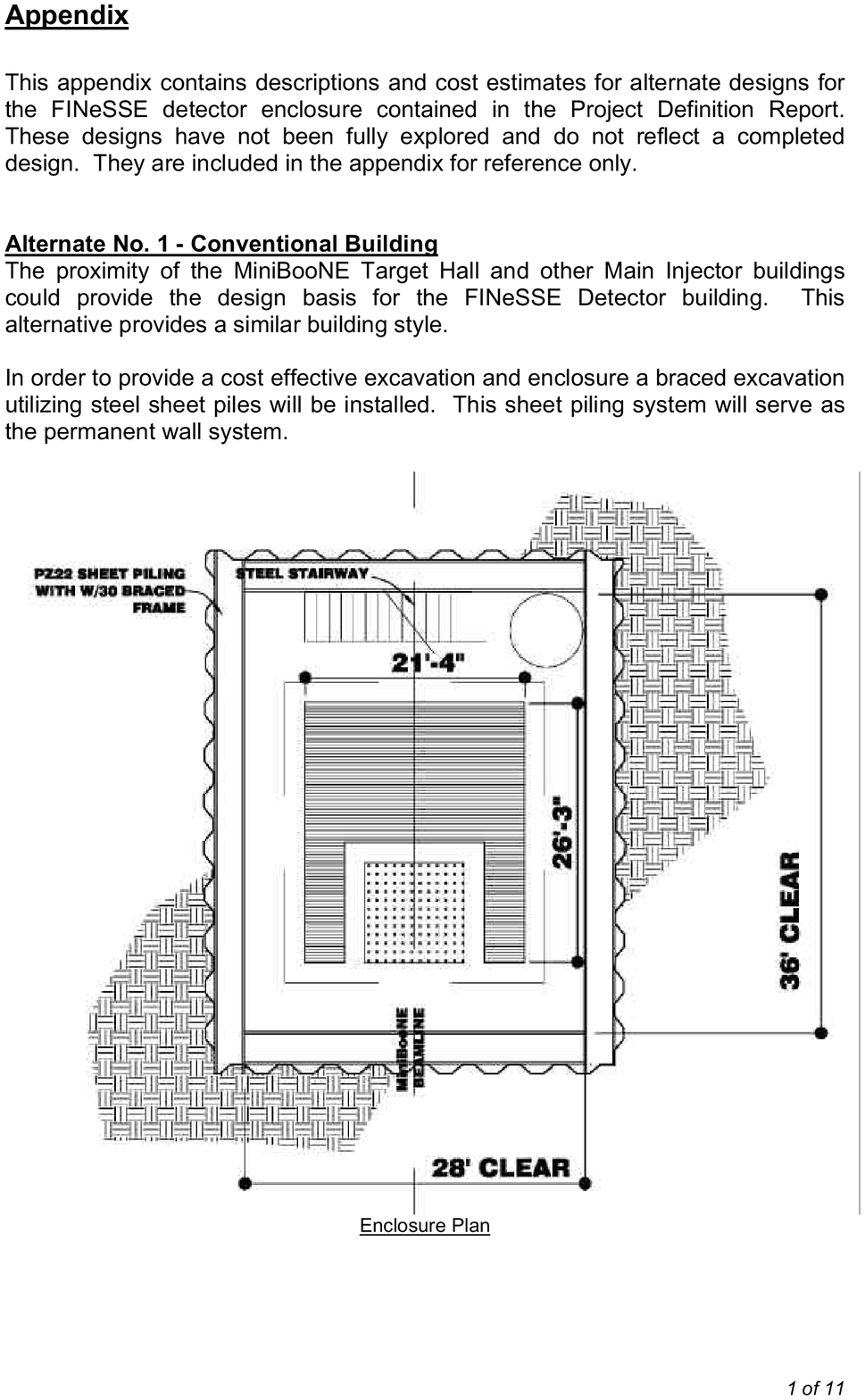}
\end{figure}

\clearpage

\begin{figure}
\centering
\includegraphics[bb=0 0 610 800,width=8.in]{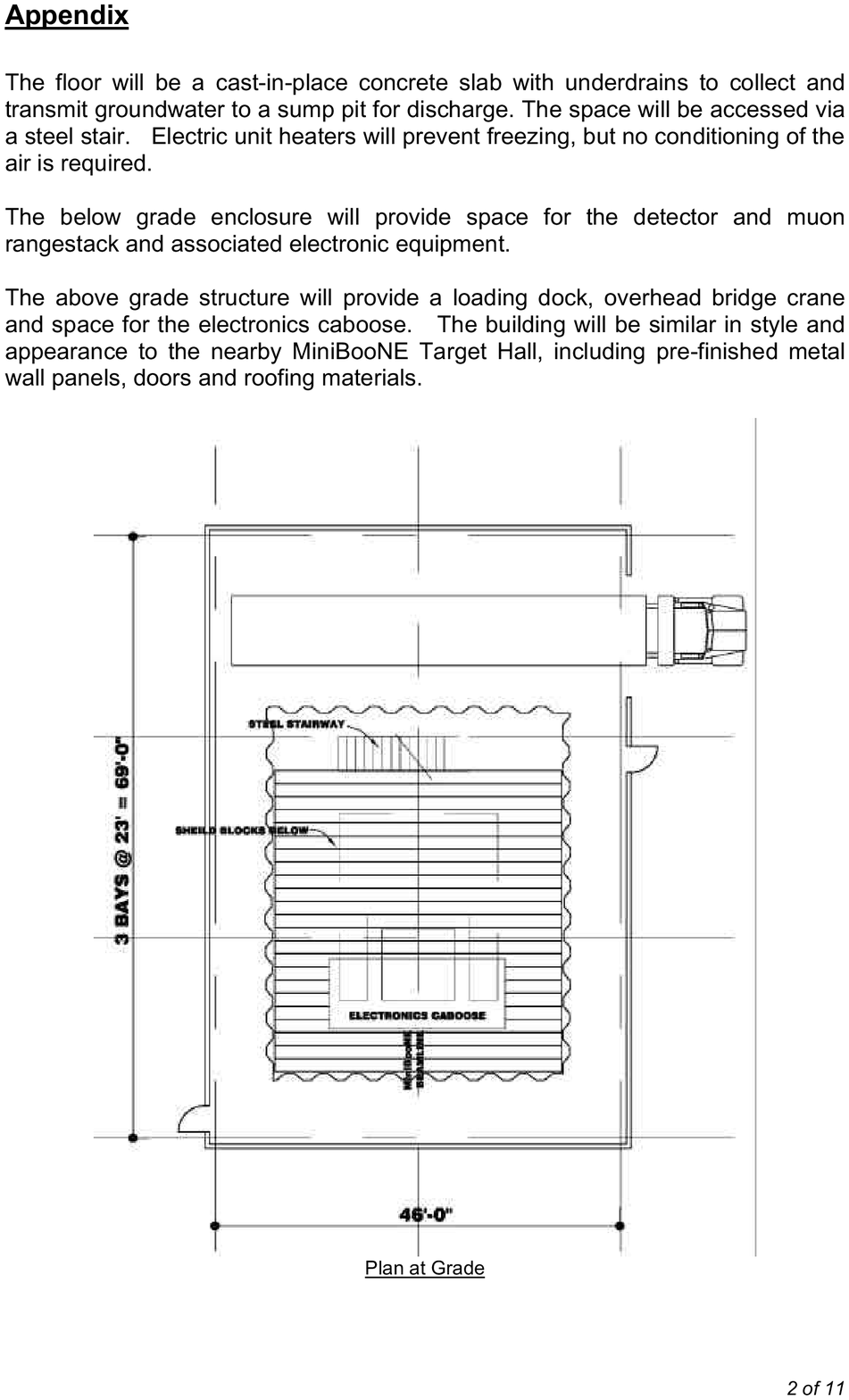}
\end{figure}

\clearpage

\begin{figure}
\centering
\includegraphics[bb=0 0 610 800,width=8.in]{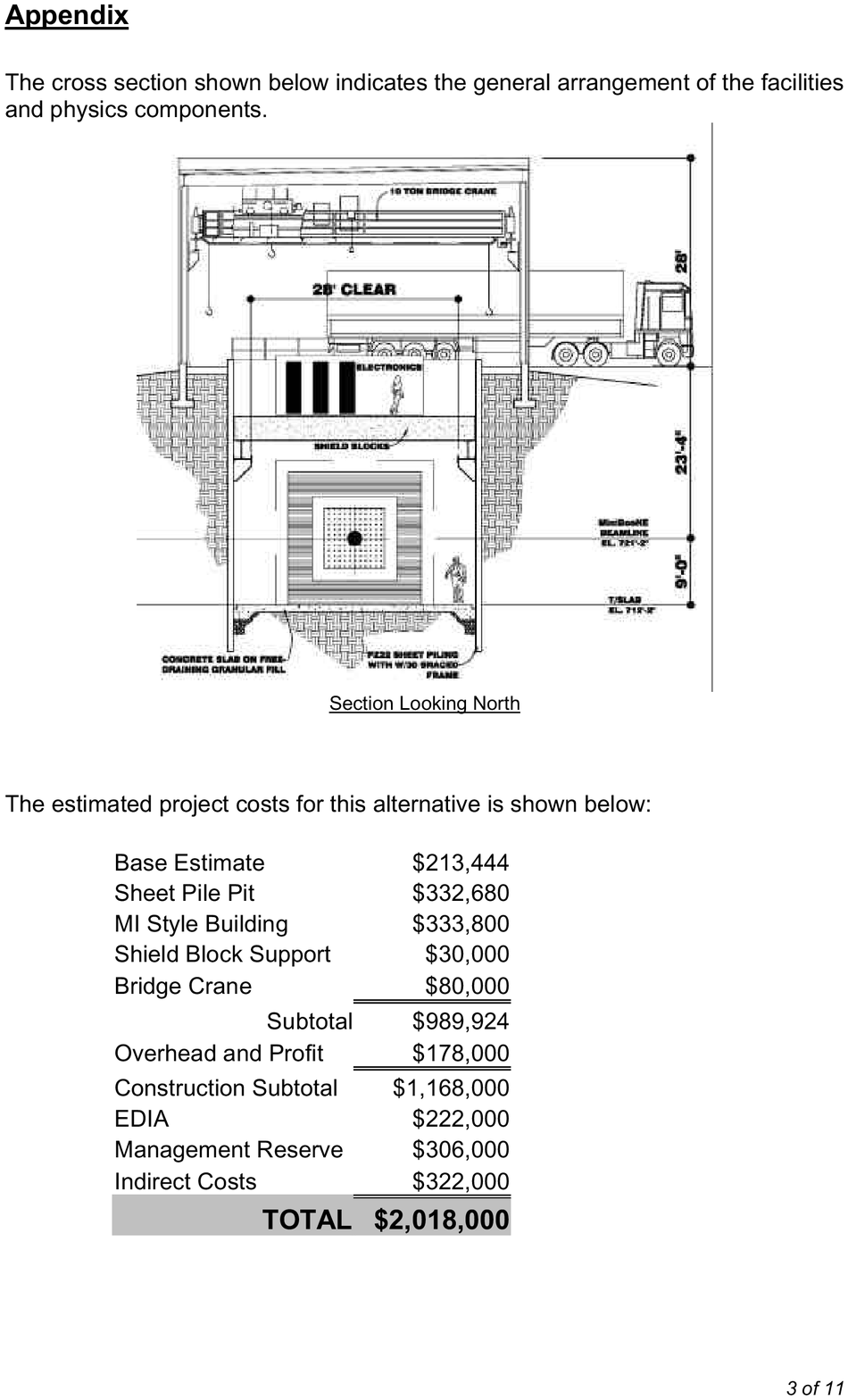}
\end{figure}

\clearpage

\begin{figure}
\centering
\includegraphics[bb=0 0 610 800,width=8.in]{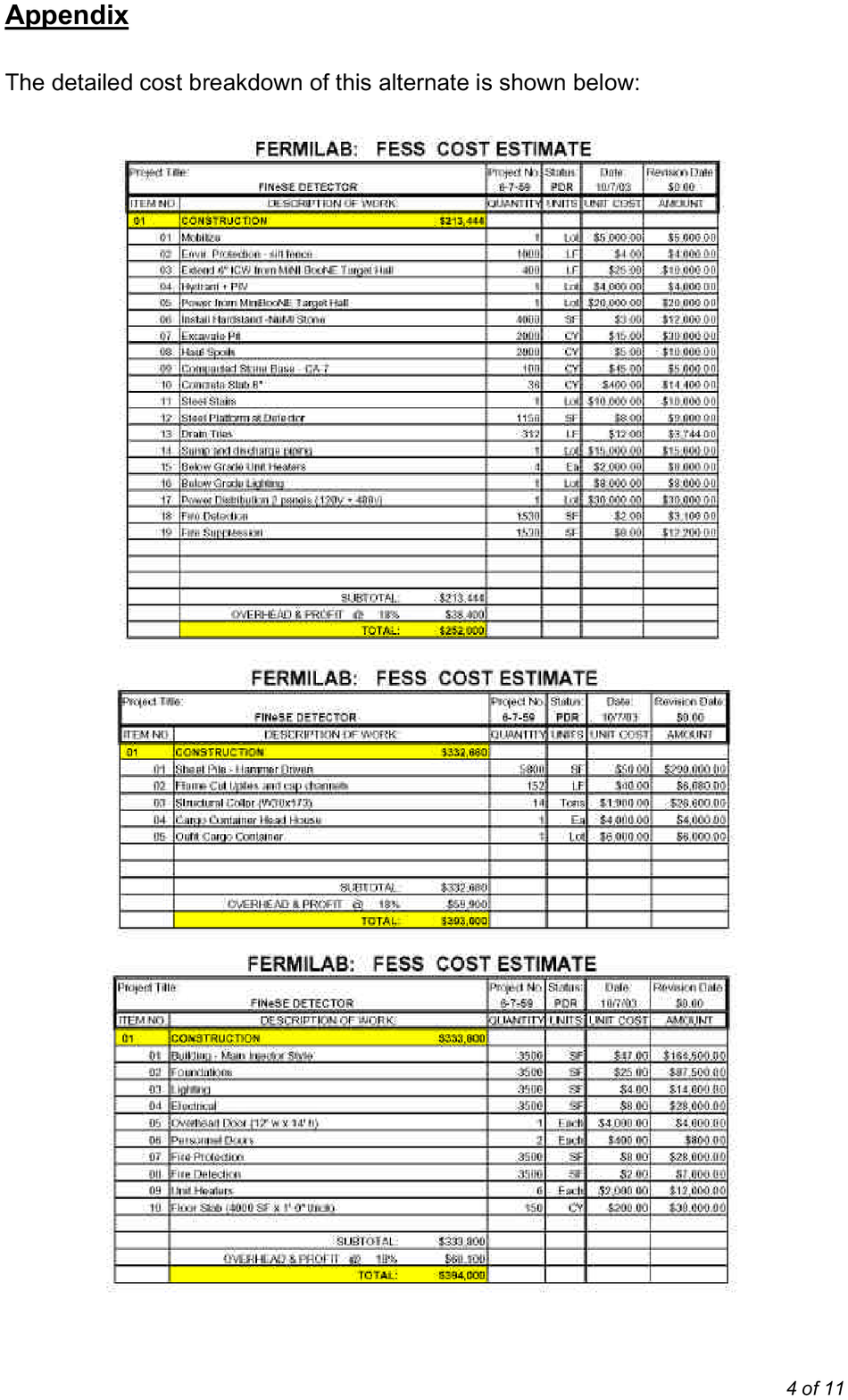}
\end{figure}

\clearpage

\begin{figure}
\centering
\includegraphics[bb=0 0 610 800,width=8.in]{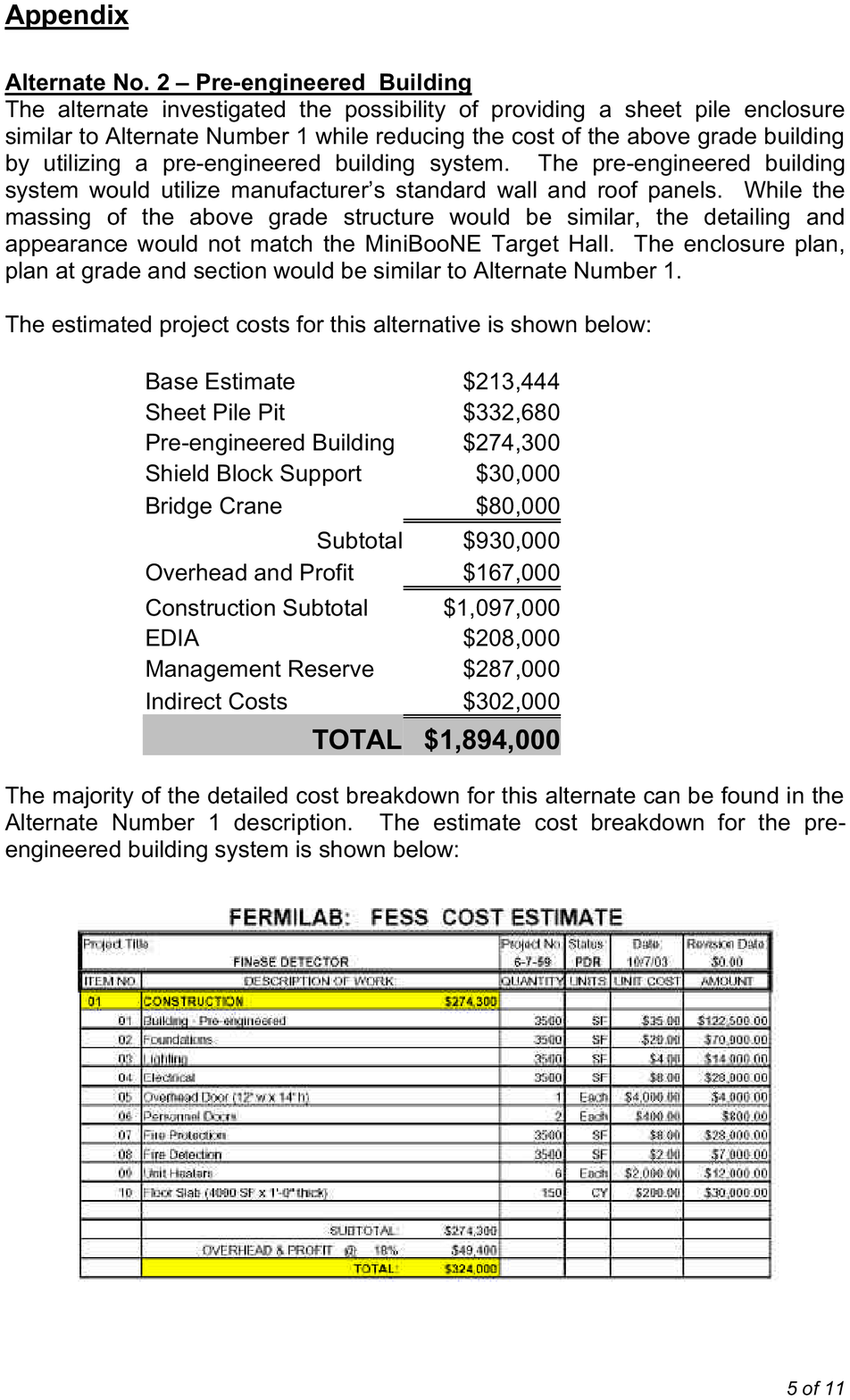}
\end{figure}

\clearpage

\begin{figure}
\centering
\includegraphics[bb=0 0 610 800,width=8.in]{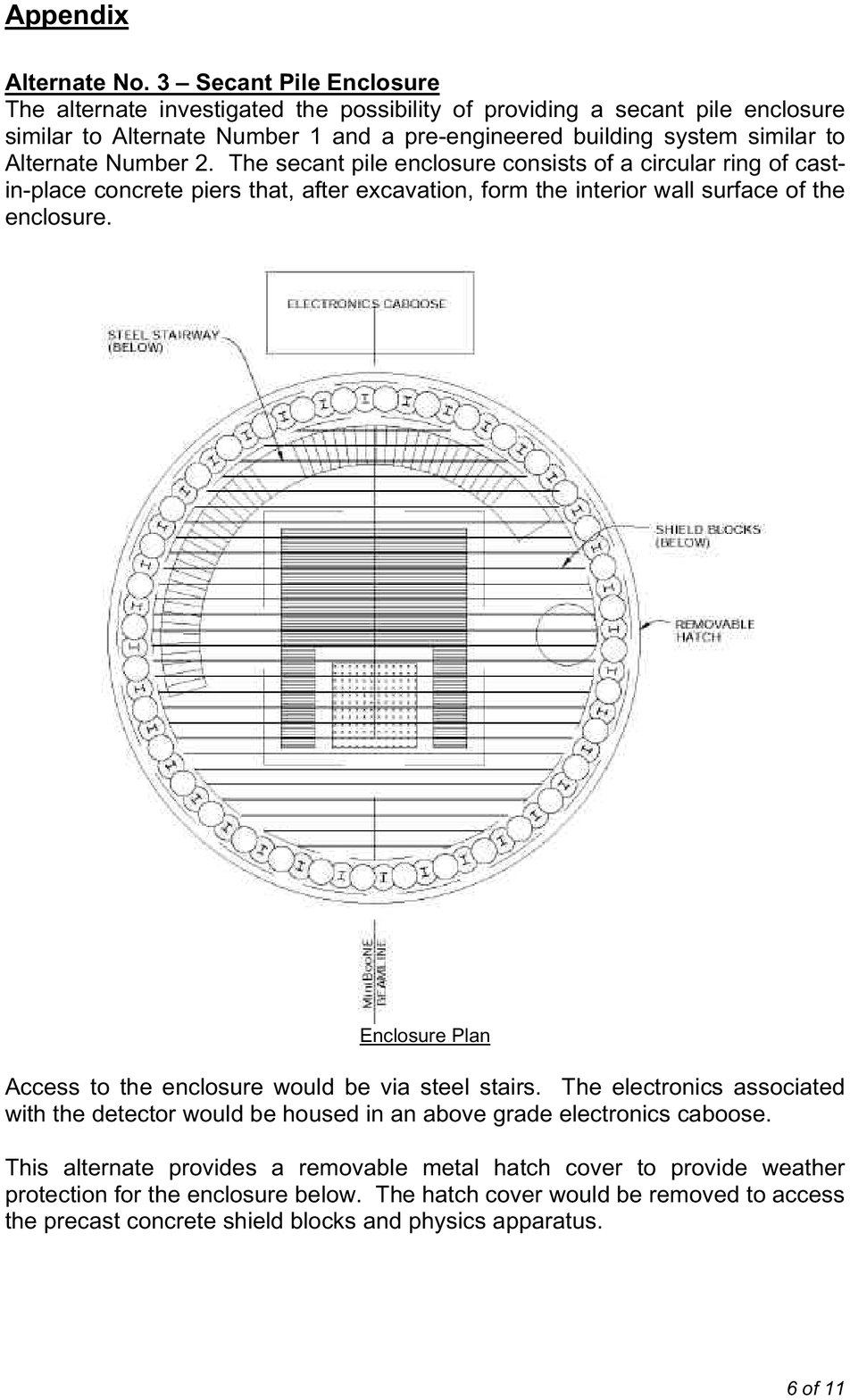}
\end{figure}

\clearpage

\begin{figure}
\centering
\includegraphics[bb=0 0 610 800,width=8.in]{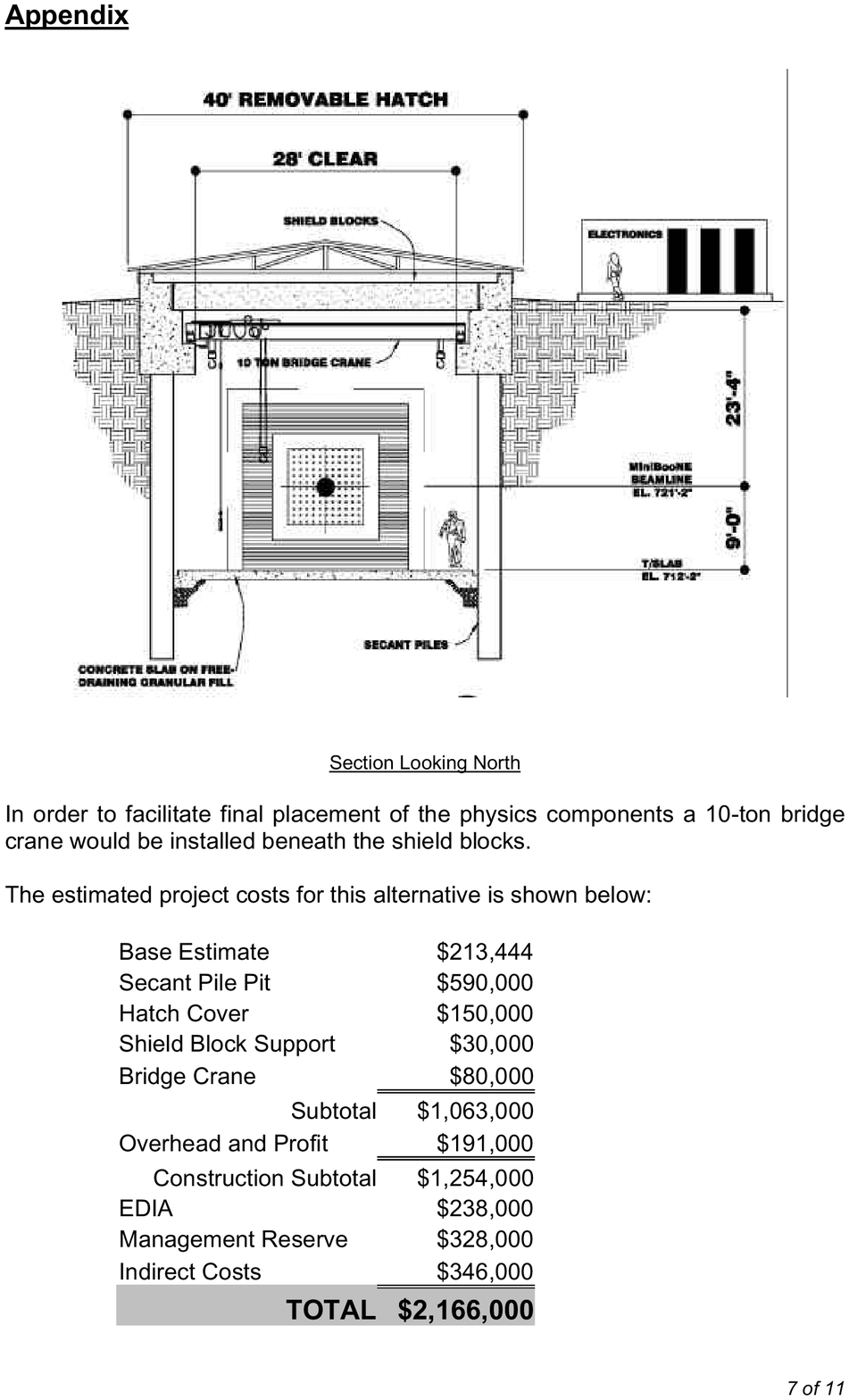}
\end{figure}

\clearpage

\begin{figure}
\centering
\includegraphics[bb=0 0 610 800,width=8.in]{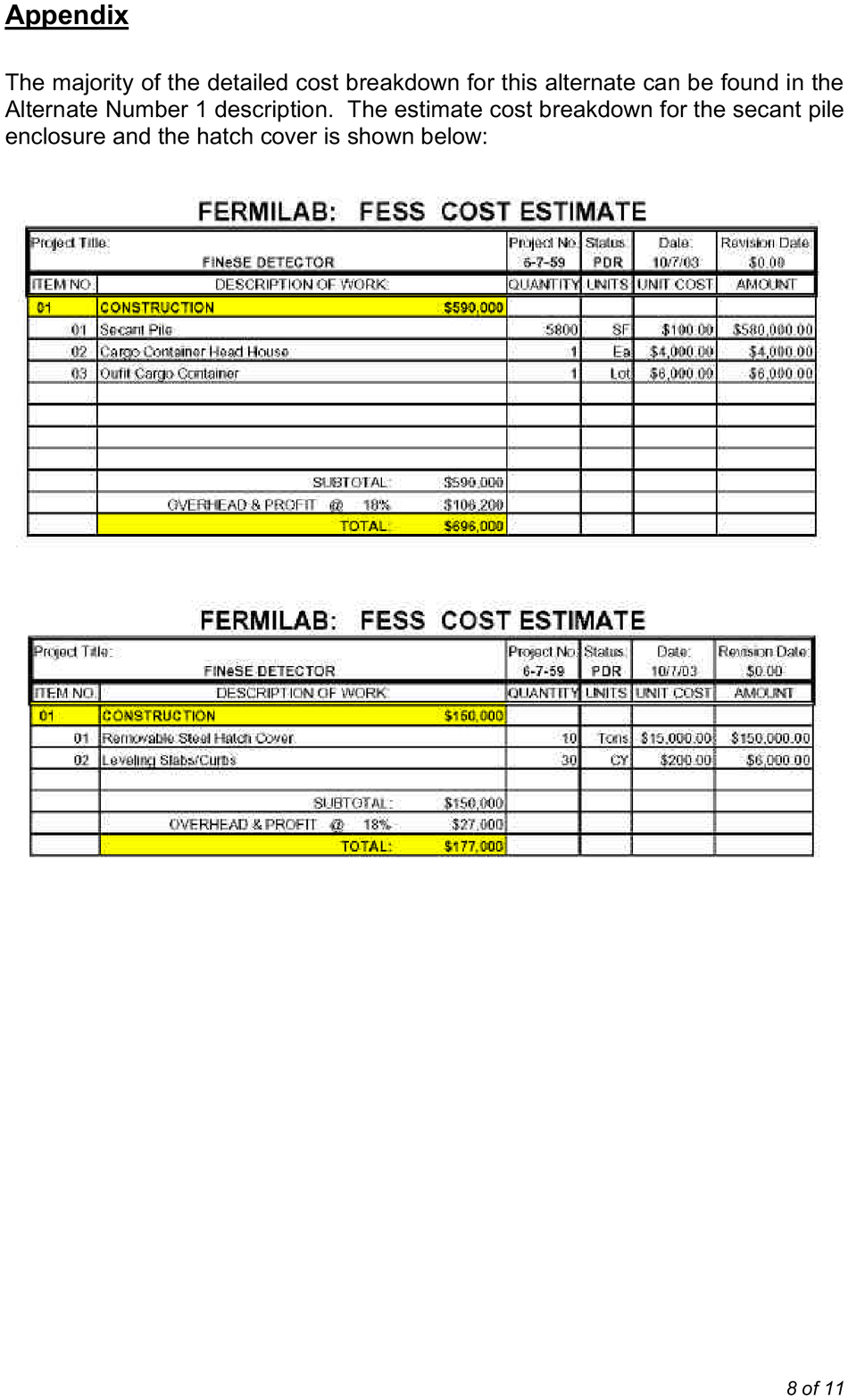}
\end{figure}

\clearpage

\begin{figure}
\centering
\includegraphics[bb=0 0 610 800,width=8.in]{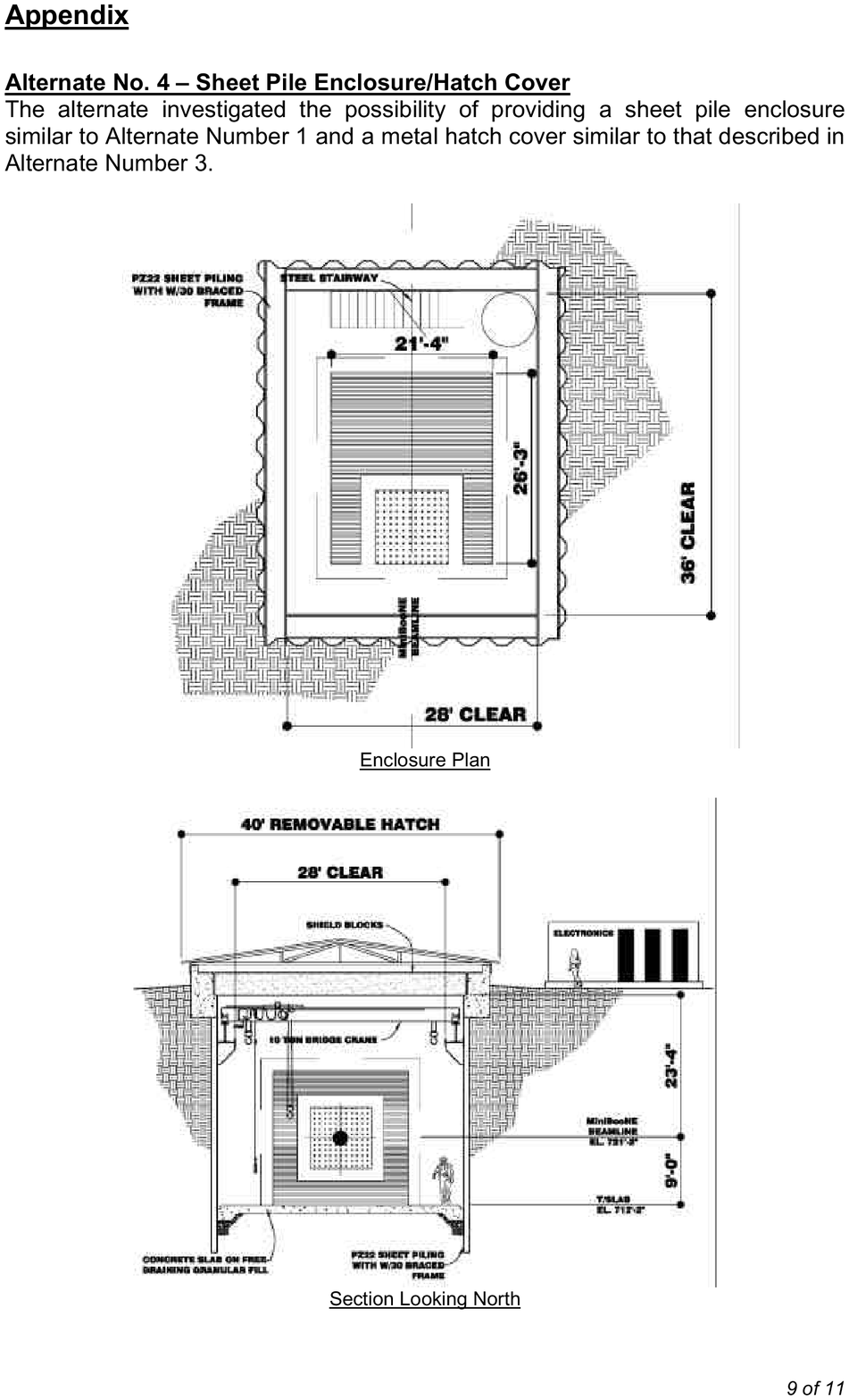}
\end{figure}

\clearpage

\begin{figure}
\centering
\includegraphics[bb=0 0 610 800,width=8.in]{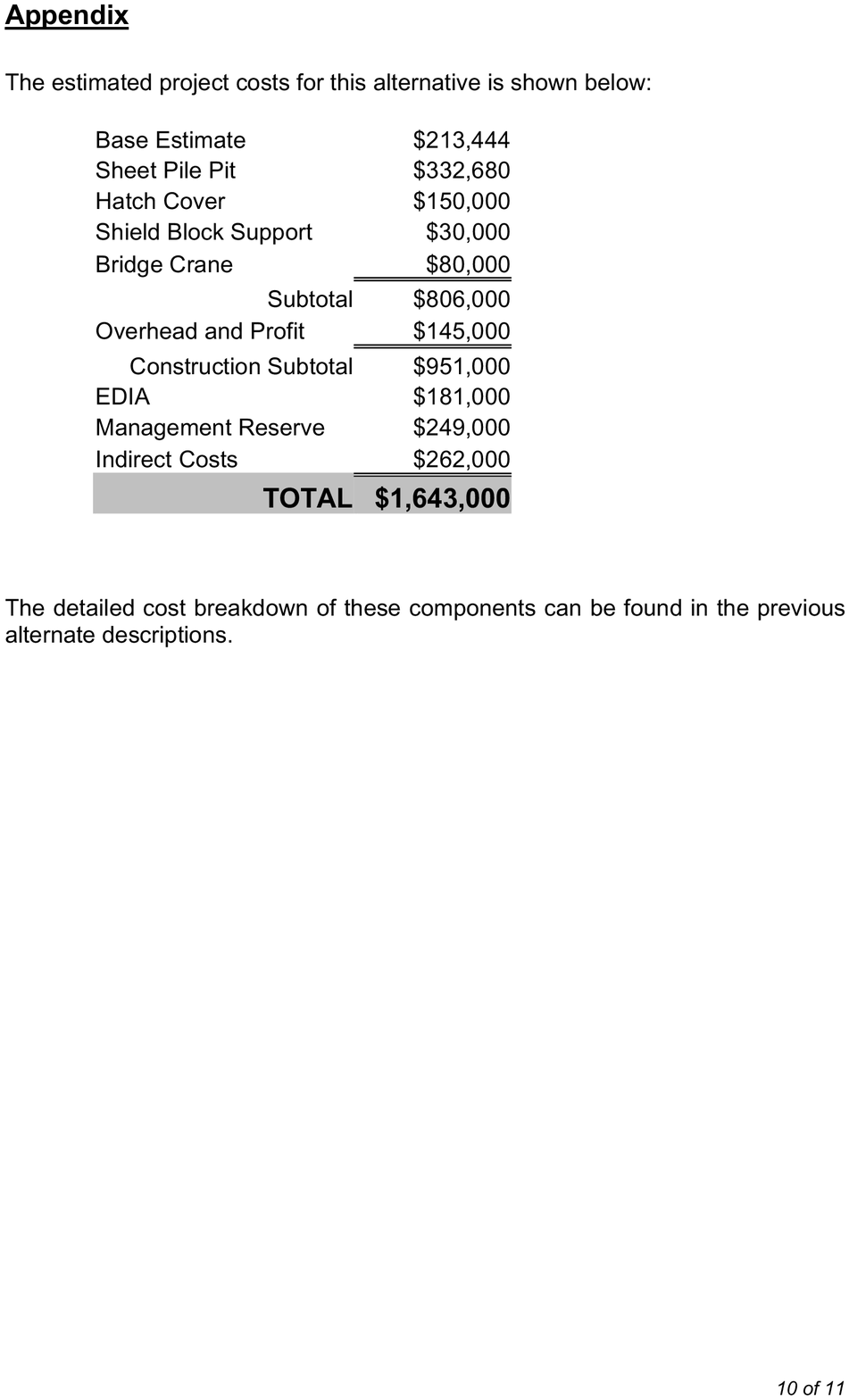}
\end{figure}

\clearpage

\begin{figure}
\centering
\includegraphics[bb=0 0 610 800,width=8.in]{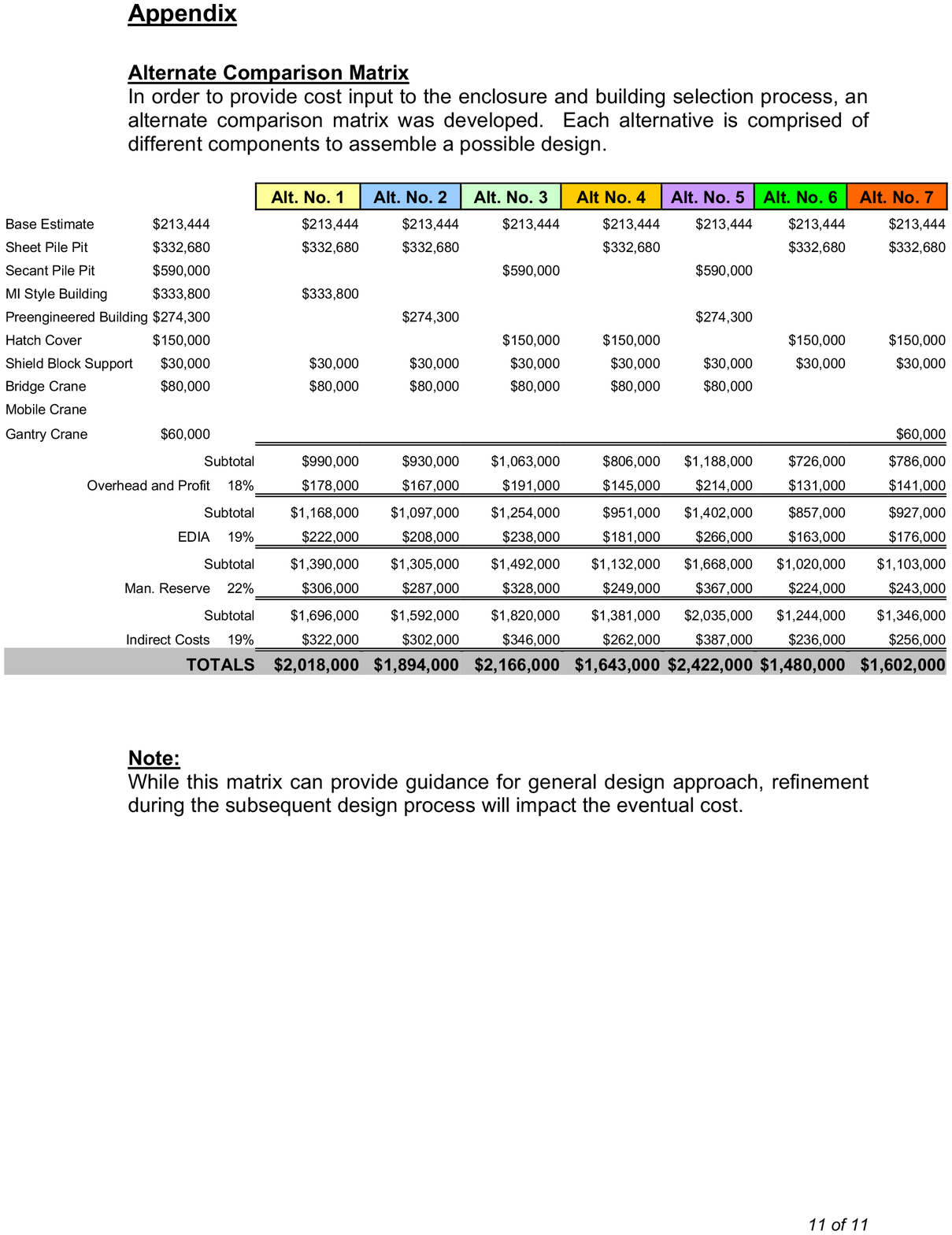}
\end{figure}

\clearpage

\begin{figure}
\centering
\includegraphics[bb=100 90 515 720,width=5.5in]{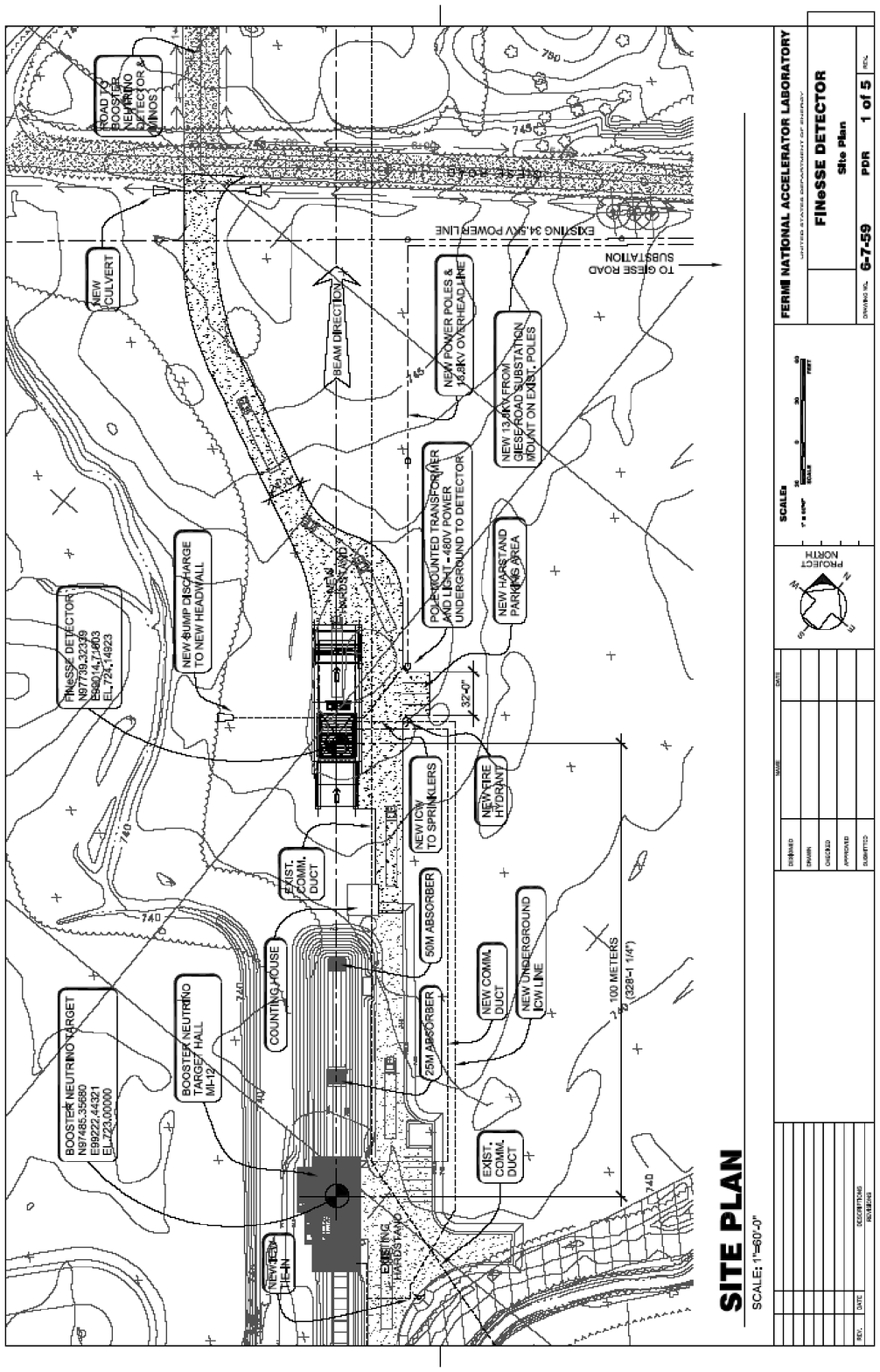}
\end{figure}

\clearpage

\begin{figure}
\centering
\includegraphics[bb=100 90 515 720,width=5.5in]{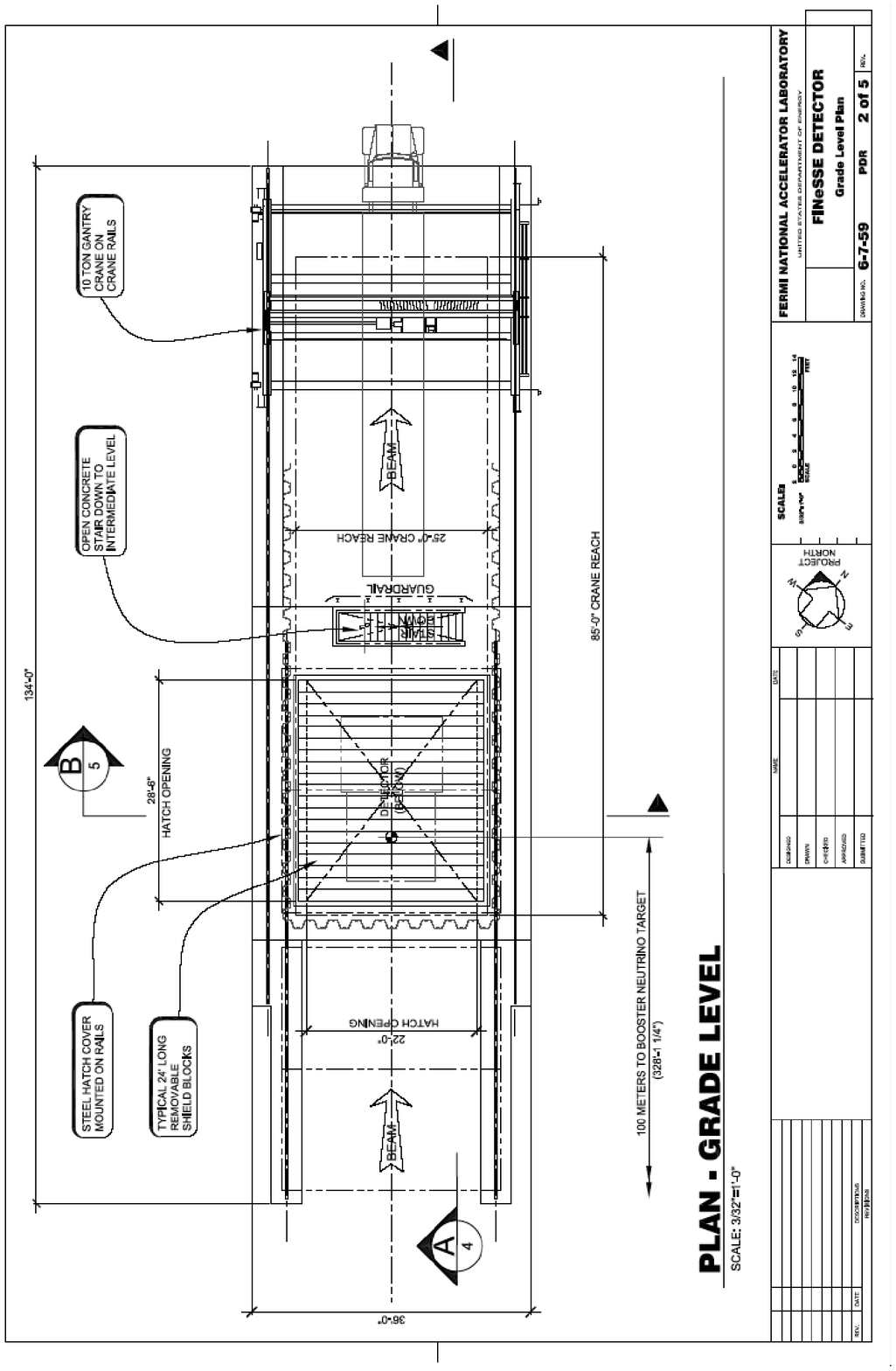}
\end{figure}

\clearpage
\begin{figure}
\centering
\includegraphics[bb=100 90 515 720,width=5.5in]{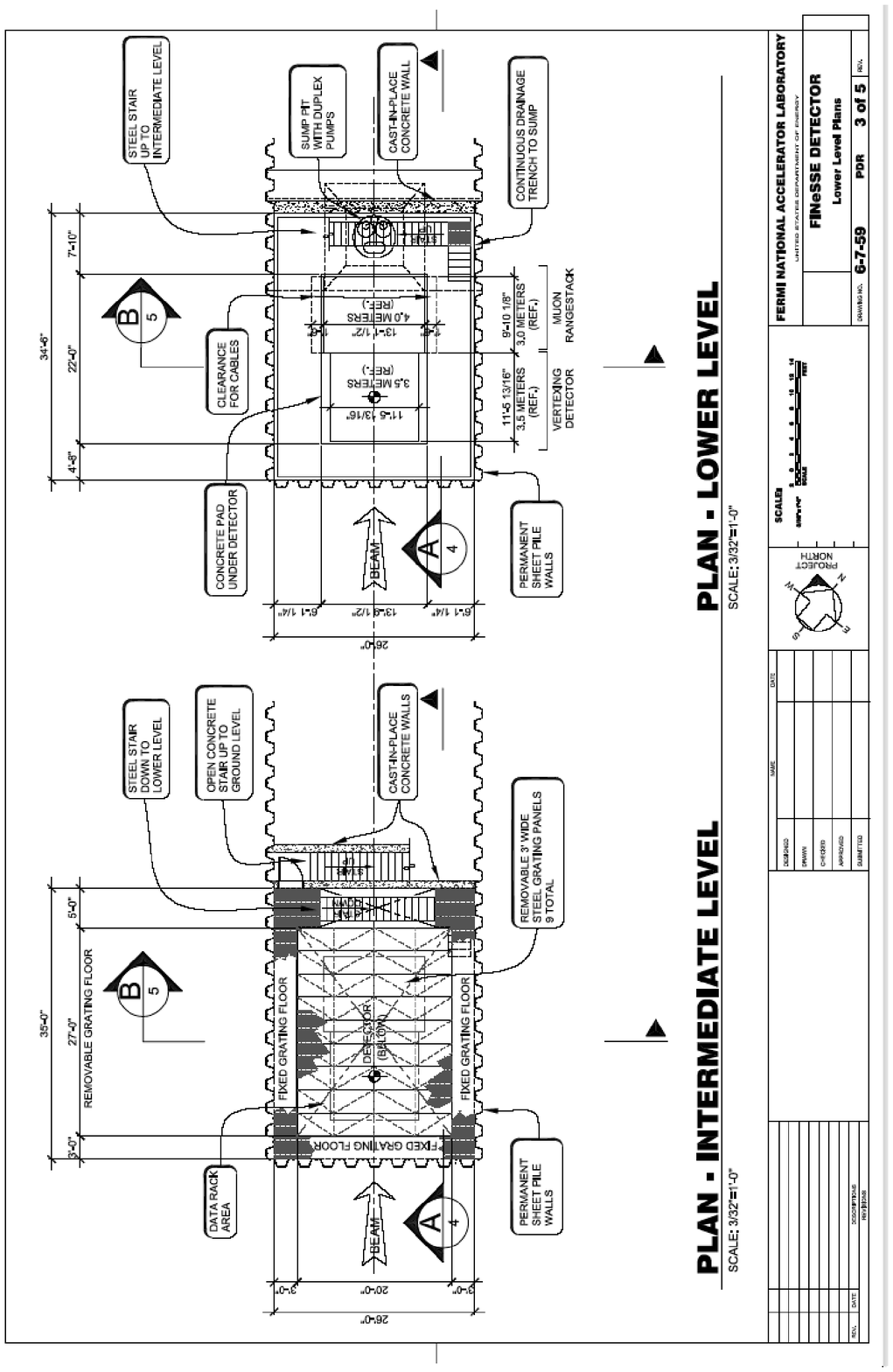}
\end{figure}

\clearpage
\begin{figure}
\centering
\includegraphics[bb=100 90 515 720,width=5.5in]{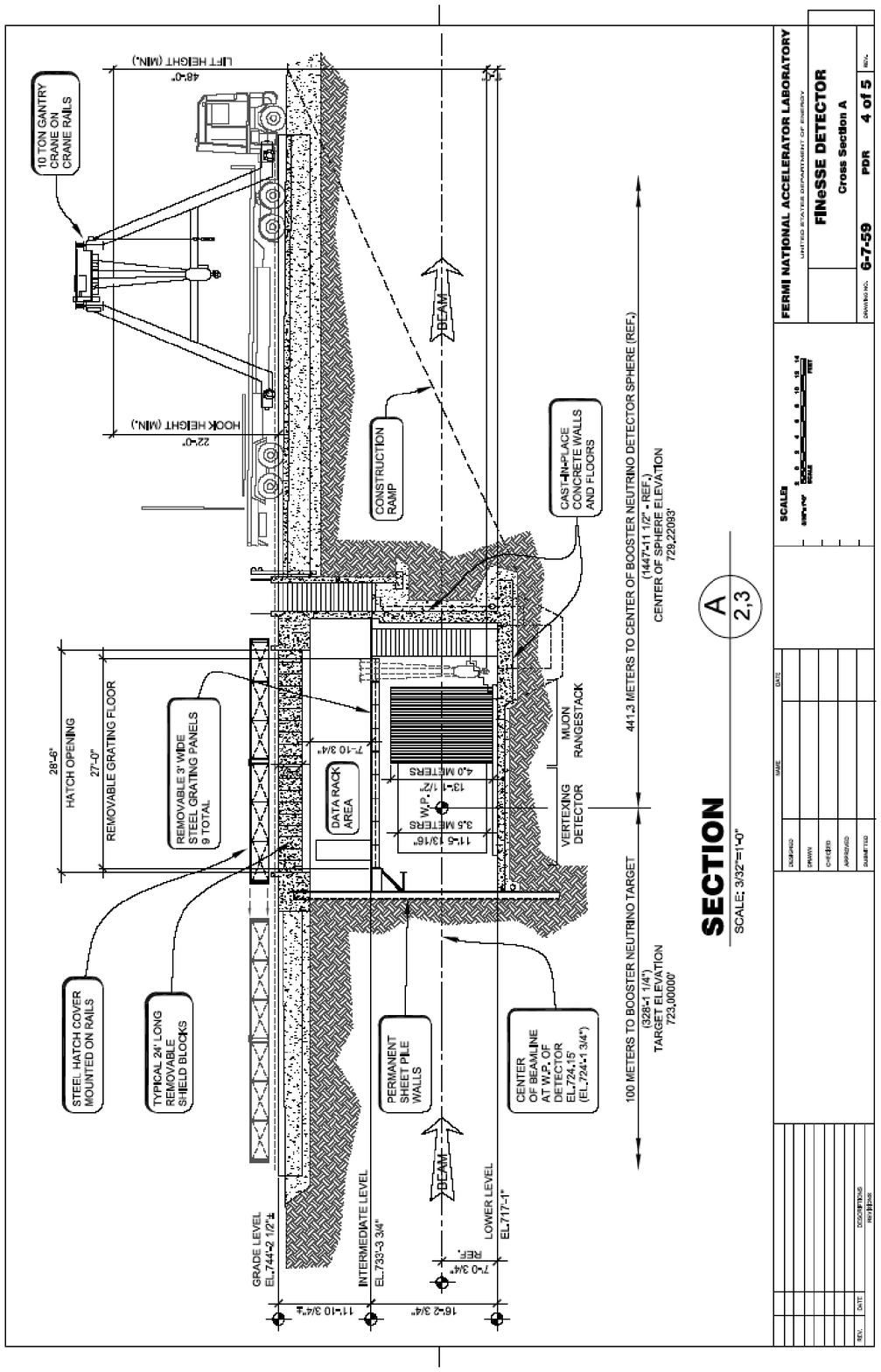}
\end{figure}

\clearpage
\begin{figure}
\centering
\includegraphics[bb=100 90 515 720,width=5.5in]{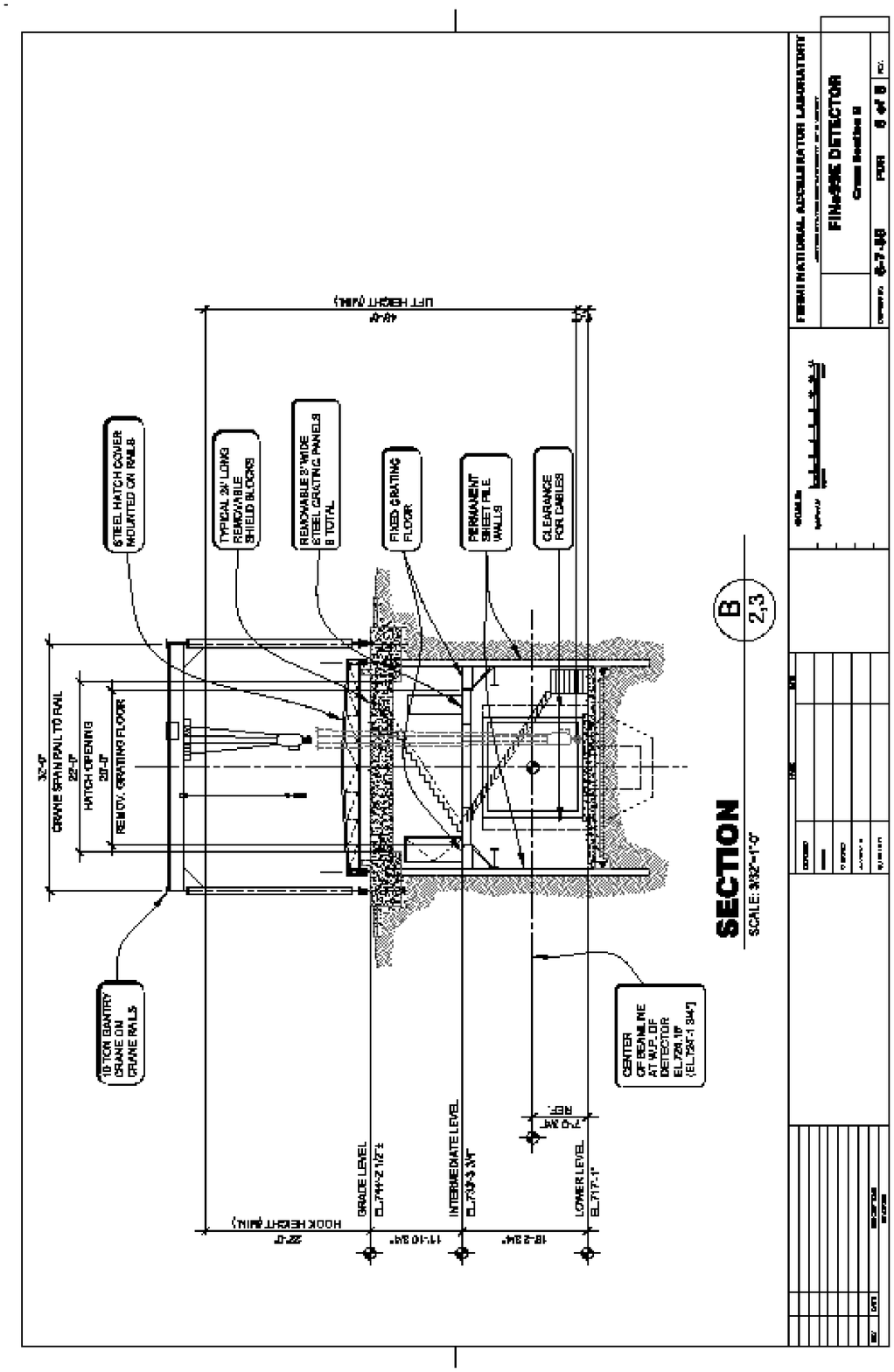}
\end{figure}

\clearpage
\begin{figure}
\centering
\includegraphics[bb=0 0 610 800,width=8.in]{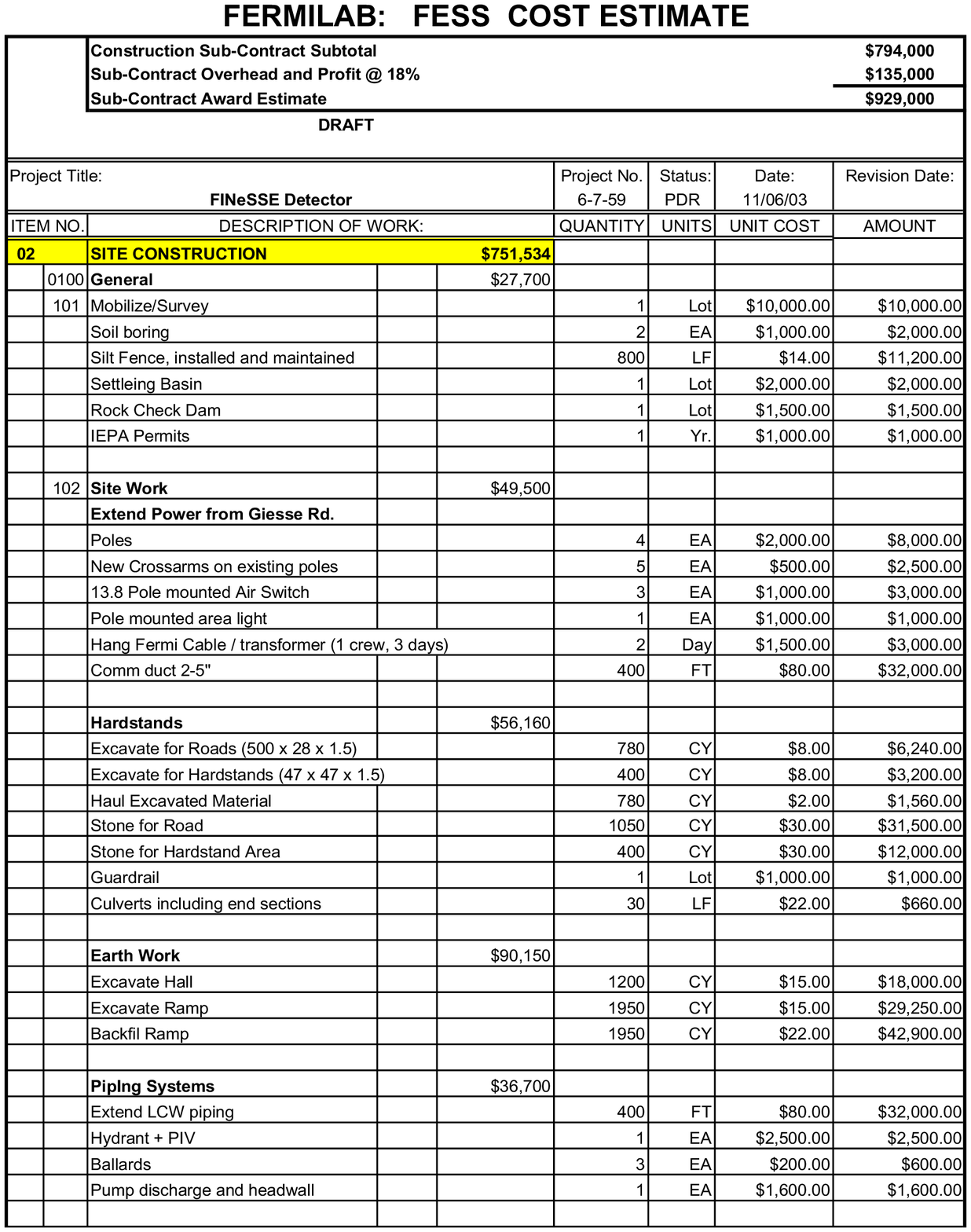}
\end{figure}

\clearpage
\begin{figure}
\centering
\includegraphics[bb=0 0 610 800,width=8.in]{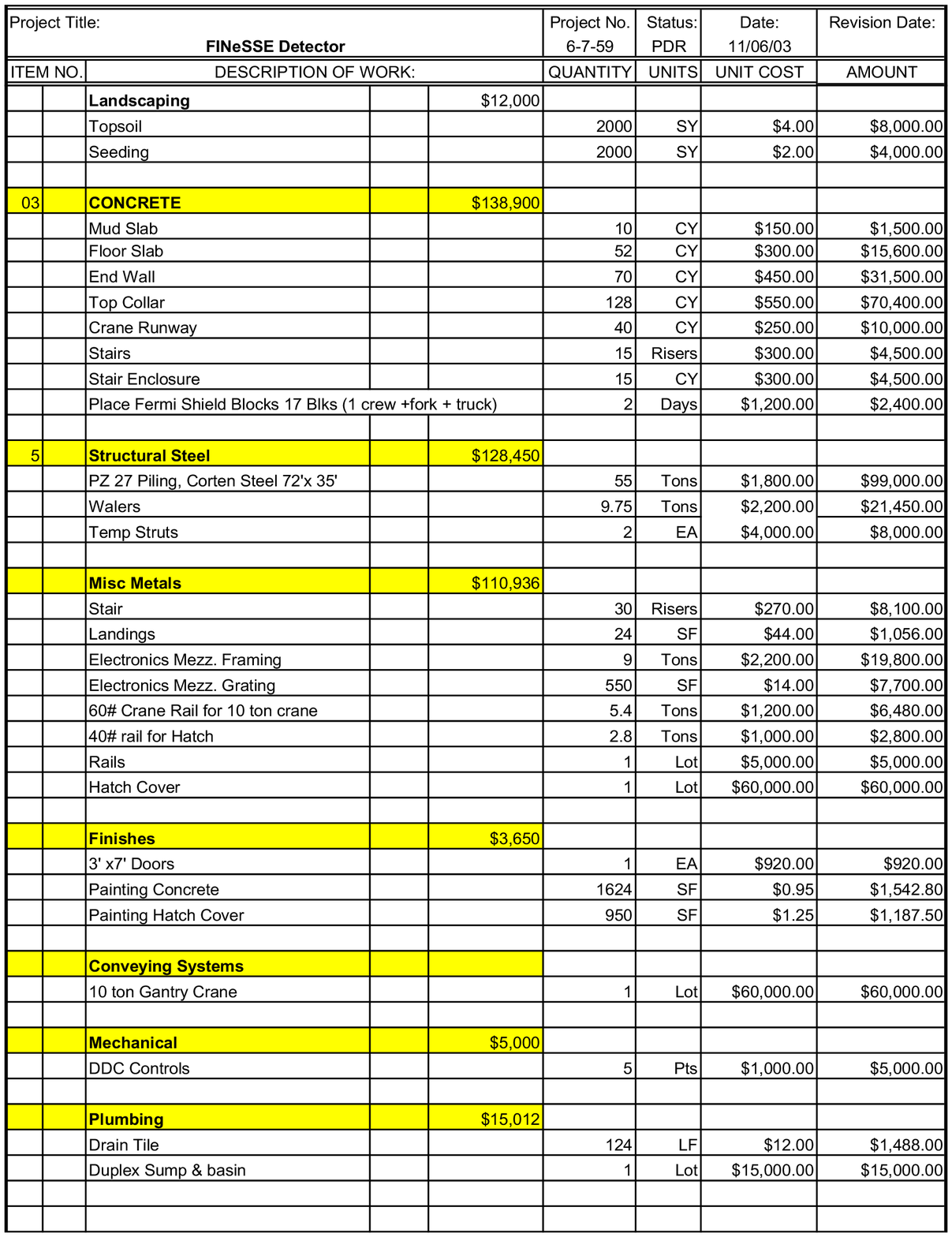}
\end{figure}

\clearpage
\begin{figure}
\centering
\includegraphics[bb=0 0 610 800,width=8.in]{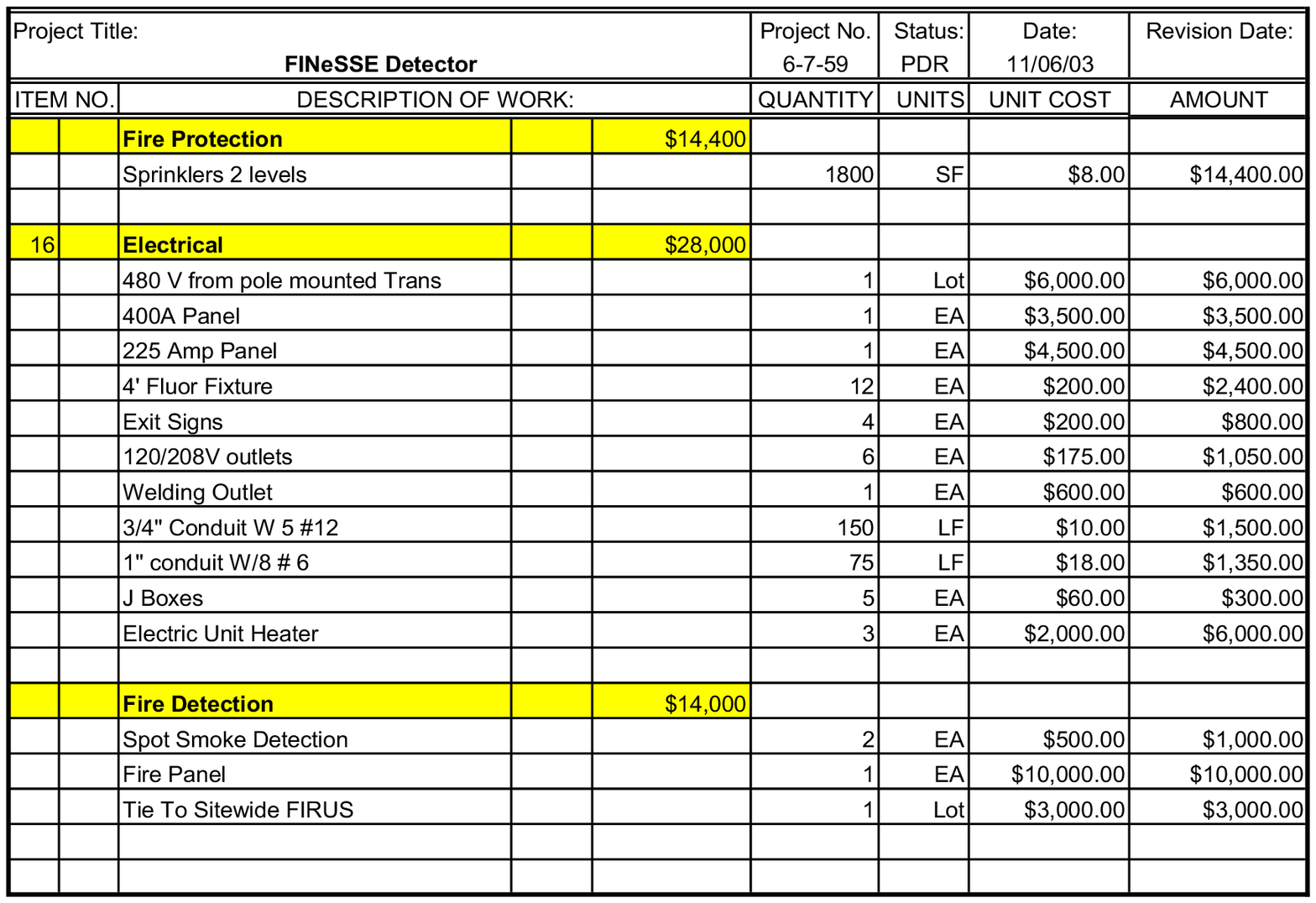}
\end{figure}

\clearpage
\begin{figure}
\centering
\includegraphics[bb=0 0 610 800,width=8.in]{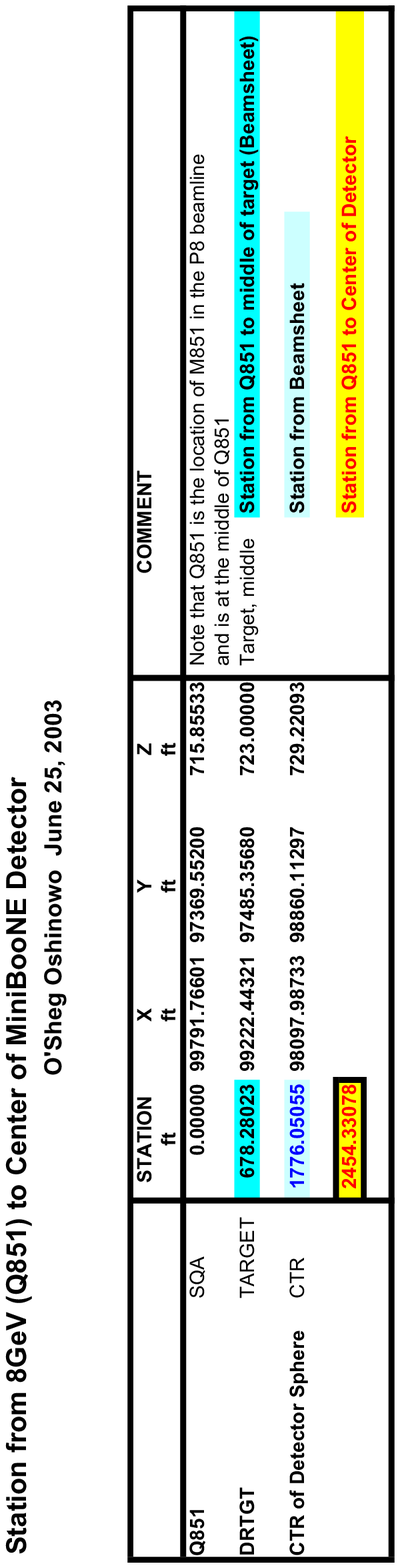}
\end{figure}

\clearpage

\end{document}